# The design of a proto-animal brain based upon spike timing

Robert Alan Brown

Mattapoisett, Massachusetts, USA

**Abstract**

A basal animal model is described as an organism attached to the sea floor living in a reproductive community. Its control system (brain model) uses logic cells (gates) to create a high frequency spike generator. Addition logic cells create a timing framework based upon Pulse Width Modulation (PWM), and create multi-cell, spike driven muscle actuators and bi-directional shift registers that serve as memories. Together, these logic cells generate a pulse train that forms a recurrent fractal in the form of coherent square waves that consist of an equal number of set pulses and reset pulses. These pulses drive the actuators that pump water and food through its shell. Some of these individuals lose their permanent attachment to the sea floor, and evolve the ability to move around using the feeding motions. This creates the hazard of getting stuck against an object or moving away from its breeding community. They evolve a sensor system and logic cells that produce a restoring motion for each avoidance motion. This keeps the animal in the specific region of an object that it encounters (hefting), and maintains the zero sum (coherence) of the fractal, and connects the organism cyberneticly to objects that are sensed in its environment. The logic units evolve to store permanent memories, and add plus and minus pulse trains together into a single pulse train that allow the animal to follow a moving object (imprinting). A problem is created if an object is encountered at the very beginning or near the very end of a pulse. This problem is overcome by the use of two pulse trains operating at the same time in quadrature. The operation of this two phase system yields insights into the operation of animal brains where multiple control systems operate at the same time, but where only some are selected to be expressed at a given time. Additional logic units evolve that allow the animal to heft using contacts with multiple objects that allow the animal to return to its point of origin (reproductive migration) using permanent memories that are created when a pulse is not reset to zero. This leaves a permanent recurrent pulse in the fractal that influences the properties of the fractal. So, information is stored in multiple logic cells instead of a connection to one logic cell, and is based upon spike timing rather than static (synaptic) connections between cells. The high frequency spike train that drives a system using PWM has a higher information transmission rate when it operates at a higher frequency. The animal brain model in this study needs a high frequency spike train of 128 Hz to provide one second motions with a 1% resolution. A high frequency spike train of 1,000 Hz and higher is needed in a higher animal brain using PWM. These higher frequencies can be obtained by using multiple, parallel, interlaced frequency generators.

**Author Summary**

It is hard to understand how the millions of neurons in the brain produce successful animal behavior. So I set out to define and investigate the simplest brain model that produces useful animal behavior. Using the Limpet as the archetype, I showed how it can graze using a specific kind of spike coding called Pulse Width Modulation (PWM). Then I show how small additions, changes, and mirror image replications of this basal model allows the animal model to avoid an object in its environment (object avoidance), to return to its home position after it avoids an object (hefting), to follow an object (imprinting), to avoid multiple objects and return to its point of origin, (reproductive migration), and learn to follow a path through objects without contacting them (maze running). This exercise yielded many insights into how the brain can function. The most striking is the importance of rhythm. Just like music, dance, poetry and spoken language, they are all

Spike timing

based upon a timing framework. The foundation of the timing framework is a high speed frequency generator (clock) that paces all of the events in these systems. All the information in these systems is contained in the set of timed pulse widths (values) stored within this timing framework. I show how this time information is stored and expressed by actuators that work like muscles with many cells that flex and extend due to spikes from the high frequency generator. Information is also stored in multi-celled, bi-directional shift register memories that use almost the same circuits as the muscles, but do not produce motion. So information is not stored at a single cell, but in a collection of cells. This explanation opens a new door into understanding how the brain works, as seen in the many examples in my paper, and places the time pulse as the foundational unit of the brain.

1. **Introduction**

The present view in neuroscience is that brain memory and plasticity are due to changes in synaptic connections in connection matrices. I show examples of these connection matrices in (Brown, 1994). While these connection systems can associate input states to output states, they lack the ability to measure and use time information, directly.

A recient study (Chen, 2014) shows that synapes believed responsible for learning a task were erased in the worm Aplysia. After new synapses were grown, the original learned behavior remained intact. This is evidence that these memories were not stored in the synapses alone.

In a paper (Bahill, Clark, & Stark, 1975), the movements of the human eye (saccades) were measured over a wide range of eyeball rotations. The best fit for the data was made using an increasing rotational velocity of the eyeball to the midpoint of the motion, and decreasing rotational velocity to the end of the motion. The time of the midpoint was attributed to a pulse width modulation controller. The question is: where did this pulse time come from? What caused the pulse? What determined when the pulse started and ended? And what is the design of this PWM controller?

I attempt to show in this paper that memory and plasticity in the brain can be due to changes in the controlled pulses stored in sets of neurons using a form of Spike Timing Dependent Plasticity (STDP) based upon Pulse Width Modulation (PWM). This form of STDP does not require changes in synaptic connections such as potentiation and depression. PWM is a powerful control system that provides the precise timing needed for most electrical heating, lighting, and electric motor speed control systems in use today, and is the foundation Morse code telegraphy. So it is ideal for storing and transmitting information as well as controlling actuators.

A PWM system consists of a train of equal spaced spikes that form a sequence of unit pulses. Each unit pulse contains an additional control spike that either starts or stops a controlled pulse within the unit pulse. The time of the controlled pulse is some fraction between 0 and 1 of the unit pulse time. The control pulse can start at the beginning of the unit pulse and stop at some time at or before the end of the unit pulse, or start at some time at or after the start of the unit pulse and stop at the end of the unit pulse. The information transmitted through the PWM system is contained in the controlled pulses transmitted through the system. Other methods of transmitting information include Amplitude Modulation (AM) where information is transmitted by variations in the amplitude of a signal, Frequency Modulation (FM) where information is transmitted by variations in the frequency of a signal, and Pulse Code Modulation (PCM) where information is transmitted as a series of binary numbers. The pulse widths in PWM can be added or subtracted directly, performing computations in real time as well as transmitting messages and controlling the motion of actuators.

The difference between animals and other life forms is the ability of animals to move themselves around (self-locomotion) in search of food and sustenance. The foundation of PWM is the use of a constant quantity over a controlled time period to produce some desired result. So when a constant velocity is employed in a motion actuator (muscle), a pulse time corresponds directly to the distance moved in that pulse time. If an animal muscle moves with an increasing velocity to the midpoint of its motion and a decreasing velocity to the end of its motion that is the mirror image of the increasing velocity, as in the Bahill paper, the distance the muscle moves using PWM is directly related to a timed pulse, also. This makes PWM particularly well suited for the control and coordination of animal muscles. I show the design of the pulse width modulation controller and a set of evolving systems that use this PWM controller to produce useful behavior in a simple proto animal model similar to a Limpet.

I propose the view that the foundational element of brain is the time pulse, and that the set of these pulses contains the information processed in the brain. The transmission of these information pulses throughout the nervous system is carried out as spike trains that operate in the PWM time framework. I show the exact time delay circuits that produce spike trains, and decode them into useful actions. These spike trains produce pulse trains in the form of recurrent fractals needed to for the locomotion of the model Limpet. The recurrent fractals also act as scanners, and form the raster of an object sensing system. I show how this timer based control system and the organism it controls can evolve into a more complex organism capable of sensing and avoiding objects, hefting (identifying objects in its environment, and using this information to remain in the location of the objects). Also, I show how the timer based control system can evolve to create temporary memories that allow an animal to imprint (follow a moving object or another moving animal), and deal with multiple objects and return to their place of hatching (reproductive migration) using time delay information alone. This process also explains the ability of some animals to learn to move through a maze of multiple objects using long term memories. The key to this process is the use of timers as the working, temporary, or long-term memory units.

A recent study (Bartol, 2015) showed that the surface area of a synaptic connection can vary. This means that there can be more docked vesicles on a larger synaptic surface capable of releasing more neurotransmitter in a given time. If a certain amount of neural transmitter is need to activate the receiving neuron, the larger synaptic connection would activate a neuron before a smaller synaptic connection with fewer vesicles, thus having a shorter time delay. I propose that these changes in synaptic time delays can be used to correct for spike transmission time delays that can occur in the PWM time framework circuits rather than serving as sources of memory.

A previous paper, (Brown, A spike processing model of the brain, 2011) describes a generalized spike timing process that transmits messages from one region to another over just a few conductors, much like a telegraph system, using pulse width information. This process meets the demands of the central nervous system where many sensors must communicate with many brain neurons through the highly restrictive paths through the spinal cord.

The following paper describes the more specific task of motor control associated with the cerebellum (motor control). It does so by investigating the simplest control system needed by a proto (archetypal) animal to move around successfully in an environment. Most of the activity of the system is the result of the internally generated fractals with very little information created by the sensors. This process contrasts with the previous paper wherein all of the activity is generated by the sensors (telegraph keys).

An assumption is made in the present paper that an animal had evolved to the point of having a single half-shell attached to the sea floor, and lived in fixed communities like coral do today. It had evolved the ability to pump water and food into a volume under its shell by means of a reciprocating (up/down) motion of its tongue. It is then able to graze on the food trapped under its shell using a series of tongue motions.

The present paper describes the control system needed to make these tongue motions, and shows a series of changes in the system that allow it to graze under its shell more effectively. It then shows how the animal can begin to move around using the motion of its tongue, and the problems of survival it then faces if the connection of its shell to the seafloor is lost. The paper shows solutions to these mobility problems that are made by small changes to the timing of elements in the circuitry, and duplications in the timing circuit and its components that are consistent with the process of evolution.

## 2. Results

The spike processing systems in this paper shows that a system of timers based upon Pulse Width Modulation (PWM) can form a viable control system for the grazing and reproductive navigational requirements of a very simple animal model. These control systems are formed by permanently connected timers and actuators that combine to create a raster of repeating patterns of motion (fractals) that are modified by spikes generated at specific times by sensors according to the location of objects in its environment. These fractals form scanning motions that allow an animal to sense objects in its environment, and remain in the field of these objects so as to remain close to its breeding population. The memory and plasticity of the control system is changed according to the spike time delays created by sensing the location of the objects it encounters, and stored within sets of memory cells. These memories are made and stored while the connections of these cells remain unchanged. This contrasts to the prevalent view in neuroscience that memory and plasticity is the result of changes in the pattern of synaptic connections.

Another indispensible element found in this study is the zero sum relationships within these fractals that maintain the orientation and behavioral structure of the fractal so that information about a given environment can be stored and retrieved. Another feature found in the study is the need for bi-symmetrical structures and circuit organization like those found in living organisms. Another feature found in the study is the usefulness of duplicating components like those found in living organisms. Another feature found in the study is the need for a poly-phase control system architecture that provides quadrature to provide object avoidance, hefting, imprinting, reproductive migration, and learning to run a maze successfully.

Also, the study found that a successful system attempts to maintain coherence (a stable recurrent fractal) by resetting itself in real time after responding to each sensed input. This restores its internal order by eliminating the buildup of greater actuator outputs than the unit output called for by the PWM protocol. This resetting process results in a changed relationship with its environment such as avoiding obstacles while maintaining its orientation so it can return to its point of origin.

It became apparent in the study that these pulse trains require parity (same number of onset spikes as offset spikes). This is a feature of Pulse Width Modulation. As seen in modern cook stoves (ranges), the temperature of the burner is determined by the "on" time compared to the "off" time (duty cycle) using PWM. If the "on" does not shut off when the burner is set for a low temperature, the system will not cycle as a wave. And the consequences to the cooking can be dire because the burner will become too hot (because it remains "on" during "off" portion of its duty cycle), unexpectedly. This holds true with the control system presented in the paper. Additions or omissions of onset or offset

Spike timing

spikes lead to the failure of coherence resulting in dysfunction. So the ability of these control systems to produce an offset spike for every onset spike, reset after every sensor pulse, restore after every action motion is critical for it to maintain its coherent, wave-like, rhythmical behavior.

The spike timing systems shown in this paper consist of four types of pulse width memories:

- Unit pulses created by the PWM timing framework.
- Working memories created by sensor spikes within the unit pulses that are set and reset to zero in four steps,
- Temporary memories created by sensor spikes that are stored for sixteen steps and reset to zero,
- Permanent memories created by sensors that are set, stored, never reset to zero, and can be read indefinitely.

The unit pulses, working, and temporary memories are like coherent waves that bottom out at zero amplitude. The permanent memories are like particles because they never have a zero value once created. Adding the time of one unit pulse period to a time value of a temporary memory pulse in a fractal wave does not allow it to be restored to zero after it is read, creating a permanent memory particle within a set of fractal waves.

Since the memory units require multiple logic units to define the length of a pulse, information is not stored at the connection to just one logic unit. Instead, a given memory is stored in the multiple memory units of bi-directional shift registers in the embodiment presented.

Also, the recurrent fractals require a zero sum of counter-clockwise and clockwise motions within a repeating cycle. These pulse trains form coherent waves like coherent light. If they fail the parity or zero sum test, they cease to act like waves, and acquire the permanent (non-zero) qualities of a particle. This can happen if a pulse is not restore after a sensor input, or if some or all of a restore pulse is not carried out. In the systems shown in the study, the failure to maintain coherence can result in dysfunction such as failing to avoid objects, heft, imprint, or migrate successfully. However, specific unrestored pulses (particles) produce the long term memory for imprinting its possible use in reproductive migration, and maze running.

So the basic principles of PWM used in the animal model became apparent, and are shown as follows:

1. The process used by the animal brain model in this paper is viewed as the creation and transmission of pulse width coded information using Pulse Width Modulation.
2. Pulse Width Modulation is based upon the expression of constant amplitude quantities (pulses) over a variable time duty cycle in repeating time periods.
3. Every event in an animal model brain based upon Pulse Width Modulation (PWM) is triggered by a spike from the High Frequency Spike Train (HFST), except for sensor produced spikes, which are then represented by a spike in the HFST.
4. The primary event in the animal model is the production of pulse fractals, which produce motion for grazing and the raster for object scanning in the environment.
5. All of the information about the environment obtained from sensor spikes is contained in the states of pulses in actuators, temporary memory registers, and permanent memory registers.

Spike timing

6. The value of the pulse is not stored by just one logic unit, but is stored at the transition between conducting and non-conducting states in the multiple logic units in bi-directional shift registers and muscle actuators.

7. For every onset spike there is an offset spike creating a pulse. For every up pulse there is a down pulse, creating a fractal. For every down fractal there is an inverted (up) fractal, creating a fractal system. For every fractal system there is an overlapping (') fractal system, forming a fractal nest. For every fractal nest there is an alternative nest, forming a specialized brain system. And for every specialized brain system there are other specialized brains systems that together form the complete brain model.

8. A logic circuit is used to connect the HFST to specific actuators and memory units that express pulses in specific fractal systems. The other fractal systems remain on hold or are unexpressed until connected to the HFST by the logic circuit.

9. In a successfully operating (coherent) PWM system, the sum of the positive onset spikes and negative offset spikes is equal to zero. The sum of the values of the positive (up) pulses and values of the negative (down) pulses is equal to zero. The sum of the positive and negative motions of down fractal and the positive and negative motions of the up fractal is equal to zero. So, at the end of the day, the sum of all of the expressed pulse values is equal to zero. This maintains the coherence of the fractal system and the original orientation of the internal fractal compass. (The purpose of sleep may be to reset any pulses left not reset without triggering new un-reset pulses.)

10. A higher frequency HFST provides a higher information transfer rate. This provides faster movements and data processing. Information transfer rates increases exponentially with spike train frequency. Higher frequency would have a huge influence on animal success and survival.

11. A HFST greater than 200 spikes per second can be created by multiple, interlaced, parallel spike trains.

The results in this paper pertain to systems with a single sensor axis. Systems with multiple scan axes need to be studied.

3. **Discussion**

The first part of this paper shows a prototypical animal that is capable of functioning in a fixed community by grazing on food pumped in under its shell. Rather than being driven by responses from sensors, its activity is driven by an internally generated recurrent fractal. Genetic variations that result in a defect in the bonding process that cements the shell to the seafloor in some animals, causes one or more animal to be set free. The already acquired grazing motion now causes the animal to move around in its environment. This causes problems of getting stuck in unfavorable locations and being separated from its fertilization and reproductive community. The rest of the paper shows how these problems can be solved by developing a withdrawal response upon contact with an object, and developing a restoring response that results in hefting (the ability to stay in a given geographical location), imprinting (the ability of following an object), reproductive migration (returning to their point of origin to produce their offspring), and learning the position of multiple objects in an environment and moving through them with making contact (maze running).

*3.1   A fixed animal with feeding motions driven by a recurrent fractal*

Many, if not all, of the activities in a living animal, such as the beating heart, breathing, blinking, etc. require a steady, repeating signal. These signals also exist in the form of brain waves (Buzsaki, 2006). These signals can be created by recurrent fractals, which automatically return to their original state and repeat themselves. This section shows how spikes,

Spike timing

timing units, and actuators can form a useful machine using a recurrent fractal.

### 3.1.1 *Immobile proto animal*

The immobile, proto-animal uses the up/down motion of its tongue to pull water in under its shell, and trap food particles under its shell, as shown in Figure 1.

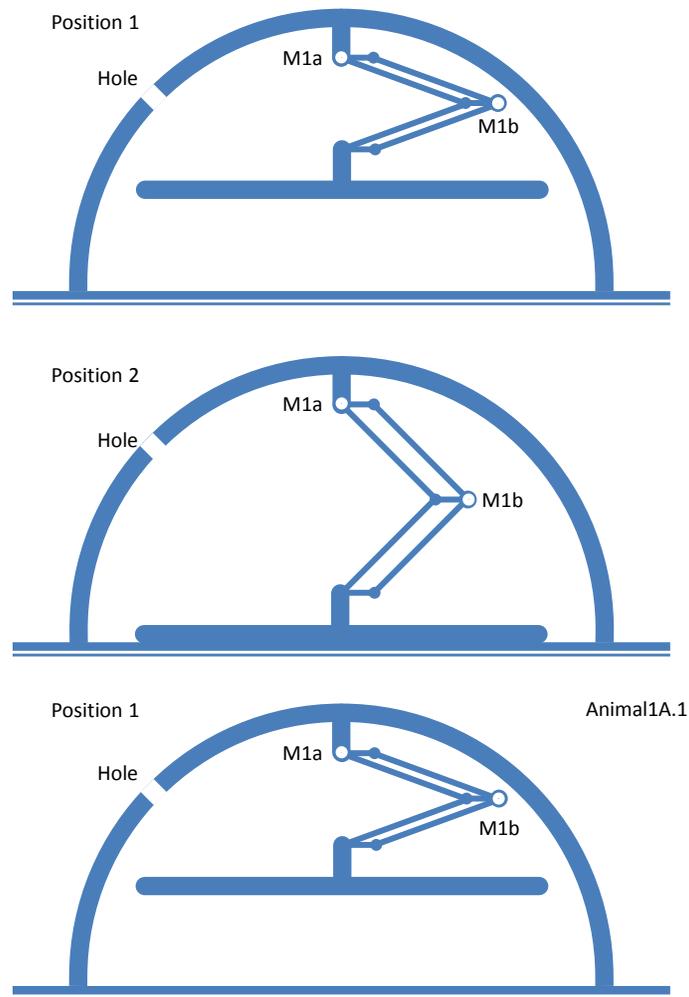

**Figure 1 A *single axis proto animal* uses just one axis to draw in food and trap it under its shell.**

The parallel, four-bar linkages can be operated by two actuators (M1a and M1b) operating at the same speed, time, and displacement. All of these motions are made by electrical spikes created by a spike generator.

### 3.1.2 *A spike generator*

A spike is short burst of electrical energy. A spike generator (Brown, A spike processing model of the brain, 2011) produces a single spike when energized, which is used to disconnect the Output Spike Terminal [4] from the power source nearly instantly, as shown in Figure 2.

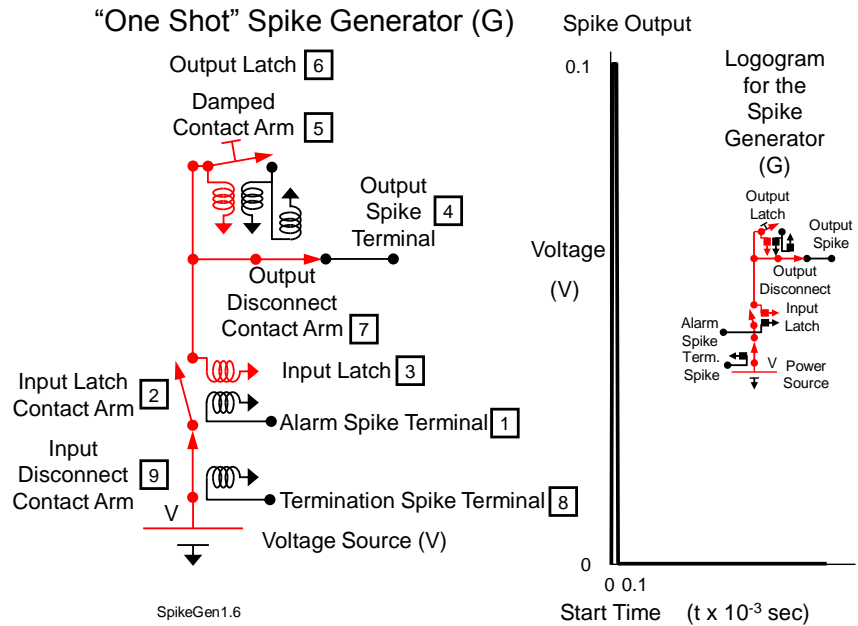

**Figure 2 An electrical *spike* can produce a powerful burst of energy that consumes very little energy if it takes place over a very short period of time.**

The spike generator (G) is turned on by a voltage at the Alarm Spike Terminal [1]. This closes the contact of the normally open Input Latch Contact Arm [2], which is held close by the Input Latch [3] as long as it is connected to the Voltage Source (V). This creates a voltage (V) on the Output Spike Terminal [4]. This voltage also closes the normally open Damped Switch Contact Arm [5] in the Output Latch [6] after a short time delay determined by the amount of damping. This opens the normally closed contact of the Output Disconnect Contact Arm [7], which disconnects the voltage to Output Spike Terminal [4], which terminates the spike.

The Output Latch [6] prevents the Spike Generator (G) from producing another spike until it is reset by a voltage signal from the Termination Spike Terminal [8], which momentarily opens the Input Disconnect Contact Arm [9]. This allows the Input Latch Contact Arm [2] to open, which releases the Output Latch [6]. (The red lines indicate conductors that are always connected to the voltage source after the Input Latch is engaged.)

The purpose of this spike generator is to create a high voltage and high current spike in a circuit with little or no resistance, and to use the high power of the voltage and current to cut off the voltage in a very short period of time so that the energy produced is only as large as needed to operate other devices at the high power level.

(The circuit shown in Figure 2 should not be used in practice without an energy integrating (timing) fuse that limits the amount of energy produced over the critical time period that would cause the wires or other components to overheat. The spike time period should be a very small fraction of this critical overload time period. Breaking a set of contacts in a circuit carrying a high current may cause the contacts to arc and weld together. If the Output Disconnect is not opened by the Output Latch, the spike will not be terminated. This may cause the circuit connected to the voltage source to heat up explosively without a properly designed time sensitive fuse since there is little or no resistance in the circuit).

### 3.1.3   *A high frequency spike generator*

A High Frequency Spike Generator (F) produces a series of output spikes separated by a preset time delay when connected to a power source, as shown in Figure 3.

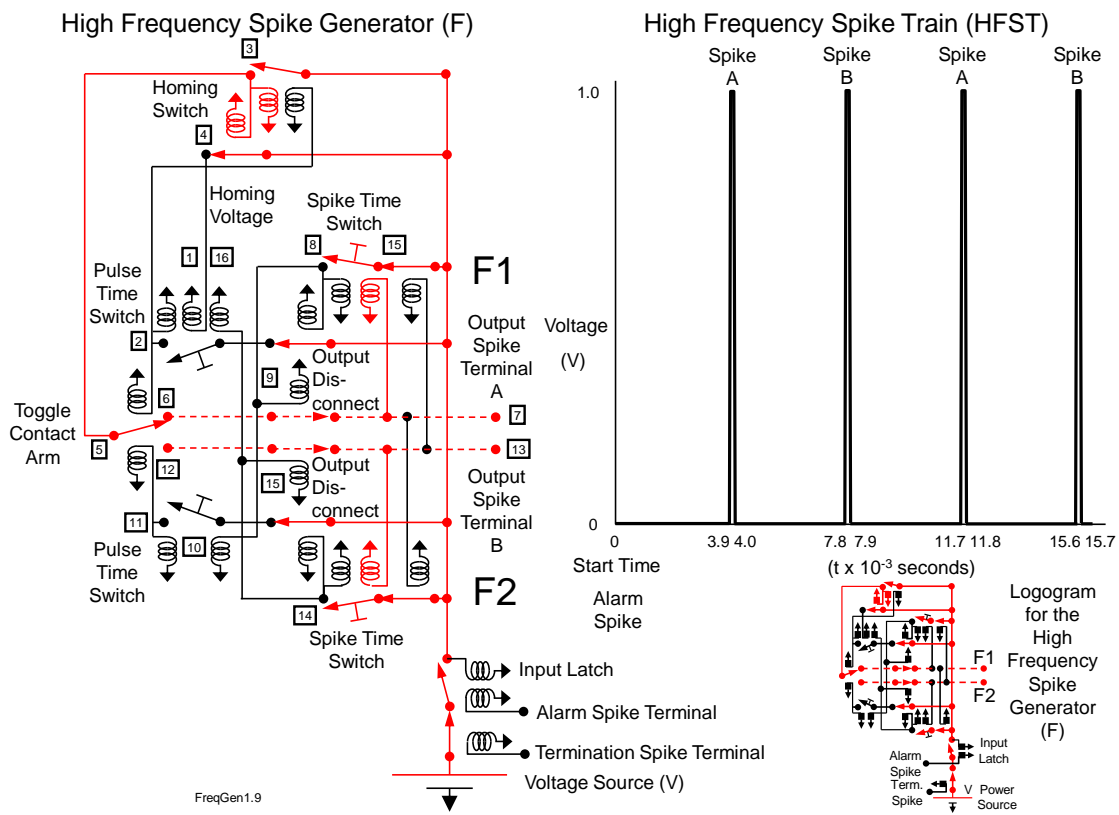

**Figure 3 A *high frequency spike generator (F)* produces a Spike A after an Alarm Spike, and produces a Spike B after some specific pulse time delay. It continues to produce a series of equally spaced spikes having equal spike width times until power is terminated.**

A spike at the Alarm Spike Terminal engages the normally open Input Latch that had been opened by a Termination Spike at some time, previously. This starts a homing voltage to the upper Pulse Time Switch [1]. This causes its contact arm to start closing. After the first pulse time delay, the upper Pulse Time Switch closes [2]. This causes the Homing Switch [3] to close, which opens the homing disconnect [4], causing a voltage to appear at the Toggle Contact Arm [5], and causing the Toggle Contact Arm to contact the upper terminal [6]. This produces a voltage at the Output Spike Terminal (A) [7]. Also, this starts the upper Spike Time Switch [8] to start closing. When contact is made after the Spike Time Delay, the output voltage at (A) is terminated by the Output Disconnect [9] ending the first Output Spike (A), and sending a voltage to the lower Pulse Time Switch [10].

This causes the contact arm of the lower Pulse Time Switch [10] to start closing.  When it contacts its terminal [11] the Toggle Contact Arm [5] is moved to the lower output contact [12], producing the start of the second Pulse Time Period at terminal [13]. This resets the upper Spike Time Switch [15], allowing it to open to its normally open position. This terminates the voltage at the lower Pulse Time Switch [10]. The voltage on the Output Spike Terminal B [13] energizes the lower Spike Time Switch [14]. This energizes the Output Disconnect [15], ending the second Output Spike (B) at [13]. This also energizes the upper Pulse Time Switch [16]. So the process is started again to generate the next Output Spike (A) at [7].

So the First Output Spike starts the series of equally spaced spikes of a given frequency that is ended only by a Termination Spike. The spike width is determined by the damping in the Spike Time Switches, and the spike frequency determined by the damping in the Pulse Time Switches. The conductor to the Toggle Contact Arm [5], the Contact Arm [5], and the spike train F1 and F2 are shown in red since they operate repeatedly after the Homing Switch [3] is closed.

Spike timing

Pulse widths are partly analogue because they represent one dimensional lengths, and partly digital because they determined by time since there are no analogue (continuous) methods of measuring time, as far as I know. Thus, time is always measured in discrete steps. So a spike frequency generator is needed to keep track of time in systems using PWM as well as in computers, cell phones, and GPS systems, where time is used as a variable.

3.1.4 *Period Timing Framework*

The actuators in the animal model in Figure 1 need to operate at specific times in a specific step sequence. Each step requires four spike times (T1, T2, T3, and T4) to complete, as shown in the Actuator Output Spike Trains in Figure 7. So the spike train shown in Figure 3 must be grouped into a set of four spikes, forming the Step Periods shown in Figure 4.

**Figure 4** The *Period Timing Framework* is created when one output (F1) of the High Frequency Spike Generator (F) is connected to seven frequency dividers and a set of Period Frequency Dividers that produce an output spike on one of four conductors, sequentially, after every 128 F1 spikes.

Each Step Period is divided into 128 Spike Times. The chart of the Period Frequency Divider Output Spike Train shows eight time units between Periods. Each division represents 16 spikes from the High Frequency Spike Generator (F). This output spike train provides the Period Timing Framework of the spike processing system used throughout this model.

### 3.1.5 *Spike driven actuator*

The actuators (M1a and M1b) in Figure 1 can be started, stopped, and reversed using the Basic Actuator Cell circuit shown in Figure 5.

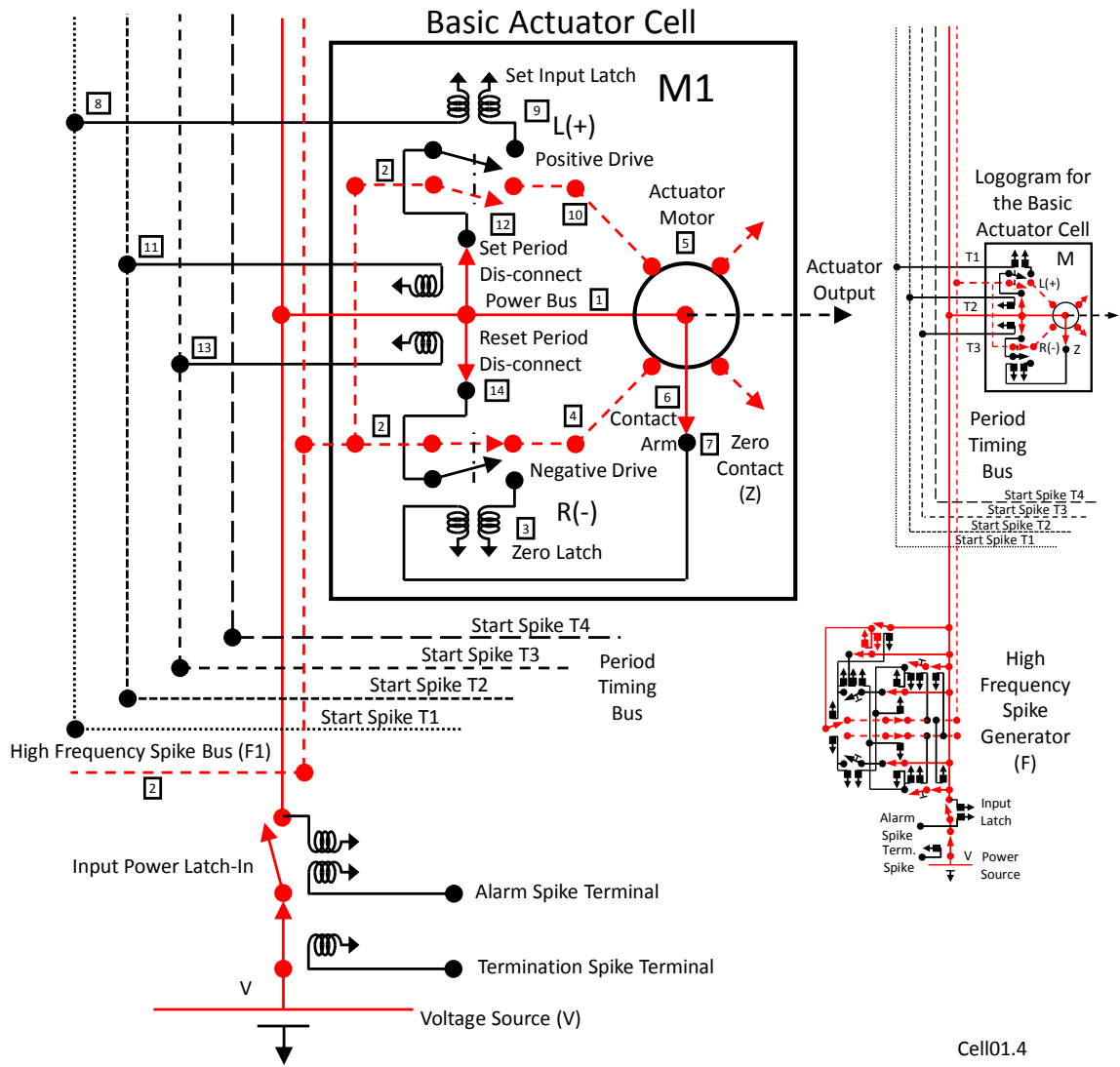

**Figure 5** The output motion of the *Basic Actuator Cell* (BAC) is started by a spike on the Set Input Latch, stopped by a spike on the Set Period Disconnect. It is reset (reversed) by a spike on the Reset Period Disconnect, and stopped by a spike through the Zero Contact.

A spike on the Alarm Spike Terminal engages the Input Power Latch. This creates a voltage on the Power Bus [1] and starts the High Frequency Spike Generator (F), which produces a high frequency spike train (F1) on the High Frequency Bus [2]. The normally closed contact in the Zero Latch [3] causes high frequency spikes to flow through the Negative Drive [4] to the Actuator Motor [5]. This causes its Contact Arm [6] to rotate clock-wise toward the Zero Contact [7] as part of the homing process that occurs before the first Start Spike T1. When the Contact Arm contacts the Zero Terminal [7], the Zero Latch [3] is engaged. This interrupts the high frequency spike train [2], causing the Contact Arm [6] to remain at Zero Terminal (Z), and causes the power side of the Zero Latch [3] to latch in the closed position.

When a spike occurs on the T1 conductor [8] of the Period Timing Bus, the Set Input Latch [9] is engaged. This causes high frequency spikes to flow through the Positive Drive terminal [10] of the Actuator [5]. This causes its Contact Arm [6] to rotate away from the Zero Terminal (Z) in the counter-clock-wise direction until there is a spike on the T2 conductor [11]. This interrupts power through the Set Period Dis-connect [12], which un-latches the Set Input Latch [9], and stops the rotation of the Contact Arm [6].

Then, when a spike occurs on the T3 conductor [13] of the Period Timing Bus, the power to the Zero Latch [3] is interrupted, which un-latches the Zero Latch causing the high frequency spike train to flow through the Negative Drive [4]

Spike timing

again until the Contact Arm [6] makes contact with the Zero Contact [7]. This re-latches the Zero Latch [3], which stops the flow of spikes to the Actuator Motor [5]. The output of the Basic Actuator Motor is shown in Figure 7.

3.1.6  *Spike driven muscle motor*

The armatures of the Actuator Motors (M1a and M1b) in Figure 1, Figure 5, and Figure 7 are turned by the spike driven muscle motor system shown in Figure 6.

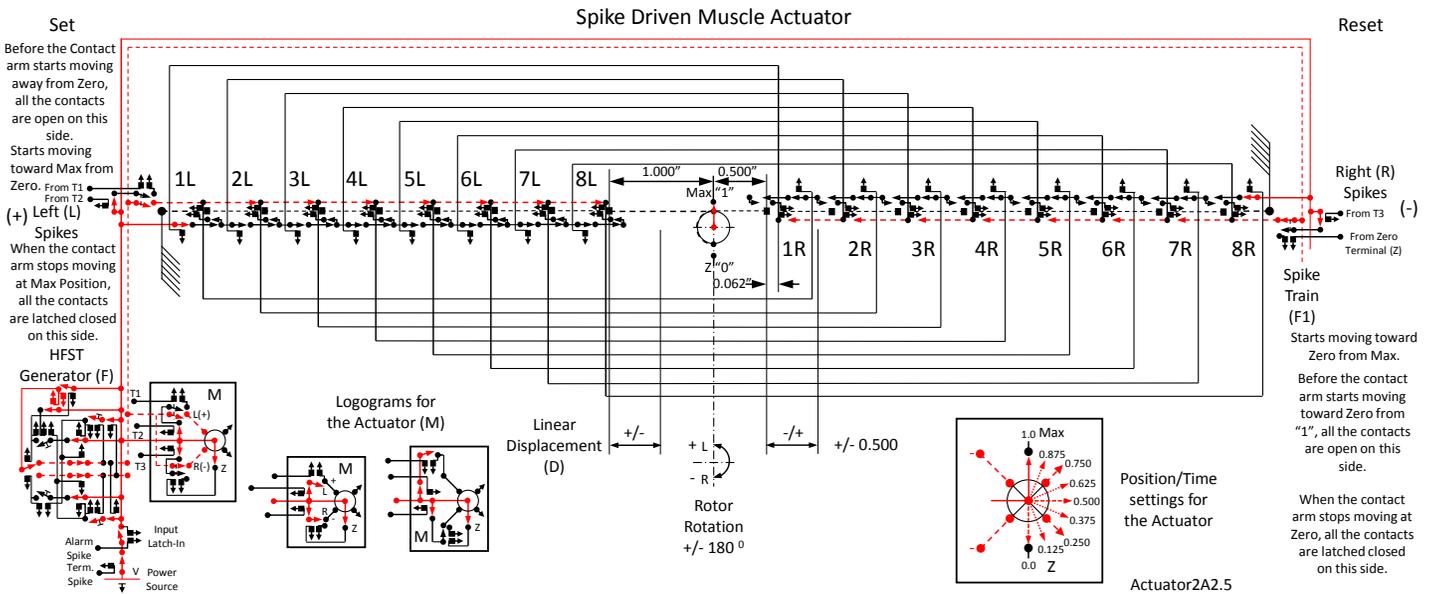

**Figure 6 A *spike-driven muscle motor* is driven by the high frequency spike train that successively contracts (flexes) a cell on one side of the rotor of actuator (M), and successively releases (extends) a cell on the other side of the rotor. These motions cause the actuator to produce a rotation in one direction or the other.**

At Start-up, a series of F1 spikes are introduced on the right side. This turns the rotor clockwise until the rotor contact arm strikes the Zero Terminal (Z), bring the rotor to it "home" position. The Zero Contact Spike disconnects the high frequency spike train and stops the rotor at the Zero (home) position, as shown in Figure 7. When there is a spike on the T1 conductor of the Period Timing Bus shown in Figure 7, a series of F1 spikes are introduced to the set of cells on the left side of the rotor causing them to become contracted (flexed) sequentially, and the set of cell on the right side to be released (extended) sequentially. This causes the rotor to turn counter-clockwise. The rotor is stopped by a spike on the T2 conductor of the Period Timing Bus shown in Figure 7 when it is fully extended to 180 degrees. In this state, all of the cells on the left side are fully contracted, and all of the cells on the right side are fully extended, as shown above. This requires that there be 128 muscle cells on both sides of the rotor. (Only eight of the 128 are shown on each side).

If two or more muscle actuators are connect in series and operated together, the velocity and displacement of the actuators increases in proportion to the number of actuators. If two or more muscle actuators are connected in parallel and operated together, the force produced by the actuators is increased in proportion to the number of actuators. Connecting multiple muscle actuators in series and parallel increases the velocity and force of the actuators. An examination of biological muscles shows these series and parallel muscle cell configurations.

The muscle motors are shift registers that can store and express pulse width information. The value of this pulse width is not stored in just one logic unit (cell). Instead, the value of the pulse width is stored by the position of the transition between open and closed contacts in the multiple logic units (cells) of the bidirectional shift register.

The normally open and normally closed relays, latches, lockouts, and actuators, frequency dividers, and summing differentials shown in these circuits are the only logic elements used in the spike timing process, except for the addition of

Spike timing

a simple shift register introduced in Figure 29.

3.1.7  *Control system for organism with a pulse generator and a single set of actuators*

An up/down motion of a feeding member, such as a tongue shown in the proto-animal in Figure 1, can be generated by the Spike Generator (G), High Frequency Generator (F), Main Timer (T), made up of the Main Frequency Divider, Period Frequency Dividers, and the actuators made up of the Basic Actuator Cell (BAC) shown in Figure 7.

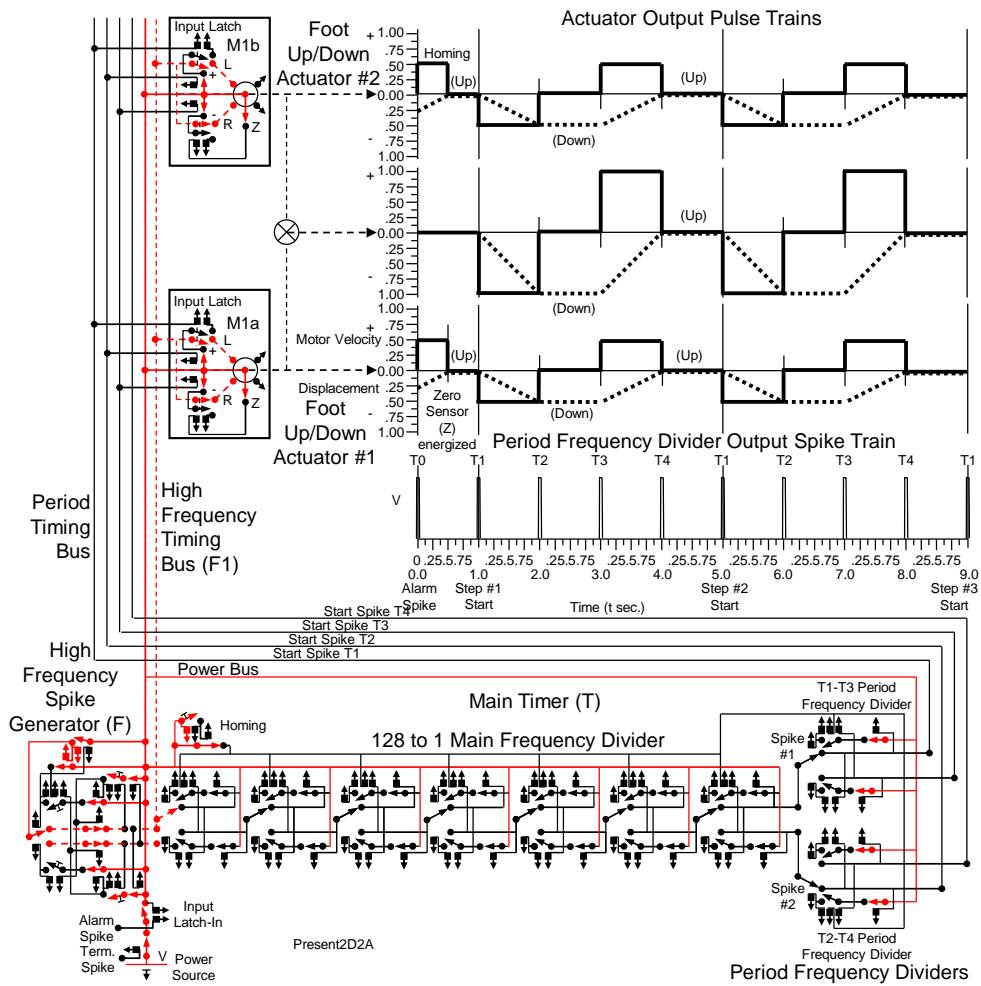

**Figure 7 A** *frequency spike generator, frequency dividers, and a pair of actuators* **can produce a recurrent fractal that generates useful behavior.**

When the Power Bus is turned on by the Alarm Spike, the High Frequency Spike Generator starts to generate spikes, as shown in Figure 3. These spikes drive home the actuators M1a and M1b clockwise to their zero positions (Z) if they are not there already, as part of the homing process. During this homing process the high frequency spikes are toggling through the Main Frequency Divider. After 128 spikes have been produced, a spike appears at the Spike #1 Terminal. This causes a spike on the T1 line of the Period Timing Bus. This closes the Input Latches in the actuators (M1a and M1b), which allows the high frequency spikes to drive the actuators (M1a and M1b) in the clock-wise direction (as shown in Figure 7) until the Main Timer (T) sends a spike on the T2 line, which releases their Input Latches. This stops the actuators until the Main Timer (T) sends a spike on its T3 conductor, which releases their lockouts and starts the reset of the actuators toward their Zero Contacts (Z). The actuators stop when their contact arms hit their Zero Contacts (Z). These motions produce the Actuator Output Pulse Trains in Figure 7. These pulse trains are recurrent fractals that continue until there is a Termination Spike.

### 3.1.8 *Two axes proto-animal*

The proto-animal with two axes allows the tongue to move in and out as well as up and down, as shown in Figure 8.

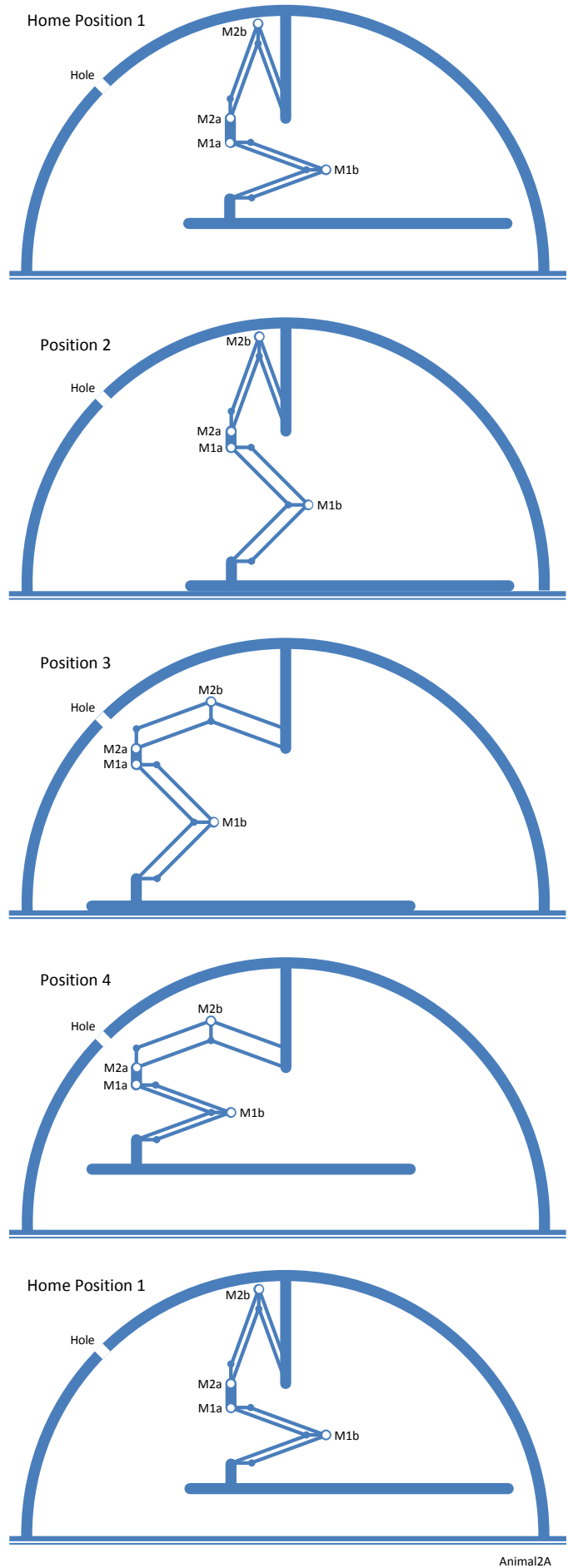

**Figure 8** The *two-axis, proto-animal* is capable of trapping food under its shell and moving it toward a mouth region.

The design of this proto-animal is similar to a present-day mollusk called a Limpet found along the shores of the British Isles attached to rocks. Some varieties of the Limpet have the hole near the top of the shell to draw in and exhaust water under the shell.

### 3.1.9 *Control system with an additional in/out axis*

The additional motion axis (Out/In) can be added to the control system in Figure 7 to drive the two-axis organism in Figure 8, as shown in Figure 9.

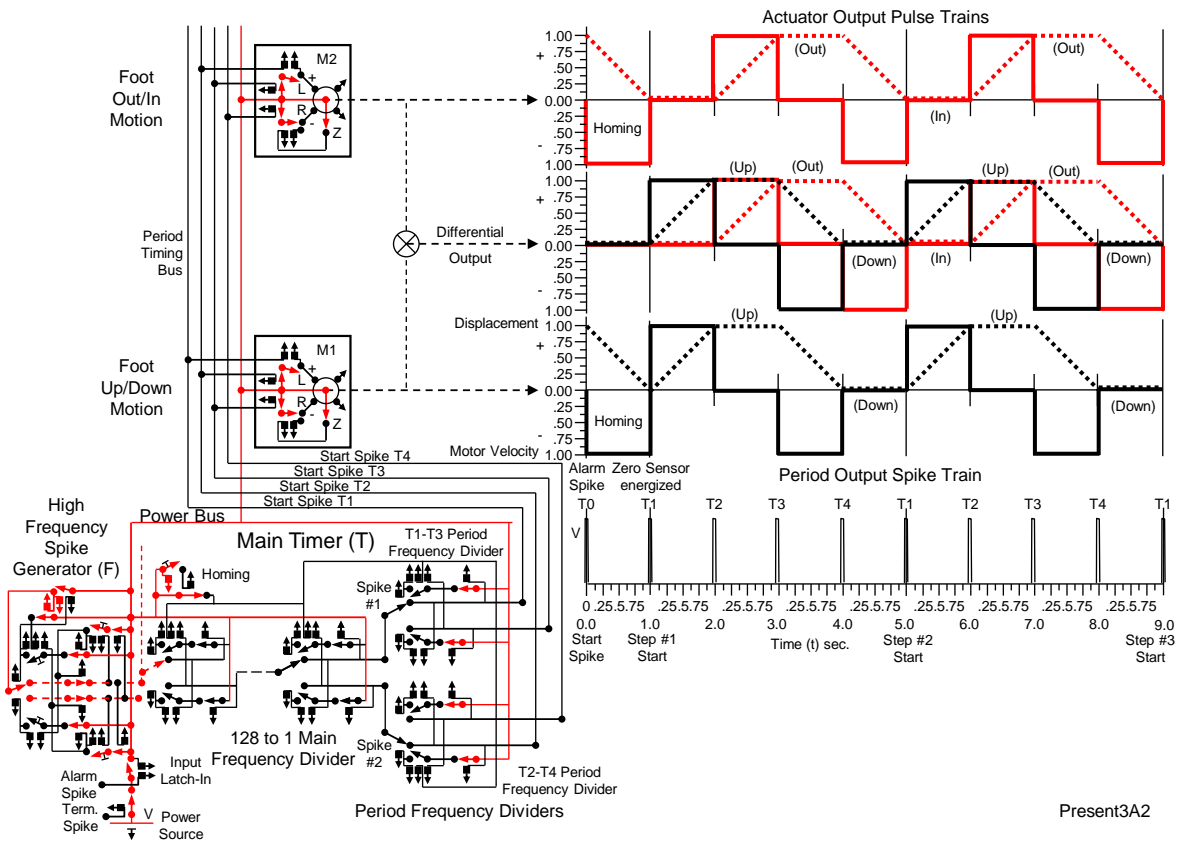

**Figure 9** **The output of the *Out/In axis* is added by the machine to the Up/Down axis resulting in a coordinated, useful motion.**

The up/down motion and an out/in motion of a tongue can be combined to pull food toward the mouth of an animal. This motion can be represented by the motion of M1 and M2 being summed in a compound mechanism, as represented by a differential. The start times of the each axis is determined by the connection made by each motor to the Timing Bus. Note that M1 starts on T1, and M2 starts on T2, and M1 stops at T2 and M2 stops at T3. The M1 resets at T3, and M2 resets at T4. Thus, the Main Timer (T) determines all of the start, stop, and reset times. An axis requires two actuators, as shown in Figure 8. To produce a given total displacement each actuator must run at half of the final speed. (To save space on the printed page, the two actuators (M1a andM1b, and M2a and M2b) are shown just as their sum M1 and M2 in Figure 9 since the (a) and (b) motors run at the same time and at the same speed. Also, the High Frequency Timing Bus is omitted since it just parallels the Power Bus, as shown in Figure 7.)

### 3.1.10 *Section summary*

The high power, low energy, and great timing ability of spikes that can be used to start and stop timers in a rhythm based control process using pulse width modulation to control the movement of an organism.

## 3.2 *A mobile animal driven by a recurrent fractal*

A genetic variation in some of these animals may have caused the shell's bond to the sea floor to weaken, and the animal finds itself set free from its fixed position. The tongue becomes a foot, and the tongue motions used for feeding now causes the animal to move across the seafloor more or less in a straight line. As it moves its shell across the seafloor, it collects food as before, but has the potential to encounter a great deal more food.

### 3.2.1 *Straight line locomotion*

Once the shell is no longer attached to the sea floor, the up/down, out/in motion of the tongue (which is now the foot) causes the animal to move in a straight line in one direction using the same control circuit in Figure 9. The newly evolved proto-animal is shown in Figure 10.

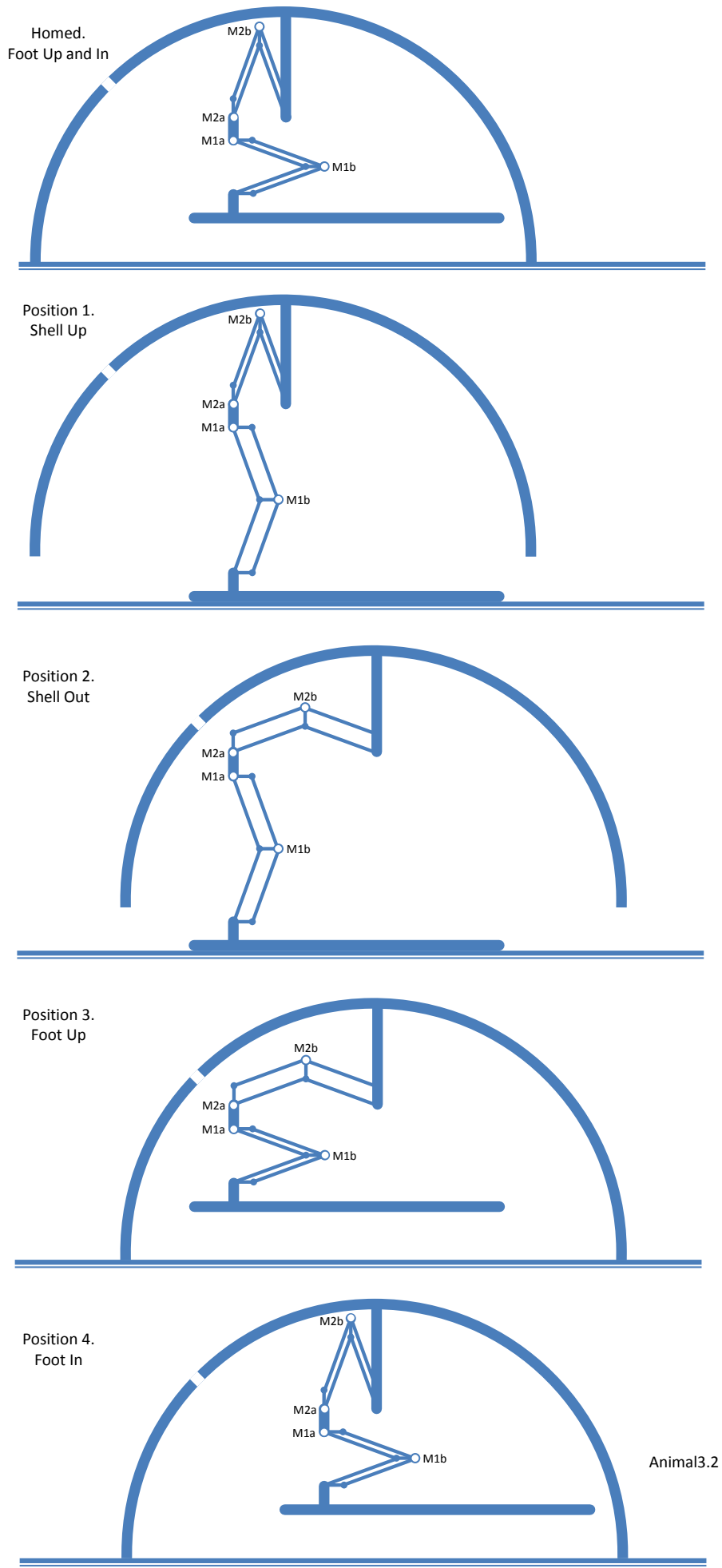

**Figure 10  The *shell and foot movement* causes the animal to move in a straight line.**

The tongue becomes the animal's foot, and allows the animal to move in a straight line so it can encounter new sources of food in each step that it takes. This greatly increases its food supply. But it also poses new threats to its survival.

Spike timing

### 3.2.2 *Diagrammatic representation of the motion*

The organism can be shown by the symbolic objects shown in Figure 11.

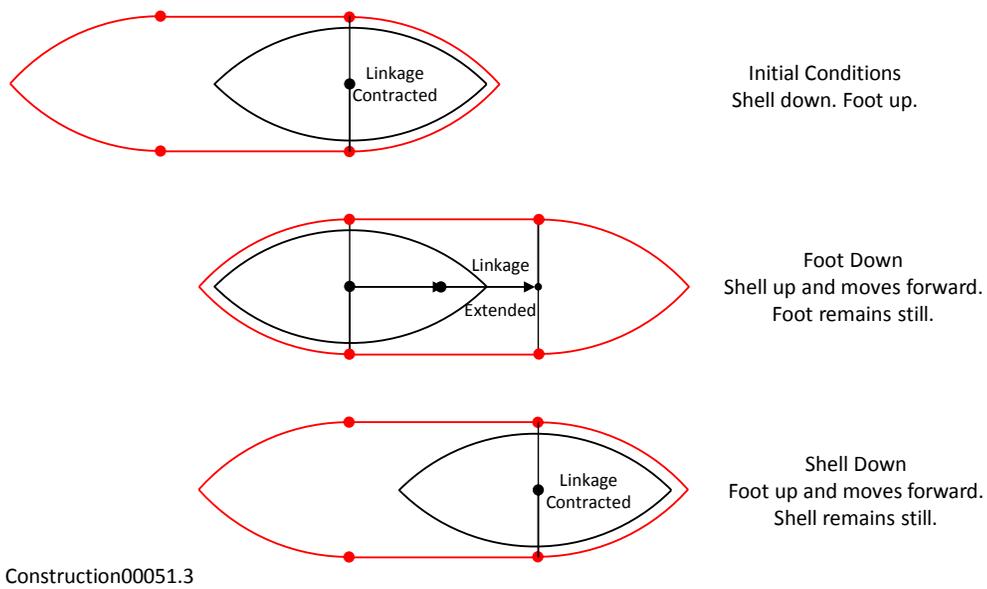

**Figure 11  The foot is *represented* by the closed pair of arcs, and the shell is represented by the lines around the foot.**

Movement is created by extending the linkage to the foot to raise the shell, extending the linkage from the foot to the shell, contracting the linkage to the foot to bring the foot up, and contracting the linkage from the foot to the shell.  But this movement in a straight line takes it away from its breeding community and into an obstacle, eventually. What is needed is a means of changing the straight line motion in to circular motion.

### 3.2.3 *Constant velocity, anthropomorphic, rotational, two-bar linkage*

When the straight line movement of the movement of the out axis is combined with a constant speed rotation axis working together, the path forms a circular arc. This can be accomplished by rotary actuators operating at a constant speed using the anthropomorphic, two-bar linkage shown in Figure 12.

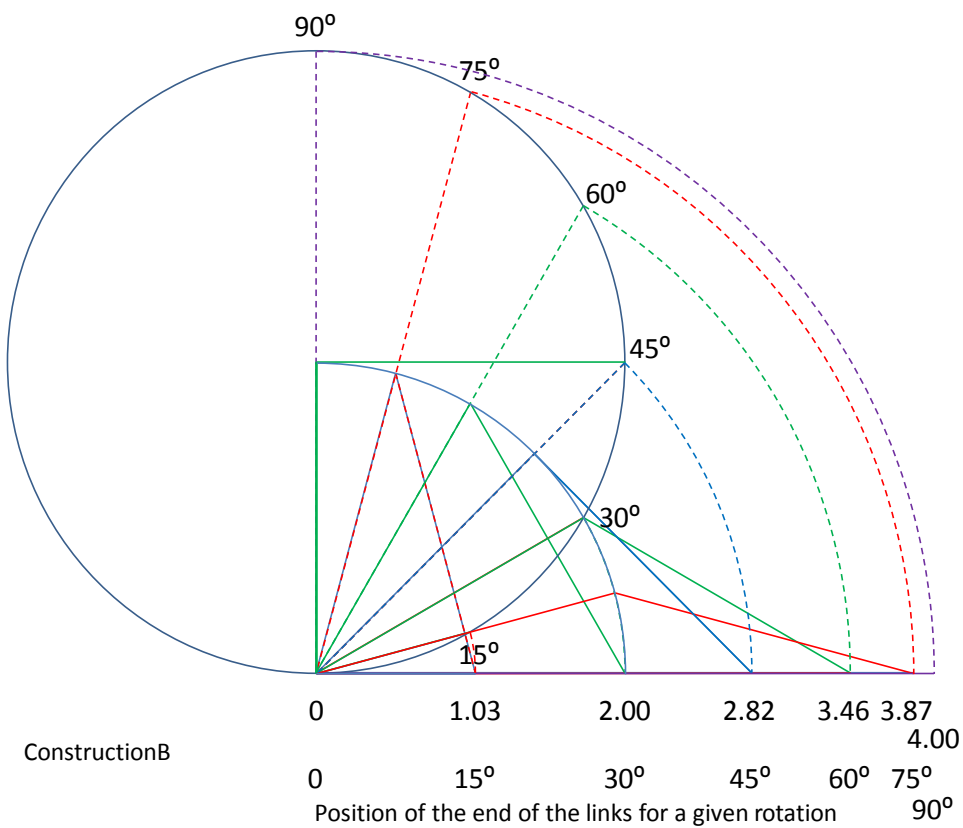

**Figure 12  A *circular arc* is formed by the end point of the linkage as the linkage extends and the linkage assembly rotates with constant angular motions.**

Spike timing

The path can continue to 180 degrees to form a complete circle. These perfect arcs are produced by the constant velocity rotary actuators characteristic of timing devices. Even though the end point is being moved at a decreasing speed by the Out/In axis linkage, end point moves at a constant speed along the circular arc when combined with the constant rotary motion.

### 3.2.4　*Kinematic model with rotation*

A third motor M3a and a slave Motor M3b can rotate the shell around the foot, and can rotate the shell an equal amount around the end of the extension linkage, as shown in Figure 13.

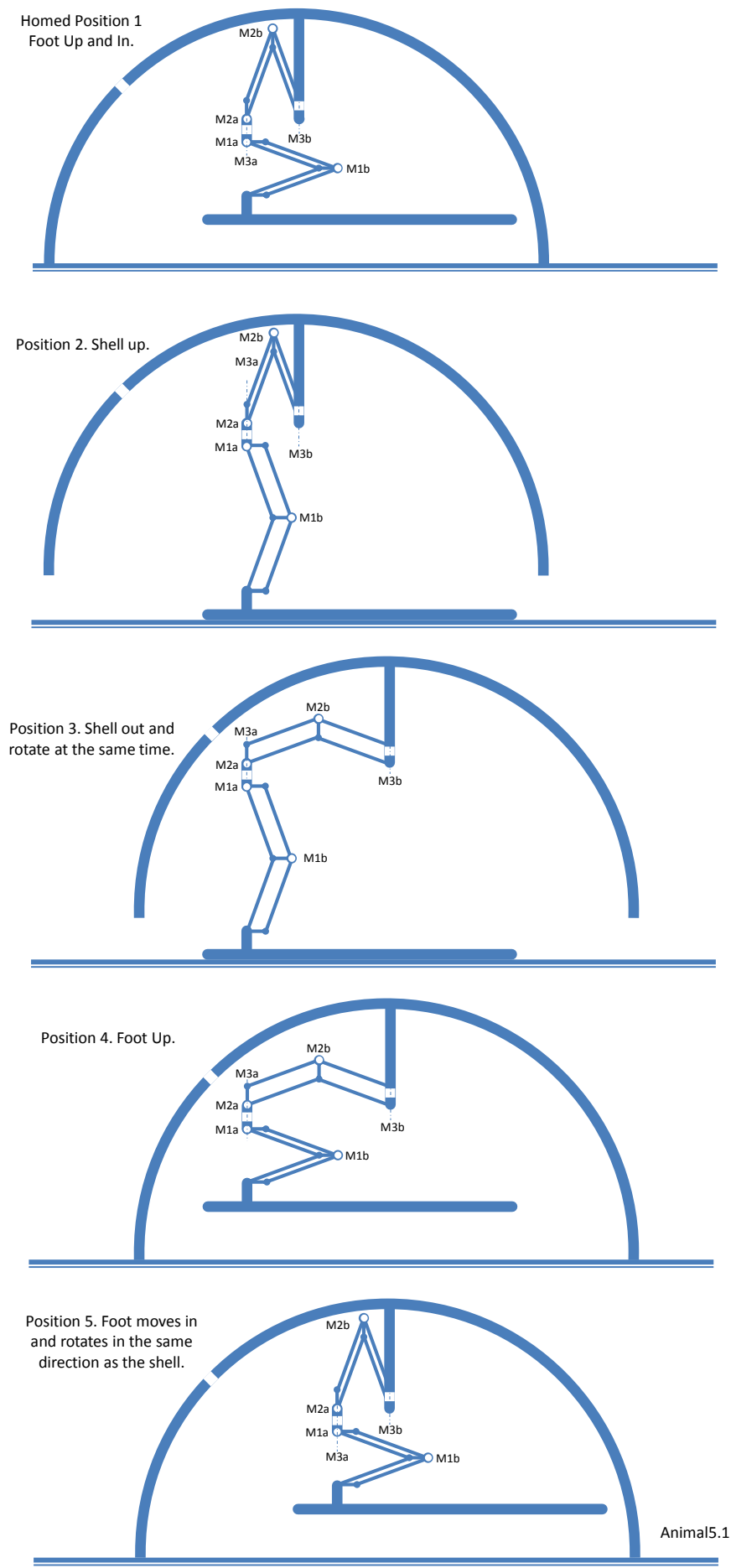

**Figure 13　The *kinematic model with shell/foot rotation* can move forward and rotate to a new position.**

Spike timing

The M3a and M3b motors allow the shell to rotation in one direction when the foot is down. But the foot must rotate in the same direction when the foot is up. This requires that the M3a and M3b (rotation axis) motors run in the opposite direct when the foot is up than it does when the foot is down, as shown on the rotation axis control system in Figure 14. So, in one complete sequence of four steps the model gets back to its original state, but with both the shell and foot the new position facing a new direction.

### 3.2.5 *Control system with a shell/foot rotation axis*

The rotation axis using motors M3a and M3b can be added to the control system in Figure 9, as shown in Figure 14.

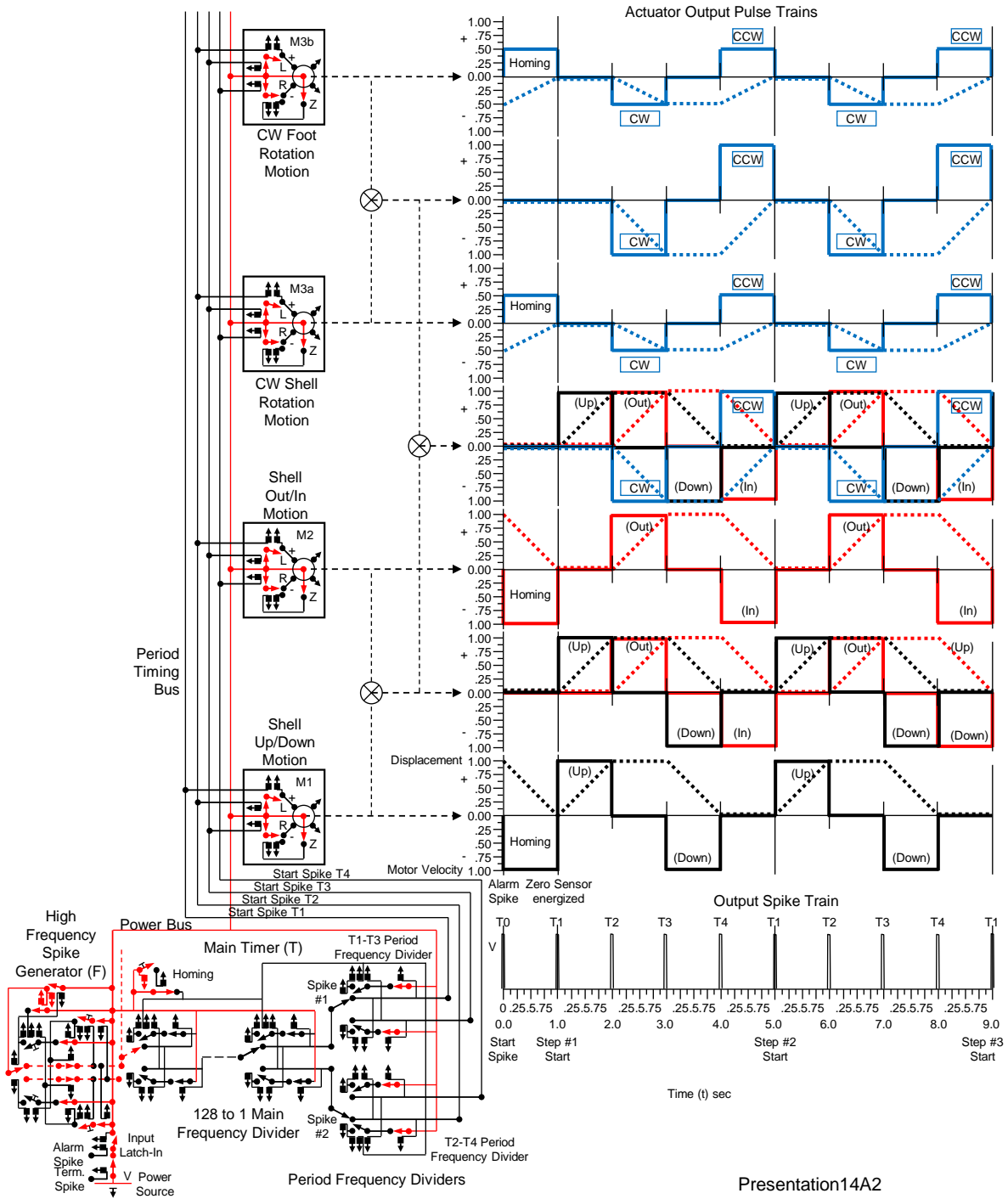

**Figure 14  The shell/foot *rotation axis* (M3a, M3b) can be added to the up/down, out/in motion control system, causing the animal model to turn and move in one direction of rotation.**

Note that the direction of rotation of the shell and foot rotation motors (3a, 3b) occur together, occur at the same speed, and have same direction of rotation. So the result in the model is the sum of these movements, as represented by the differential.

### 3.2.6 *Trajectory of the animal model*

A constant speed movement of the Out Axis and the Rotation Axis together creates a constant speed circular arc, as shown in Figure 12. The degree of rotation of the shell is determined by the angle or rotation of the extension linkage connected

Spike timing

to the foot and the angle of rotation of the linkage to the shell. So a full 90 degree rotation of the shell with respect to the foot is produced by a rotation of 45 degrees of the linkage at the foot using M3a, and 45 degrees at the shell using M3b. So the sum of M3a and M3b is 90 degrees when the extension linkage rotates 45 degrees, as shown in Figure 15.

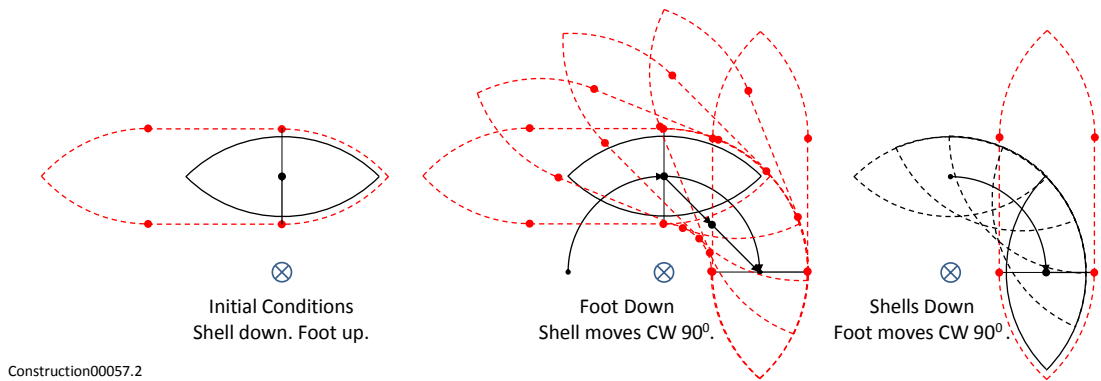

**Figure 15  The straight, rigid shell will not be able to lift the foot after a rotation of 90 degrees because the shell would come down on its own foot.**

If the shell is straight, as shown in Figure 11, the shell would not be able to lift the foot after a rotation of 90 degrees, as shown in Figure 15. So the shell would need to be permanently curved in the direction of rotation to encircle the foot, as shown in Figure 16.

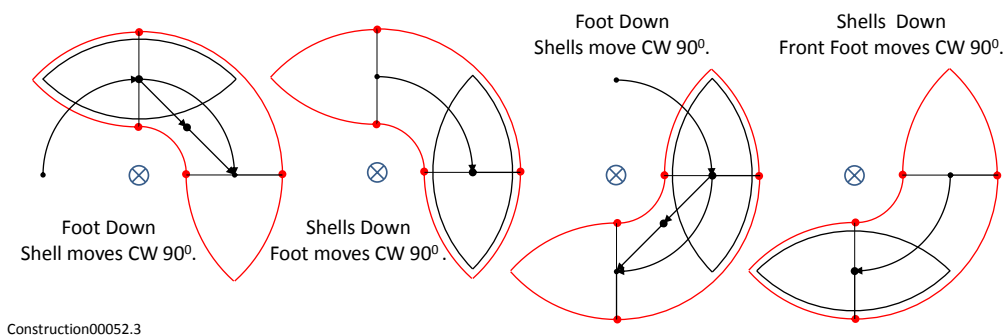

**Figure 16  The *curved rigid shell body* moves forward when the foot is down, and the foot moves forward when the shell is down.**

Even then, the bulk of the weight of the organism needs to be in the foot to maintain balance. The alternating motion of the M3 motors allows the animal to move forward rotating in one direction, only.

3.2.7    *Circular motion in one direction*

The tongue motions of what is now the foot can be converted into a movement in a circle by repeating the down/up, out/in, and the rotary motion in one direction using the control system shown in Figure 14, as shown in Figure 17.

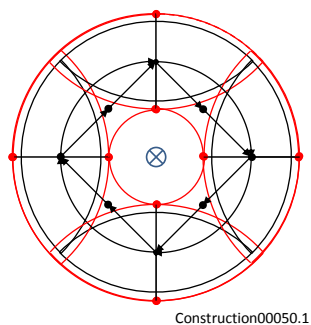

**Figure 17  A *full 360 degree turn* is made by repeating four 90 degree right-hand steps in just one direction of rotation.**

These 90 degree arcs can be repeated indefinitely to produce a continuous movement in a circle in the clockwise direction. This motion allows the animal to find encounter food sources, and keeps the animal close to its breeding community. But it limits the size of its grazing area.

Spike timing

### 3.2.8 *Section summary*

The time based system evokes its own behavior plan by means of its fractal structure that causes the organism to move around in its environment.

## 3.3 *Rotation in both directions*

The size of the grazing area can be increased by adding a second rotation axis motor so the shell can rotate in both directions. The rotatory axis motors need to operate in steps that are one-half the frequency of the Down/Up and Out/In axes.

### 3.3.1 *Two rotation axis motors*

For the shell to rotate in both directions alternatively requires that the shell and the foot rotation be in the same direction in one Step Period, and rotate in the opposite directions in another Step Period. This requires two rotation axes motors, as shown in Figure 18.

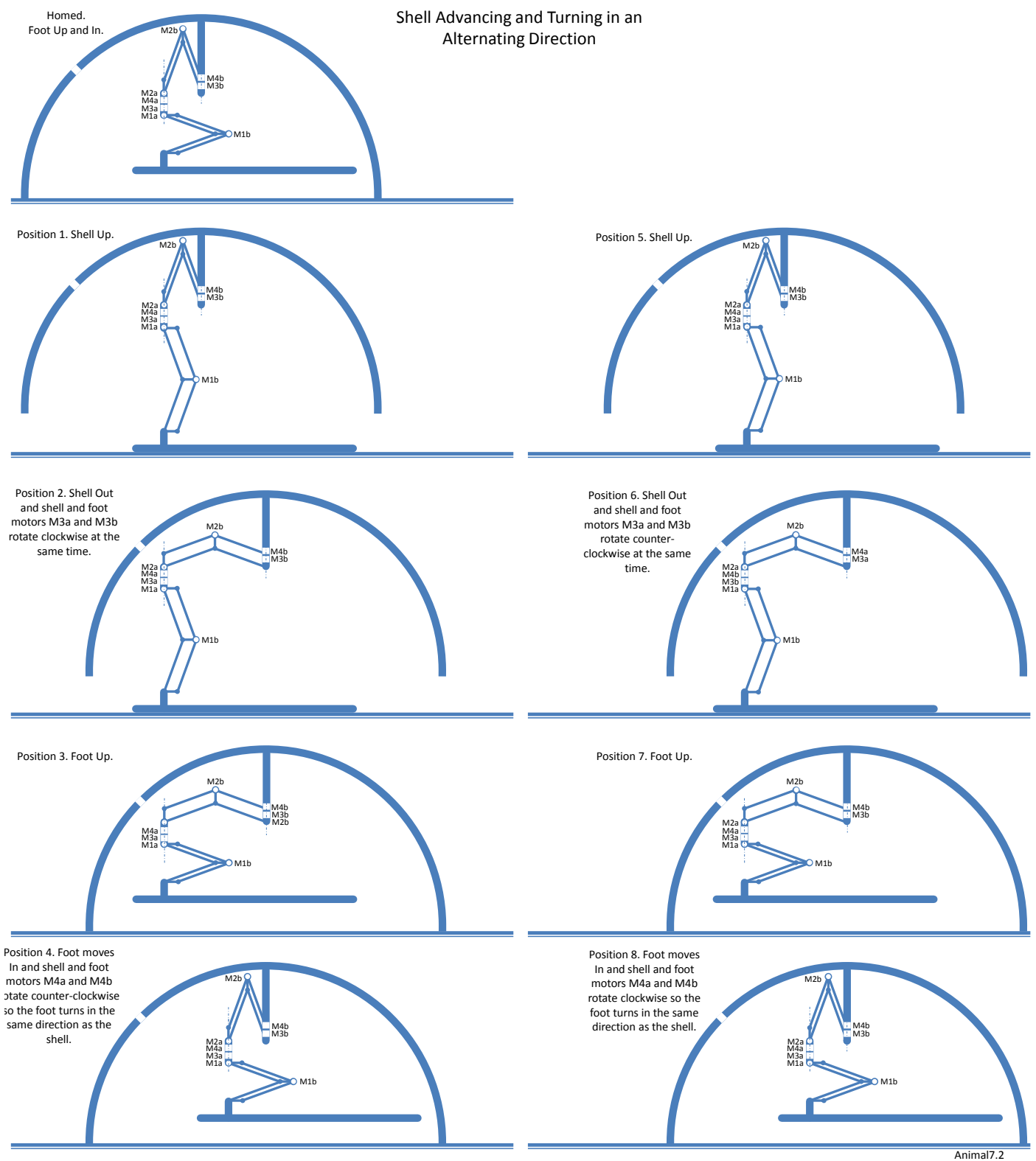

**Figure 18** A *second set of shell and foot rotation axis motors* **(M4a and M4b) are required for the shell to rotate in both directions.**

Spike timing

For the shell to turn in the opposite direction in Step 2 from Step 1, it needs to rest in Step 2 instead of Step 1. This requires that there be a separate foot rotation axis, as shown in Figure 19

### 3.3.2 *Control system with a second rotation axis*

The effectiveness of the grazing patterns can be increased by allowing rotation in both directions. This is accomplished by adding a second rotation axis (M4a, M4b) to the control system shown in Figure 14, as shown in Figure 19.

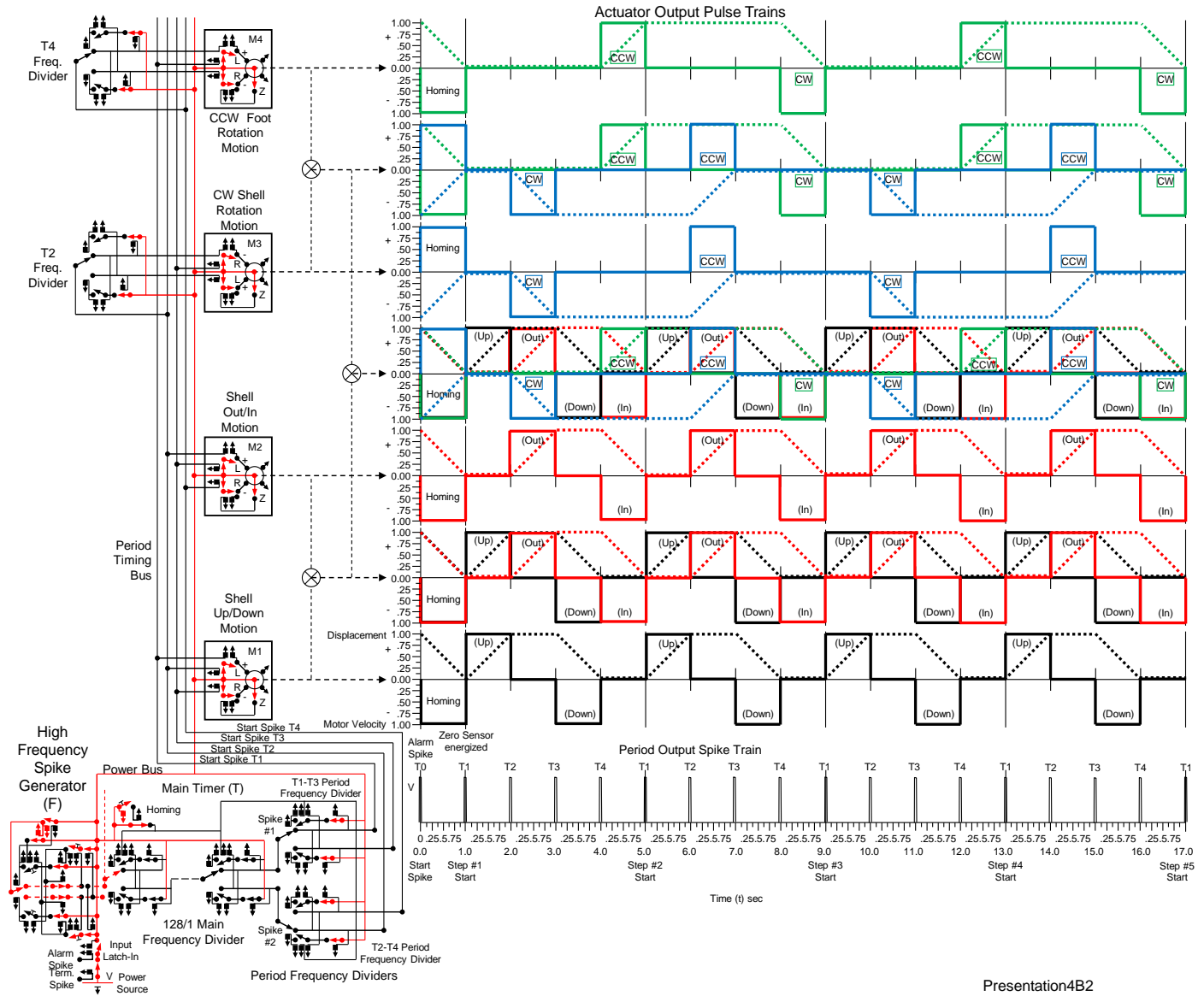

**Figure 19** The *shell and foot rotation axes* need to operate at half the frequency of the Up/Down, Out/In axes.

The T2 frequency divider starts M3 at the first T2, and resets it at the second T2. The T4 frequency divider starts M4 on the first T4, and resets M4 on the second T4. This causes the rotation axes to operate at half the frequency as the other axes. This simple control system would work for an organism with a flexible shell, as shown in Figure 20.

### 3.3.3 *Trajectory possible with a flexible shell*

The flexible shell allows the organism to turn in both directions, as shown in Figure 20.

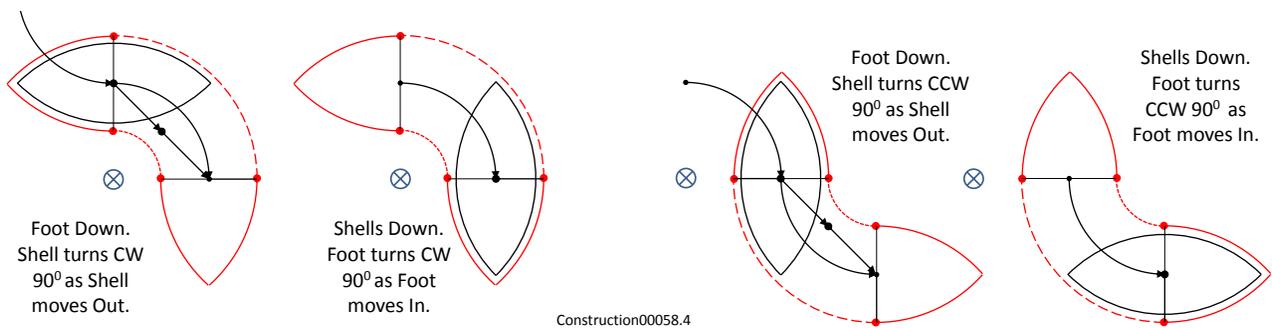

**Figure 20** The *side walls of the shell* must stretch in only one direction, like an accordion made with rigid folded plates.

The back part of the shell is dragged forward when the front part of the shell moves forward. A shell like this might be too flimsy to be lifted and placed accurately by the foot.

Spike timing

### 3.3.4 *Rotation in both directions by splitting the shell into a front shell and a back shell*

The rotation in both directions requires that the shell be flexible, or that there be two shells, a front shell and a back shell. The two separate shells increase the ability of the animal to lift its foot. The space between the two shells can be filled by a flexible membrane, or be left open, as shown in Figure 21.

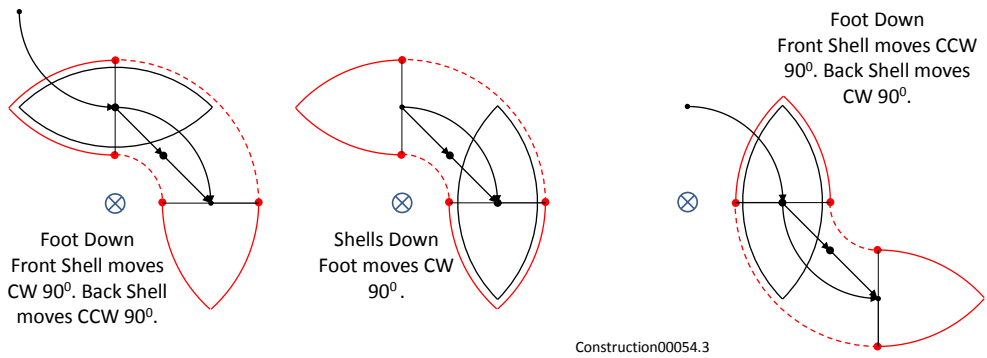

**Figure 21** The *back shell rotation axis* operates at the same time as the front shell, but in the opposite direction when the animal rotates in alternate directions.

The ability to rotate and move forward in either direction is the basic requirement for an animal if it is to range freely in its environment.

### 3.3.5 *The design of the organism with a front and back shell*

Splitting the shell into two halves requires another set of in/out and rotation axes, as shown in Figure 22.

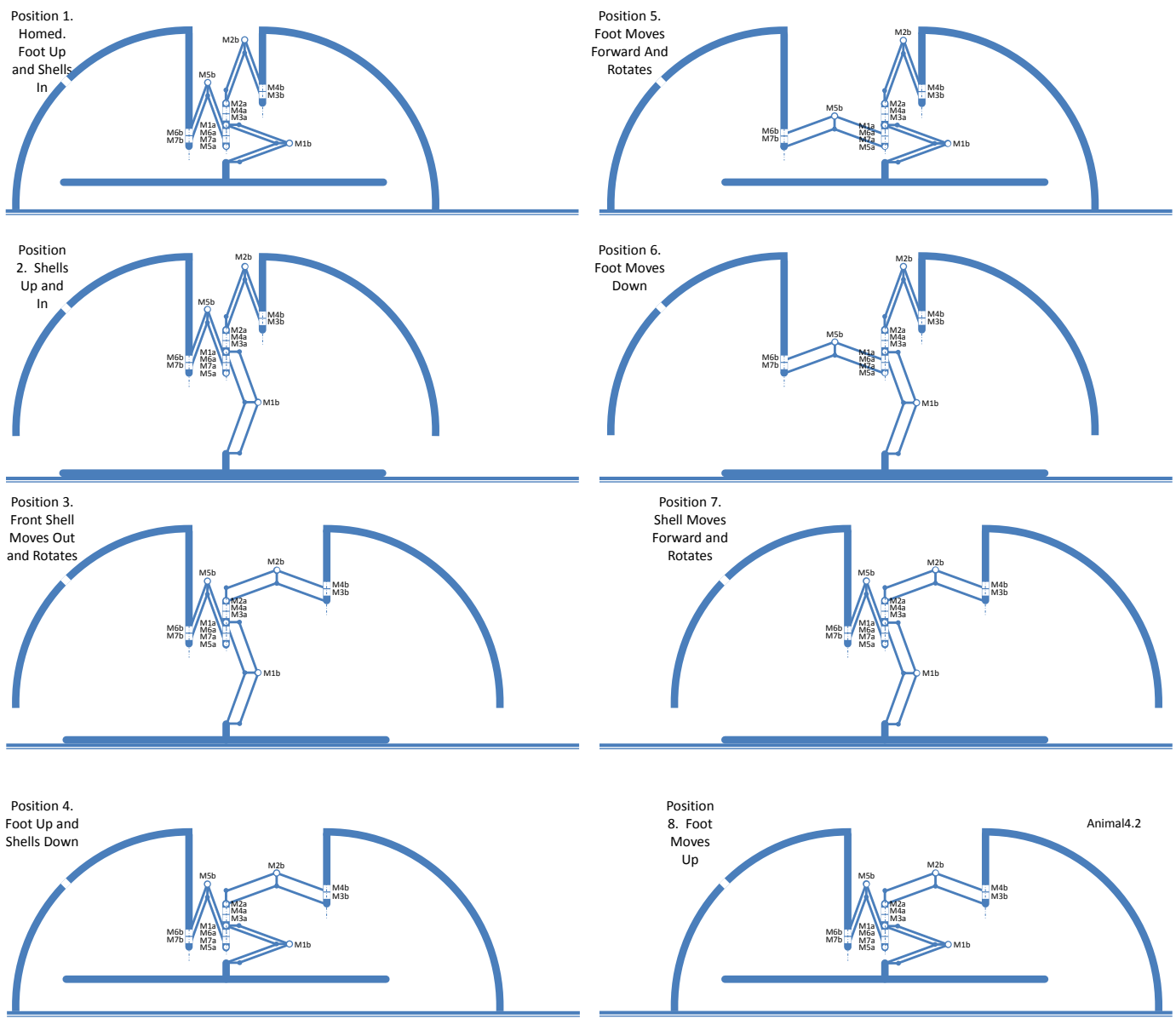

**Figure 22** The *two shells operate at the same time*, but the direction of rotation of the back shell must lag one step behind the front shell.

Spike timing

The two extension axes are not shown completely contracted. This is because it would not show the individual links when contracted. In the execution of the motions, the links would be extended 45 degrees, and fully contracted to make the 90 degree rotation movement shown on Figure 12.

3.3.6 *Control system needed to produce rotation in both directions in an animal with two shells.*

The control system need to produce rotation in both directions with two shells requires three more Basic Actuator Cells, the Back Shell In/Out (M5), the Back Shell Rotation (M6), and the Back Shell/Foot Rotation (M7), as shown in Figure 23. The three new axes are basically a duplication of the previous three axes, Front Shell Out/In (M2), Front Shell Rotation (M3), and Front Shell/Foot Rotation (M4). They both operate at the same time, but in the opposite direction.

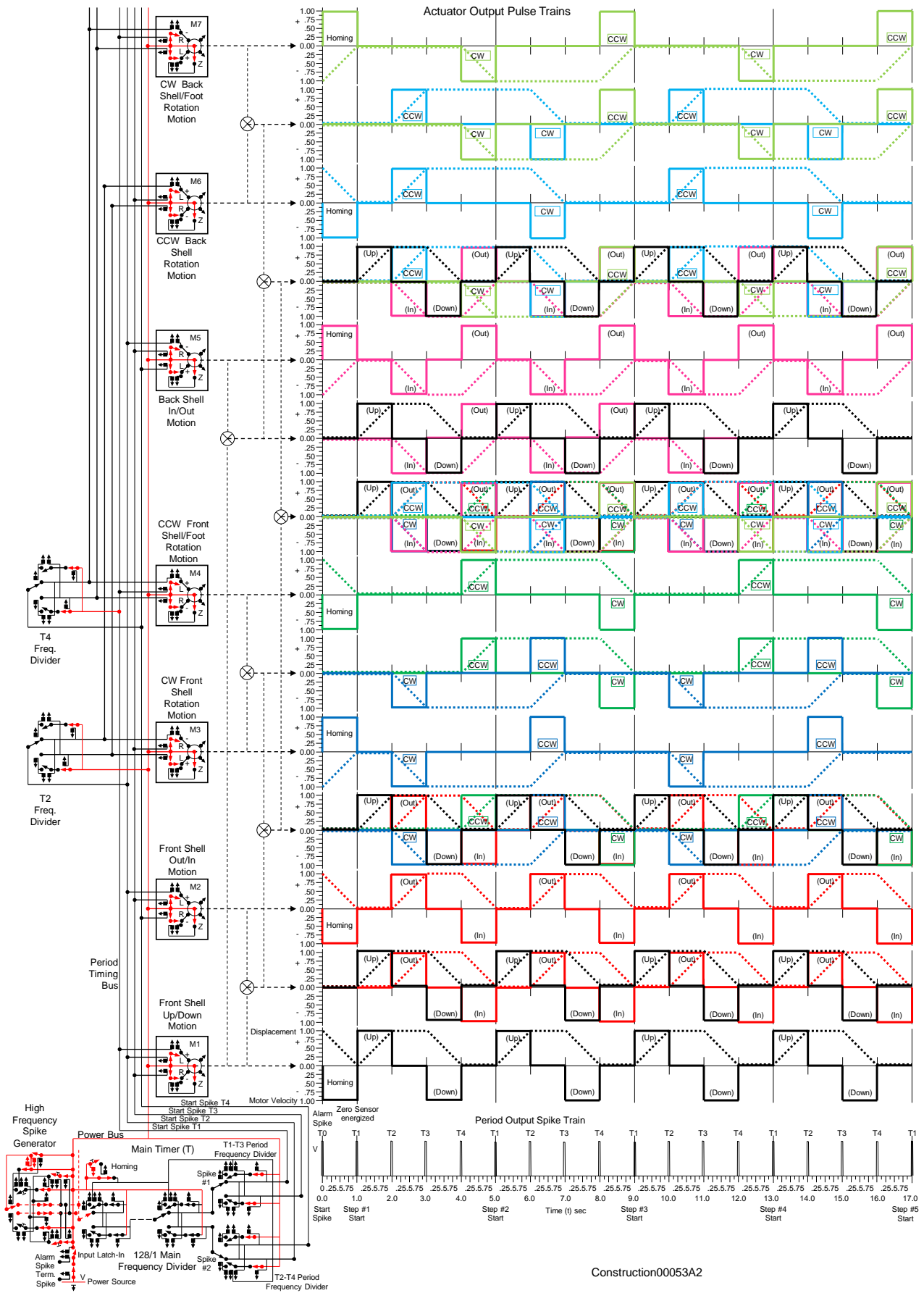

**Figure 23** The *back shell rotation axis* operates at the same time as the front shell, but in the opposite direction when the animal model alternates its direction of rotation.

Spike timing

So an organism with a front shell and a separate back shell would need six pairs of actuators (axes) plus the shell Up/Down axis.

### 3.3.7 *Sequence of Front Shell/Foot and Back Shell/Foot motions*

The sequence of Front Shell and Back Shell motions is shown in Figure 24.

ALTERNATING (STRAIGHT-LINE) MOTION

Step 0, T0-T1
Initial Conditions (Homing).
Foot moves up.
(Shells move down).
Front and Back Shells move in.

Step 1, T1-T2
Foot moves down.
(Shells move up).

Step 1, T2-T3
Front Shell moves out CW $90^0$.
Back Shell already in.

Step 1, T3-T4
Foot moves up.
(Shell moves down).

Step 1, T4-T1.
Shells remain stationary.
Front Shell/Foot Axis moves in CW $90^0$.
Back Shell/Foot Axis moves out CW $90^0$.
Foot moves CW $90^0$.

Step 2, T1-T2
Foot moves down.
(Shell moves up).

Step 2, T2-T3
Front Shell/Foot Axis moves out CCW $90^0$.
Back Shell/Foot Axis moves in CW $90^0$.
Front Shell moves CCW $90^0$.
Back Shell moves CW $90^0$.

Step 2, T3-T4
Shells move Down
(Foot moves up).

Step 2, T4-T1
Shells remain stationary.
Front Shell/Foot Axis moves in CCW $90^0$.
Back Shell/Foot Axis moves out CCW $90^0$.
Foot moves CW $90^0$.

Step 3, T1-T2
Foot moves down.
(Shells move up).

Step 3, T2-T3
Front Shell/Foot Axis moves out CW $90^0$.
Back Shell/foot Axis moves in CCW $90^0$.
Front Shell moves CW $90^0$.
Back Shell moves CCW $90^0$.

Step 3, T3-T4
Shells move Down.
(Foot moves up).

Step 3, T4-T1
Front Shell/Foot Axis moves in CW $90^0$.
Back Shell/Foot Axis moves out CW $90^0$.
Foot moves CW $90^0$.

Step 4. T1-T2
Foot moves down.
(Shell moves up).

Step 4, T2-T3
Front Shell/Foot Axis moves out CCW $90^0$.
Back Shell/Foot Axis moves in CW $90^0$.
Front Shell move CCW $90^0$.
Back Shell moves CW $90^0$.

Step 4, T3-T4
Shells move Down
(Foot moves up).

Step 4, T4-T1
Shells remain stationary.
Front Shell/Foot Axis moves in CW $90^0$.
Back Shell/Foot Axis moves out CW $90^0$.
Foot moves CW $90^0$.

Construction00060.1

**Figure 24** The *motions of the Front Shell* determine the path of the animal.

The Front Shell moves out clockwise in Step 1, counter-clockwise in Step 2, clockwise in Step 3, and counter-clockwise in Step 4, as shown above, creating motion in an alternating straight line, as shown in Figure 25.

### 3.3.8 *Path created by rotation in alternating directions*

Using alternate rotations, the animal can move in a nearly straight line movement by repeating the down/up and out/in motions, and rotation axes shown in Figure 23. This alternating motion is shown in Figure 25.

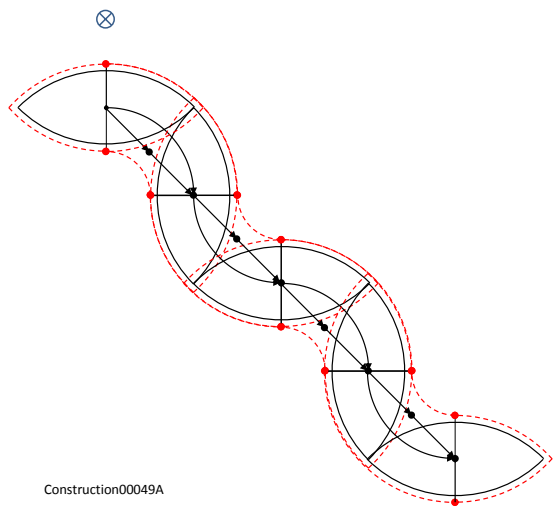

**Figure 25  The *straight line motion* will take the animal away from its fertilizing community, and cause it to run into some obstacle or cull de sac, eventually.**

However, the ability to rotate in both directions creates the potential for the animal to move in a figure-eight, which requires that the animal be able to turn around in a complete circle in both the clock-wise and counter-clockwise directions.

### 3.3.9 *Section summary*

Multiple fractals acting together are needed to produce motion in both directions.

## 3.4 The figure-eight recurrent fractal

One of the simplest recurrent motion fractals is in the form of a figure-eight. It provides a simple way for an animal to extend its grazing area and stay in a given location. So far, the organism operated using a set of simple, alternating fractal. To create motion in a more complex pattern of a figure-eight requires the creation of a more complex control system. This requires the addition of more semi-axes motors operating at different times to synthesize the more complex fractal.

### 3.4.1 *A recurrent fractal*

A figure-eight can be drawn as shown on Figure 26.

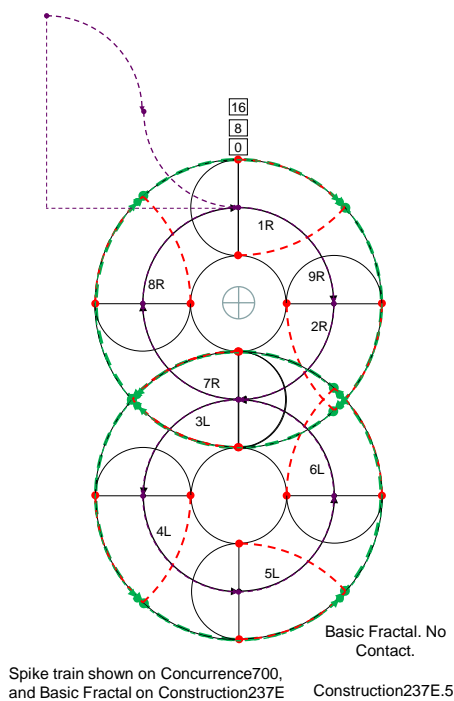

**Figure 26  A *figure-eight* is an example of a recurrent fractal.**

Spike timing

The movements are described as steps labeled 1R, 2R, 3L, 4L, 5L, 6L, 7R, and 8R. Obviously there is a change in direction of rotation after Step 2R and Step 6L. This is a better grazing path than going around in a circle because the animal grazes over a larger area.

3.4.2  *Problems with this two-shell animal turning in a figure-eight*

Moving in a figure eight in eight steps requires that M3 produce up to four repeated turning arcs in one direction, and four repeated turning arcs in the other direction, as shown in Figure 27.

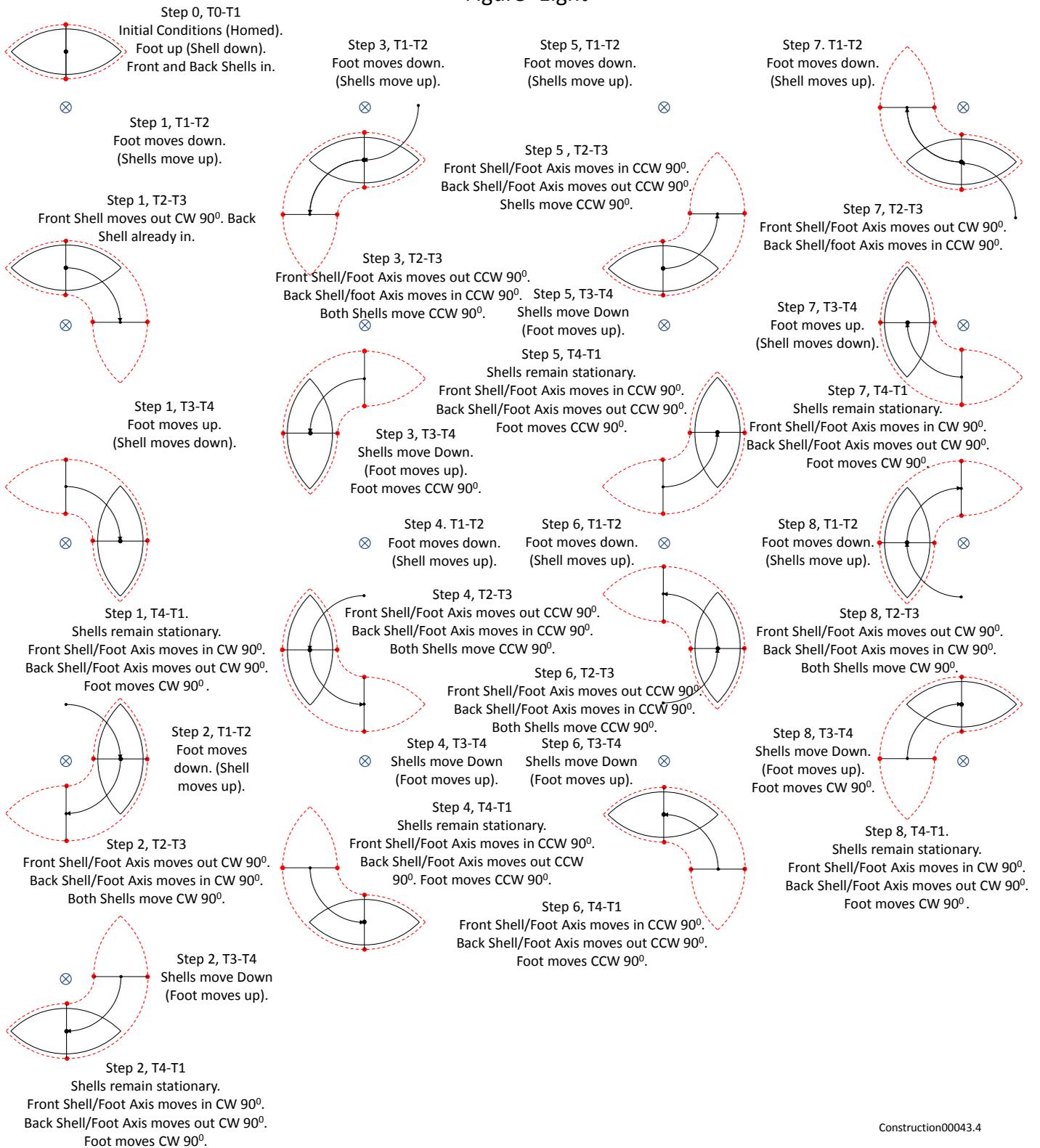

**Figure 27  A *figure-eight motion* requires repeated motions of the Front Shell Rotation Axis in both directions.** However, each motor must reset after it makes a motion. So making a complete circular motion in both directions is not possible with the control system shown in Figure 23. The more complex control system shown in Figure 29 is needed to make a complete circle in both directions.

### 3.4.3 *Actuator Cell needed to form a figure-eight*

The foot and shell actuators have been modified so that they are enabled by the Step Bus (SP), and are expressed (activated) by T2 or T4 signals on the Step Period Bus in Figure 29, as shown in Figure 28.

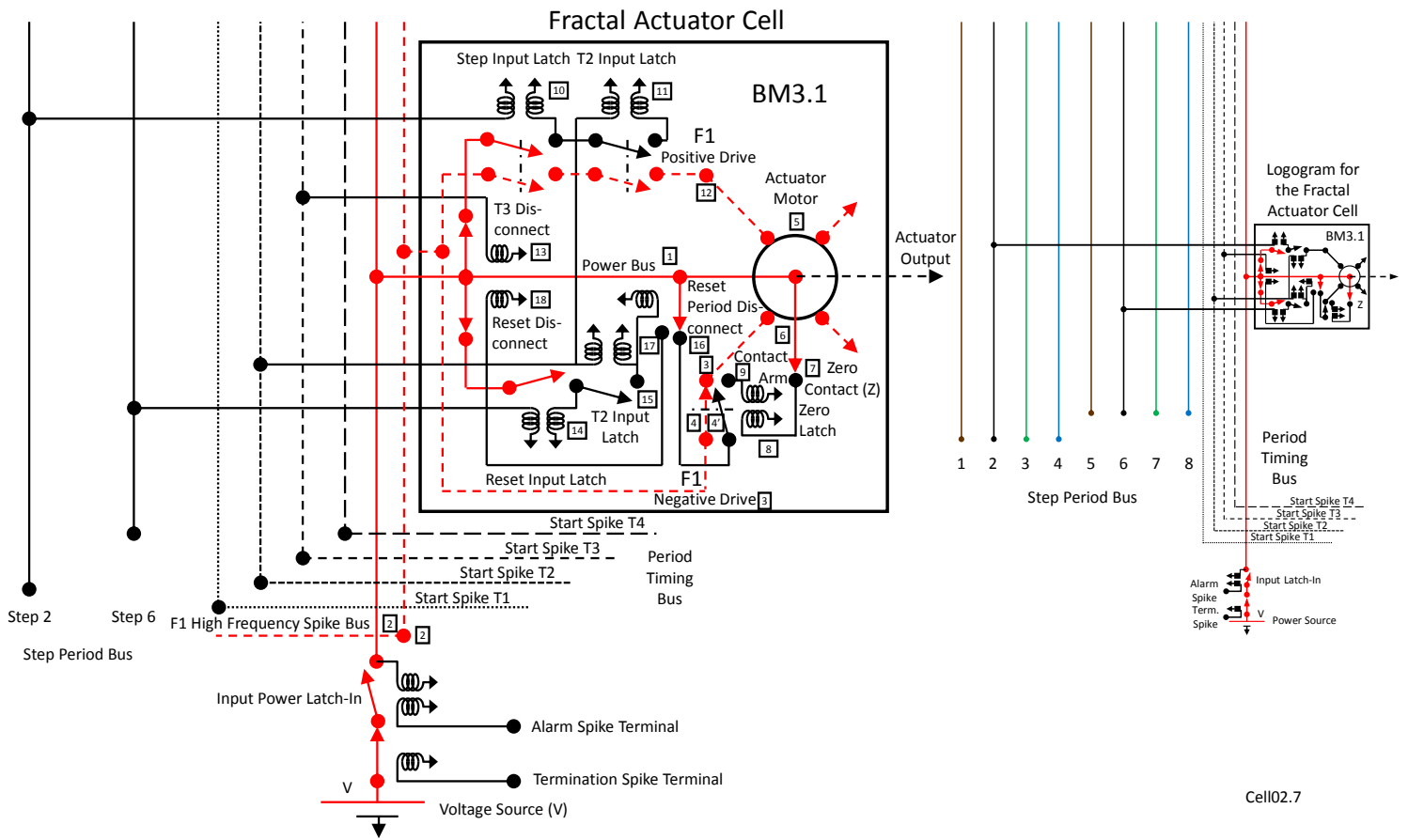

**Figure 28 The *Fractal Actuator Cell* (FAC) is started by a spike on the Step 2 conductor in the Step Period Bus and then a spike on the T2 conductor in the Period Timing Bus.**

Operation of the Fractal Actuator Cell

Homing: The Power Bus [1] is energized by a spike at the Alarm Spike Terminal. This causes a voltage on the Power Bus [1] and spikes on the High-Frequency Spike Bus [2] to flow to the Negative Drive [3] through the spike contact [4] of the Zero Latch [8] to the Actuator Motor [5]. This drives its Contact Arm [6] in the negative (Clock-Wise) direction toward the Zero Contact [7]. When the Contact Arm arrives at the Zero Contact [7] the Zero Latch [8] is engaged. This causes the Negative Drive Contact Arm [4'] to contact the latch terminal [9]. This interrupts the High Frequency Spike train to the Actuator Motor, leaving its Contact Arm [6] at the Zero Terminal.

Set Spike: When a spike occurs on the conductor in the Step Period Bus (Step 2) connected to this Actuator Cell, its Step Input Latch [10] is engaged. Then, a short time later the spike on the T2 conductor engages the T2 Input Latch [11]. This causes power and the high frequency spike train to flow through the Positive Drive [12], causing the Contact Arm [6] of the Actuator Motor [5] to rotate in the positive (Counter-Clock-Wise) direction until it is stopped by the spike on the T3 conductor of the Timing Bus. This engages the T3 Disconnect [13], which unlatches the Set Input Latch [10] and the T2 Input Latch [11]. This process produces an output pulse of one period.

Reset Spike: Four steps later, a spike on the Step 6 conductor engages the Reset Input Latch [14], and another spike on the T3 conductor of the Timing Bus engages the T3 Input Latch [15]. This causes the Reset Period Disconnect [15] to break the power flow to the Zero Latch [8]. This allows the Negative Drive Contact [4'] to return to its normally open position. Also, power flows through Contact [17] to the Reset Disconnect [18], which dis-engages the Reset Latches [14] and [15]. This also allows the Reset Period Contact [4] to return to its normal position. This causes the High

Spike timing

Frequency Spike Train [2] to flow through the Negative Drive [3], which causes the Contact Arm [6] of the Actuator to rotate in the negative (CW) direction until it contacts the Zero Contact [7]. This re-engages the Zero Latch [4'], which opens Contact [4] stopping the Actuator Arm at the Zero Contact [7], and closes Contact [4']. This engages the Zero Latch [8] so the Contact Arm 6 remains at the Zero (Z) position. This process produces an output Reset Pulse that is equal to and opposite from the Set Pulse.

Thus, the basic Fractal Actuator Cell (FAC) is not started by a spike on the T2 Bus alone, as in Figure 7. Since there are eight Step Periods in the figure eight, the Step Period Bus (SPB) operates at one-eighth of the frequency of a conductor in the Start Spike Bus (T). So a given FAC is started every eight Steps. In the example shown, the Set Spike occurs at T2 in Step 2, and the reset occurs at T2 in Step 6 since the reset occurs four Steps after the start (Set) spike. So four FACs are needed to produce the eight-step fractal.

### 3.4.4 Control system need to form a figure-eight

An additional frequency divider and a new circuit called a shift register are needed to change the direction of rotation of the animal to form the figure-eight. The two-level frequency divider and eight element shift register shown in Figure 29 can provide the unique signals need to change direction of the arc in an eight-step pattern. The shift register produces a spike output at ¼ of the frequency of the main timer (T). This kind of circuit will be used repeatedly because of its ability to provide spike outputs after many different evenly-spaced time delays created by the system timer (T).

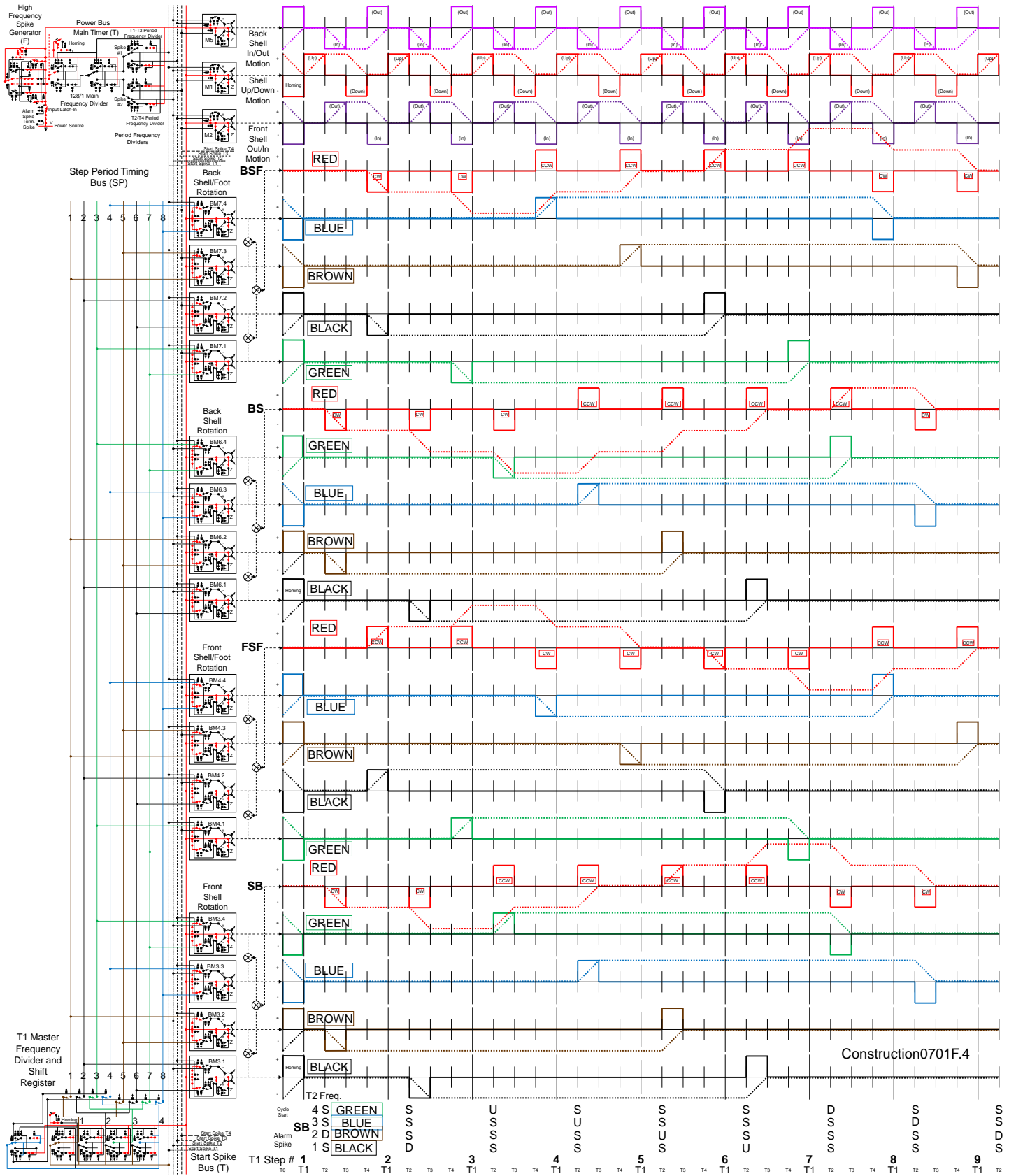

**Figure 29** A T1 *Master Frequency Divider and Shift Register* provides a lower frequency Step Timing Bus (SP) than the Start Spike Bus (T). It is driven by the T1 Start Spike, and defines the individual Step Periods 1 through 8.

Four motors are required for each of the Front Shell Rotation, Front Shell Foot Rotation, Back Shell Rotation, and Back Shell Foot Rotation functions instead of the original one motor for each function shown in Figure 23. The sum of these motions produces the fractal (S) that produces the needed motions for these functions at a given time so the animal can move in the figure-eight. These fractals (SB), (FSF), (BS), and (BSF) are similar to sine waves that repeat after eight

Spike timing

steps. This could be the source of the Alpha waves in the human brain.

Eight different equally spaced spike times are signaled by the output lines 1-8 of the Frequency Divider and Shift Register Circuit shown in Figure 29. Each output line representing a specific step. This creates four 90 degree steps in one direction of rotation, and another four steps in the other direction of rotation, forming the figure-eight shown in Figure 26.

Notice that the direction of rotation of the Back Shell lags one step behind the direction of rotation of the Front Shell. So both shells turn in the same direction except after the change in direction of rotation of the front shell in the middle of the figure-eight.

Moving in a figure-eight fulfills one of the basic requirements of the spike-timing process, which is to undo everything that is done. Every set has a reset. This keeps the actuators from extending beyond their limits, a fundamental requirement in robotics.

3.4.5  *An inverted figure-eight*

The figure-eight in Figure 26 can be inverted, as shown in Figure 30.

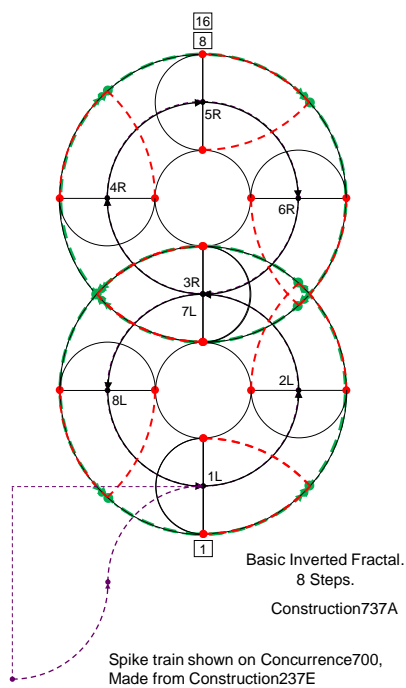

**Figure 30  The *inverted fractal* can have the same starting point as the basic fractal, but each step rotates in the opposite direction.**

The inverted fractal can be connected to the basic fractal at their common starting point. And one fractal can continue into the other fractal by a single change in direction at their connection point.

### 3.4.6  *Control system for the inverted fractal*

The control system needed for the inverted fractal is just an inverted version of the control system for the basic fractal, as shown in Figure 31.

**Figure 31  The *inverted control system* creates rotation motions in the opposite direct as the Basic (bottom) control system.**

A map of the motion of the animal is encoded in the pulse train (BSF) created by the control system.

3.4.7  *Fractal forming a figure-eight and an inverted figure-eight*

The grazing area of the organism can be doubled by connecting an inverted figure-eight to the end of the first figure-eight, as shown in Figure 32.

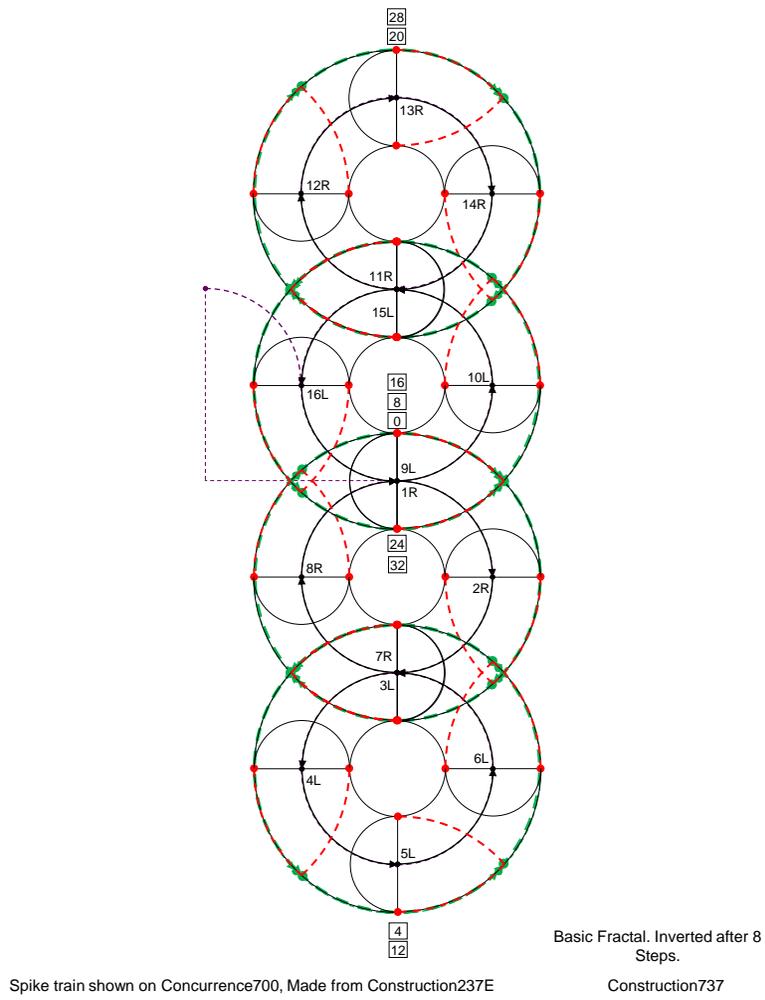

**Figure 32  Putting the *basic fractal and inverted fractal* together nearly doubles the size of the grazing area.**

The control system for the inverted fractal is symmetrical with respect to the original control system, typical of the symmetry found in animal structures.

3.4.8  *Control system that is bi-symmetrical*

The control system for the basic fractal in Figure 29 and inverted fractal in Figure 31 can be joined together, as shown in Figure 33.

Spike timing

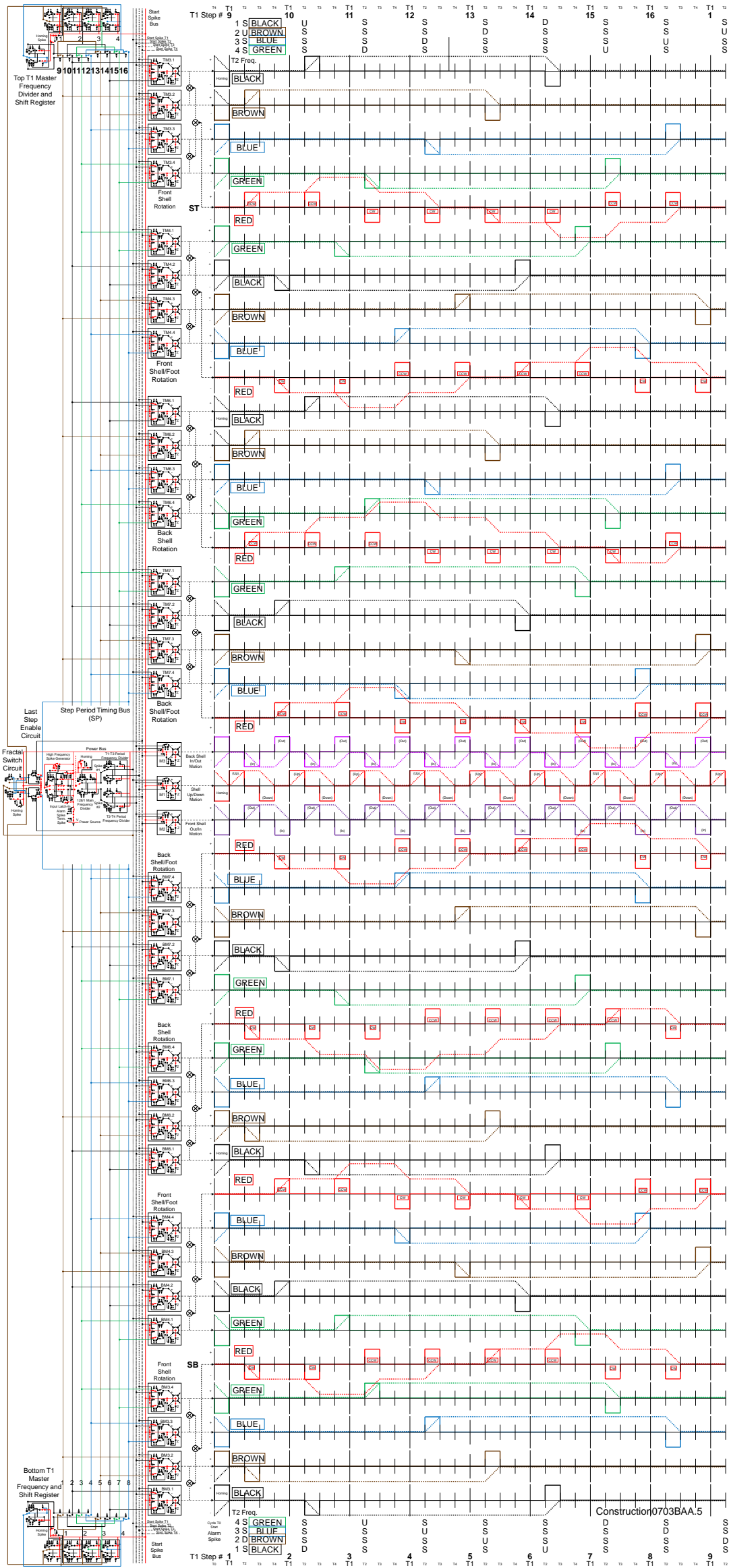

**Figure 33** The bi-symmetric control system has a *Fractal Switch Circuit* between them that switches control from one control unit to the other unit to form the double figure-eight.

Spike timing

The T1 Start Spike from the Timer (T) is first sent to the Frequency Divider and Shift Register for the bottom control system when the system is homed at start up. This causes bottom fractal to be active. When it reaches Step 8, the Fractal Switch connects the T1 Start Spike to the Frequency Divider and Shift Register in the top control system by a spike from T4. This stops the activity of the bottom control system, and causes the top control system to become active. When the top control system reaches Step Sixteen, the Fractal Switch re-connects the T1 Start Spike Train to the Frequency Divider and Shift Register in the bottom control system, as shown in Figure 33.

3.4.9    *Close up view of the Fractal Switch circuit*

A closer look at the Fractal Switch is shown in Figure 34.

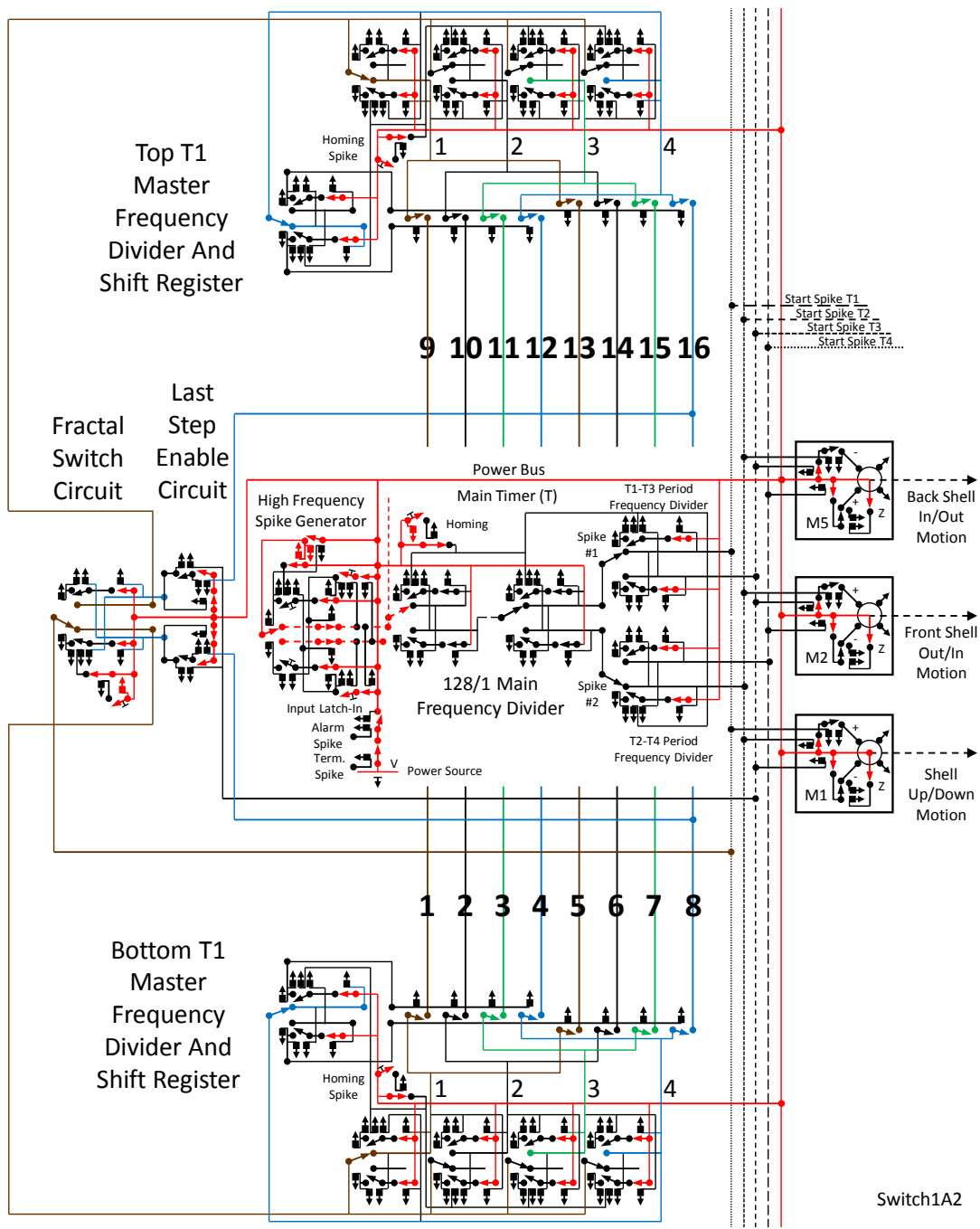

**Figure 34 A *Fractal Switch* sends the T1 Start Spike to the bottom control system, and then switches it to the top control system after eight steps. Then the Fractal Switch sends the T1 Start Spike Train back to the bottom control system after another eight steps.**

The Last Step Enable Circuit senses Step 8 or Step 16, and triggers the Fractal Switch Circuit after the step pulse is over at T3. So the basic and inverted fractals operate alternately. This very simple Fractal Switch toggle circuit increases the grazing area, and keeps the animal in a given region. It provides the most basic level of behavior of a proto-animal. It gives the animal a chance to survive by roaming freely in an environment in search of food. There are some other simple changes that can enhance the animal's grazing pattern.

3.4.10 *Thirty-two-step fractal*

A slight change in the control system allows the animal to go twice around the figure eight before it switches to the inverted fractal shown in Figure 35.

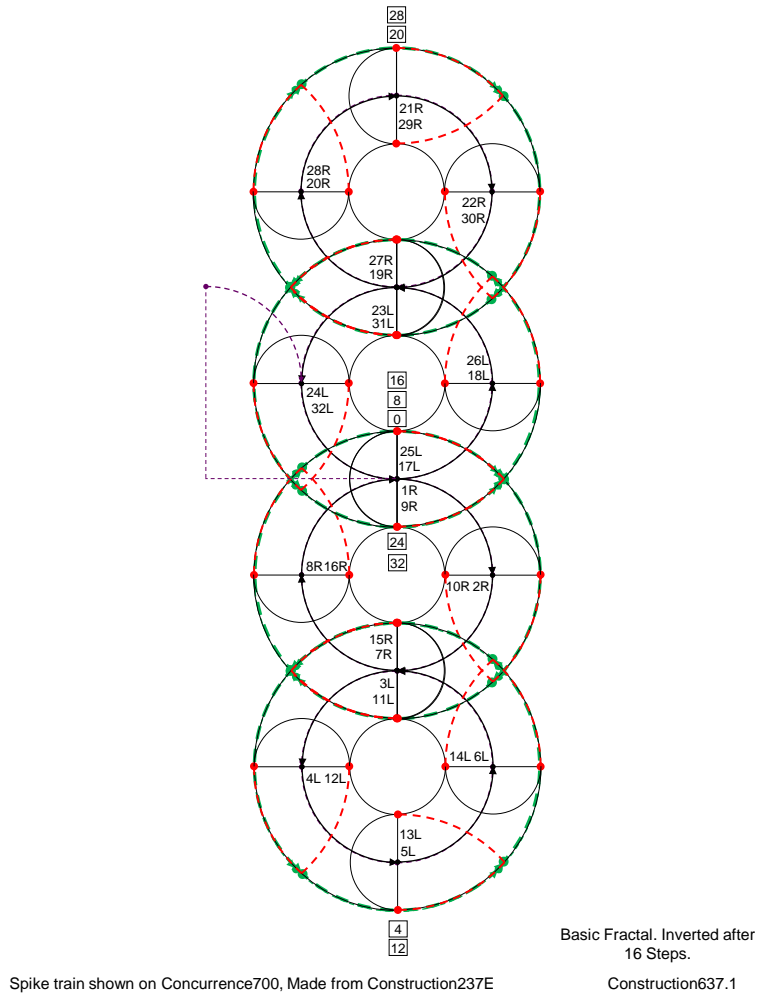

Spike train shown on Concurrence700, Made from Construction237E     Basic Fractal. Inverted after 16 Steps.
Construction637.1

**Figure 35  The *32 step fractal goes around twice* in one direction before it changes to go around twice in the other direction.** The additional second run allows the animal to pick up residual food without encountering new obstacles. Also, this fractal is necessary for the animal to be able to restore its orientation after it changes its orientation after making contact with an object. This is discussed in the next Section 3.5 *Objection sensing and avoidance*.

3.4.11 *Control system for the thirty-two step fractal*

The thirty-two-step path shown in Figure 35 can be created by the two-level Master Frequency Divider and eight-unit Shift Register Circuit shown in Figure 36.

Spike timing

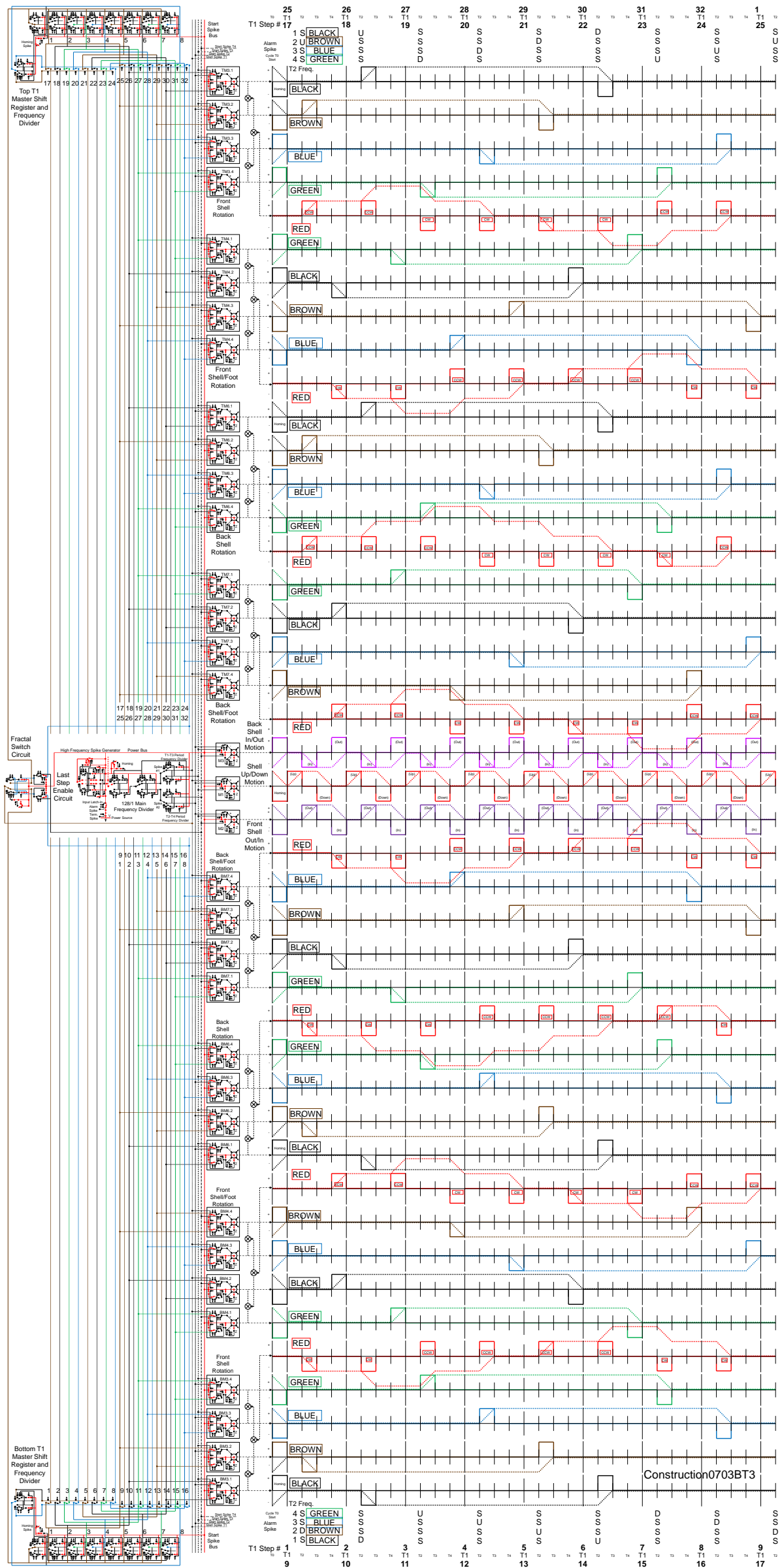

**Figure 36** The *basic and inverted sixteen step control system* produces thirty-two-step fractal.

Spike timing

This is accomplished by using a two-level Frequency Divider and eight-step shift register that switches the direction of the fractal using the Fractal Switch shown enlarged in Figure 37 after sixteen steps rather than at eight steps.

3.4.12 *A close up view of the sixteen step Fractal Switch circuit*

A closer look at the sixteen step Fractal Switch circuit is shown in Figure 37.

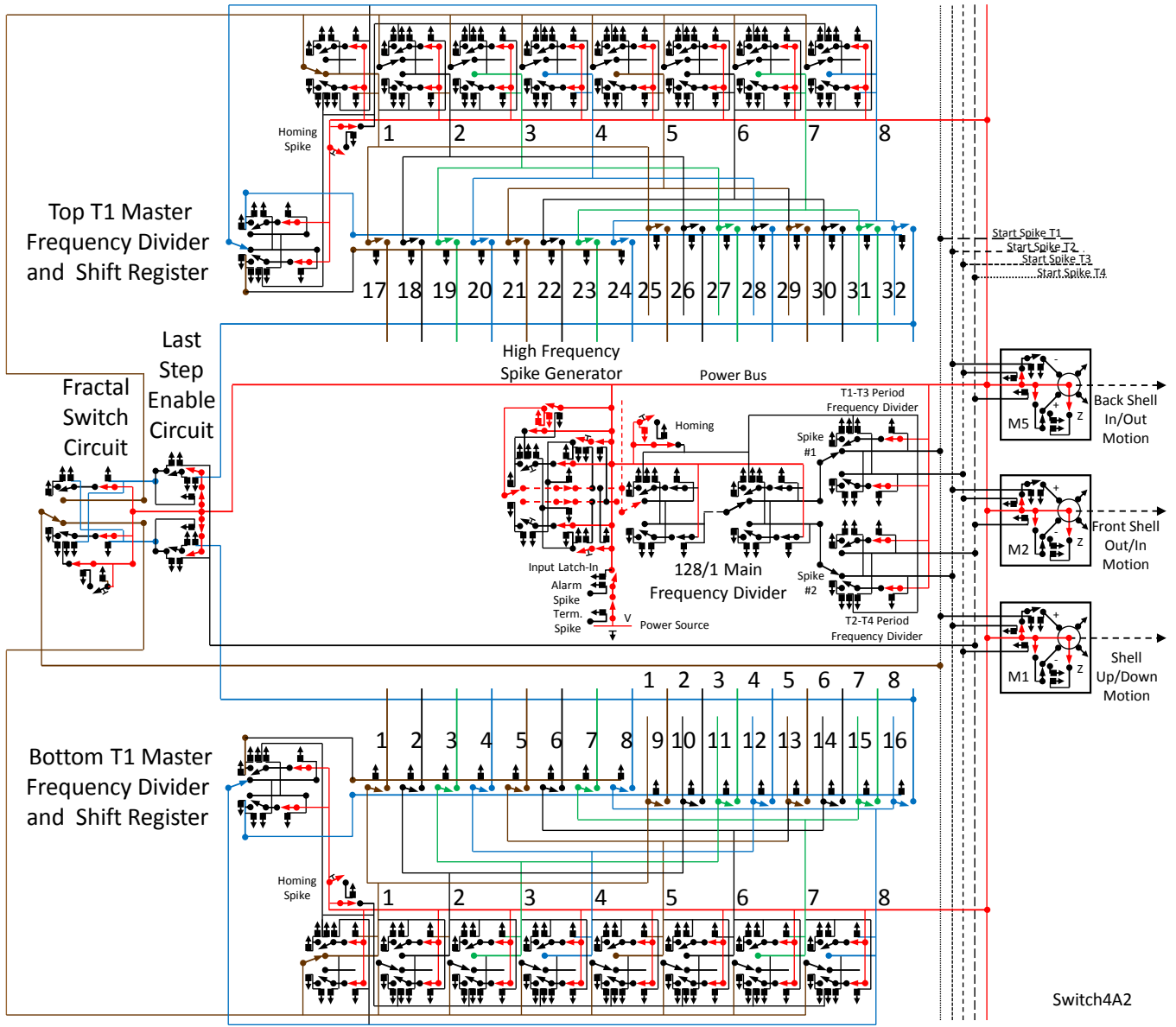

**Figure 37** The *frequency divider and 8 step shift register* **switches the fractal after sixteen step times.**

The Fractal Switch switches the T1 Spike from the bottom frequency divider to the top frequency divider, and switches from the top frequency divider to the bottom frequency divider. So it takes thirty-two steps to repeat the step cycle. These additions contribute to its ability to graze. And they set up a control system that allows it to react to objects in its environment that are contacted after Step 4, as shown in the next section.

3.4.13 *Section summary*

An organism driven by multiple fractals evokes a figure eight behavior plan that moves its presence into a larger area of its environment.

## 3.5 Object sensing and avoidance

Now that the organism has been set free to roam and graze in the world, the greatest challenge it faces is to avoid getting trapped against an object, or getting separated from its breeding community. This section shows how the problem of getting stuck can be reduced by a system that senses an object in its environment and produces an avoidance reflex.

### 3.5.1 Avoidance path

One way for an animal to reduce the chances of getting stuck against an object is to reverse its direction of rotation when it contacts an object, and continue to move in a new arc away from the object, as shown on Figure 38.

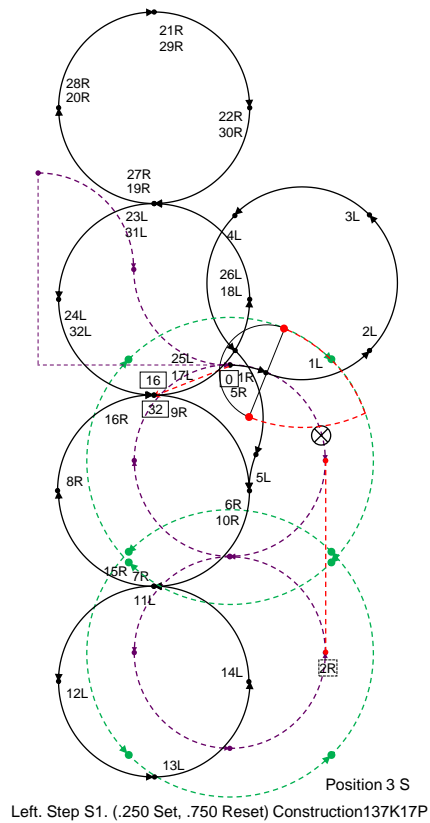

**Figure 38  An *avoidance path* takes the animal away from an object in its path, and keeps it away from the object.**

The organism also needs a second change in direction to restore the orientation of the original figure-eight. This can be created by producing a change in direction in Step 5 that is the complement of the contact angle in Step1. This complementary change in direction occurs in the fourth step after the step in which contact is made with the object. The sixteen-step fractal allows for the complement restoring pulse to occur even if the contact occurs in the eighth step. This allows the fractal to be restored to its original orientation (zero-sum of rotations) regardless of when it encounters an object. Even though the end of the fractal is displaced from the original start position of the fractal, the orientation of the fractal is maintained.

The line connecting the origin [0] of the original fractal and the start [16] of the new fractal is a Displacement Vector ($v_1$) that defines the change in position of the fractal after contact is made with an object at a particular time in a particular step.

### 3.5.2 Mechanism needed to turn away from an object when it is contacted

Two additional extension and rotation axes need to be added to allow the organism to turn away from an object when it is encountered, as shown in Figure 39.

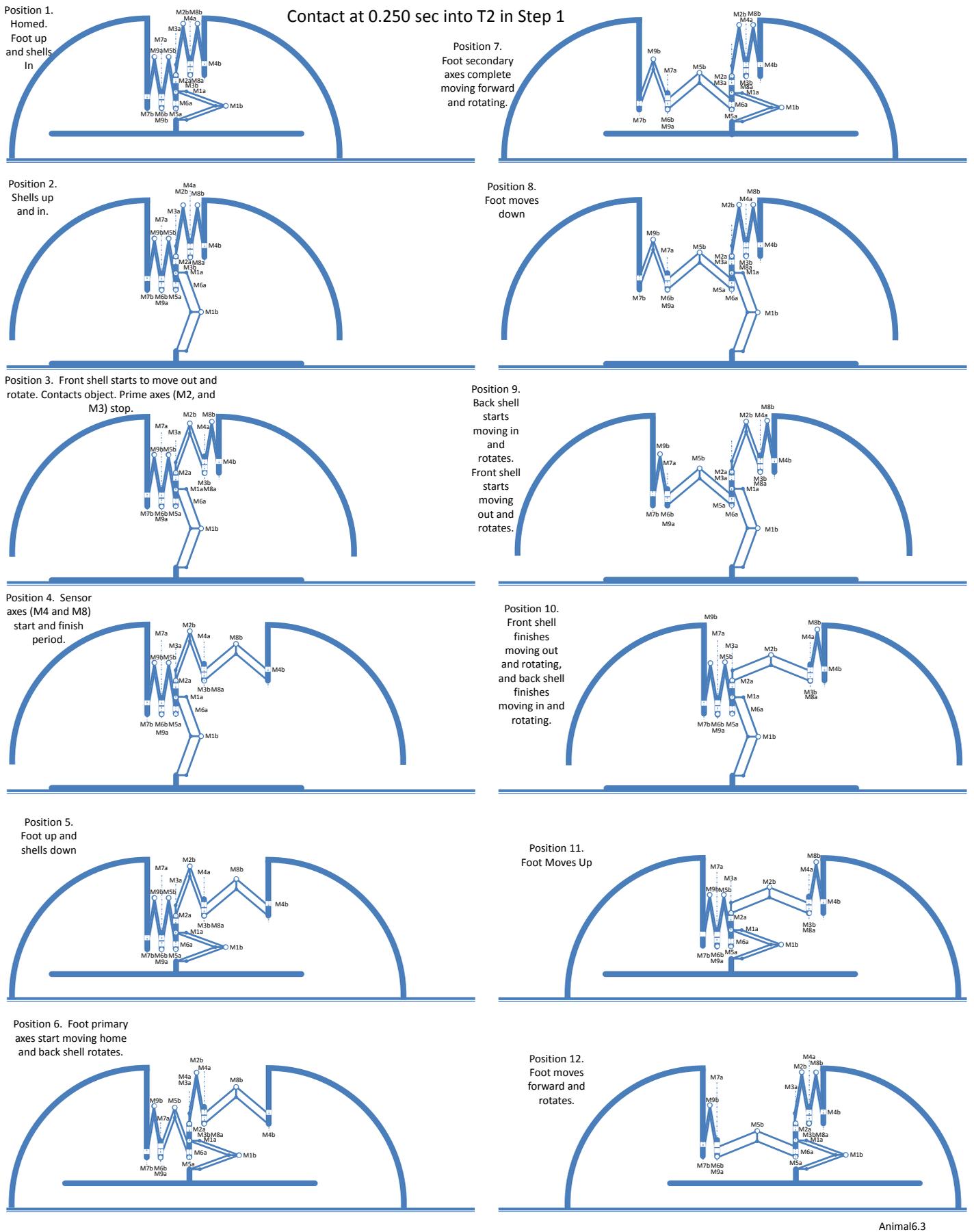

**Figure 39** *Additional out and rotation axes* are activated when the organism contacts an object.

When contact is made with an object, the primary Out Motor (M2a) and primary Rotary Motor (M1a) axes are stopped, and the secondary Out Motor (M2b) and secondary Rotary Motor (M1b) axes are activated, as shown in Figure 47. If contact is not made with an object, just the primary shell out and shell rotation axes operate in each Step Period, as shown in Position 7 through Position 10 above.

### 3.5.3 Two nodes in the avoidance motion

The path followed by the organism is described by circular curve fitting with a constant radius and a variable placement of

Spike timing

the nodes (places of reversals or changes in radius). Two nodes are required to carry out the avoidance. The first node is the position of the foot at the start of the step. The second node is at the position of the foot at the moment of contact with an object. These nodes are shown in Figure 40 with respect to the time axes shown in Figure 47 for contact made 0.250 sec into Step #1.

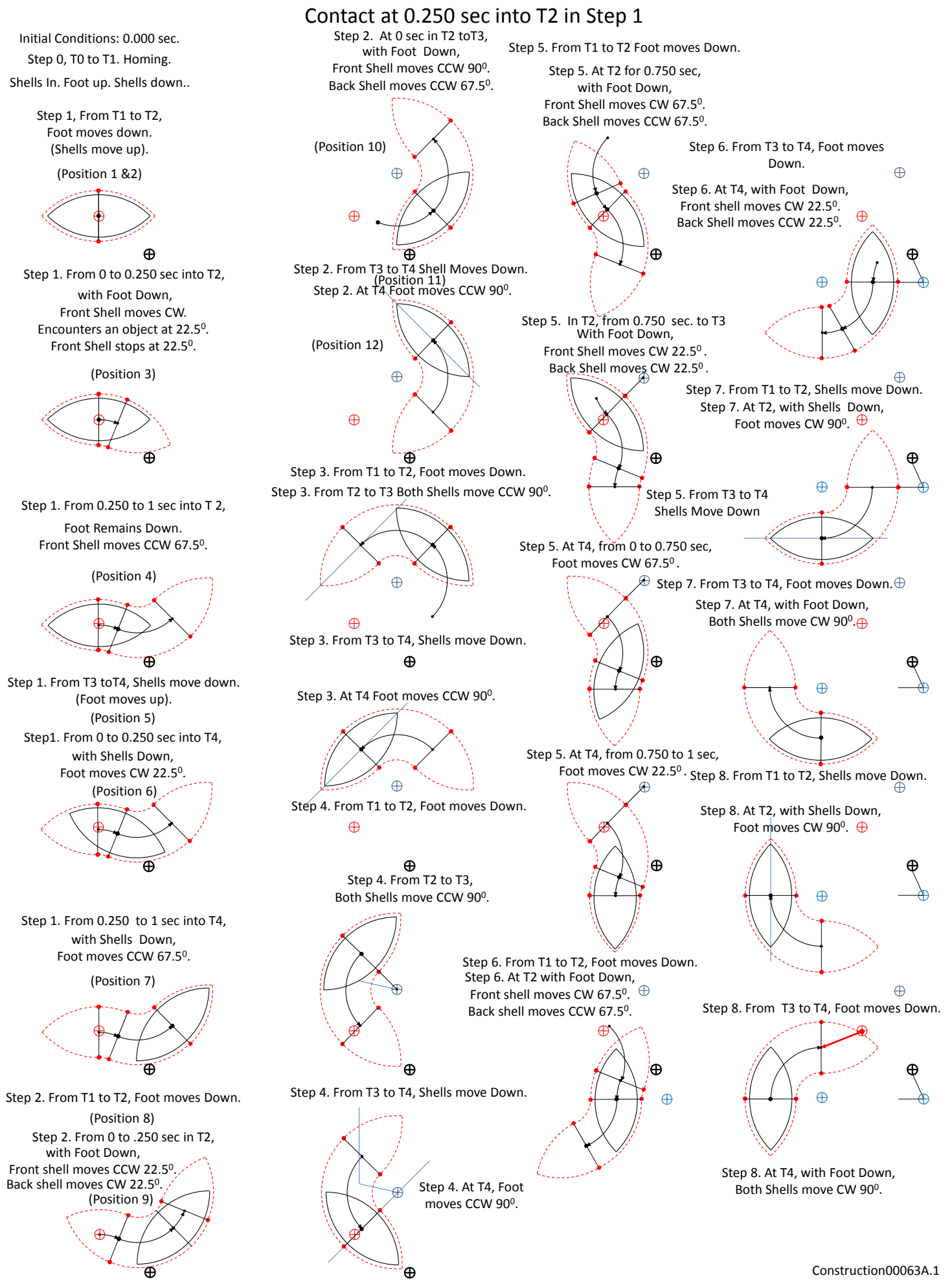

**Figure 40** The *direction of rotation of the front shell is reversed when contact is made with an object* **0.250 sec into T2 in Step 1. The out motion continues until the end of its step period at T3.**

Spike timing

At the beginning of a step period, the foot is down, and the shell out motion and rotation motion begin. When contact is made with an object, the direction of rotation reverses and the out motion continues to the end of the step period. This causes the organism to move away from the object. Four steps later, the complement of the avoidance pulses is created by the control system that returns the fractal to its original orientation. Then the control system causes it to continue indefinitely in a double figure-eight in a new position away from the offending object in Step 8 defined by the Displacement Vector ($v_1$), as shown in Figure 38.

### 3.5.4 *Object contact*

An object is contacted by a switch armature (whisker) (RK) on the right leading edge of the front shell, as shown in Figure 41.

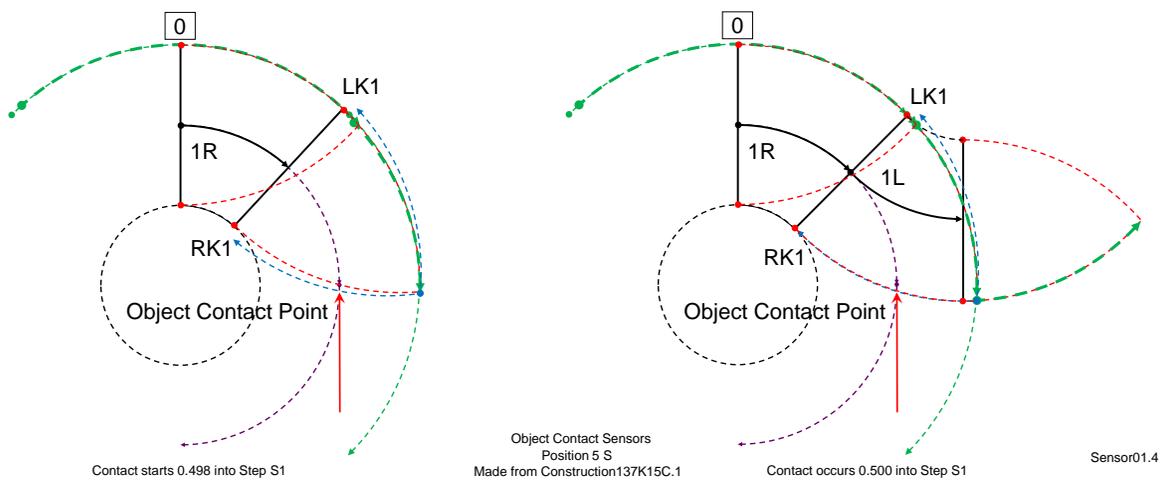

**Figure 41 The front shell has a *left and right object sensor switch* (LK1 and RK1) that is closed when the edged of the front shell contacts an object.**

The animal model has whiskers (LK1 and RK1) directed backwards from the front tip of its front shell. According to the direction of rotation of the front shell, a left (LK1) whisker or right (RK1) whisker is moved against the front shell when contact is made with an object. This closes a switch contact that is used to generate at spike, as shown in Figure 42.

### 3.5.5 *Object Sensor*

The object sensor is a single spike generator (encoder) that is intended to operate while the front shell is moving out and rotating clock-wise or counter-clockwise, only. If it were to operate at any other time, this would not indicate a contact with an object caused by the movement of the front shell. For example, the sensor may be struck by a moving object while the shell is at rest. This requires a different response than a contact made as a result of the movement of the shell. So a series of disabling latches and enabling disconnects are required, as shown in Figure 42.

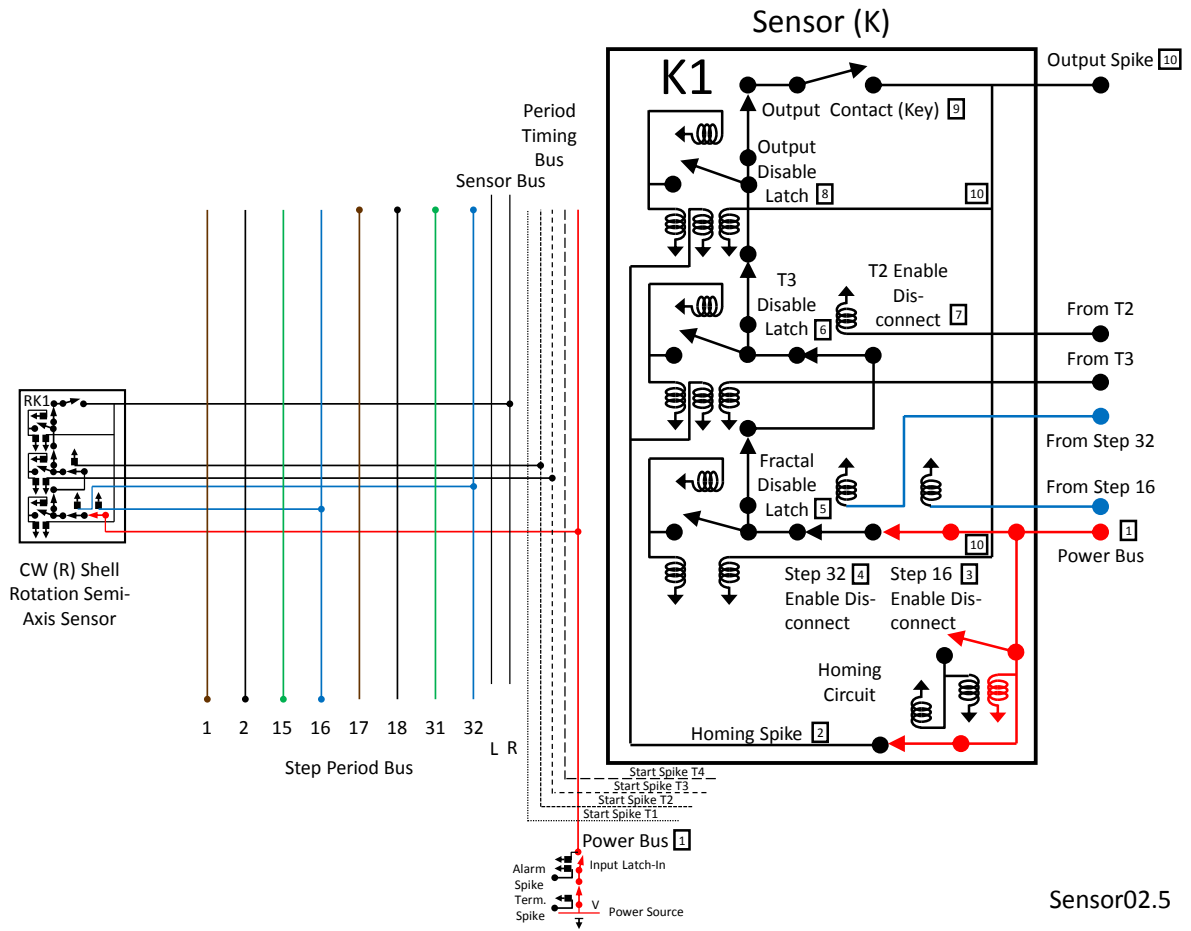

**Figure 42  The *object sensor (K)* is a one-shot spike generator that operates when it contacts an object only when the front shell is moving forward and rotating.**

Power [1] flows to the Homing Circuit when the Power Bus [1] is energized by an Alarm Spike. This produces a Homing Spike [2] that engages the T3 Disable Latch [6] and the Output Disable Latch. This prohibits power to the Output Contacts [9] until T2 Period of the first step (S1), when the T2 Enable Disconnect [7] releases the T3 Disable Latch. Then the T3 Disable Latch is reengaged at the start of the T3 Period, cutting power to the Output Contact. This allows power to the Output Contacts [9] to occur only between the start of the T2 Period and the start of the T3 Period when the Front Shell can be moving.

If an object is contacted during this time period, the Output Contact Key [9] is closed, creating a voltage on the Output Terminal [10]. This engages the Output Disable Latch [8] and the Fractal Disable Latch [5], which shuts off power to the Output Contact [9]. This completes (terminates) the Output Spike [10], and prohibits any more Output Spikes until power to the Output Disable Latch [8] is turned off and then on again. This would release Output Disable Latch [8], allowing power to flow to the Output Contact [9] again. Power to the Output Disable Latch [8] is restored by a spike from the end of the first fractal (Step 16) or from the end of the second fractal (Step 32), and a spike to the T2 Enable Disconnect [7]. (The sensor is labeled "K" because it corresponds to the operator key in telegraphy.)

### 3.5.6 *Actuator offset time controlled by a sensor*

A separate rotation actuator is needed for the first part of the rotation trajectory, as shown in Figure 40. It is represented by the motor (M3) in Figure 39. It is started by a spike on the Step 1 conductor in the Step Bus and by s spike on the T2 conductor in the Period Timing Bus, as shown in figure 43.

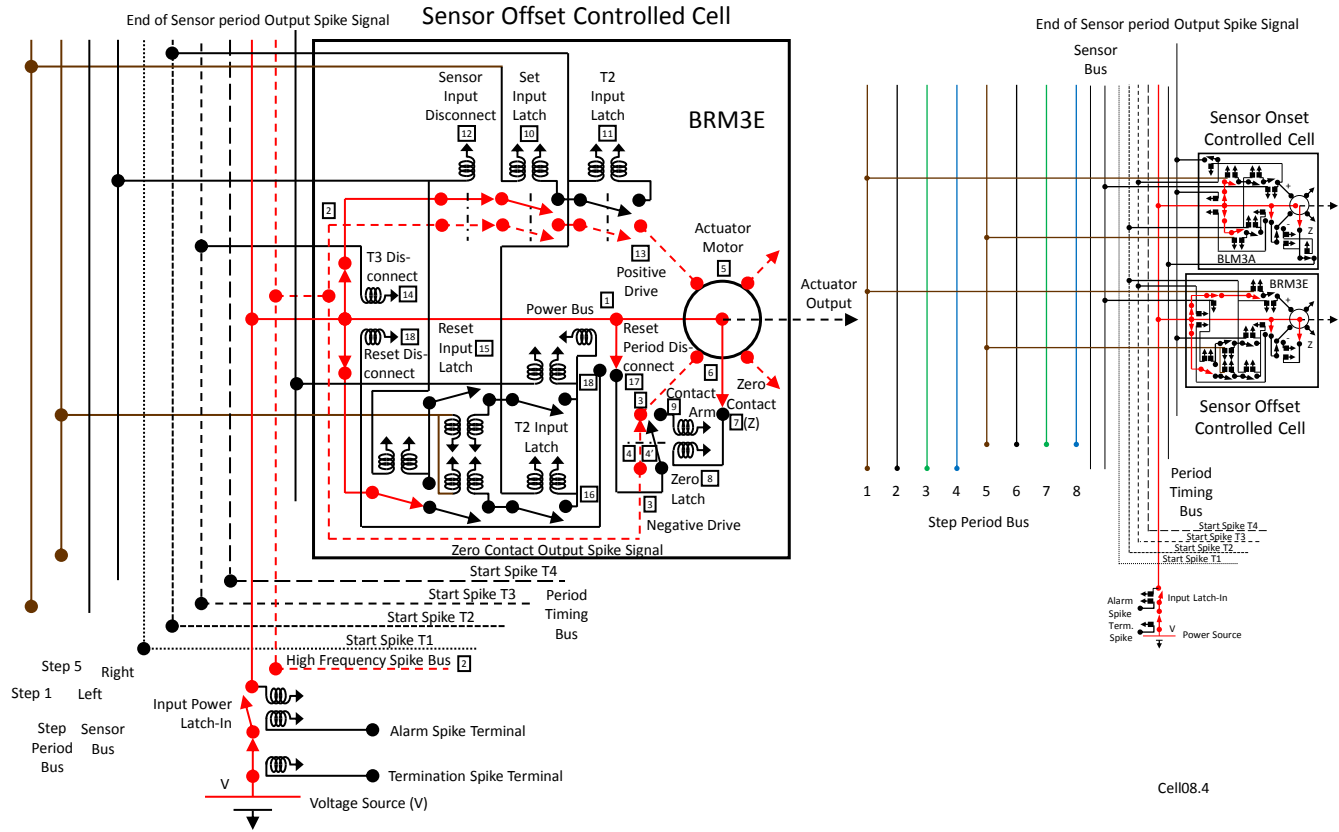

**Figure 43 A *Sensor Offset-Controlled Actuator* starts at the beginning of a specific step period, and is stopped by a sensor spike. If there is no sensor spike, it produces a full pulse that ends at T3.**

The Sensor Offset-Controlled Actuator (BRM3E) is the same as the basic Fractal Actuator (M), except that the output pulse is stopped by the sensor contact spike. Also, if there is a sensor contact spike, its reset is started by the Step 5 spike and the Zero spike at the end of the pulse created by the Sensor Onset-Controlled Actuator (BLM3A) in Figure 44. If there is no sensor controlled spike, the Sensor Offset-Controlled Actuator (BRM3E) produces a full pulse that is in the same direction of rotation (right) as called for by the Basic Fractal in Step 1.

### 3.5.7 *Actuator onset time controlled by a sensor*

For the animal model to sense and avoid an object, it must have a second rotary actuator (decoder) that is controlled by the sensor. It is started by the right sensor spike from Sensor (RK). It is represented by the motor (M4) in Figure 39. Its output causes the front shell to rotate in the opposite direction of the Basic Fractal, causing it to turn left and move away from the object. The Sensor Controlled Actuator (BLM3A) needs a Sensor Input Latch [12] that is enabled by the Set Input Latch [10] from the Step Bus and the T2 Input Latch [11], and engaged by a Sensor Output Spike in Figure 42, as shown in Figure 44.

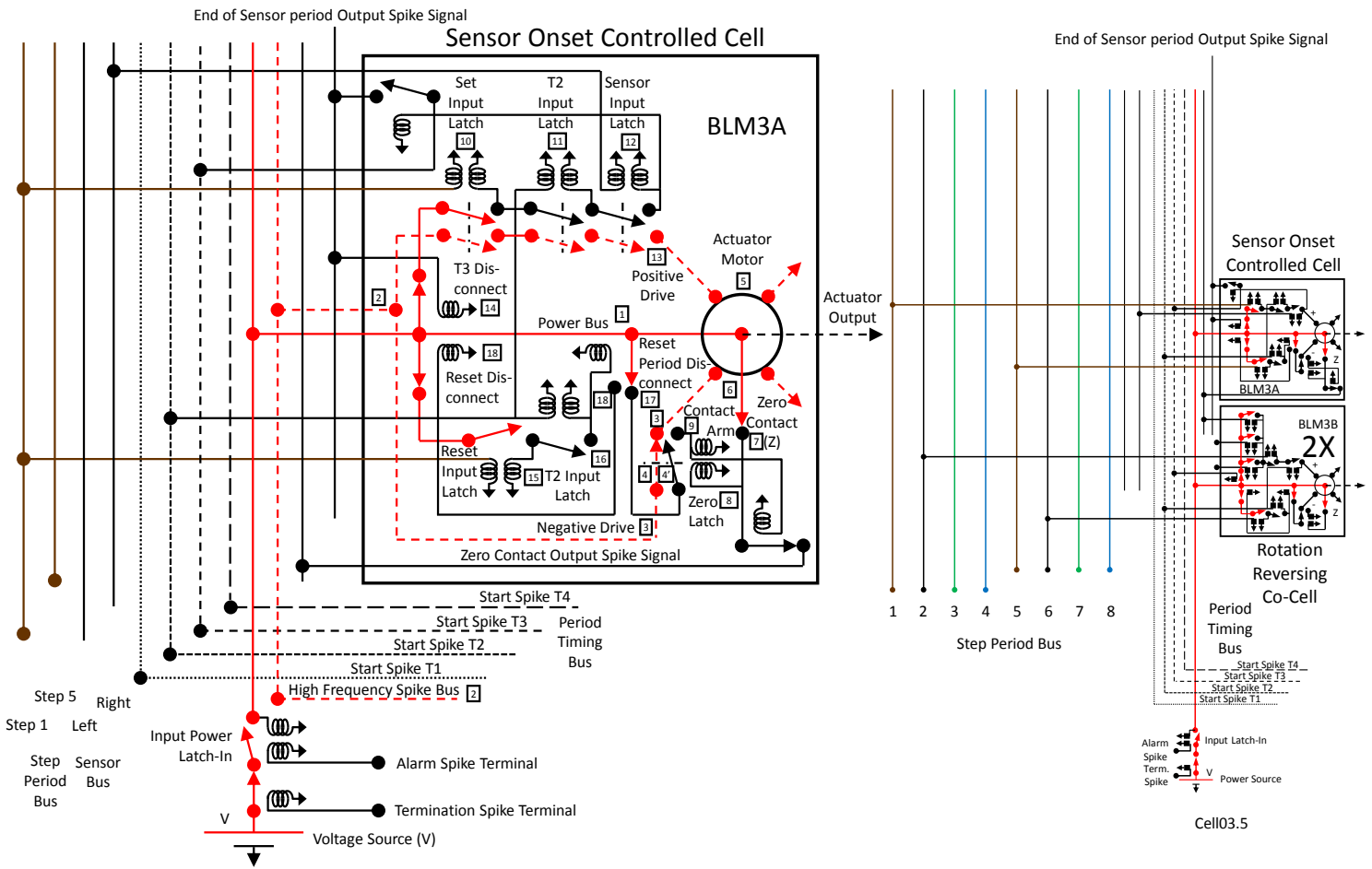

**Figure 44 The *Sensor Onset-Controlled Cell* starts to produce an actuator output pulse when an object is contacted in its Step Period.**

The Sensor Onset-Controlled Cell (BLM3A) is enabled by a spike on the Step Period Bus (Step 1) and a spike on the Period Timing Bus (T2). But it does not produce an output unless there is a spike on its Sensor Bus in that Step Period. If it is started, its output is stopped by a spike on T3 in this example. This leaves the Contact Arm [6] at some position from zero less than the full 180 degree rotation of the Fractal Actuator Cell shown in Figure 28. So the Contact Arm [6] travels less than 180 degrees when it is reset back to zero, as shown in Figure 47. If there is no Sensor Contract spike, there is no output pulse created by the Sensor Onset-Controlled Actuator. And there is no reset pulse required.

### 3.5.8  Rotation reversing pulse generator

Also, an additional 2X pulse is required to reverse the polarity of the basic fractal pulse. Its purpose is to maintain the zero-sum property (no greater rotations in one direction than the other after a complete fractal cycle) of the new fractal created by the contact with an object. It is produced by the (BLM3B) actuator in the next step after contact is made in Step 1, which is Step 2, as shown in Figure 45.

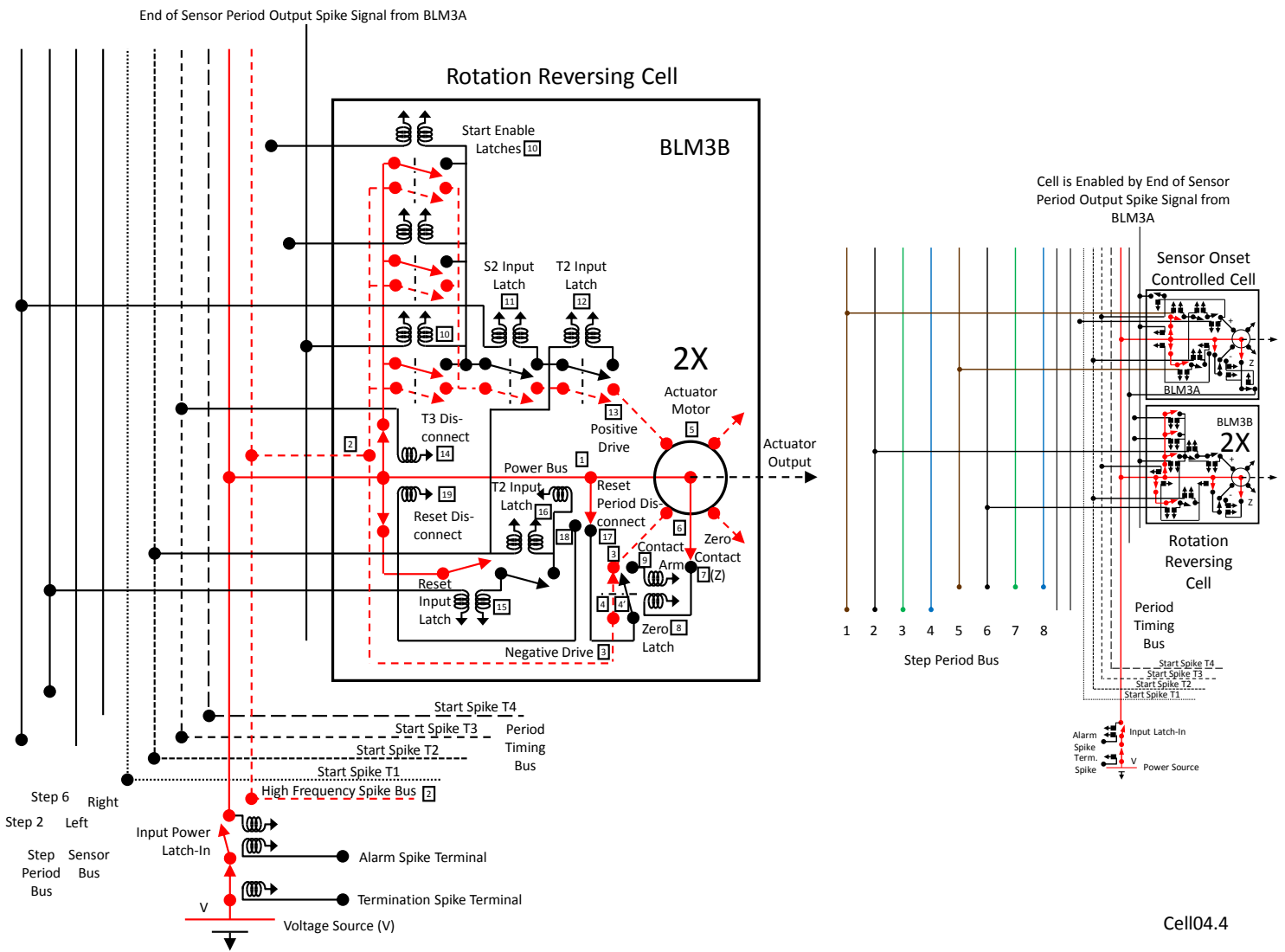

**Figure 45** The *Rotation Reversing Cell* produces a pulse that ends at T3, and is twice the amplitude of the Fractal Actuator. The Sensor Controlled Co-Cell (BLM3A) sends a spike signal to the Rotation Reversing Cell at the end of the T2 time period if it has responded to the contact sensor. This sets one of the Start Enable Latches [10] in the Rotation Reversing Cell. In the next Step Period (2), its S2 Input Latch and T2 Input Latch are engaged, causing its Actuator Motor [5] to start rotating until disconnected by the T3 Start Spike. This produces the two-times-normal-amplitude pulse output that lasts for one Step Period.

This cell may be enabled by other Sensor Controlled Cells, which will require is to become active only after two or even three Steps. So the T3 Disconnect [14] is not enabled in a Step Period unless the cell has been made active.

If contact were made in Step 2, no extra rotation reversing pulses are required for contacts in Step 2. Three extra pulses are required for contact in Step 3. Two extra pulses are required in Step 4. One extra pulse is required in Step 5. No extra pulses are required in Step 6. Three extra pulses are required in Step 7. And two extra pulses are required in Step 8. These extra pulses are required to maintain the original orientation of the fractal after contact with an object. They are generated by actuators connected to the cell activated by the sensor, as shown in Figure 47.

Note that the fractal pulse train (S1.250) in Figure 47 is returned to the original pulse train (S) by the restore pulses in Step 5 after being displaced by the sensor pulse in Step 1. The top matrix does not receive a spike from the

Spike timing

sensors. So it executes the double figures-eight when it is activated by the Fractal Switch, as shown by the trajectory in Figure 38.

The Rotation Reversing Cell is shown to have twice the amplitude of the Fractal Actuator so it can reverse the sign of the pulse. However, one of the basic features of Pulse Width Modulation is the use of constant amplitude Unit Pulses throughout the system. So, in practice, the double amplitude can be created by just adding the output of two Rotation Reversing Cells, each having the Unit Amplitude of the system.

3.5.9   *Out/In actuator cells*

Also, two Out/In axis motors are needed because a primary Out/In axis (M1X) must stop when the object is sensed, and a secondary Out/In axis (M2X) must start and complete the out motion in that period while the rotation is moving away from the object.

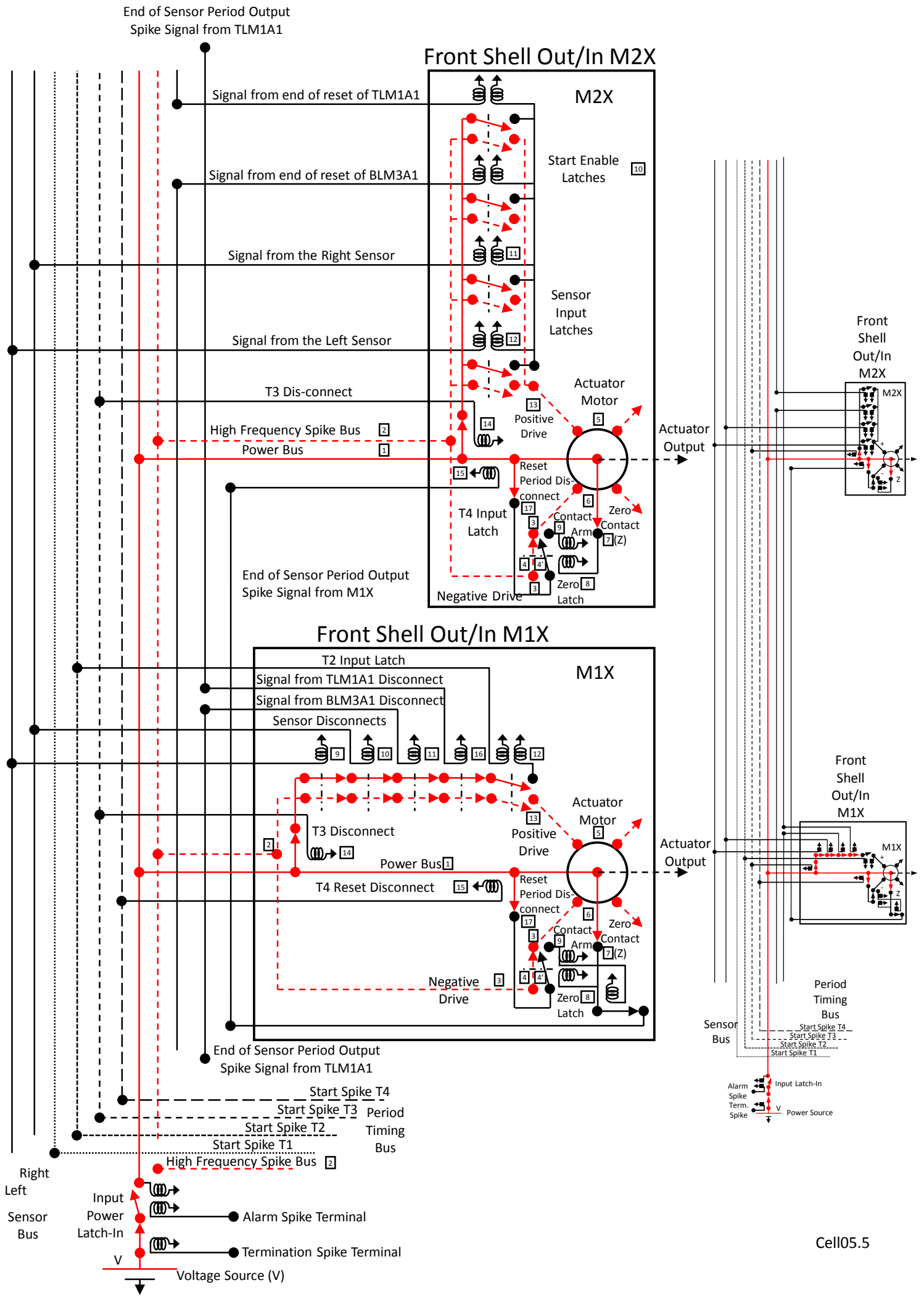

**Figure 46** The *Out/In Actuator Cells M1X and M2X* are identical, except that Cell M1X has multiple inhibitory input disconnects, and Cell M2X has multiple excitatory input latches.

The Out/In Actuator Cells M1X and M2X correspond to motors M2 and M8 in Figure 39. If there is no contact with an object, Actuator M1X carries out all of the Out/In motion. If there is contact with an object, Actuator M2X carries out the avoidance Out/In motion. The trajectories of this two-stage, Out/In motion, are shown in Figure 47.

### 3.5.10 Control system with an object sensor

Two sensors (LK1 and RK1), the Sensor Controlled Cells (BRM3E and BLM3A), and the Out/In Actuator Cells (M1X and M 2X) can be added to the system shown in Figure 36, forming the Sensor Controlled Matrix in Figure 47. One sensor (RK1) senses contact with an object while the organism is rotating and moving forward in the right-hand (CW) direction, and the other sensor (LK1) senses contact with an object while moving in the left-hand (CCW) direction. Each sensor is connected to a new right (R) spike line or a new left (L) spike line in the Sensor Bus that extends to the top and bottom matrices, as shown in Figure 47.

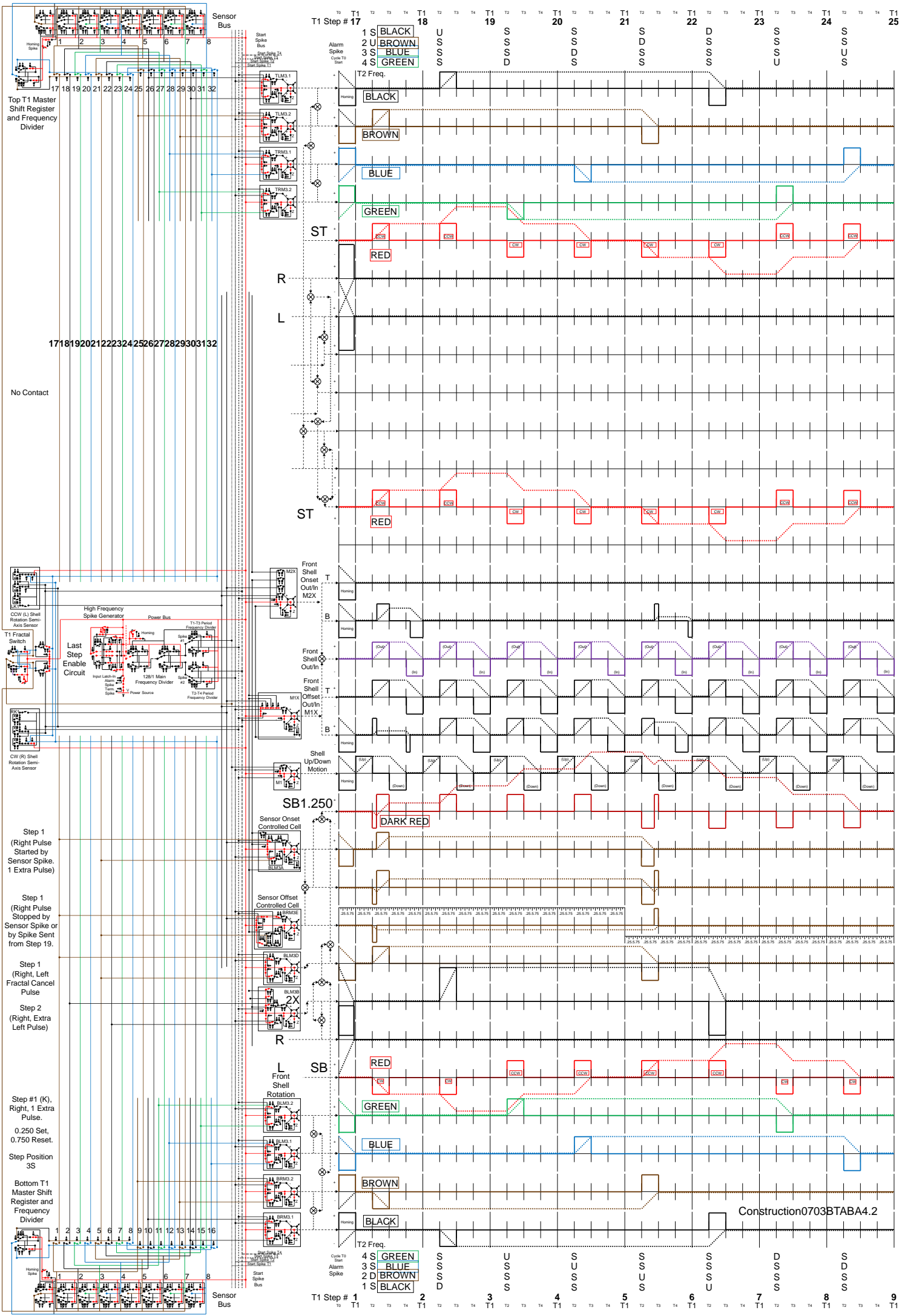

**Figure 47** The *object sensor* (LK1) produces a spike at 0.250 sec into Step 1 that causes the Sensor Controlled Cells (BRM3E and BLM3A) to produce a compound pulse train that causes the organism to turn away from the object and establish a figure-eight trajectory in a new location away from the object.

Only the Front Shell rotation axis is shown because of the page-space limitations. The operation of the Foot and Back

Spike timing

Shell were last shown in Figure 36. They both make the same motions. But the Back Shell motions occur midway between the Front Shell motions, and are in the opposite direction of rotation. So, from this point on, only the Front Shell will be presented.

At the start of Step 1, the Pulse Generating Actuator (BLM3D) starts producing a CCW pulse that is in the opposite direction of rotation of the CW pulse being generated by the Basic Fractal Actuator (BM3.2). This cancels the pulse from the Basic Fractal in this embodiment. At the same time, the Sensor Controlled Actuator (BRM3E) shown in Figure 43, corresponding to motor M3 in Figure 39, starts producing a CW pulse that is stopped (interrupted) by a contact spike at 0.250 sec into Step 1 from the Sensor (RK1). This contact spike starts the left turning Sensor Controlled Actuator (BLM3A) shown in Figure 44, corresponding to motor M4 in Figure 39. This creates a 0.750 sec CCW pulse in Step 1 that ends at T3. The 0.250 sec CW pulse from (BRM3E) is added to the 0.750 CCW pulse from (BLM3A). This causes the reversal in direction of rotation shown in SB1.250 caused by a spike from Sensor (RK) 0.250 sec into Step 1, starting the new S1.250 fractal shown in Figure 47.

The complementary pulse is started at the beginning of Step 5 by the reset motion of the actuator (BLM3A). The reset of the (BRM3E) actuator is started when the contact arm of (BLM3A) reaches zero. So the complementary pulse is created in Step 5 by the reset action of the two actuators, as shown on the S1.250 fractal in Figure 47.

Also, the Out /In motion of the Front Shell also needs to be carried out by the two Out/In Actuators (M1X and M2X) in Figure 46, corresponding to the two actuators M2 and M8 shown in the animal model in Figure 39.

The sequence of actions of the Out/In axes that occur due to contact with an object at 0.250 sec into Step 1 is shown below:

1. The Out/In Actuator (M1X) starts on every T2 spike, and is stopped (interrupted) by a Sensor spike or a T3 spike if there is no Sensor spike.
2. The Out/In Actuator (M2X) is started by a Sensor spike, and is stopped by the T3 Disconnect. If there is no Sensor spike, (M2X) remains motionless.
3. The reset of the Out/In Actuator (M1X) is started by the next T4 spike, and is stopped by the contact with its Zero Terminal.
4. The reset of the Out/In Actuator (M2X) is started by the spike made by the Zero contact of the (M1B) above, and is stopped by the contact made with its Zero Terminal.
5. The complementary restore of (M1X) is started by a T2 spike in Step 5, and is stopped by the Zero spike at the end of the reset of (BLM3A).
6. The complementry restore of (M2X) is started by the Zero spike at the end of the reset of (BLM3A), and is stopped by T3.
7. The reset of the complementry restore of (M1X) is started by T4, and is stopped by its Zero spike.
8. The reset of the complementry restore of (M2X) is started by the Zero spike from (M1B) above, and is stopped by its own Zero spike.

The result is that the motion of the Front Shell and Out/In axes of the S1.250 pulse train cause the organism to turn away from the object, to move away for three steps, and then to restore itself to the original orientation in a new location in the fourth step later, as shown in Figure 38 and Figure 40.

Spike timing

### 3.5.11 *Object in the field of the top fractal*

An object may lie in the field of the top fractal instead of the bottom field. An example is shown in Figure 48.

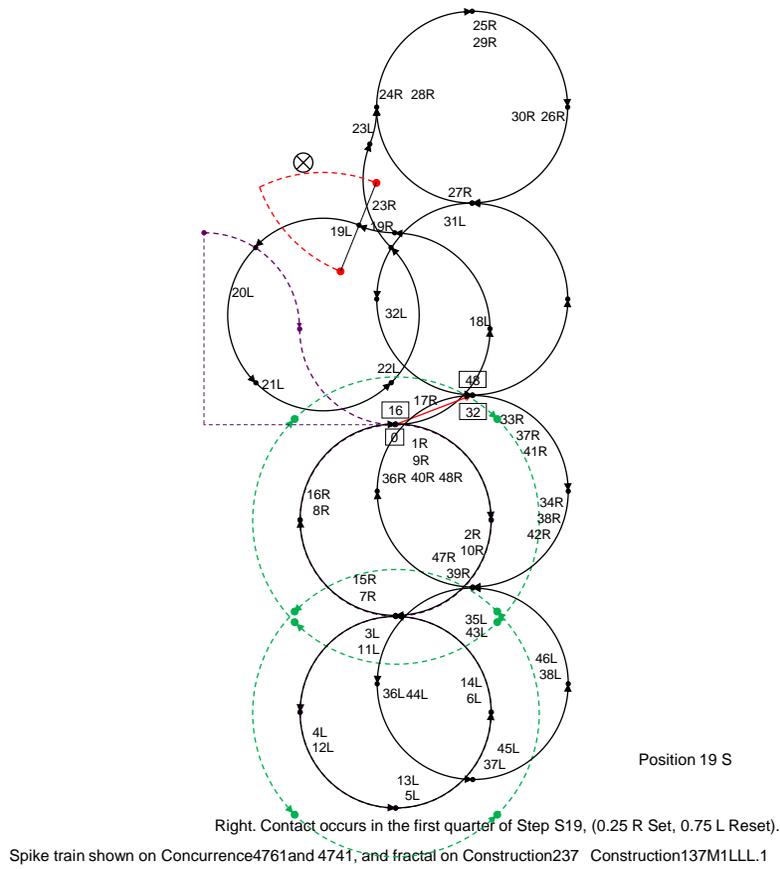

Right. Contact occurs in the first quarter of Step S19, (0.25 R Set, 0.75 L Reset).
Spike train shown on Concurrence4761and 4741, and fractal on Construction237   Construction137M1LLL.1

**Figure 48  The *object is contacted at 0.250 sec into Step 19R*, causing the animal to turn left for the remainder of Step 19.**
The animal then produces the complement of the motion in the fourth step after the object is contacted. This complement motion is made in Step 23. This complement motion restores the orientation of the fractal to the original orientation, but displaces the fractal from its original fractal position at the beginning of Step 17 at its end position at the end of Step 32. This Displacement Vector ($v_2$) from [16] to [32] is determined by the Step Period and the time of contact in the Step Period.

### 3.5.12 *Control system needed to deal with contact in the inverted fractal*

The control system needed to deal with an object contact in the inverted fractal in Step 19 is shown in Figure 49.

Spike timing

**Figure 49** The *control system for the inverted fractal* brings the inverted fractal (ST) to a new position by the end of Step 32 in response to a contact at 0.250 sec into Step 19.

Even though the fractal extends to 16 steps in the bottom fractal, and to 32 steps in the top fractal, the illustration in Figure 49 extends to the end of Step 12 in the bottom fractal and to Step 28 in the top fractal because of the space limitations on the printed page. The remaining space does not contain new information since it is just the continuation of the basic figure-eight fractal.

The contact in Step 19 requires three extra double pulses in Steps 20, Step 21, and Step 22 to maintain the original orientation of the fractal. And it needs the complementary restoring pulses in Step 23 along with the restoring pulses in

Spike timing

Steps 24, Step 25, and Step 26 to maintain its original orientation.

Again, after the organism makes contact with an object it moves away from the object and continues indefinitely in the double figure-eight path in the new position away from the object defined by the displacement vector (v), as shown in Figure 48, while maintaining its original orientation. This grazing motion would allow the animal to avoid an object in its environment, but may cause it to drift away from its breeding community if it slips during its motion. If the object moves, the animal will act to avoid the object. This will cause the animal to move to other new locations.

3.5.13 *Movement of the organism between two objects*

The motion shown in Figure 38 and Figure 48 are combined to show the movement between an object contacted in Step 1 and an object contacted in Step 19, as shown in Figure 50.

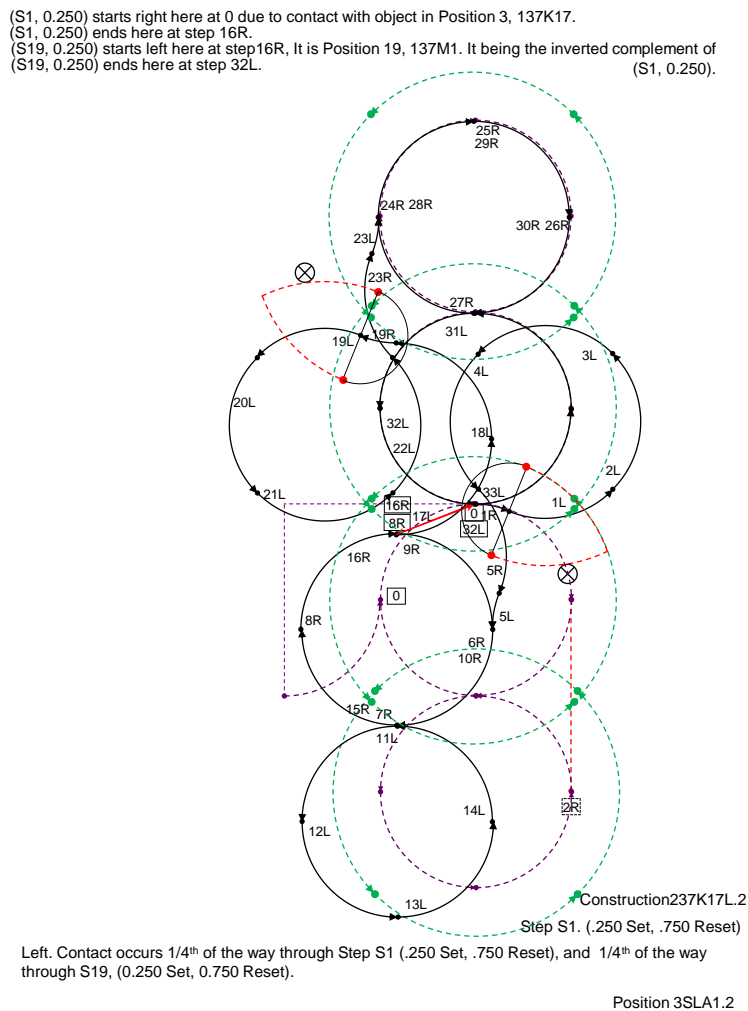

**Figure 50 The *motion caused by making contact with two objects in two particular locations* causes the organism to remain fixed in a new location, alternating back and forth between the two original objects, and coming back to the starting point.**

The magnitude and direction of the Displacement Vector ($v_1$) of the fractal in Figure 38 is equal and in the opposite direction of the Displacement Vector ($v_2$) of the fractal in Figure 48. So the object in Figure 48 causes the fractal to return to the starting point of the fractal in Figure 38, where it would encounter the first object in that location, as shown above. So the fractals repeat their motions back and forth between these two objects, and maintain the original orientation assuming there is no slippage in the footing of the organism, and that the two objects remain stationary.

These trajectories are produced using computer graphical synthesis. The arc lengths are produced using angular dimensions in software with a resolution of one degree. Since a step rotates 90 degrees, one degree represents a resolution of 1.1%. This is sufficient to draw these trajectories that move away and return to their starting point, accurately. However, less resolution would cause some trajectories to miss the return point. So the higher resolution of the 128 spike pulses should be adequate to bring the animal back to its starting point.

Spike timing

### 3.5.14 *Control system dealing with a different object in each fractal*

The contact time settings in Figure 47 and Figure 49 can be combined, as shown in Figure 51.

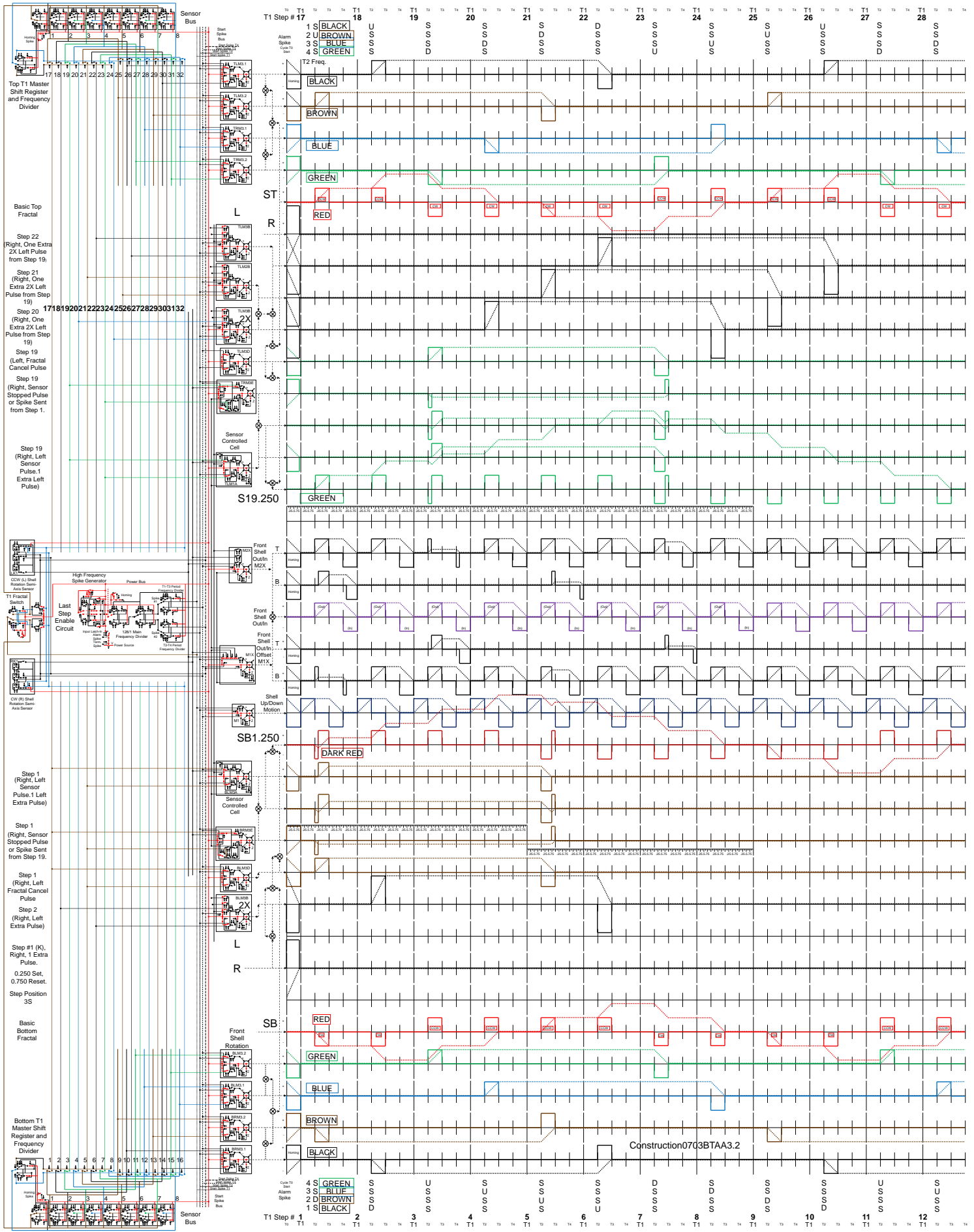

**Figure 51** **If *two objects are contacted at the same time* (0.250 sec) in Step 1 and in Step 19, the two fractals will cycle between these two objects.**

For every location of an object in one fractal there is a location of an object in the other fractal that returns the organism to the other object, as shown in Figure 50 for Step 1 and Step 19. The Constellation Diagram in Figure 58 in the next Section 3.6 shows what steps will restore the fractal after different contact positions in every step.

3.5.15 *Special condition for Step 2*

If the animal model contacts a point object in Step 2, the object will lie near the intersection of the two circles of the figure eight some time in Step 2, as shown in Figure 52. The trajectory will be able to move away from the object in eight steps. But it will contact the object a second time if the trajectory goes to the full sixteen steps, as shown in Figure 52.

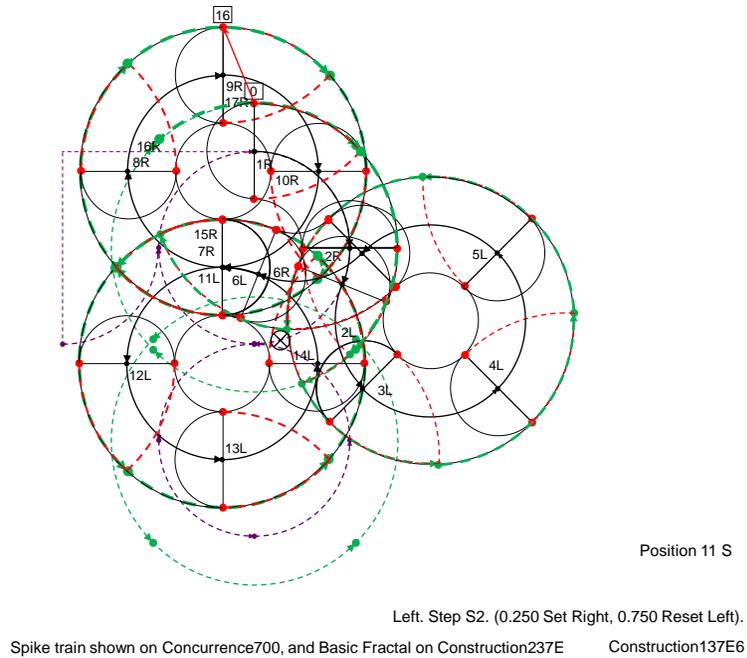

Position 11 S

Left. Step S2. (0.250 Set Right, 0.750 Reset Left).

Spike train shown on Concurrence700, and Basic Fractal on Construction237E    Construction137E6

**Figure 52** *If the object is contacted in Step 2*, **it will be contacted a second time late in Step 13.**

So a special circuit is needed that switches the fractal after Step 8 if an object is contacted in Step 2, as shown in Figure 53 in the next section. Step 18 in the top fractal has the same problem that is solved by switching back to the bottom fractal in Step 24 instead of Step 32.

3.5.16  *Trajectory that can move away from an object contacted in Step 2*

An example is shown in Figure 53 of how a contact made in 0.250 sec into Step 2 in the bottom fractal jumps to the top fractal after Step 8 without contacting the object a second time, and returns to the starting point of the bottom fractal if it contacts a second object at 0.250 sec in Step 20.

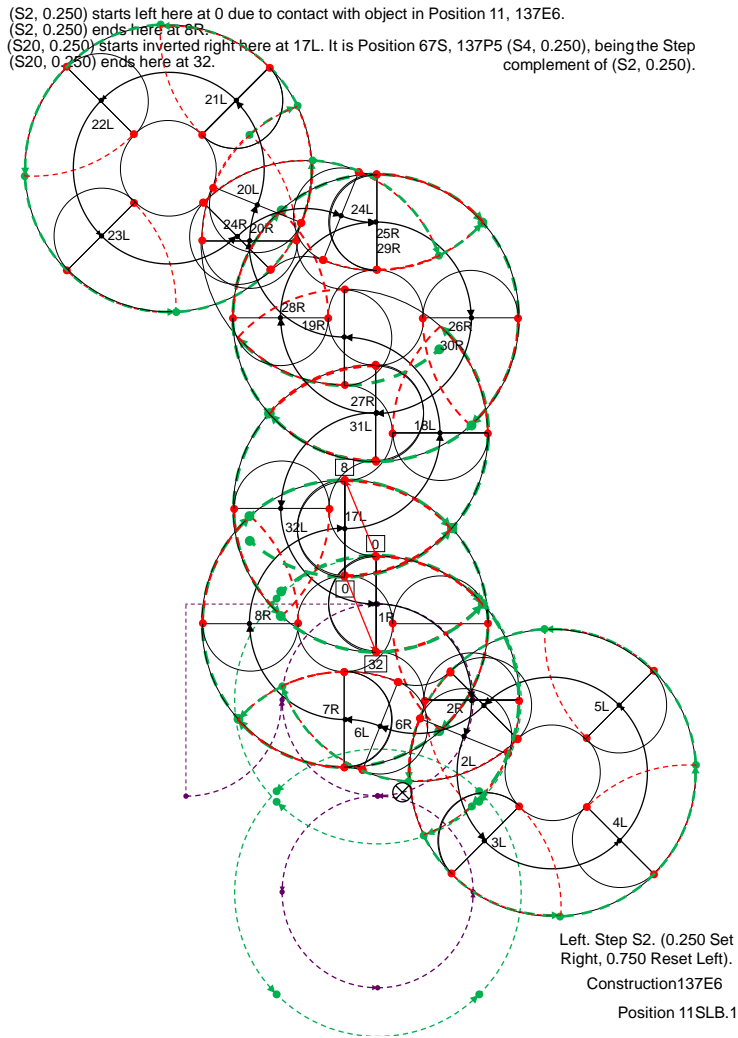

**Figure 53 Contact with an object any time in Step 2 requires that the bottom fractal switch to the top fractal at the end of Step 8.**

Notice that Step 8 jumps to Step 17, and Step 20 changes from a right-hand turning to a left hand turning at 0.250 sec into Step 20.

### 3.5.17 *Special enable circuit that switches the fractals after eight steps*

A logic circuit can be added before the Fractal Switch in Figure 53 that senses when Step 2 actuator TRM4A or Step 18 actuator BLM4A are made active by the object sensors LK1 or RK1, as shown in Figure 54.

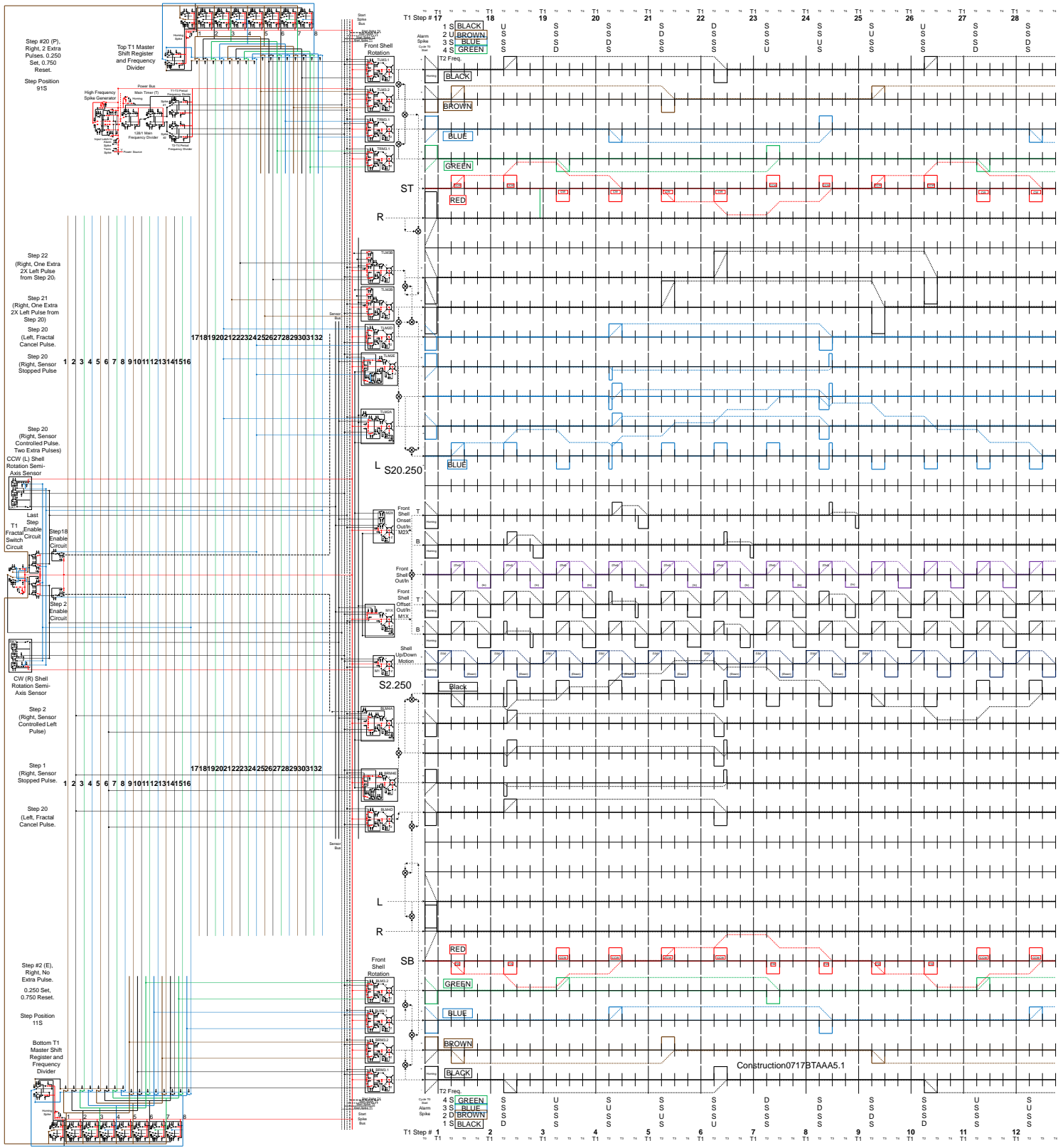

**Figure 54** When the Step 2 Object Sensing Cell (BLM4A) is made active by a sensor contact, it sends a spike to the *Step 2, 18 Enable Circuit* that causes the Fractal Switch to change from the bottom fractal to the top fractal on Step 8 instead of Step 16. Likewise, an object contacted in Step 18 (in the top fractal) will cause the top fractal to contact it a second time if it continues to Step 32. So the top fractal also needs to switch to the bottom fractal at Step 24 instead of Step 32. This trajectory is shown in Figure 56.

3.5.18  *Close-up view of the Step 2, Step18 fractal switch enable circuit*

An AND circuit can be placed in front of the Last Step Enable Circuit in figure 54 that produces an output spike if there is an object contact in Step 2 or Step 18, as shown in figure 55.

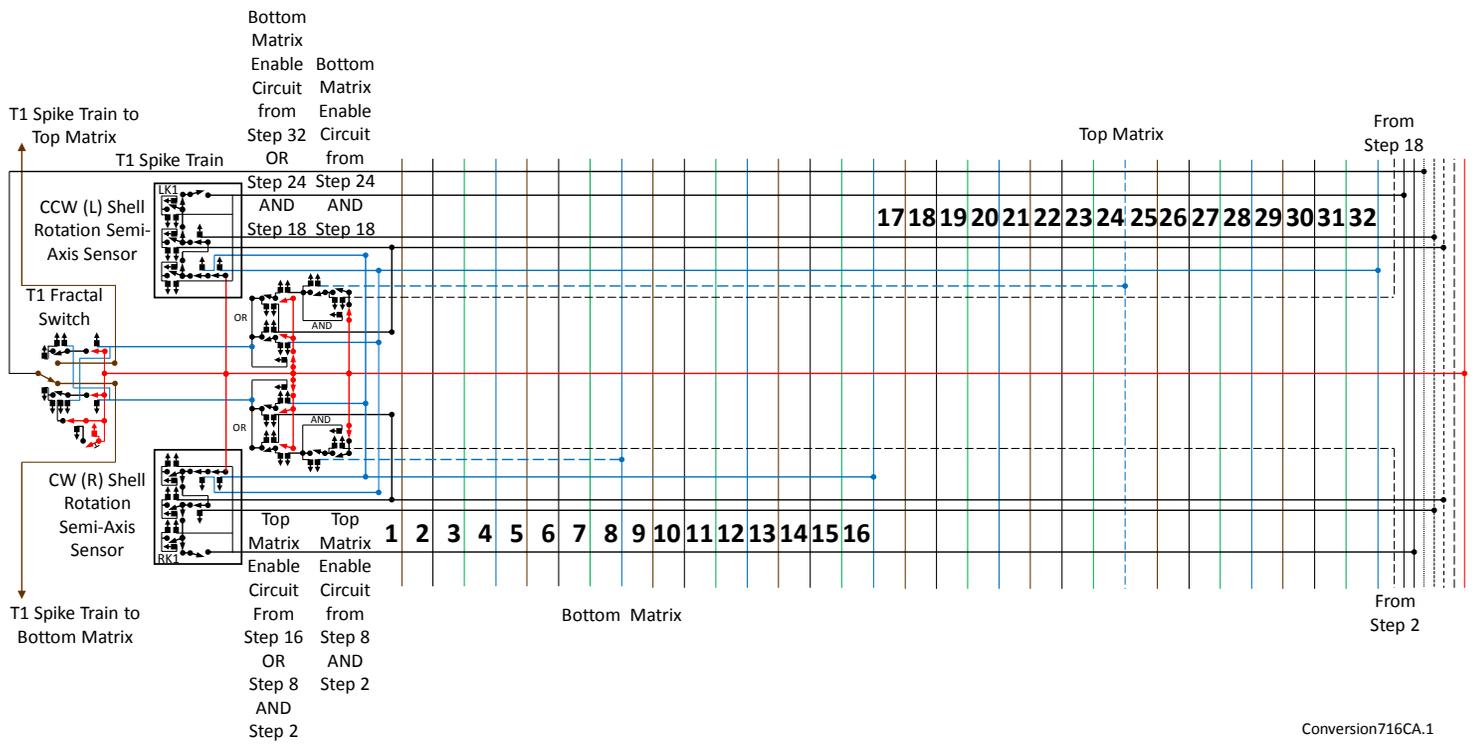

**Figure 55** The *Fractal Switch Enable Circuits* from Step 2 AND Step 8, and Step 18 AND Step 24 cause the T1 Fractal Switch to the Top or Bottom Matrix, respectively.

The Last Step Enable Circuits that switches the T1 Fractal Switch are modified so that they produce an output spike when the Top or Bottom fractal ends at Step 16 or Step 32, OR when there is an output spike from the Step 2 AND Step 8 Enable Circuit, OR when the is an output from the Step 18 AND Step 24 Enable Circuit.

3.5.19  *Trajectory that switches after eight steps in the top fractal*

The top fractal also switches after eight steps if it contacts an object in its second step (Step 18), as shown in Figure 56.

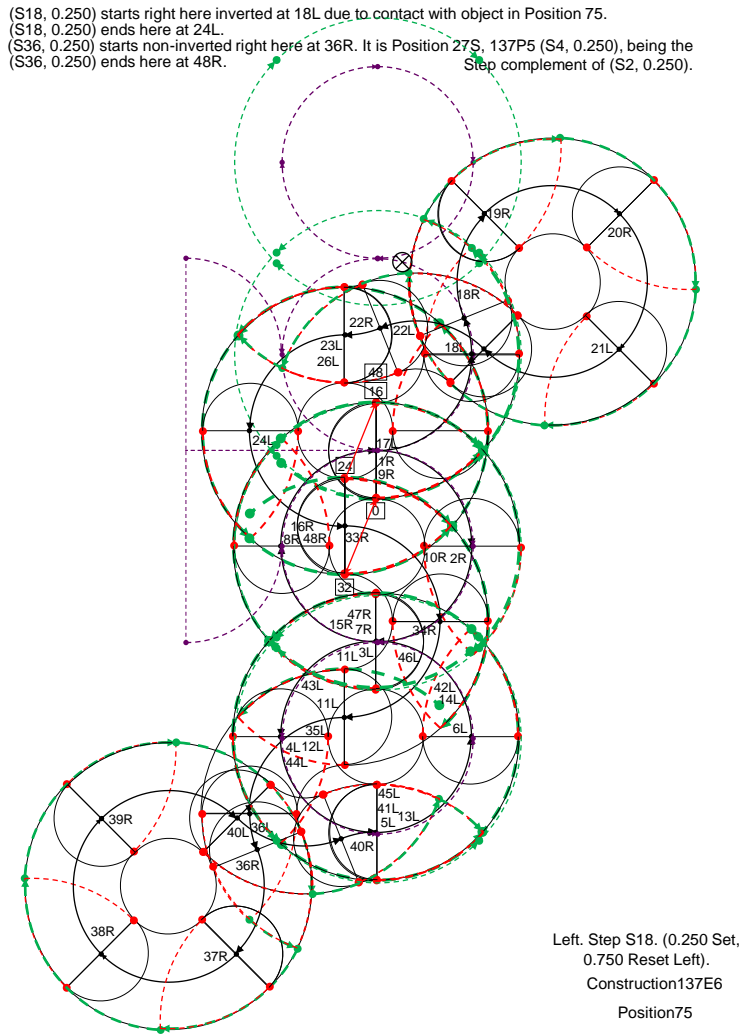

**Figure 56 If the *object is contacted in Step 18* of the top fractal, it must switch to the bottom fractal in Step 24 instead of Step 32 to avoid contacting the object a second time.**

The top fractal ends at Step 24. The bottom fractal starts at Step 33 and restores the orientation and location of the trajectory when there is a 0.250 set spike in Step 36, which is four steps (S4) into the bottom fractal.

3.5.20 *Control system that switches after eight steps in the top fractal*

The control system that is needed to create the trajectory in Figure 56 is shown in Figure 57.

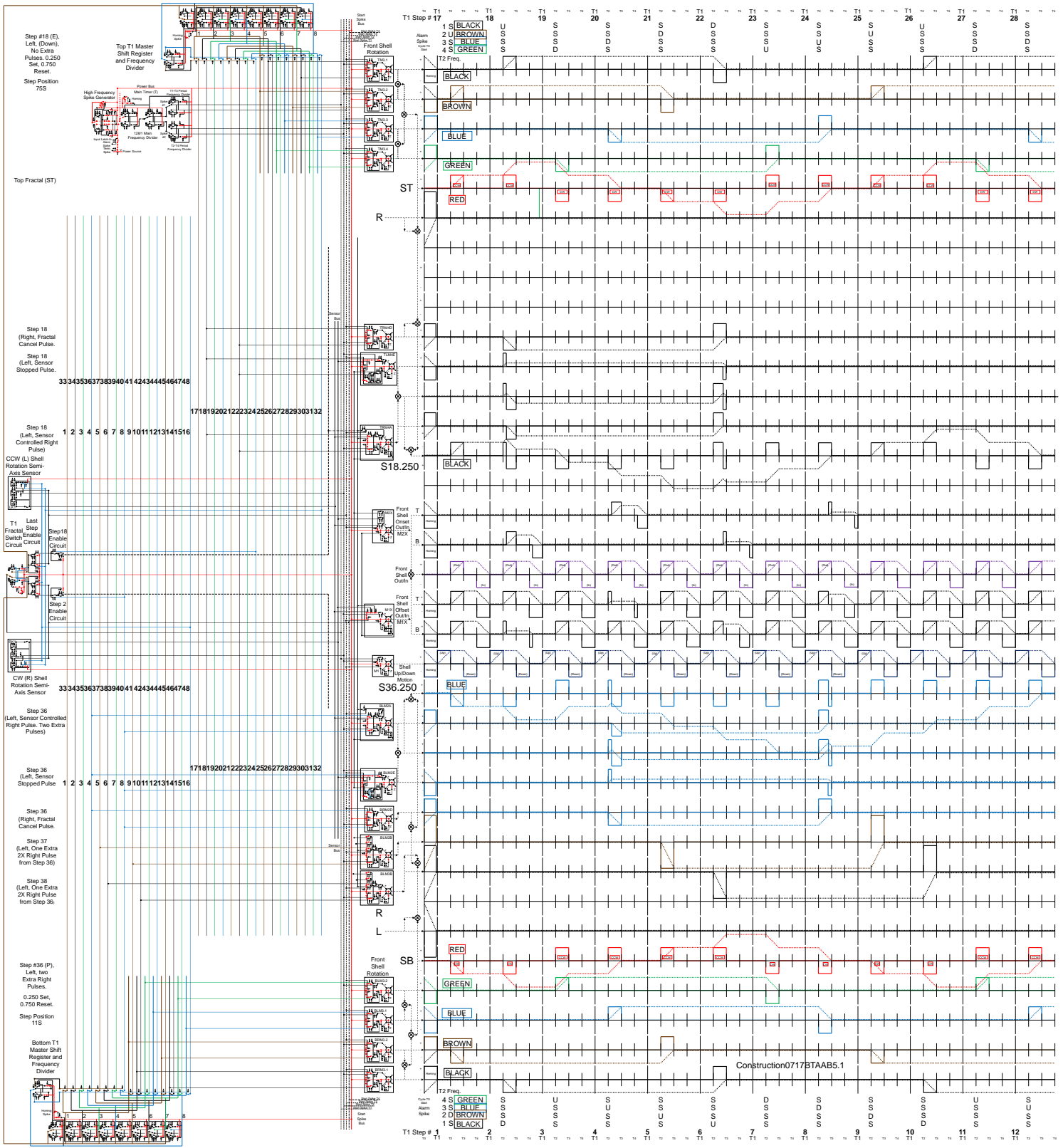

**Figure 57 The *Sensor Cell (TRM4A)* in Step 18 sends a spike to the Step 18 Enable Circuit that causes the top fractal to switch to the bottom fractal in Step 24 instead of Step 32.**

So the fractals are switched at the end their last step by the Last Step Enable Circuit except when an object is contacted in the second step in either fractal.

3.5.21 *Section summary*

The fractal system initiates interactions with its environment like a scanner causing changes in the behavior of the organism.

## 3.6 The attachment of an organism to an object in a specific location

Hefting is the process of becoming locked into a given location by means of sensing fixed objects in that location. In fact, there is a different position of an object in the field of one fractal that will produce this hefting effect for every position of an object in the field of the other fractal. This section shows how the control system can produce the effect of the second object when it encounters a first object, resulting in the organism remaining in the region of the first object that it encounters.

### 3.6.1 Single-phase constellation diagram

As shown in Figure 51, there is a specific contact time in one fractal that brings the animal back to the original position when there is contact in the other fractal. Some other examples of these contact times are shown in the Single-Phase Constellation Diagram in Figure 58.

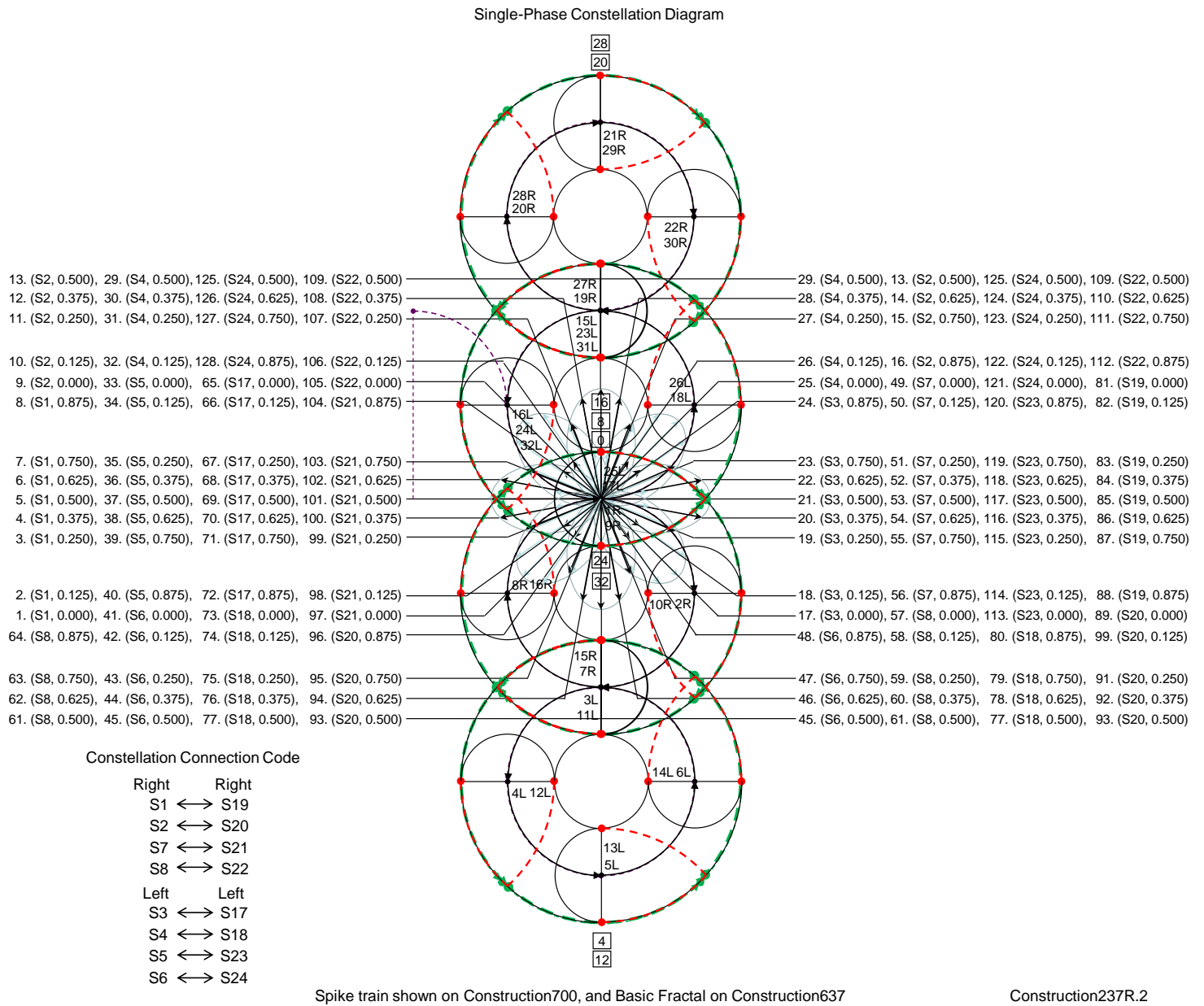

**Figure 58** The *Constellation Diagram* shows examples of contact times in each step that restore the system to its original position.

The constellation diagram shows the set of vectors created by each of eight equally spaced contact time samples in each step for Steps 1 through 8 in the bottom fractal, and for Steps 17 through Step 24 in the top fractal. The Connection Code shows which steps contain vectors that are equal and opposite for the same value of contact time. In the example of the trajectory shown in Figure 50, the 0.250 sec contact time in Step 1 (Position 3) results in a vector ($v_1$) that is equal and in the opposite direction to the vector ($v_2$) that is produced by the 0.250 sec contact time in Step 19 (Position 83).

Spike timing

### 3.6.2 *Modification of the Sensor Controlled Offset Cell*

The Sensor Controlled Offset Cell in Figure 43 needs to be modified so that it is disconnected (offset) by a spike from the Virtual Send Cell, as shown the BRM3EV cell in

**Figure 59**.

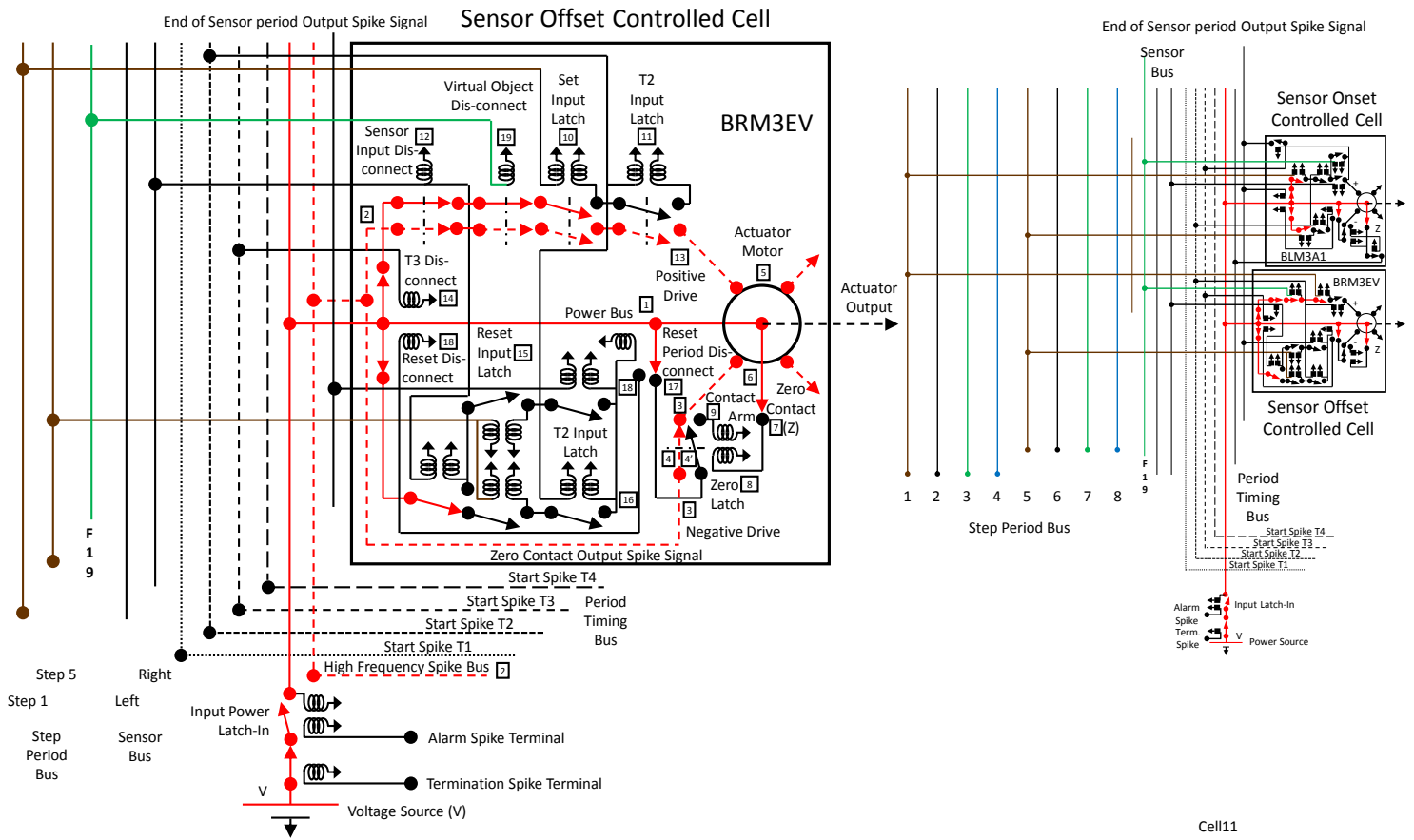

**Figure 59  The *Sensor Controlled Offset Cell* is disconnected by a sensor spike or a spike from the virtual send cell determined by the Constellation Diagram.**

A Virtual Object Disconnect [19] is added in series with the Sensor Disconnect [12]. This terminates its pulse, which is reset to zero four steps later.

### 3.6.3 Modification of the Sensor Controlled Onset Cell

The Sensor Controlled Cell in Figure 44 can be modified so that it responds to a signal in Step 19 from a Memory Send Co-Cell in the other matrix by adding a Remote Input Latch, as shown in **Figure 60**.

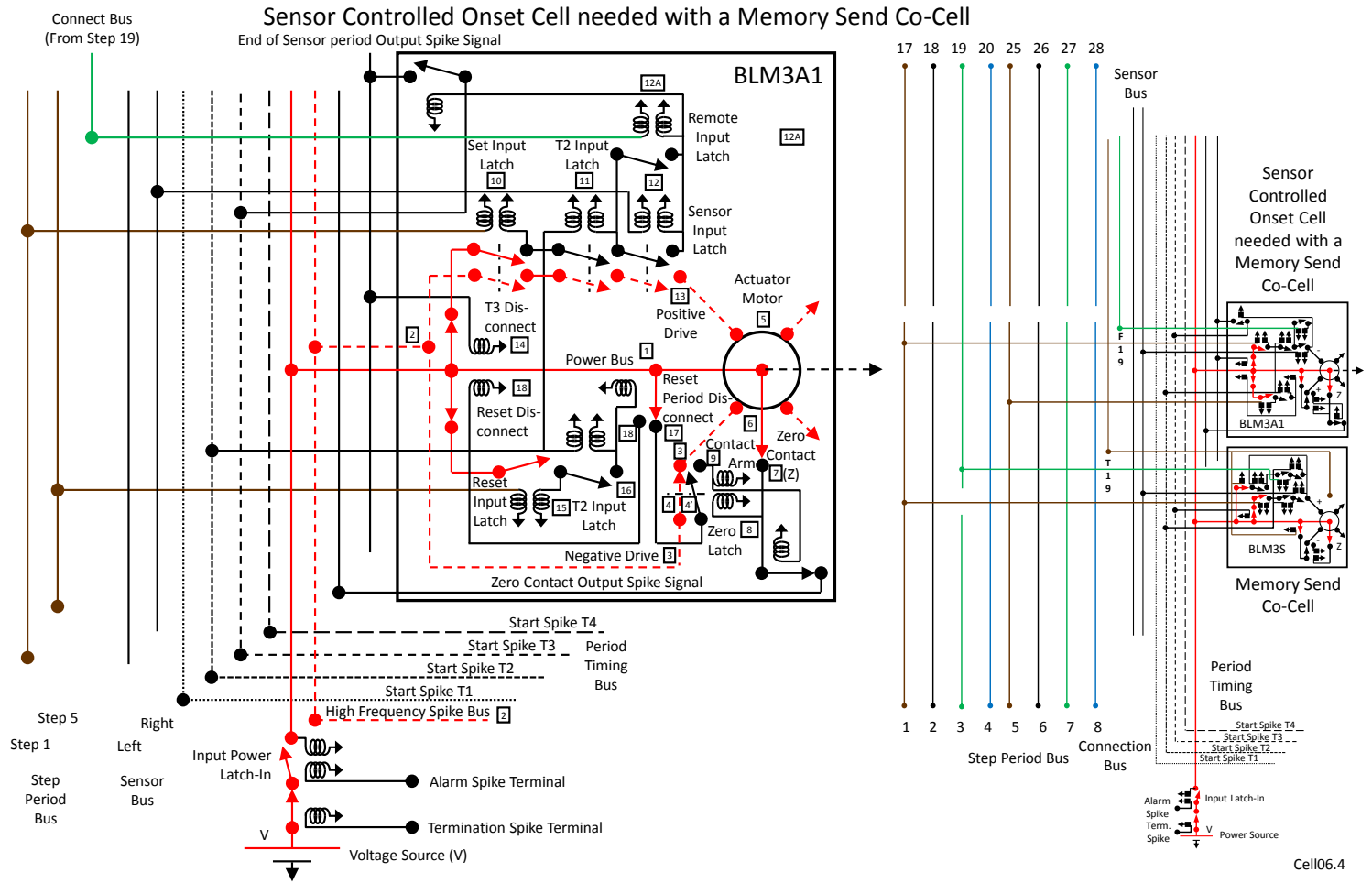

**Figure 60 The *Remote Input Latch* starts the Sensor Controlled Onset Cell as if it had sensed a contact with an object.**

The Remote Input Latch [12A] is connected in parallel (the logic "or") with the Sensor Input Latch. This causes it to act as if contact had been made with an object or by a signal from Step 19 using the Memory Send Co-Cell (BLM3S) in the next section.

### 3.6.4 *Memory Send Co-Cell*

The Memory Send Co-Cell in Figure 61 is connected to the same Step Bus conductor as its Sensor Controlled Cell. So it receives the same sensor contact time information. It holds this information, and then transmits the recorded contact spike time to the restoring Sensor Controlled Cell in the other matrix according to the Connection code shown in the Constellation Diagram in Figure 58.

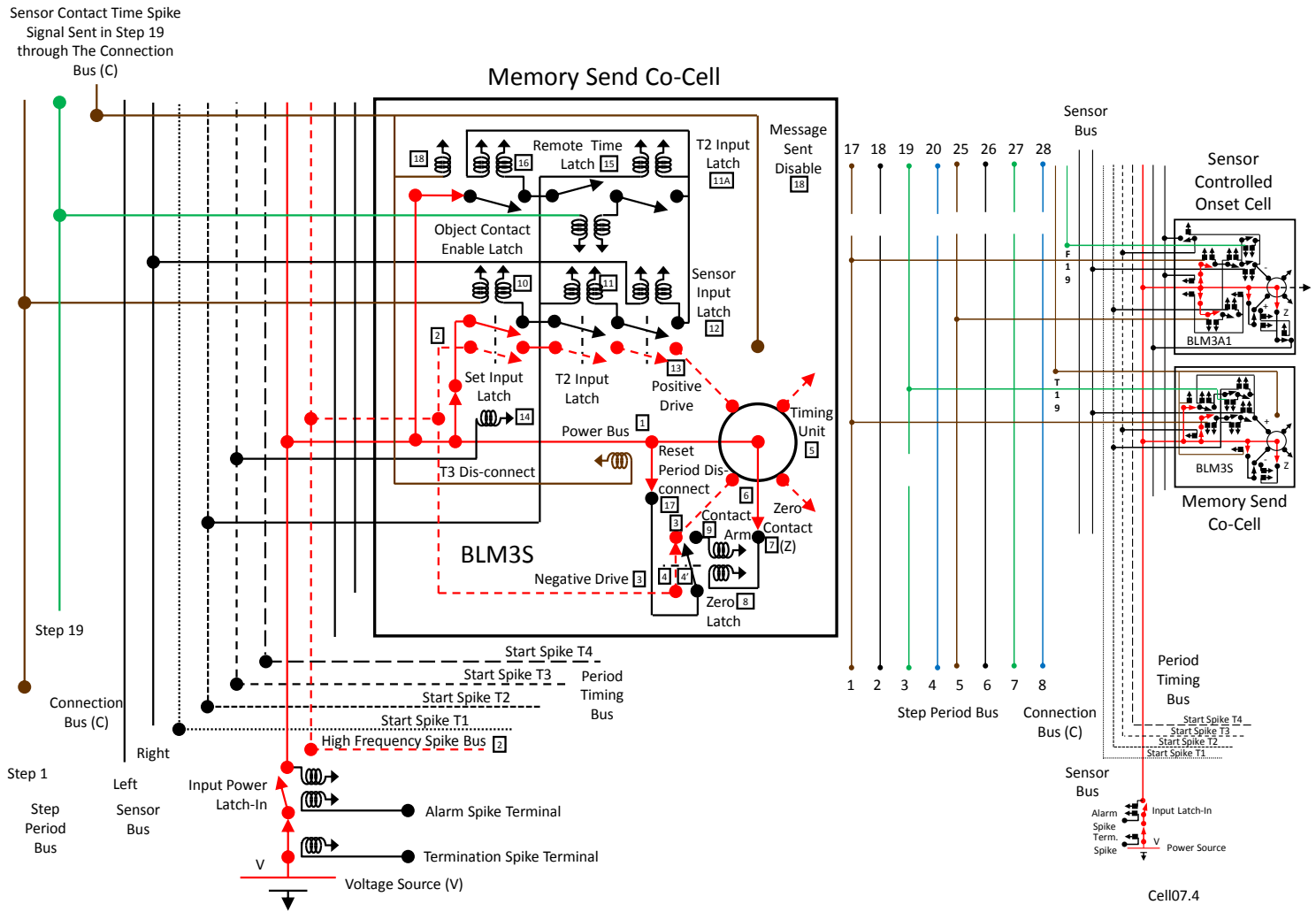

**Figure 61 The *Memory Send Co-Cell* uses a Timing Unit to store the time of the object contact within a Step Period.**

The Memory Send Co-Cell holds the object contact time in its Timing Unit [5] because it is not reset in the contact Step Period. It is restarted in the next repeat of the original Step Period, which restores the fractal to its original position. A spike corresponding to the original contact time is produced when the Contact Arm [6] contacts the Max Contact [18]. This spike is terminated by the Message Sent Disable disconnect [19]. This releases the Contacted Enable Latch [20], and starts the reset of the Timing Unit through the Reset Period Disconnect [17]. This brings the Contact Arm [6] back to the Zero Contact [7] and re-engages the Zero Latch [8].

### 3.6.5 *Shift register as a motionless timing unit*

The Send Co-cell does not need to produce motion to produce a spike at the Max Position. It merely needs to record and store the time of the spike made by contact with an object in a given Pulse Period, and send it to another Action Cell in the Step Period of the other Action Cell. This can be accomplished by the Bi-directional Shift Register shown in Figure 62.

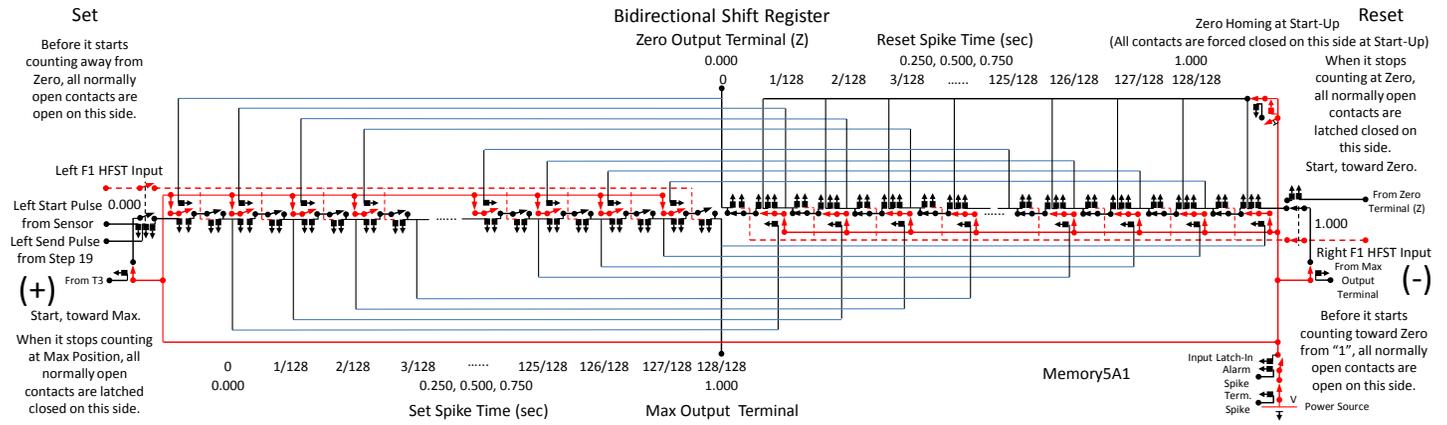

**Figure 62 The *bi-directional shift register* is homed to zero, like the actuator motors, and starts counting away from zero when the left side high frequency spike train and power circuits are energized. It starts counting back toward zero when the right side circuits are energized.**

The bi-directional shift register produces an output spike at the Max Output Terminal when the last contact on the left side is closed, and produces an output spike at the Zero Output Terminal when the last contact on the right side is closed. The bi-directional shift register will continue to represent time by the symbol for the actuator motor and its contact arm, as shown in Figure 61, even though it has become a digital timing unit.

The value of the send pulse is not stored by just one logic unit, but is stored at the transition between open and closed contacts in the multiple logic units in the bi-directional shift register.

### 3.6.6 *Action diagram of the Sensor Cell and its Memory Co-Cell*

The Sensor Cell and Memory Co-Cell both measure the time of contact. The Sensor Cell produces a set and reset motion. But the Memory Co-Cell advances to its Maximum Position Terminal during the step period determined by the Connection code, and sends the value of the original contact time to the Sensor Cell in that Step Period, as shown in Figure 63.

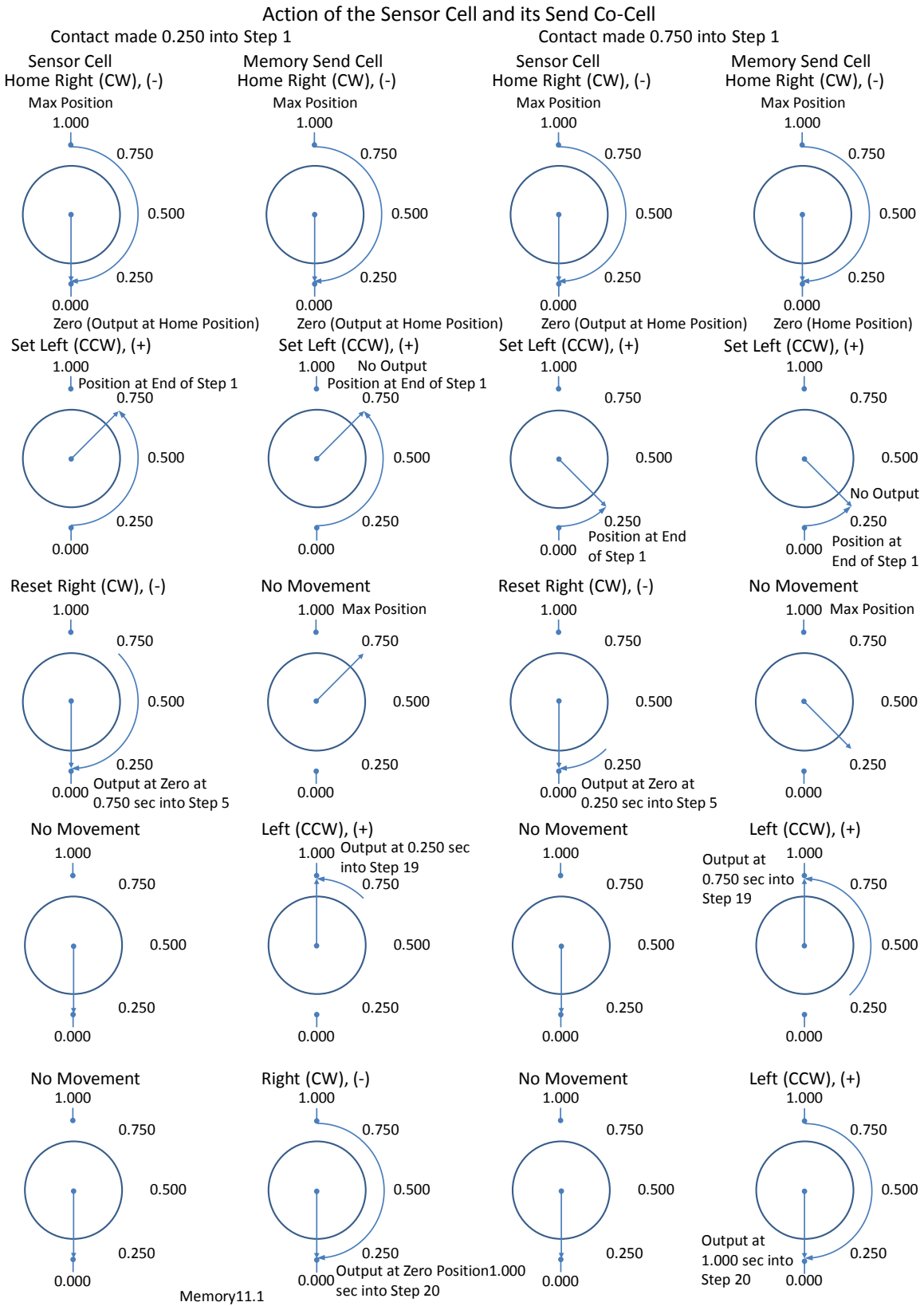

**Figure 63 The *Send Co-Cell* creates a spike at the original contact time.**

The illustration in Figure 63 shows the movement of a contact arm in the Co-Cell to represent time like an analog clock. However, the spike outputs are produced digitally by the bi-directional shift register in Figure 62 without the need of a moving contact arm.

### 3.6.7 Control system with an internal connection matrix

A Connection Bus (C) connects the output contact spike time stored in a Send Co-Cell in the matrix that contacts an object to the step in the other matrix that starts the restoring motion in the other matrix using this same contact spike time. The connections made by the Connection Bus are determined by the Constellation Diagram in Figure 58. In this example, the Send Co-Cell in Step 1 is connected to the Action Cell in Step 19, and the Send Co-Cell in Step 19 is connected to the Action Cell in Step 1, as shown in Figure 64.

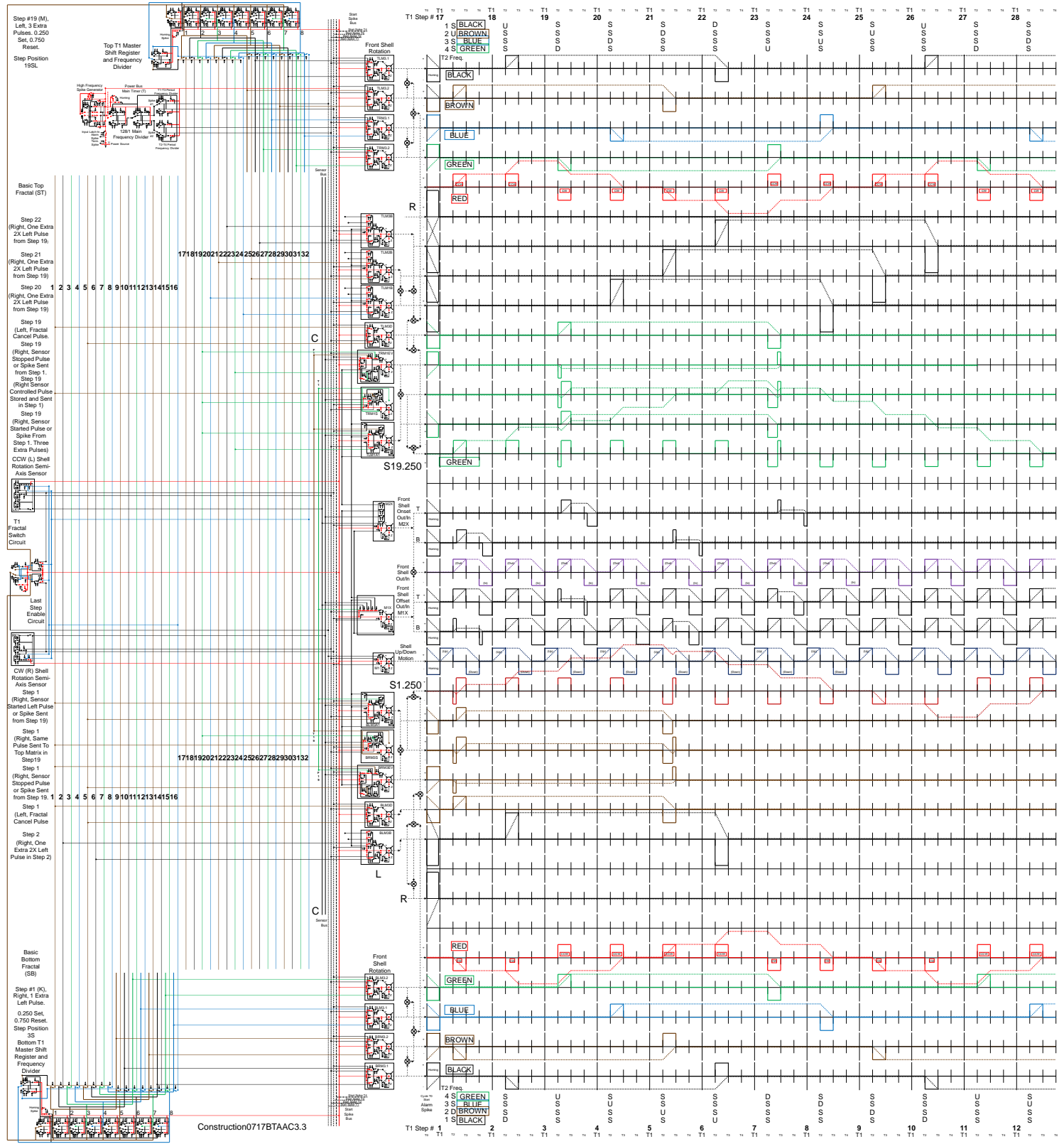

**Figure 64 The *internal connection matrix* (C) sends a spike time signal from a Send Co-Cell in one matric to an Action Cell in the other matrix at specific moments in time that allows an organism to move away from, and come back to a single object in either field of the two fractals.**

The Send Co-Cell (BLM3S) in Step 1 records a 0.250 sec. contact time in Step 1 in the bottom matrix. It then reproduces that 0.250 sec. spike time to the Action Cell (TLM1A) in Step 19 in the top matrix through the Connection Matrix (C). This causes the organism to return to the original position and orientation of the bottom fractal, where it would make

Spike timing

contact at 0.250 sec. in Step 1 of the bottom matrix again. Thus, the organism would become attached to a specific geographical location around the object.

So, when contact is made by the left hand sensor at 0.250 sec. into Step 1, the BLM3A cells is activated. This starts a 0.750 pulse to be created, which cancels the (SB) fractal pulse for the remainder of Step 1. This enables the BLM3B cell, which is activated by the Step 2 start spike. This creates a double values pulse for the whole Step 2 unit period. This reverses the direction the of the SB pulse during the Step 2 unit period.

If the TLM1A cell had been contacted at 0.250 sec. into Step 19, it would produce a pulse that cancels the remained of the inverted fractal in Step 19. It would also enable the TLM1B, TLM2B, and the TLM3B cells, which would be activated in Step 20, Step 21, and Step 22. The TLM1S cell would also be activated also. So it stores the 0.250 time setting until the next Step 1, at which time it send the 0.250 time setting over the Connection Bus (C) to the BLM3A cell. This activates the BLM3A cell, which cancels the bottom fractal for the remained of Step 1, and activates the BLM3B cell so it reverses the fractal in Step 2 as before.

The output of the BLM3A and the TLM1A cells are added to the basic fractal. This creates the response behavior needed by the organism to reverse its direction of rotation when contact is made with an object, and then restore its original orientation, as shown in Figure 50.

### 3.6.8 *Close-up view of virtual object circuit*

In the example shown Figure 65, the Memory Send Co-Cells connected to Step 1 sends the measured object contact time to the Sensor Controlled Cell connected to Step 19 in Step 19 over the Connection Bus (C), or the Sensor Controlled Cell connected to Step 1 receives the measured object contract time from the Memory Send Co-cell connected to Step 19 in Step 1 over another conductor in the Connection Matrix (C). This will restore the system to its original location, as shown in the trajectory shown previously in Figure 50, using a virtual object created by the Send Cell.

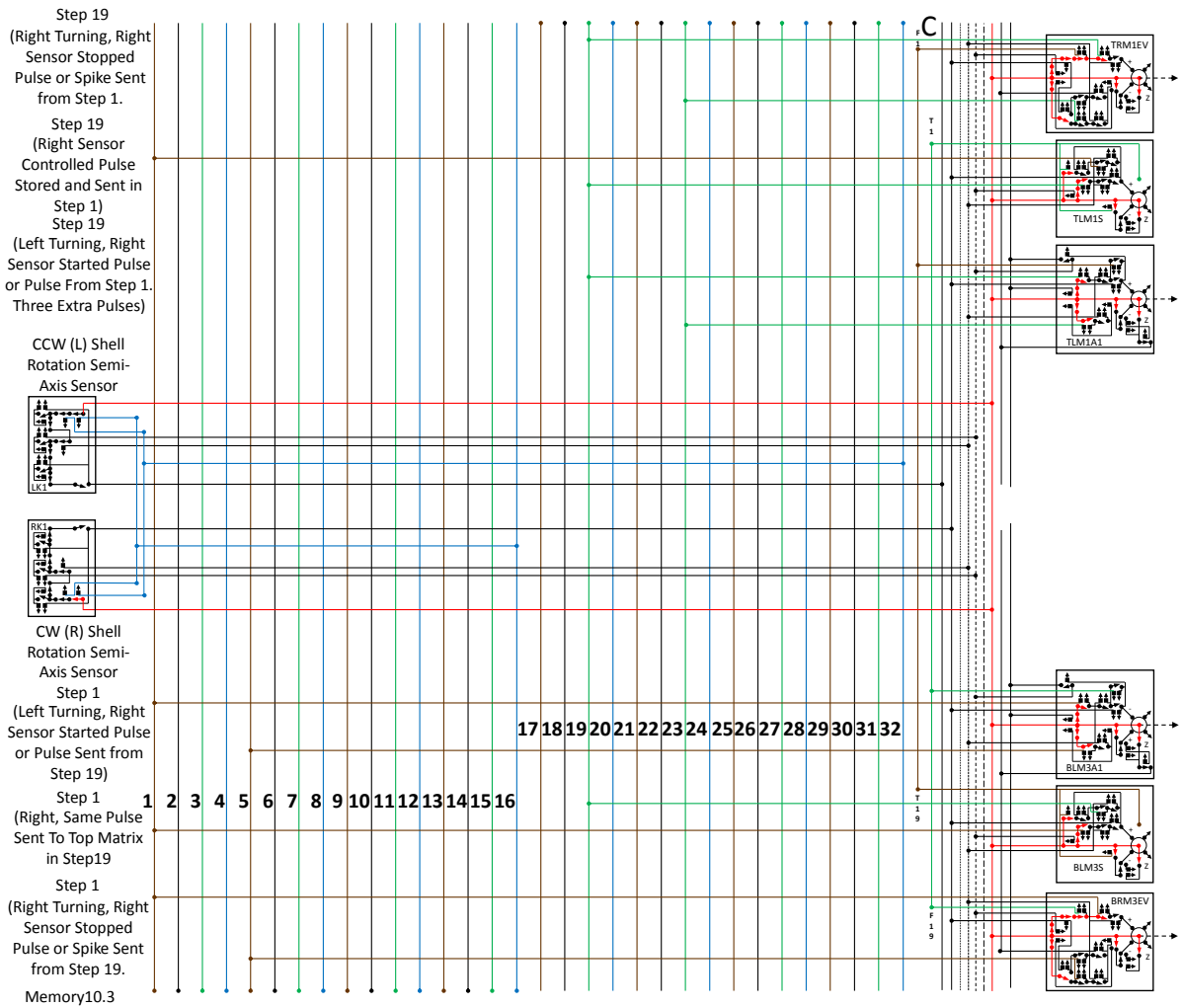

**Figure 65 The *Co-Cell* (BLM3S) stores the contact spike time in Step 1 as a virtual object, and sends its contact time to Sensor Cell (TLM1A) in Step 19.**

Its armature is stopped by the T3 pulse at the end of Step 1 in a position three quarters of the way to the Maximum Stop Contact, and remains in that position until it is restarted by the start spike at the beginning of Step 19. This causes the armature to continue toward the Maximum Stop Contact. It will reach the Maximum Stop Contact at 0.250 sec. into Step 19. This causes a spike to be sent to the TLM1A cell at Step 19 along the Connection Bus (C), which starts its actuator at 0.250 sec. into Step 19 just as if it had contact a second object at this time. So it would produce the trajectory shown in Figure 50.

The time spike sent to the sensor cell (TLM1A) in Step 19 does not occur on the sensor bus. So its Co-Cell (TLM1S) is not activated. When the trajectory returns to the first step in the bottom fractal, now Step 33, the original object would be encountered again, and the process is repeated.

### 3.6.9 *Connecting all of the Basic Steps in both matrices according to the Constellation Diagram*

A unique set of cells belonging to Step 1, shown in Figure 47, and Step 19, shown in Figure 49, are required to produce the reaction response and recovery pulse trains shown in Figure 64. The unique set of cells required for all eight steps of the top and bottom fractal is shown in Figure 66. To save drawing space the whole change in direction is shown in just one step even though the change in direction at the second node is due to the motions of two axes, sequentially, as shown

Spike timing previously.

**Figure 66** All of the sensor circuit cells shown in both matrices are needed to produce the *reaction and recovery response* to the contact with an object in any of the eight steps in the bottom or top fractal.

Spike timing

This figure shows the different pulse trains in all of the basic steps that are created when contact is made at 0.250 sec. in a step. The avoidance pulses are shown as being created by a single, double unit amplitude Sensor Controlled Onset Actuator (BLM3A) instead of the single unit amplitude offset and onset actuators (BRM3E and BLM3A) shown previously. This representation is used because of the limited page size of the graphics software. It adequately represents the effect of the object contact, and will be used for the rest of the discussion.

The control system shown in Figure 66 connects the front shell rotation axis in a specific step in one matrix to a the front shell rotation axis in specific step in the other matrix, as shown in the Single-Phase Constellation Diagram in Figure 58. So there are only sixteen connecting lines needed in the Connection Matrix (C) for all of the infinite number of possible contact times in all of the steps. A different pulse train and animal path is produced for every different contact time encountered in every different step. A commonality occurs between all of the pulse trains in the bottom and top matrix since there is an inverted version of each pulse train the other matrix.

Note that there are 23 cells in both the bottom and top matrices that are required to produce an avoidance response in the shell rotation axis for contact with an object in any of their eight steps. So, 92 cells would be needed for the Front Shell, Front Shell/Foot, Back Shell, and Back Shell/Foot axes shown in Figure 36.

3.6.10  *Attachment to a given location when the object moves*

An interesting feature of this location attachment system that produces hefting is that a moving object will pass through the fractal without changing the location and orientation of the fractal. The fractal will return to the original point of contact after each thirty-two steps, and then establish a new point of contact when it encounters the object in its new position. This process will continue while the object is in the field of the object, and the fractal will return to its original location and orientation after each contact. When the object moves out of the field of the fractal, the original figures-eight will reappear in their original location and orientation.

3.6.11  *Attachment to a given location when the organism slips*

If the organism slips during the execution of it fractal, the object will be contacted in a new position. So the organism will be in a different position with respect to the object. If the orientation is changed by the slip, the original orientation will be lost. The next section shows how the animal can return to the original position if the orientation is preserved.

3.6.12  *Section summary*

This fractal based system can become attached to its environment by producing an internal spike time that returns the organism to its point of origin in a process described as hefting.

*3.7  Attachment to an object that moves along the centerline of the fractal*

It turns out that imprinting is an intrinsic aspect of hefting. An animal that is imprinted is inclined to follow another moving object or moving animal. Some additions to the control systems shown above will cause the position of the figures-eight to adjust itself so as to remain in contact with the object at the same time and in the same period in a new position of the object. This process causes the location of the trajectory and the animal to move with the movements of the object. It also corrects for slippage in the movement of the organism if orientation is maintained. The object can move in either direction along the centerline of the fractal, or move at right angles to the centerline of the fractal. Or, the object can move in some combination of in-line and at right angles to the centerline. This section deals with the movement of the object in either direction along the centerline of the fractal.

Spike timing

### 3.7.1 *Path of the fractal due to the movement of an object along the centerline of the fractal*

When the object moves downstream on the path of the trajectory, the path of the trajectory will move in the same direction by the same amount if the imprinting process shown in Figure 67 is followed.

1. (S1, 0.250) starts right here at [0] due to contact with object in Position 3, 137K17.
   (S1, 0.250) ends here at [16].
   (S19, 0.250) starts left here at [16], It is Position 19, 137M1LL. It being the inverted value of (S1, 0.250).
   (S19, 0.250) ends here at [32].
   The object is moved <u>Down Stream</u> 1/8 Unit to new position.
2. (S33, 0.375) Starts right from [32] due to contact with the object in the new position. It is Position 4, 137K19.
   (S33, 0.375) ends here at "a" [48].
   (S51, 0.500) starts left here at "a" [48]. (See calculations). It is Position 85, 137M12L.
   (S51, 0.500) ends here at [64}. It is the new starting location for the fractal.
3. (S65, 0.250) starts right at [64]. It is a repeat of the original reaction.
   (S65, 0.250) ends at [80]R.
   (S83, 0.250) starts right at [80]. It is a repeat of the original .
   (S83, 0.250) ends at [96]R.

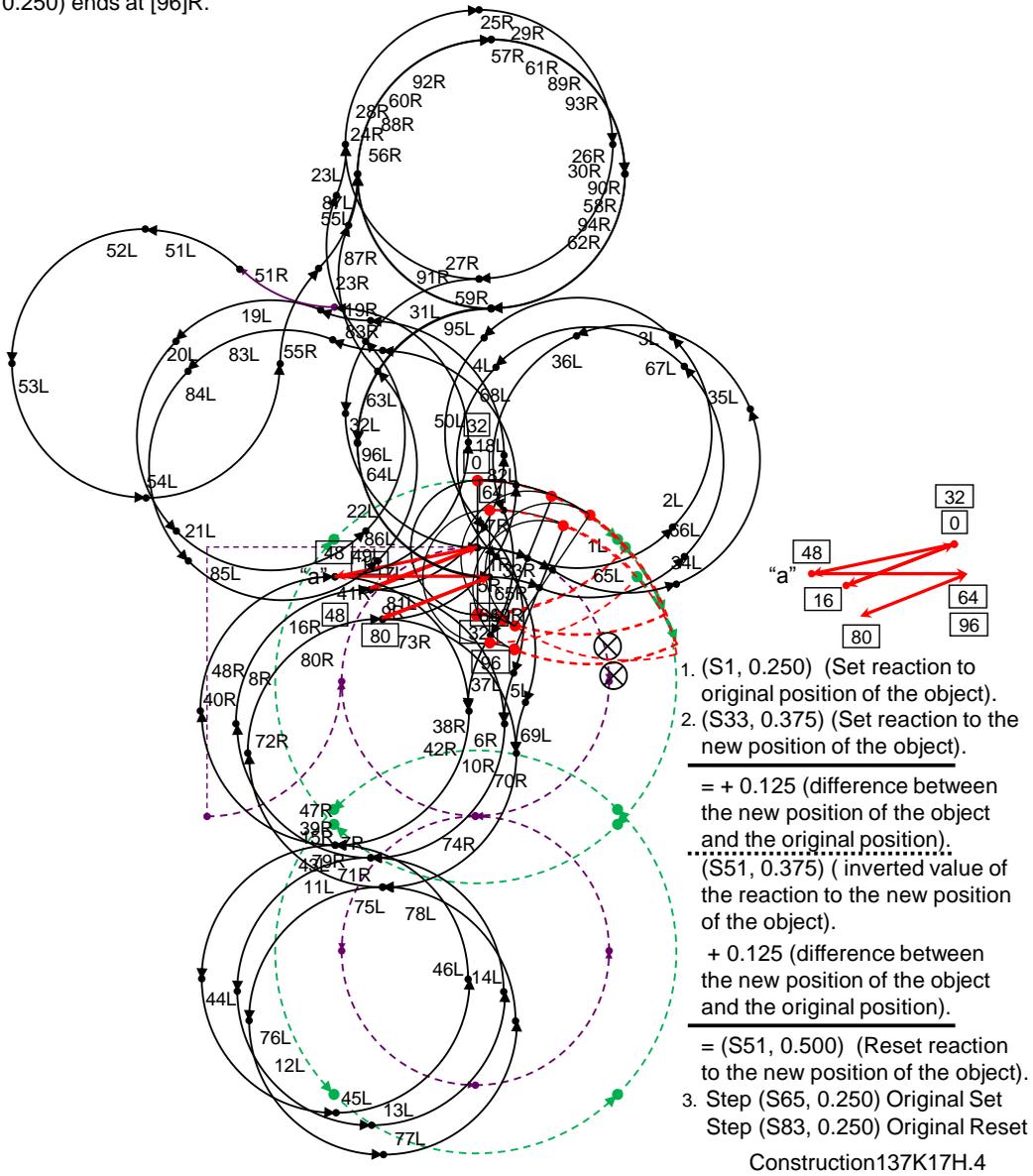

1. (S1, 0.250) (Set reaction to original position of the object).
2. (S33, 0.375) (Set reaction to the new position of the object).

= + 0.125 (difference between the new position of the object and the original position).
(S51, 0.375) ( inverted value of the reaction to the new position of the object).

+ 0.125 (difference between the new position of the object and the original position).

= (S51, 0.500) (Reset reaction to the new position of the object).
3. Step (S65, 0.250) Original Set
   Step (S83, 0.250) Original Reset
   Construction137K17H.4

**Figure 67  In the *imprinting process*, the figure eight formed by the first contact with an object follows the object if the object moves.**

The process of following the object around takes six trajectories, each consisting of 16 steps, for a total of 96 steps, as shown above. This imprinting process involves a computation using time (delay) measurements. The difference between the new position of the object (S33, 0.375) and original position of the object (S1, 0.250) is + 0.125. This value is added to the inverted value of the reaction to the new position of the object (S51, 0.375), which is equal to S51, 0.500). This value is used as the restore reaction to the new position of the object. The result of this sample calculation is shown in Figure 67, above. Four additional cells are required to carry out this calculation: three Sensor Spike Time Decoder Cells and a Dual Input Memory Send Co-Cell.

### 3.7.2 *The Sensor Spike Time Decoder Cell*

The Sensor Spike Time Decoder Cell measures the contact time with an object using the F2 High Frequency Spike Train, and decodes it into an F2 spike train of a specific length (pulse width) that represents the contact time rather than a motion of a specific length, as shown in Figure 68.

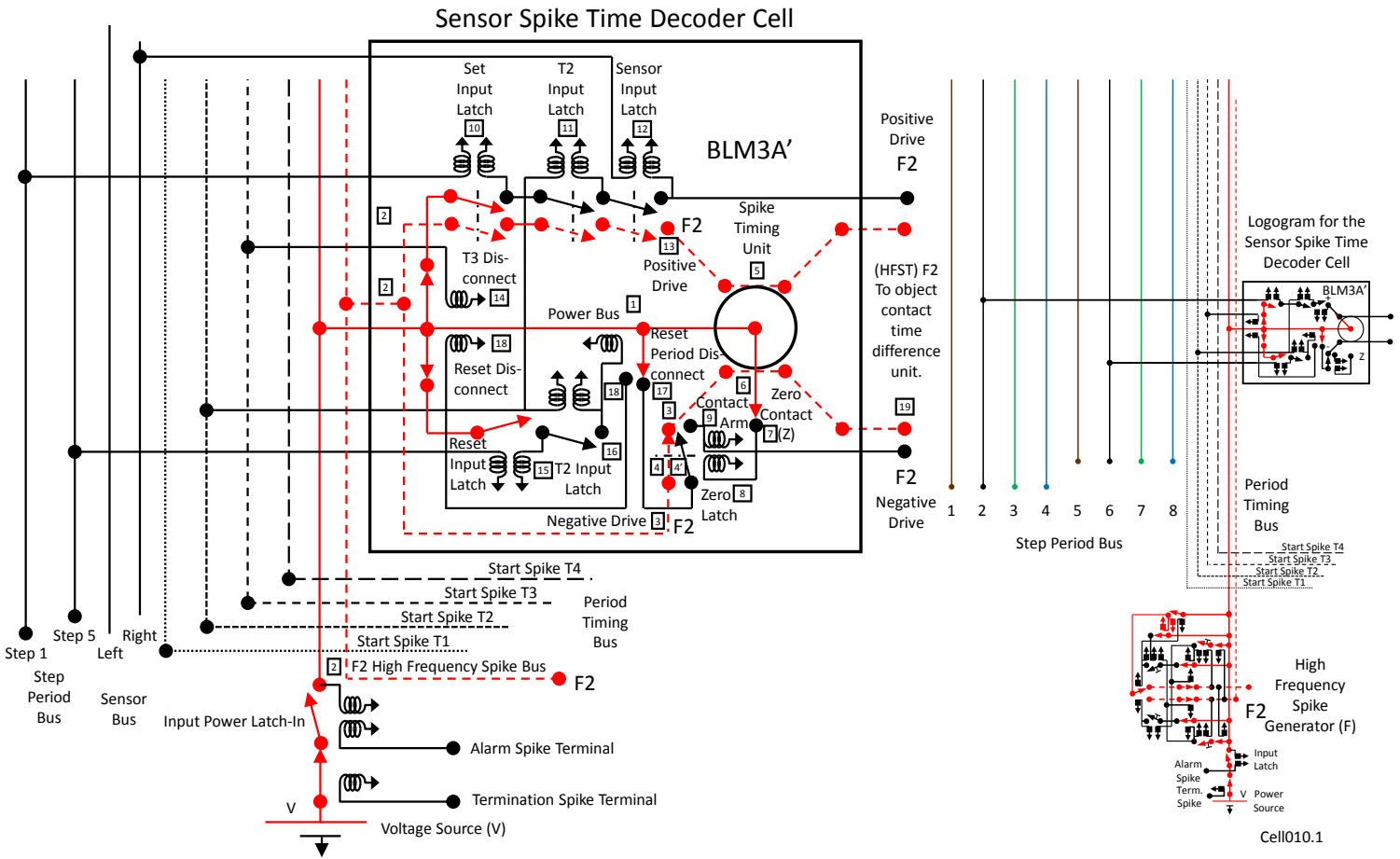

**Figure 68 The *Sensor Spike Time Decoder Cell* uses a Spike Timing Unit to store a contact time, and produces an F2 output spike train containing a specific number of spikes according to the measured contact time.**

One Sensor Spike Time Decoder Cell (BLM3A') records and outputs the positive value of a first contact time. The second Sensor Time Decoder Cell (BLM3A") records and outputs the value of the next contact time. A third Sensor Spike Time Decoder Cell (BLM30) records the first contact time, and outputs the inverted (negative) value of the first contact time.

They all operate using the F2 terminal of the High Frequency Spike Generator (F) instead of the F1 terminal used by the rest of the system. Since the F2 spike train is 180 degrees out of phase with the F1 spike train, the spike value in the Dual Input Memory Send Co-Cell in Figure 70 using the F1 spike train can be modified by the outputs of the Sensor Spike Time Decoders that use the F2 spike train, as shown in Figure 72, Figure 73, and Figure 74.

### 3.7.3 *Spike driven differential*

The spike outputs of the decoder cells in Figure 68 can be added or subtracted to one another in the same manner as the mechanical differentials shown previously by using the "not and gates" NAND shown in Figure 69.

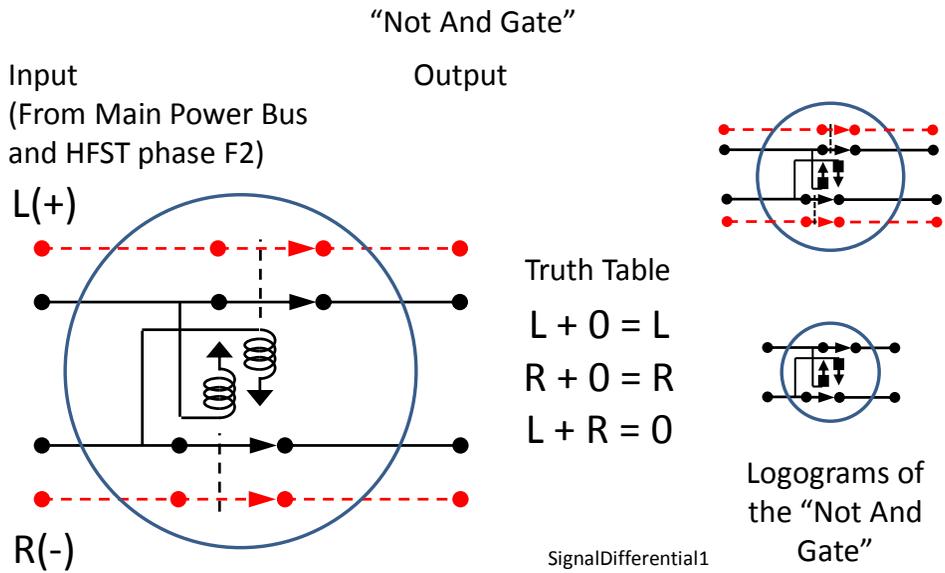

**Figure 69** The *spike driven differential* produces no output when plus and minus signals are present.

The spike driven differential works the same way as the mechanical differential since one input is a plus signal and the other input is a minus signal, both of equal magnitude (the unit amplitude in PWM systems). The decoder cells determine the pulses from the Main Power Bus that are used to switch the F2, HFST signals.

### 3.7.4 *The Dual Input Memory Send Co-Cell that computes the restore reaction that causes the fractal to follow a moving object*

The Send Co-Cell shown in Figure 61 needs to be modified so that it adds the difference between the original contact time and the new contact time, to the new contact time. It sends this computed time to the Action Cell in the other matrix in the correct step that restores the fractal to the new position that produces the original contact time, as shown in Figure 70.

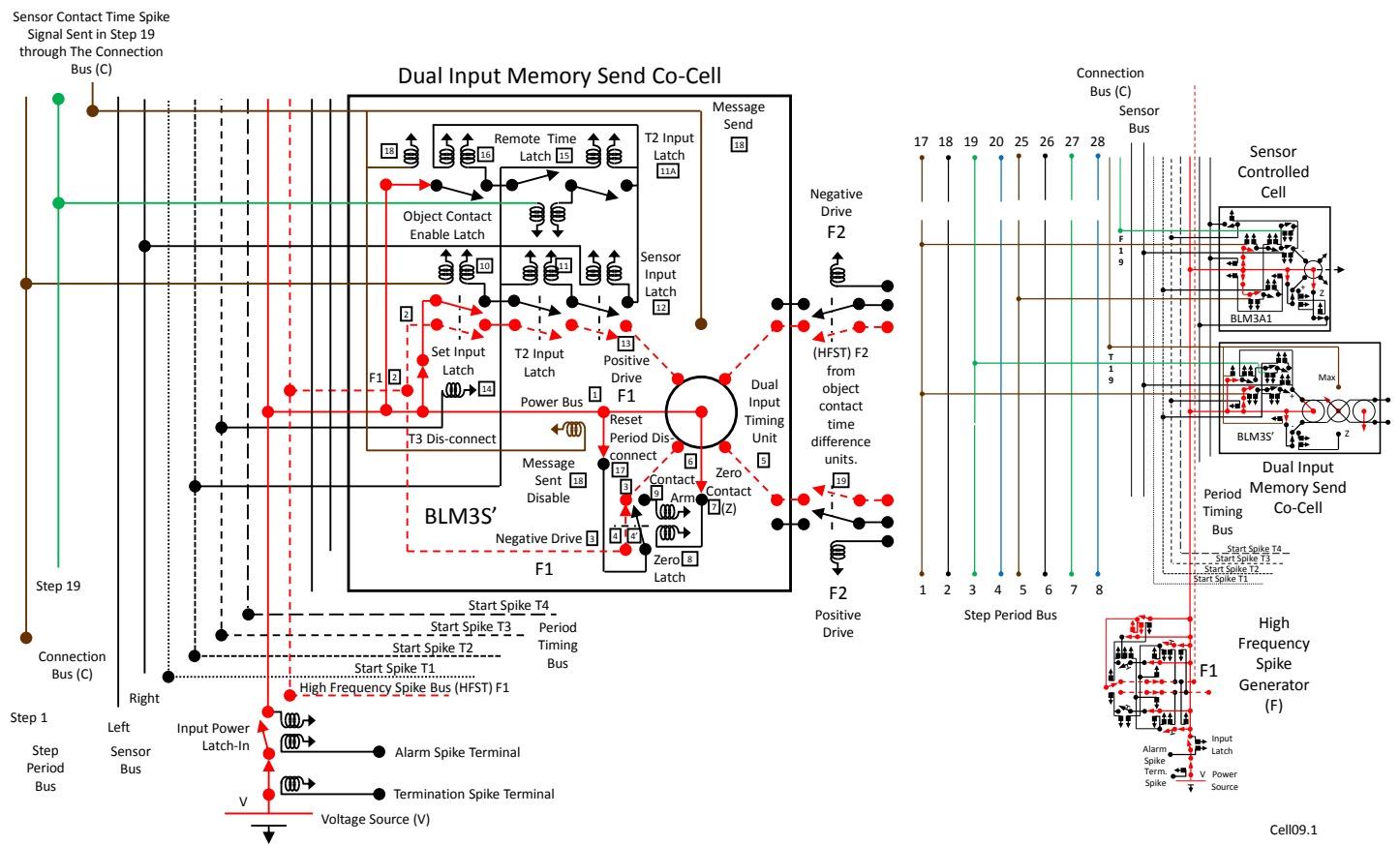

**Figure 70** The *Dual Input Memory Send Co-Cell* adds the difference between the pervious contact time and the present contact to the present contact time, and sends the result to its restore actuator cell.

The sensor contact time is recorded in the Dual Input Time Unit [5] by the number of spikes from the F1 High Frequency

Spike timing

Spike Train through terminal [13]. This number is increased or decreased by the number of spike from the F2 High Frequency Spike Train [19]. The resulting count determines when the output time of the Message Send [18]. Since the spikes in F2 occur between the spikes in F1, positive driving spikes from both F1 and F2 and negative spike from both F1 and F2 are added or subtracted together, and allow the positive and negative spikes from both F1 and F2 to cancel each other out.

3.7.5  *The Dual Input, Bidirectional Shift Register*

The Dual Input, Bidirectional Shift Register is similar to the Bidirectional Shift Register shown in Figure 62, except it contains the second High Frequency Spike Train F2 on Terminal B shown in High Frequency Spike Generator (F) in Figure 3. F2 is a separate spike train having spikes that occur 180 degrees out of phase ( in between) the spikes produced by the High Frequency Spike Train F1, as shown in Figure 71.

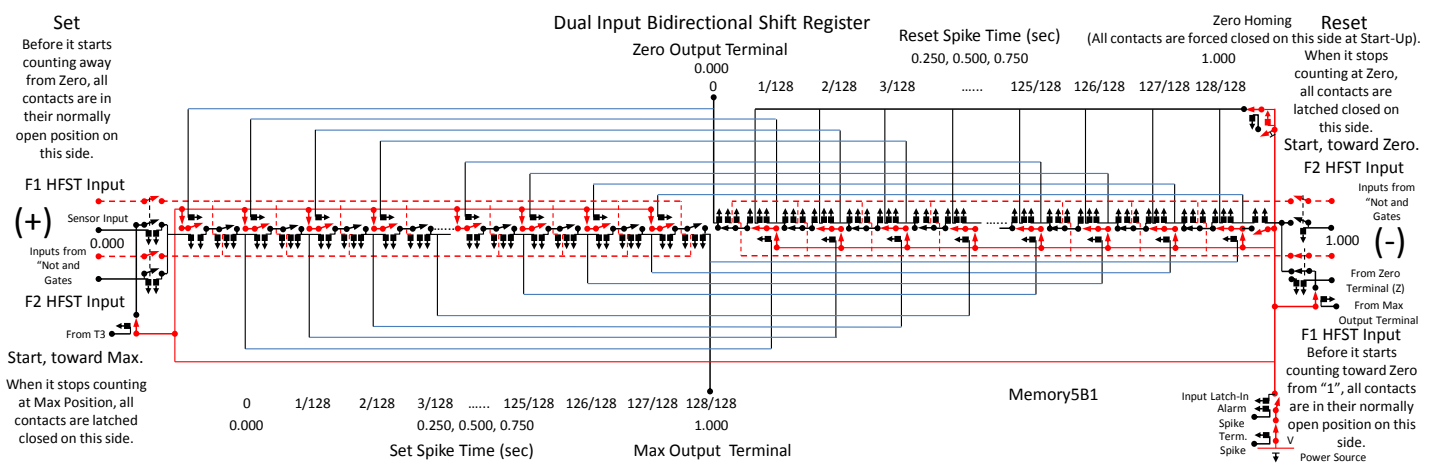

**Figure 71 The *Dual Input, Bidirectional Shift Register* adds spike inputs from two interlaced spike trains of opposite polarity.**

The second set of inputs from the F2 spike train adjusts the count in the Dual Input Bidirectional Shift Register so that it produces an output on its Max Output Terminal that reflects the difference between a first and second contact time with an object. The value of the computed pulse is not stored by just one logic unit, but is stored at the transition between open and closed contacts in the multiple logic units in the bi-directional shift register.

3.7.6  *Measuring the movement of the object along the path of the animal, and shifting the location of the path of the animal according to the movement of the object*

For the animal to follow the movement of an object along the path of the trajectory (imprinting) requires the four additional types of actuator/timers in each step position, as shown in Figure 72 through Figure 74. The first object contact has been selected to be 0.250 sec. into Step 1, as shown in Figure 72. The imprinting actuator/timers needed for Step 1 are shown in the bottom matrix as: BLM3A' (first positive pulse), BLM3S' (sender), BRMA0 (permanent memory of the first contact time), and BLM3A" (second negative pulse). In the first step, the timer of the actuator (BLM3A1), the sender (BLM3S') are set to the first contact time of 0.250 sec. BLM3S' keeps a permanent record of the first contact time. It is the only memory not reset by the process, except when there is a power shutdown, which releases its power input latch.

Spike timing

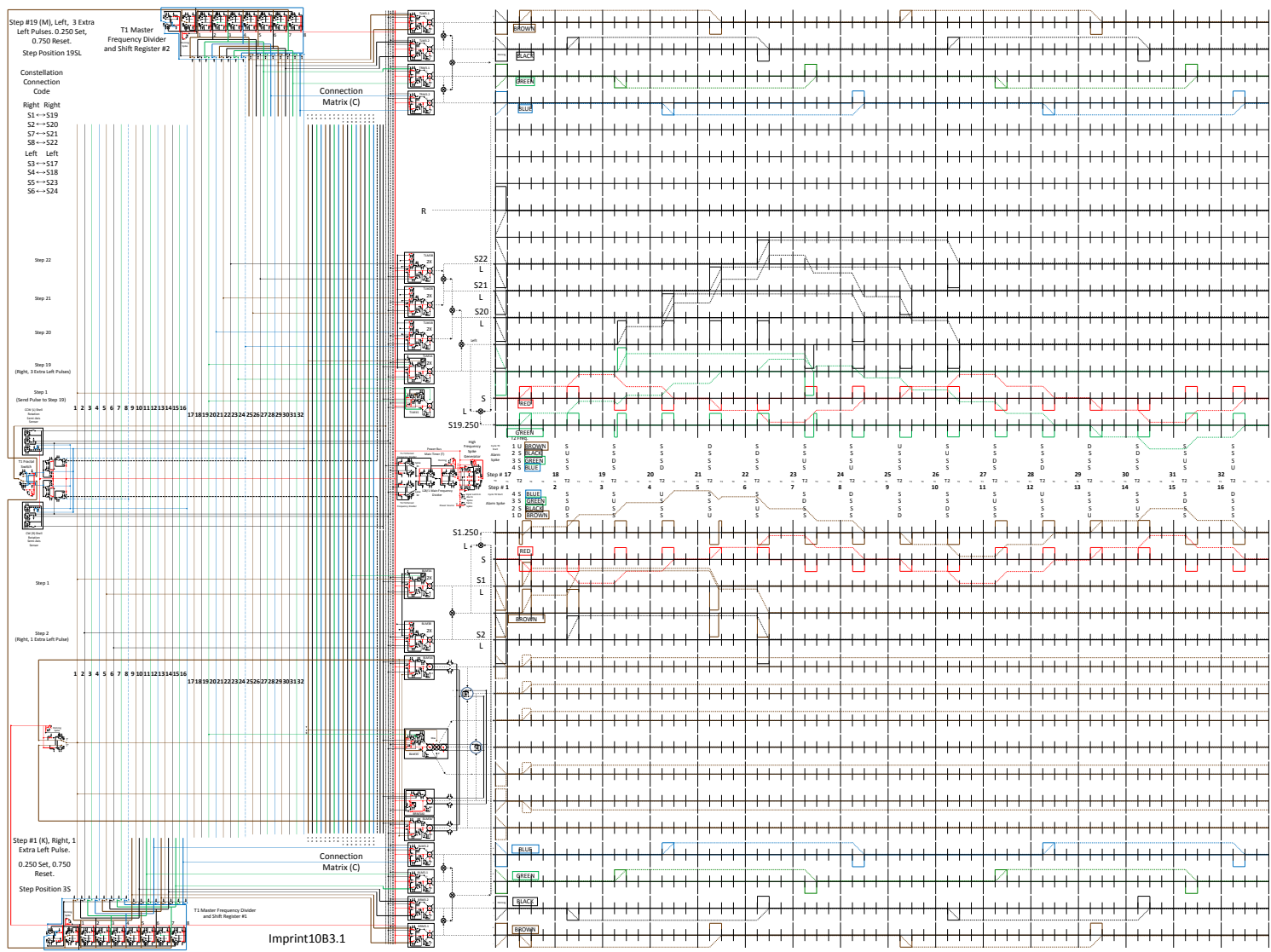

**Figure** 72 **The *contact time with the first object* is recorded in the bottom matrix in Step 1, and is used to restore the orientation of the fractal during the operation of the top matrix in Step 19.**

The sense cell BLM3A senses the presence of an object at 0.250 sec. into Step 1, and produces a 0.750 sec. 2X pulse that stops the right turning rotary motion, and starts a left turning rotary motion for the remainder of the spike period. The BLM3A' cell produces a 0.750 sec 1 X pulse, and the BRM3A0 cell produces a 0.750 sec 1X pulse in the opposite direction of the 0.750 pulse from the BLM3A' cell. The result is used to set the timer in the BLM3S' so that it will send the original contact time (0.250 sec.) of the object to the co-cell TLM1A in the other matrix when it is made active. This will cause the organism to return to the original start position, and contact the original object again if it does not move.

3.7.7  *Second contact with the object*

If the object remains stationary, and the organism does not slip during its motion, the organism will return to the object at the same time (0.250 sec) into Step 33 and the BLM3A actuator will repeat its original behavior. The BLM3A" actuator will start producing a positive pulse and send a spike to the BLM3A' actuator, simultaneously, causing BLM3A" to reset in the negative direction. This cancels the positive BLM3A", exactly. Since BLM30 has been rendered inactive, no change is made in the value of the send actuator BLM3S'. So the organism will repeat its original behavior.

If the object advances along the centerline of the fractal by 1/8 of an inch, the organism will contact the object slightly later at 0.375 sec. into Step 33, as shown in Figure 73. This will cause BLM3A" to produce a pulse that starts later at 0.375 sec. This second positive BLM3" (0.625) pulse is subtracted from the first longer negative (0.750) reset pulse of BLM3A'. This results in a 0.125 sec. pulse that is the difference in time between the original contact and the second contact. This result is used to increase the time setting of the BLM3S' cell to 0.500 sec, which is twice the time difference

Spike timing

the first and second contact, as shown in Figure 73.

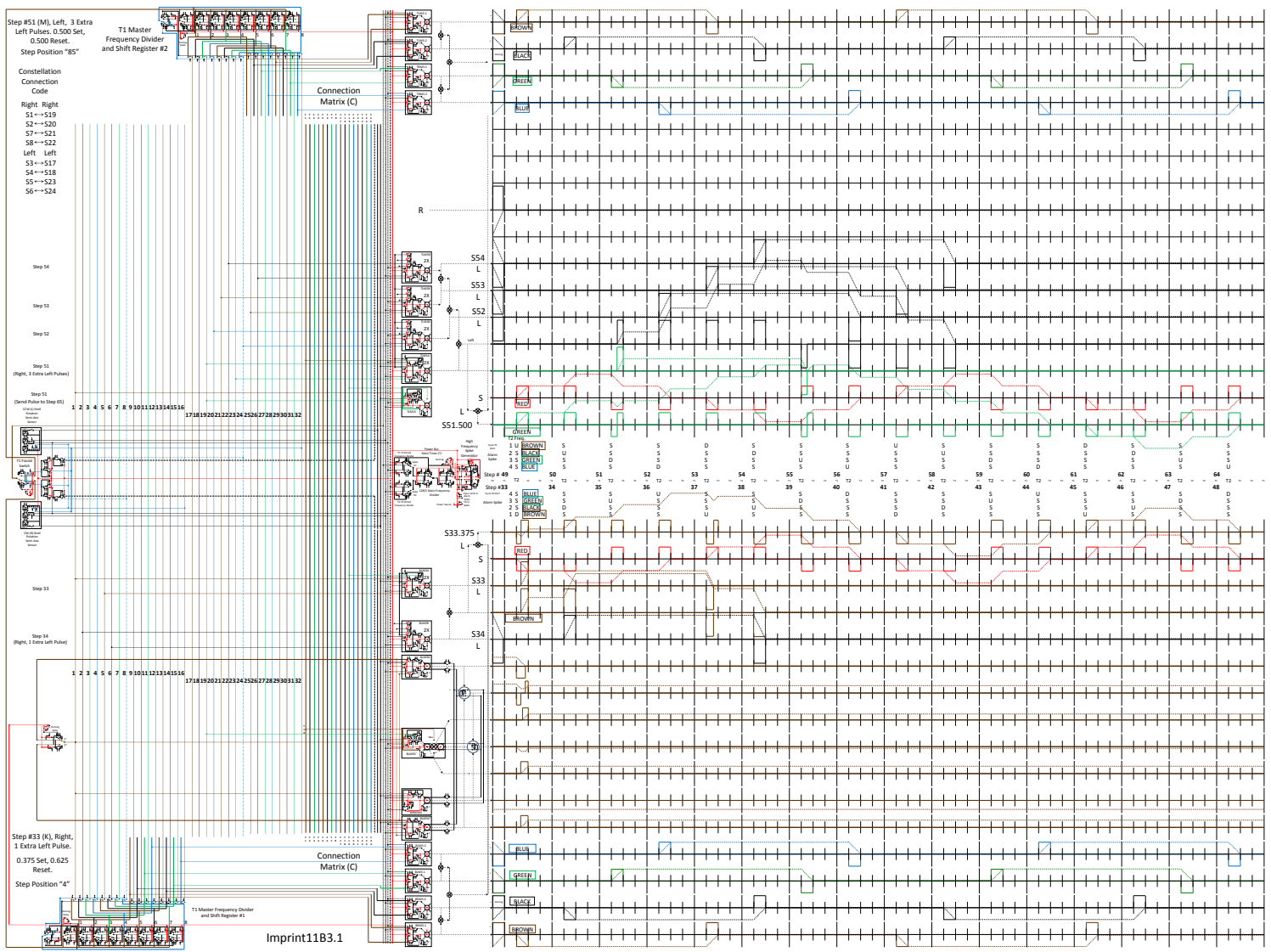

**Figure 73** The *second contact with the object* **causes the return spike time to be twice the difference between the first and second contact time.**

The send actuator BLM3S' causes the TLM1A actuator in the top matrix to produce a spike at 0.500 sec. in to Step 51. The trajectory of the return fractal brings the organism to the new position determined by the movement of the object.

3.7.8   *Third contact with the object*

If the correction is carried out successfully, the organism will contact the object at the same time within the spike period as with the original (0.250 sec.) contact. This will cause the new trajectory to be same as the original trajectory. But it will take place in the new location of the fractal determined by the new position of the object. The time setting of the send cell BLM3S' will be returned to the original setting, as shown in Figure 74.

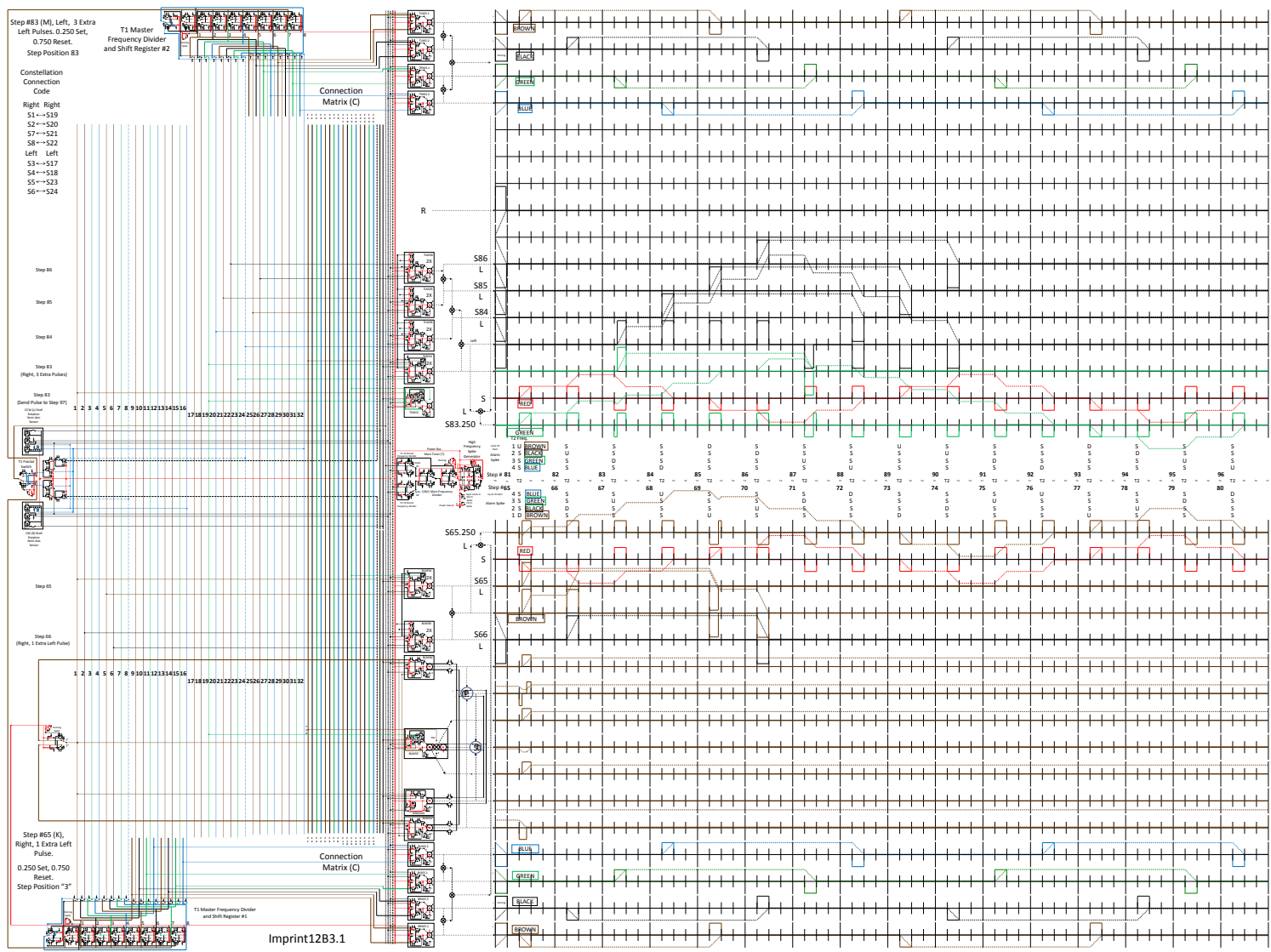

**Figure 74 After the *third contact with the object*, the organism executes the original trajectory in the new position of the object.** The timing of the send cell BLM3S' is restored to the original (0.250) value because the increases positive value of the BLM3A' pulse of 0.750 sec. is greater than the reset pulse of BLM3A'' of 0.625 sec. This increases the value of the pulse in BLM3A' to 0.750sec from 0.625 sec. (The value of the pulse is one minus the object contact time). Also, this process is used to correct for any slippage in the footing of the animal, so the animal always comes back to the same geographical location around the original object. If the object doesn't move, the animal will come back to the same location in Steps 1, 33, 65, and 97, …, as well as following a moving object it has contacted previously.

If the object disappears, the system will continue to make the 0.250 restoration response without the contact set reaction. This will cause the animal to move repeatedly in the direction of the restoration vector (searching for the lost object).

### 3.7.9   Section summary

A system can follow a moving object in its path by adding the contact time difference between the new position of the object and original position of the object to the inverted value of the reaction to the new position of the object in a process called imprinting.

### 3.8   Attachment to an object moving at right angles to the path of the fractal

The object may move at right angles to the path of the trajectory, or the animal may slip causing the path of the trajectory to move at right angles away from the object. This requires some additional sensors and actuators that move the trajectory so the object is contacted in the middle of the sensor face, which lies on the centerline of the basic fractal (See Figure 76).

### 3.8.1 *Motion of the fractal when the object moves at right angles to the fractal*

When contact is made with an object that is not on the centerline of the fractal, small additions to or subtractions from the Shell Out/In motions can be carried out to the end of the fractal that increases or decreases the magnitude of the Displacement Vector (v), as shown in Figure 75.

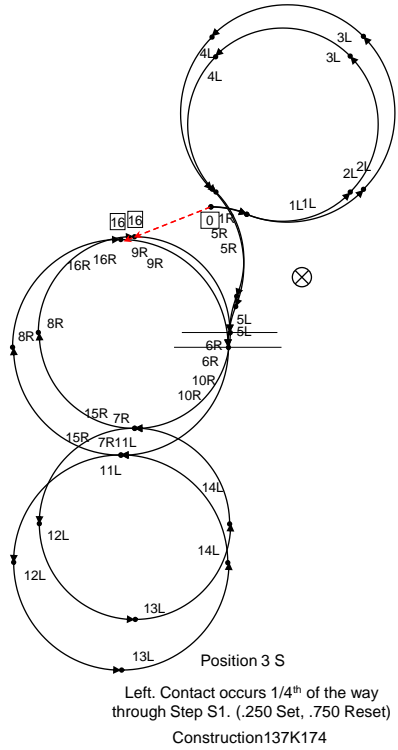

**Figure 75** An *increase or decrease in the Out/In motions* increases or decreases the magnitude, but not the direction, of the **Displacement Vector.**

Increasing the one inch radius of the arcs by 1/8 of an inch changes the length of the Start /End Vector by 1/8 of an inch. Since the return vector, after sixteen steps, will remain the same, the location of the fractal will shift 1/8 inch to the left. This will cause the object to be contacted closer to, if not exactly on its centerline.

### 3.8.2 *Sensing the position of the object at right angles to the path of the animal*

A second set of sensors can be added to both the left and right side of the organism that can be used to determine the position of an object away from the centerline of the fractal, as shown in Figure 76.

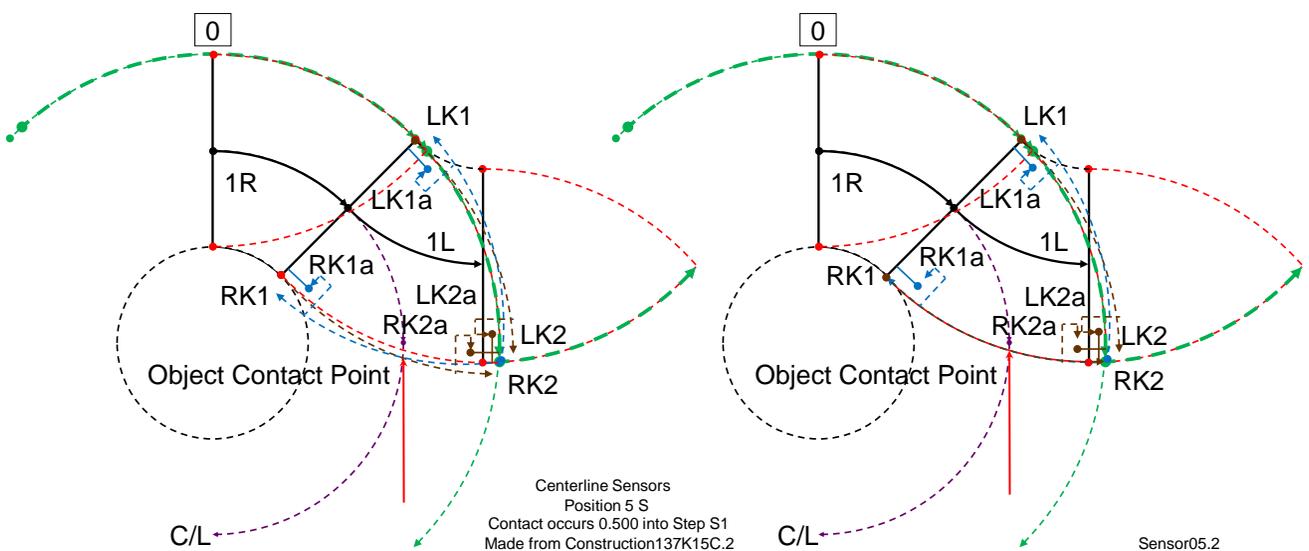

**Figure 76** A *second set of make" sensors* (RK2 and LK2) are added along with four "break" sensors (RK1a, LK1a, and RK2a, LK2a) are added to help determine if the object is contacted inside of the centerline of the trajectory path or outside of the trajectory path.

If the object is contacted inside of the centerline of the path, the Inner Contact Arm (RK1a) will break first. If the object is contacted inside of the centerline, the Outer Contact Arm (RK2a) will break first. If the object is contacted right on the

Spike timing

centerline, both contact arms will break at the same time. The contact break causes the Contact Arm Switch Contacts to open instantly, starting the timers shown in the circuit diagram below. The timers will run until the contact arms stop moving (at the same time) when the Main Contact Arms (RK1, RK2, or LK1, LK2) senses the presence of the object, and stop the forward movement and rotation of the shell.

### 3.8.3 *Circuit needed to measure the offset from the centerline of the path of the fractal*

The measured time between the break and make of the Contact Arm Switch is recorded by the timers shown in Figure 77.

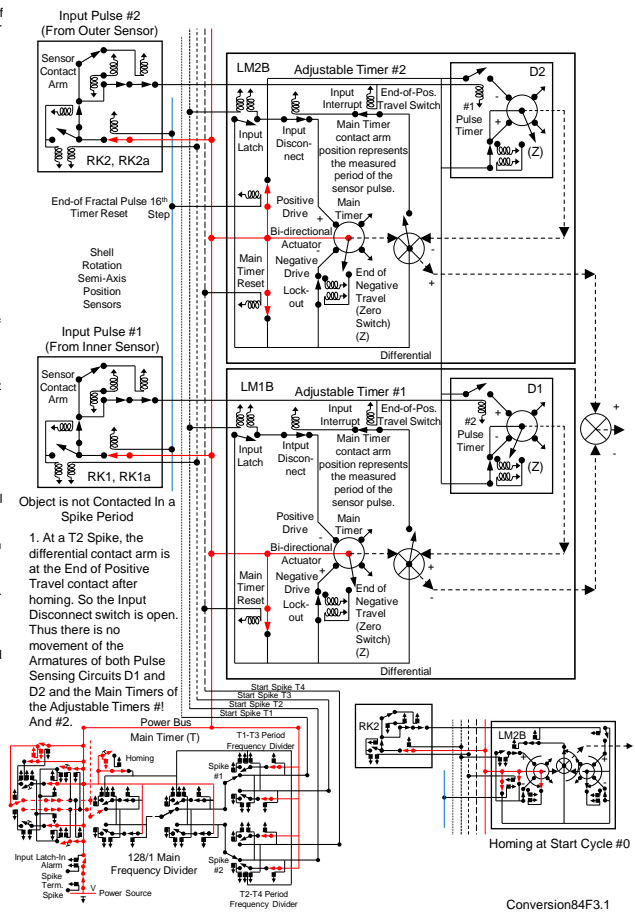

**Figure 77** The *path size is changed* by the circuit that measures the time difference between the Outer and Inner Contacts when they encounter an object, and produces a small pulse correction to the Out/In Axis in each step.

The time values stored in the Pulse Timers #1 and #2 determine the position of the contact arms of the differentials. This causes the Main Timers to produce motions proportional to the stored pulse times. The Output Differential produces a net motion equal to the differences in the pulse times produced by the Main Timers. These small motions are added to or subtracted from the Shell Out/In Axis, which changes the magnitude of the Shell Out/In Axis. This changes the size of the figure-eight.

### 3.8.4 *Measuring the deviation of the position of an object from the centerline*

The difference between the start of contact with the Outer Sensor and Inner Sensor is shown on Figure 78 as the pulse on the middle graph. The width of the pulse is determined by the degree of offset of the object from the centerline of the fractal, and the polarity of the pulse is determined by the position of the object outside or inside of the centerline.

Spike timing

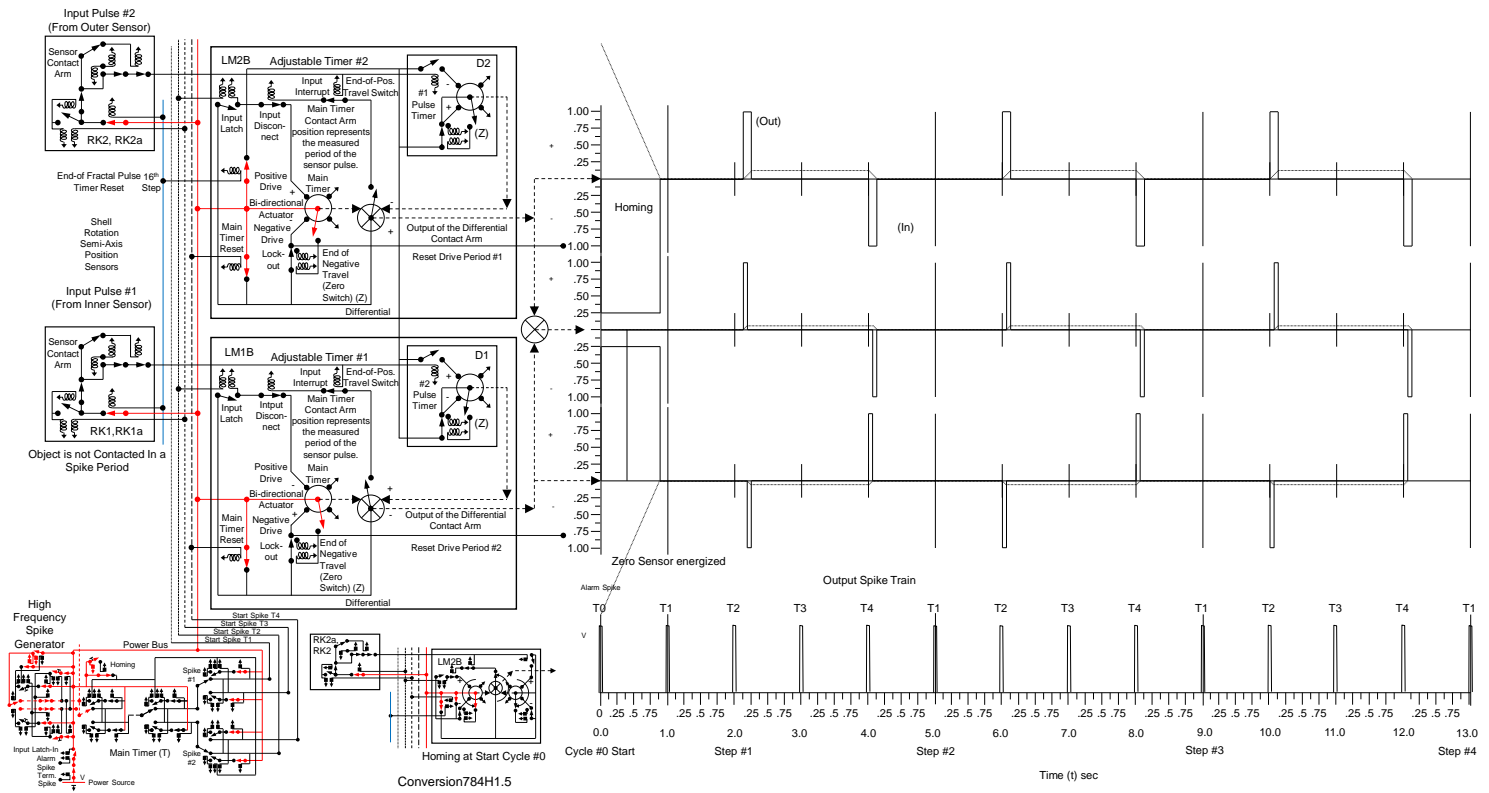

**Figure 78** The *output of the differential contacts* can be seen as two pulses of opposite polarity.

The difference between the two pulses is shown by the output of the differential in the middle graph, which represents the *deviation of the object* from the centerline of the fractal.

3.8.5  *High frequency operation of the deviation measuring circuit*

If the deviation sensing circuit is made to operate at a higher frequency, a series of pulses can be generated during the period of the "Out" and "In" motion, as shown in Figure 79.

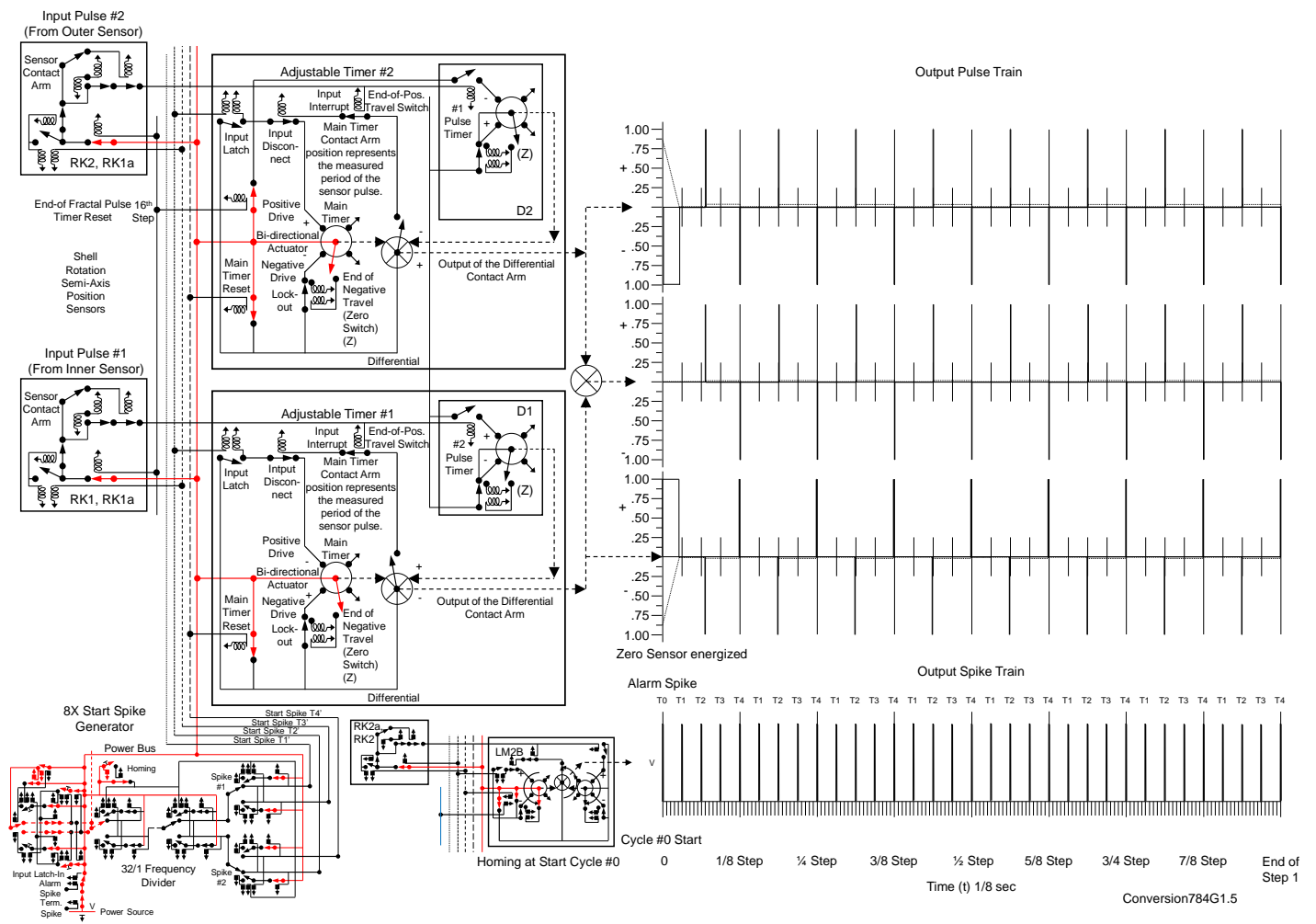

**Figure 79** *Multiple pulses can be generated in a single step period* by operating the control system at a higher frequency than the step frequency.

Spike timing

When the frequency divider is run at one-eighth the frequency as the standard 128-to-one used before, eight "Out Pulses" and eight "In Pulses" are generated in each step period. Each of the eight pulses is $1/8^{th}$ of the original deviation pulse shown in Figure 78. So, eight of these pulses are equal to the original deviation pulse.

### 3.8.6 *Using the pulse generators to interrupt the Out/In pulse*

The pulse generators can be rectified by using their reset signal to interrupt the In/Out pulse as shown in Figure 80.

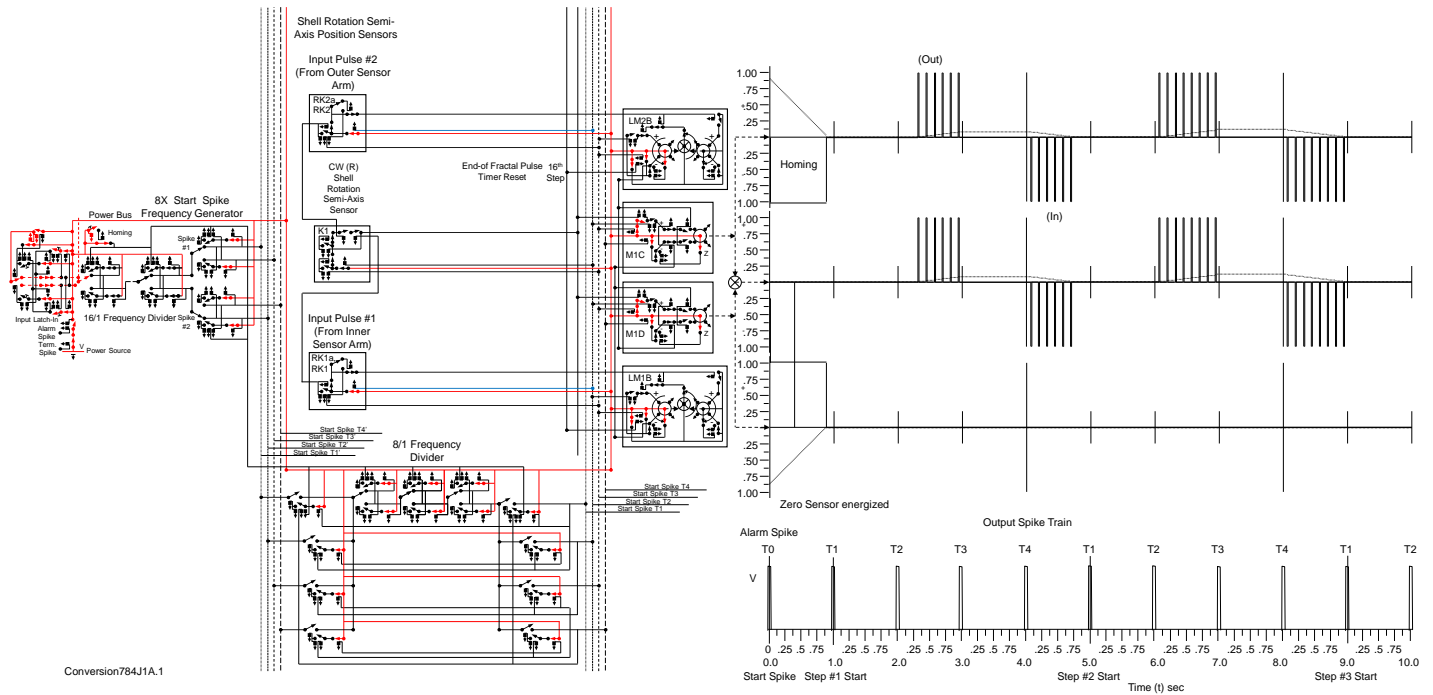

**Figure 80** A *series of high frequency pulses generate a displacement* that is the product of the pulse width times the number of pulses.

The sum of the pulse width times the number of pulses in a step period is used to add to or subtract from the Out/In motion for the remainder of the fractal. In this example, six pulses are produced in the first step because the pulses start ¼ of the way into this Step Period. And eight pulses are produces in all of the other step periods. This changes the size of the fractal, and changes the magnitude of the Start/End Vector (v) according to the measured offset of the object from the centerline.

### 3.8.7 *Control system needed to relocate the fractal so it contacts the object on its centerline*

The memory cells shown in Figure 80 can be placed into the control system diagram that includes the resulting pulse trains, as shown in Figure 81.

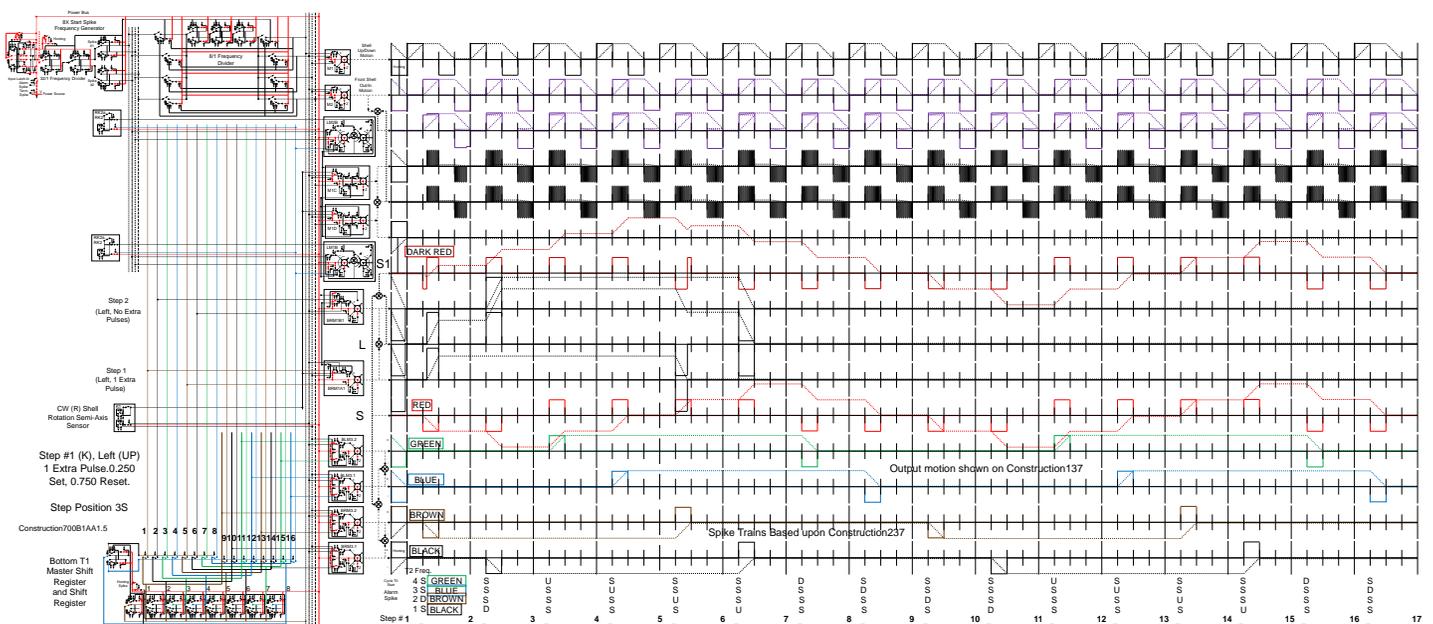

**Figure 81** The *output of the pulse generator* is added to or subtracted from the output of the Out/In Axes after contact is made with an object according to which side of the centerline contact is made.

Spike timing

The pulse generator operates at eight times the frequency of the (S) spike train. When an object is encountered, the RK2a sensor is energized before the RK1a sensor if the object lies outside of the centerline. This causes a positive pulse to be produced with a pulse width determined by the time difference between the pulses produced by RK1a and RK2a. The spikes that produced these pulses are sent to the Out/In actuators M1C and M1D.

If the RK2a sensor operates first, these spikes enable M1C to generate a short pulse coming from LM2B. Then RK1 starts LM1B to generate a pulse that disables M1C and the pulses from LM2B disable M1D so no short spikes are produces by M1D. The result is that a short pulse is produced M1C that is equal to the time difference between the start of the pulses from RK1a and RK2a since both pulses end at the same time when and end contact is made by both RK1 and RK2, simultaneously. Since the time base of the RK sensors and the LM2B and LM1B pulse generators is eight times the frequency as the Out/In axis, eight pulses are produced in a full Step, and fewer pulses are produces according to the object contact time. This produces a total displacement equal to the magnitude and direction of the deviation of the path of the animal from the centerline of the original trajectory. This process is basically a negative feedback servo loop using Pulse Width Modulation (PWM).

3.8.8  *Section summary*

A system can re-center itself along the centerline of its original trajectory by creating a series of small pulses determined by the measured time difference between the offset distances on both side of the centerline, and use these small pulses to adjust the Out/In motion to relocate the fractal trajectory.

3.9  *Quadrature*

The control systems shown so far provides avoidance, hefting, or imprinting for all of the possible contact times in all of the possible eight steps in the double figure eights when the contact occurs between around 0.250 to 0.750 sec. within a Step Period. If contact occurs before 0.250 sec. or after 0.750 sec. in a given Step Period, the organism may run into the object again before it completes its fractal trajectory. If the organism makes contact with an object very close to zero seconds into a step, it will not be able to get away from the object at all. It will essentially return to the object in just four steps. In other words it can't get started on the object avoidance, location attachment, or object following process. This timing problem is common in the physical world, such as in starting reciprocating engines, electric motors, and the synthesis of voice signals using digital information. The solution to this problem is to use two phases that are 90 degrees apart, sense when a start event occurs, and select the phase that will work best with the timing of the event.

### 3.9.1   *Objects contacted very early or very late in a Step Period*

If an object is contacted very early or very late in a Step Period, the response will be nearly a complete circle, as shown in Figure 82.

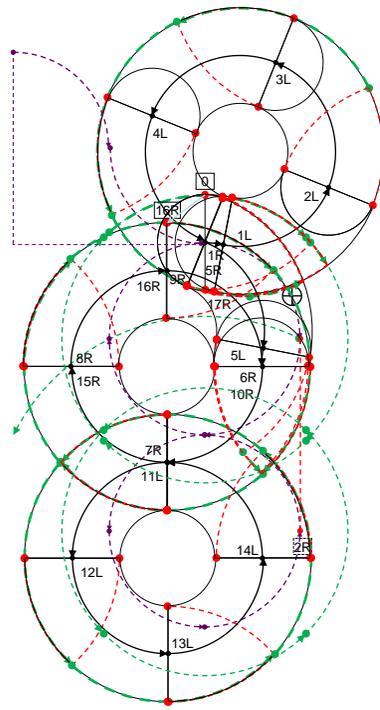

Left. Contact occurs near the beginning of Step 1. (.125 Set, .875) Reset

Spike train similar to Concurrence9751, 752, 753, and fractal on Construction237    Construction137K18P

**Figure 82  The *response to an early or late spike* does not take the organism far enough away for the object it encounters.** The steps following an early or late contact time cause the organism to return to the same point, nearly. If contact is made right at the start of a step, the organism will return to the contact point, exactly, as shown in Figure 83.

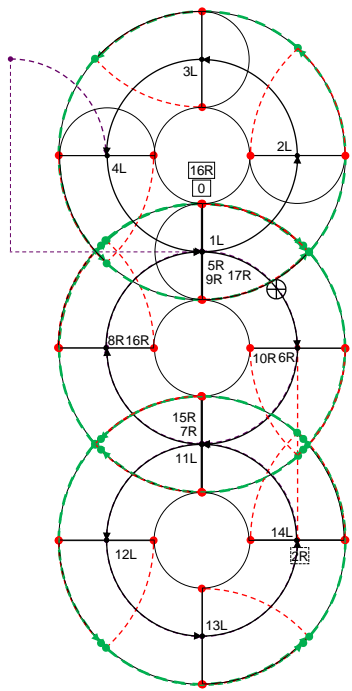

Left. Contact occurs at the beginning of Step 1. (0.000 Set, 1.000) Reset

Spike train similar to Concurrence9751, 752, 753, and fractal on Construction237    Construction137K23P

**Figure 83  When *contact is made at the very start or very end of a step*, the organism returns to the same point, exactly.** This is similar to the problem created when the pedals of a bicycle are directly above and below on another. Pushing down on either pedal does not produce a rotating motion. So the pedals need to be backed around 90 degrees to get the bicycle moving. Because the organism is not able to move away from an object contacted at the very beginning or end of a step, some means must be found to get around this problem if the organism is to thrive by moving in its environment without

Spike timing

getting stuck at one point.

### 3.9.2 *Poly-phase constellation diagram*

A second "Q" system can be added that operates a half step before or after the original system (S). The second system is often called the "Q" phase because it is 90 degrees out of phase (in quadrature) with another system. The original single-phase constellation diagram in Figure 58 can be modified to include the (Q) phase by replacing the time periods between 0.750 sec. in each step to 0.250 sec. in the next step in the original (S) phase by times between 0.250 sec. to 0.750 sec. in the Q phase, as shown in the poly-phase constellation diagram in Figure 84.

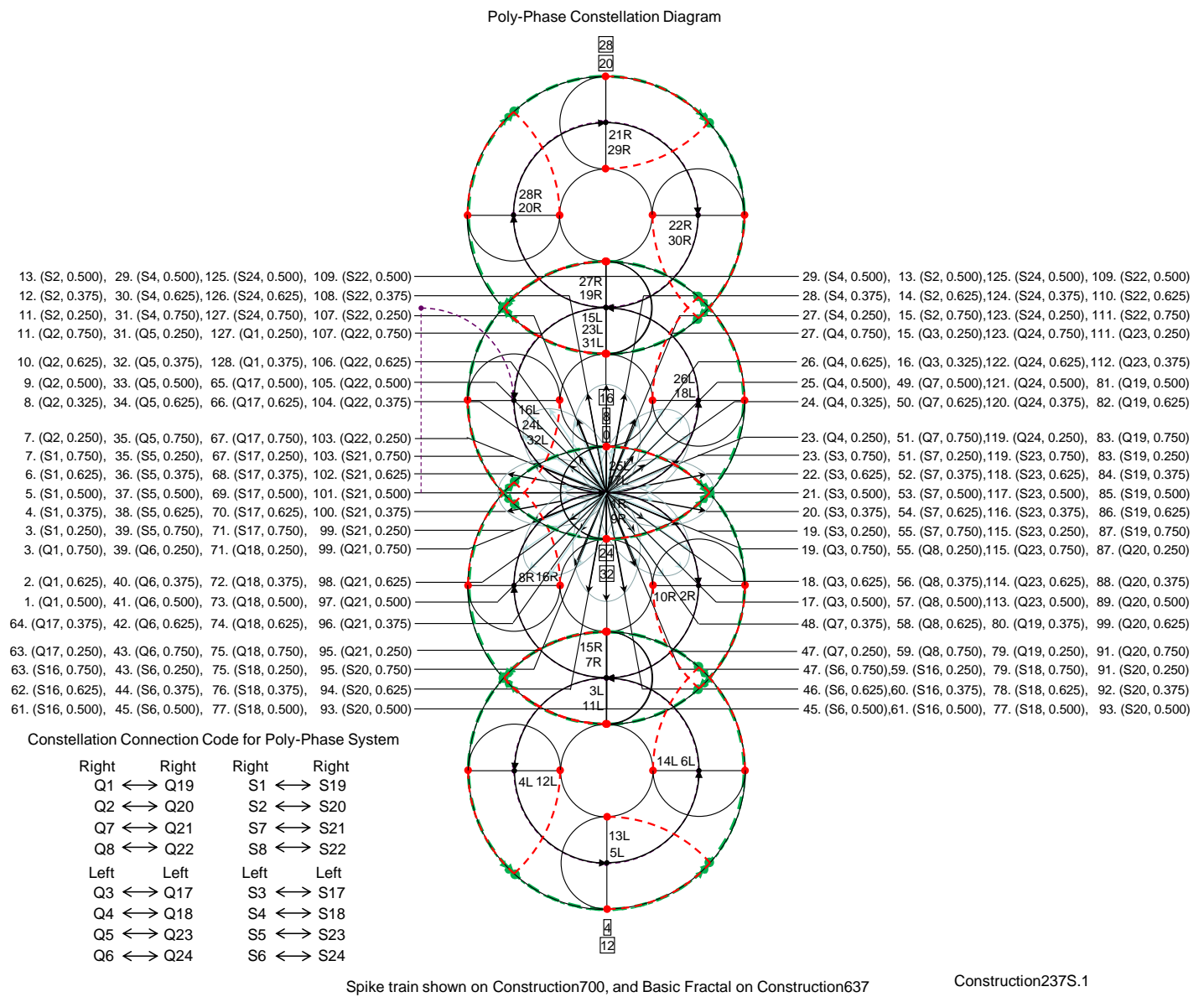

**Figure 84 The *poly-phase constellation diagram* shows the times when the Q or S phase must operate.**

The Q phase leads the S phase. Note that both (S1, 0.250) and (Q1, 0.750) occupy Position 3 above, the same as in Position 3 in the single-phase constellation diagram in Figure 58.

### 3.9.3 *Path after contract in an early Q phase position*

The time of contact of 0.250 sec. into Step 1 in the S phase would occur at 0.750 sec. in the Q phase. This corresponds to Position 3Q, 0.750 on the Poly-Phase Constellation Diagram in Figure 84. The Q phase leads the S phase because the Start Time of the system, which is set to be time 0.000 in Step 1 in the S phase, is 0.500 sec. into the Q phase. This contact time is will result in a hefting value in the Q phase even if contact were to occur right at the Start time of the system, which is one (Homing) step after the Alarm Spike. An example of the path after a contact at 0.750 sec. in the Q phase is shown in Figure 85.

Spike timing

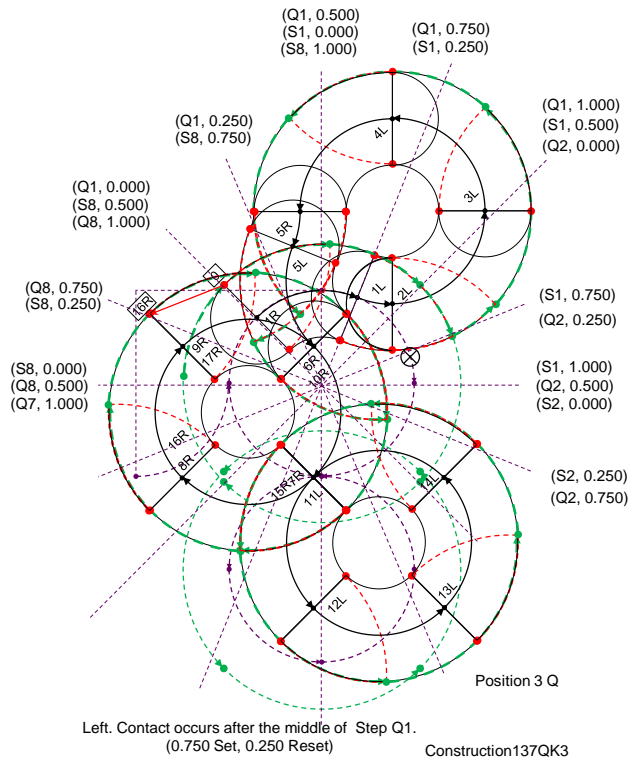

**Figure 85** The *axis of the path of the Q phase generated motion* is tilted 45 degrees CCW from the axis of the path of the S phase using the eight-step figure-eight shown in this paper, but the displacement vector ($v_1$) remains the same in the Q phase as in the S phase for an object in the same position.

Note that the contact point at 0.750 sec. in Step Q1 is in the same position with respect to the original figure-eight as the contact point at 0.250 sec in Step 1 in the S phase shown above. A control system must be designed that selects the Q phase if contact is made slightly before 0.750 sec in the Q phase (0.250 sec in Step 1 in the S phase). And the S phase will be selected will be selected if contact is made just after 0.750 sec in the Q phase. (The trajectory is shifted 45 degrees because the step rotation is 90 degrees. The term quadrature comes from the 90 degree shift in phase, which is one quarter of a revolution in mechanical crank system that rotates 360 degrees. This is one half of the 180 degree stroke, the same ratio as the 45 degree shift in the 90 degree fractal step system.)

### 3.9.4  *Path generated in the Q phase*

The path generated by a contact being made in 0.750 sec. in Q1 (in the Q phase) is shown in Figure 86.

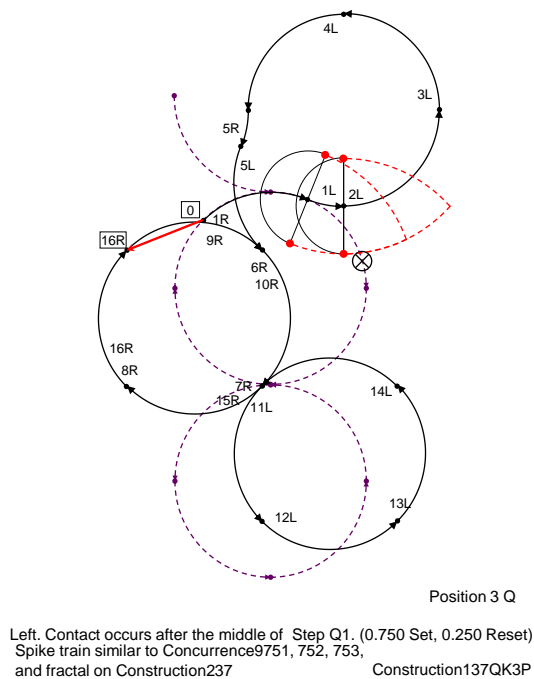

**Figure 86** The *path of motion in the Q phase* is different from the path in the S phase even though contact is made by an object in the same position.

Spike timing

The Q phase allows an avoiding response when contact is made too early or too late in the S-phase.

3.9.5 *Poly-phase control system*

The poly-phase control system requires a timing system with a higher resolution (higher frequency) to allow either Q or the S phase to operate within a given step period. This is because there are six possible start times (T) in each Step Period. The Q or the S phase is expressed depending upon when contact is made with an object in a particular step, as shown in Figure 87.

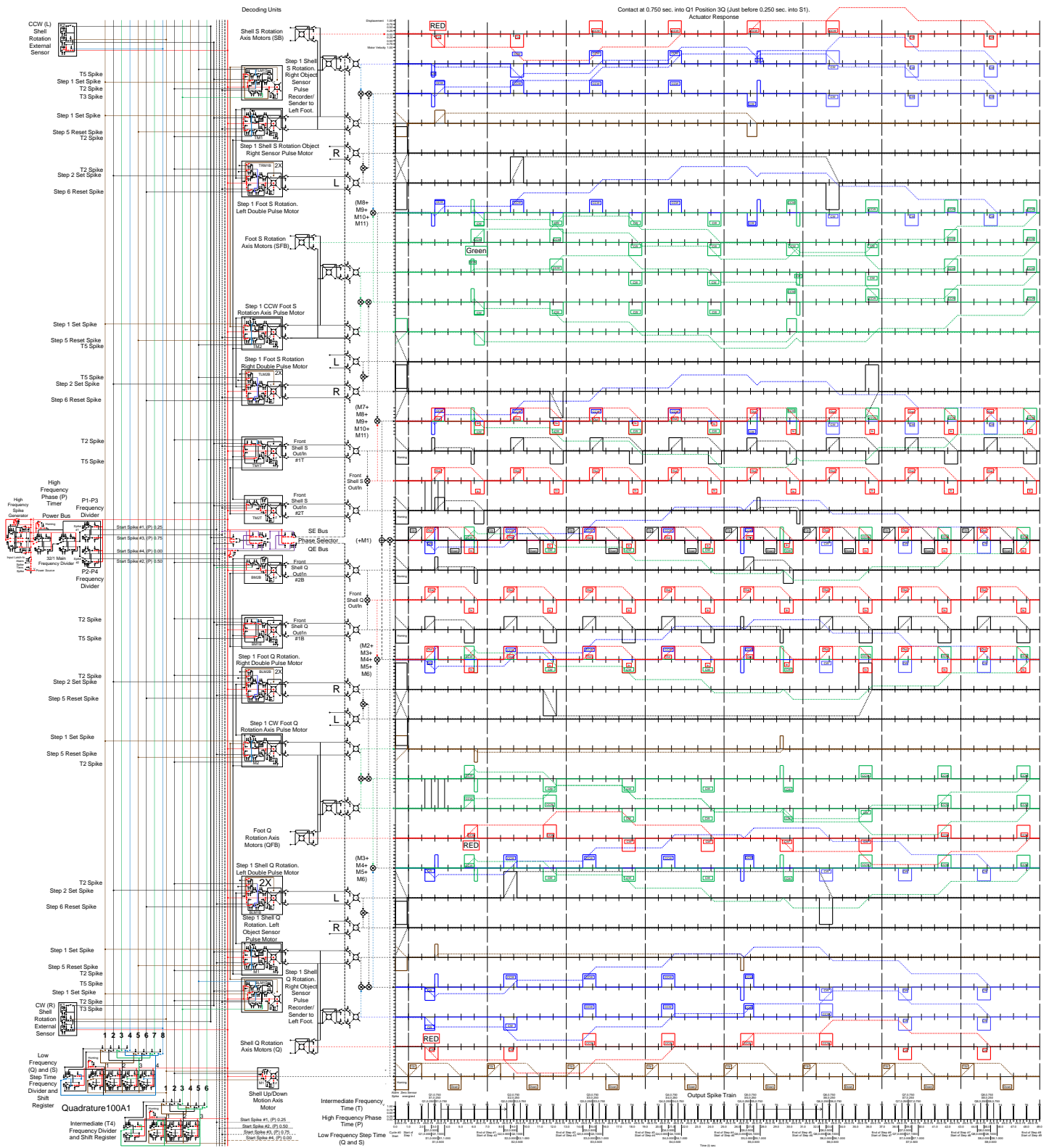

**Figure 87 The *poly-phase* control system selects the Q or S expression bus to produce behavior.**

The Phase Selector measures the contact spike time, and determines which phase has its 0.250 sec and 0.750 sec time slot during that contact spike time. It then activates the expression bus (QE or SE) belonging to that phase. The example shown in Figure 87 assumes the object is sensed just before 0.750 sec in Step 1. The homing process selects (defaults to) the Q phase. So the Q phase is expressed, as shown. If the object had been contacted just after 0.750 sec, the S phase would have been selected. And the behavior shown in the S-phase would have been expressed with a contact time of 0.250 sec in the S-phase instead of a contact of 0.750 sec shown in the Q phase. Once the system selects a particular

Spike timing

phase, it stays on that phase until it encounters an object that requires that it change phase.

The control system needed to move the foot if contact is made with an object is shown in Figure 87 for the first time since the control system shown in Figure 29. As shown on the control system in Figure 29, the foot requires a separate control circuit from the shell. This is shown above as operating in the T5 period instead of the T2 period, and its direction of rotation is in the opposite direction of the Shell. A Sensor Recorder/Sender is required to measure the contact time with the object in the T2 period, and send it to the foot actuator in the T5 period.

The control system for the Back Shell is not shown also because of drawing space limitations. However, the Back Shell motion follows the Front Shell motion by one step, as shown on Figure 29.

### 3.9.6 *Path of the restoring motion from the inverted fractal*

According to the poly phase Constellation Diagram, the restoring spike time is 0.750 sec. in Step Q19, which is Position 83. The sixteen step path of the eight step figure eight with this contact time is shown in Figure 88.

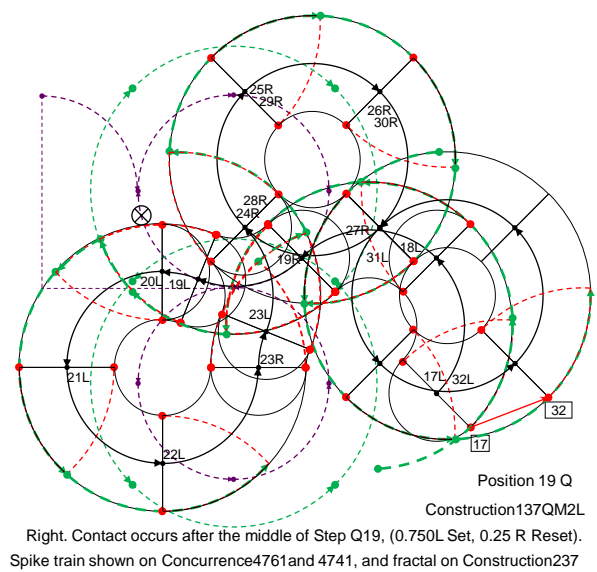

Position 19 Q
Construction137QM2L
Right. Contact occurs after the middle of Step Q19, (0.750L Set, 0.25 R Reset).
Spike train shown on Concurrence4761and 4741, and fractal on Construction237

**Figure 88  The inverted version of Step Q3 produces Step Q 19.**

The vector (v) formed by the starting point Step 17 to the end point Step 32 of the fractal formed by the contact in Step (Q19, 0.750) is the same magnitude, but in the opposite direction of the vector (v) formed by the starting point 0 and the end point of Step16 of the fractal formed by contact in Step (Q1, 0.750). So an object encountered at (Q19, 0.750) will bring the animal back to the starting point of the path were it encountered an object at (Q1, 0.750), as shown below in Figure 89.

### 3.9.7 *Path of the animal when contact is made in Step Q1 and then returns to its starting point by a spike signal made by the control system in Step Q19*

The end point of the path in Step (Q1, 0.750) is connected to the starting point of (Q19, 0.750) by the control system, as shown in Figure 89.

Spike timing

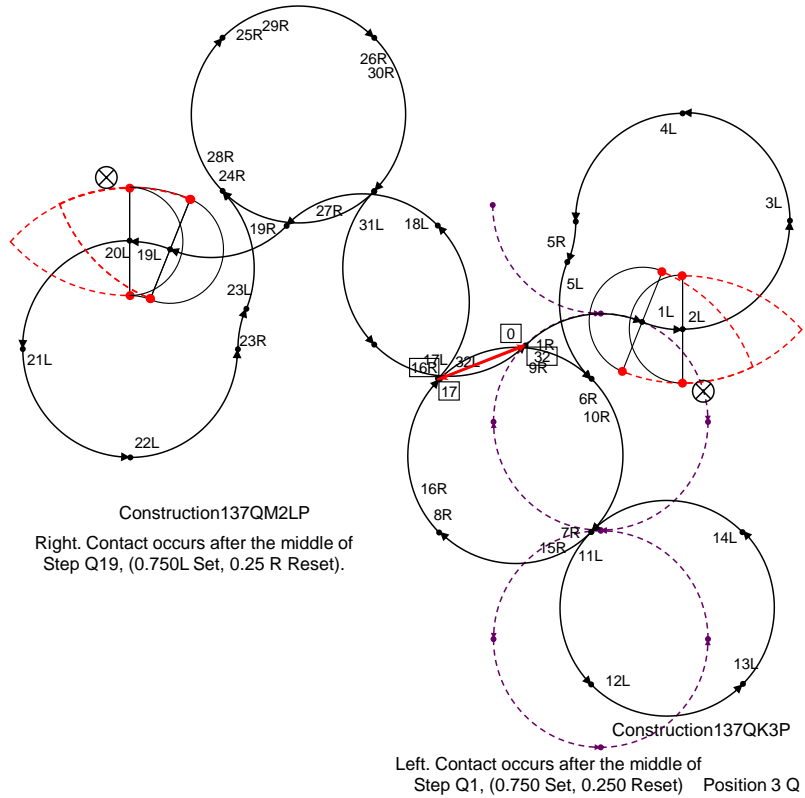

**Figure 89** The *path of the animal after contact is made takes the animal away from the object*, **and the virtual object created by the control system brings it back to the original object.**

The purpose of the control system is to connect Step Q1 to Step Q 19 so that any spike time value between 0.250 sec. and 0.750 sec. within Q1 will produce the same restoring spike time value in Q19. The control system needed to drive the poly-phase system is created by combining the single phase control systems shown previously with the poly-phase control system in Figure 87. Because of the limitations of the paper size in a printed presentation, the remaining circuits will show the S-phase only. This assumes that contacts are made at time between 0.250 sec. and 0.750 sec. in each step in the examples given.

### 3.9.8 *Change of phase*

Since the Q-phase is set at the start by default, the system must change phase if contact is made just after 0.750 sec in the Q-phase. The trajectory for a contact at 0.250 sec. in Step 1 in the S-phase is shown in Figure 90.

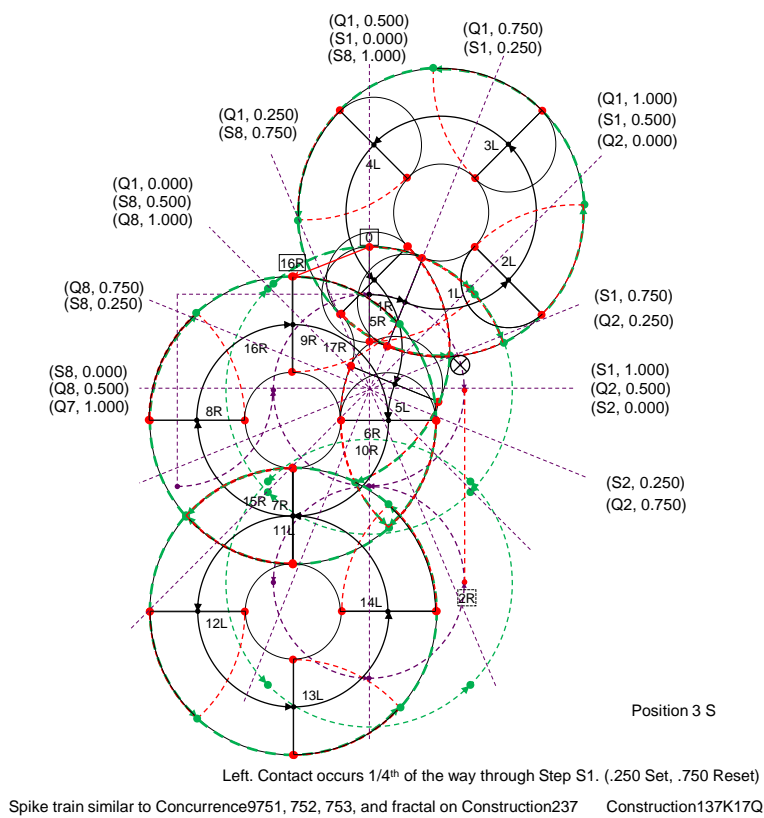

**Figure 90** The *system switches to 0.250 sec in the S-phase* when the object is just passed 0.750 sec in Step Q1.

Spike timing

The object is nearly in the same place. But the trajectory ends in a different position.

### 3.9.9 *Pulse train during a change in phase*

The pulse train produced by the control system in Figure 91 shows how it changes with a change in phase.

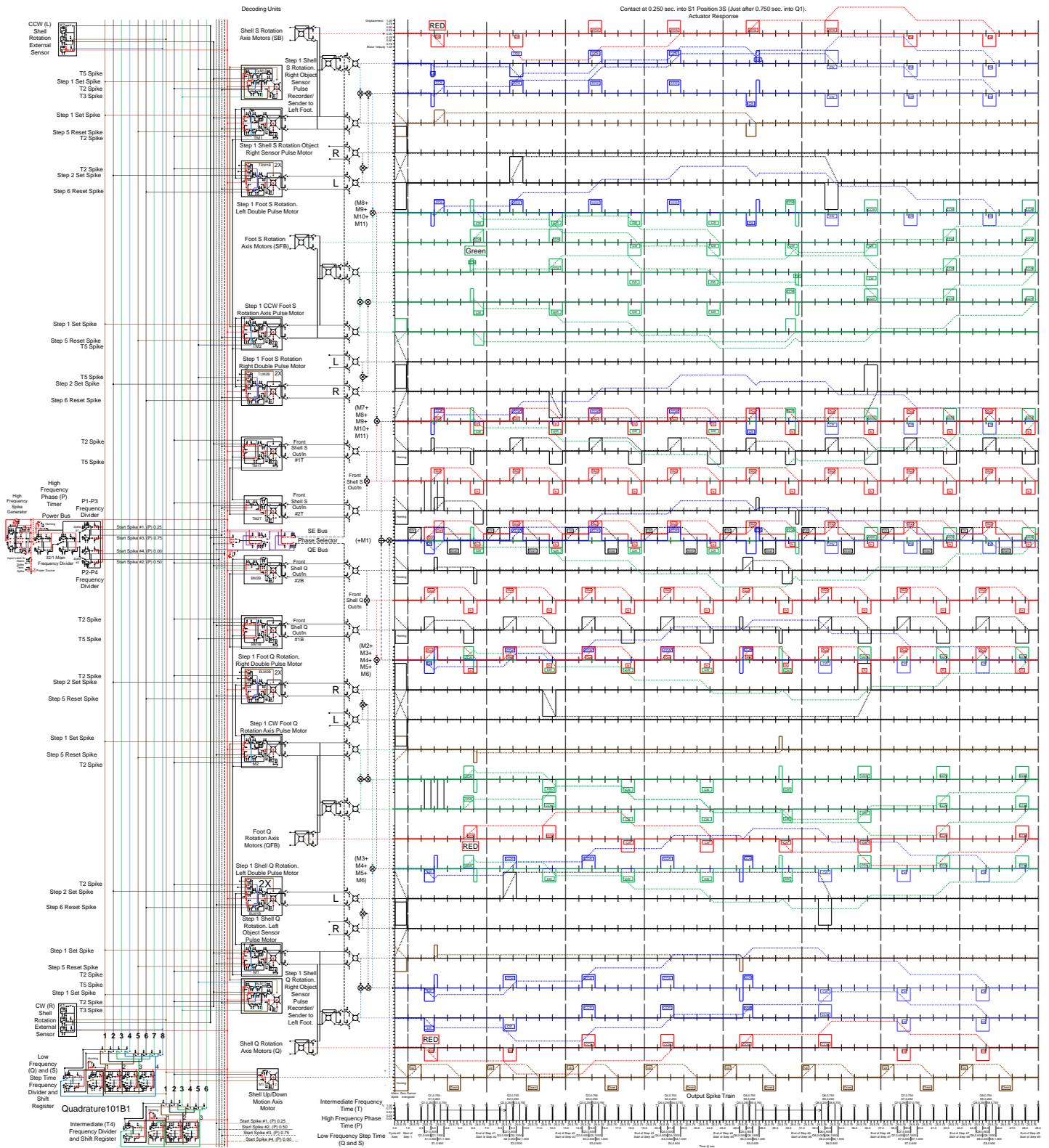

**Figure 91** The pulse train in the very center of the graphs shows how the actuators must respond to a *change in phase*.

One effect of the phase change is that the Out axis has to be able to travel 50% longer during a change from the Q phase to the S phase. This means that 50% more over-travel must be provided in the Out/In axis in the poly-phase system.

### 3.9.10 *Section summary*

Operating two fractal systems that are 90 degrees out of phase allows the system to select the phase that has a contact time nearer the middle of its unit pulse time so the system can avoid making contact second time with the object before it completes its avoidance motion.

*3.10  Dealing with multiple objects*

The control systems shown so far can avoid, heft, and imprint after making contact with just one object during the course of a sixteen step fractal. With only one point of contact, the orientation of the axis of the trajectory can rotate around the point due to foot slippage over time. However, adding two more fractals to the control system makes it possible for the first initial fractal to be put on hold when a second object is contacted before the initial fractal is completed. Then the first of the two added fractals can respond to the second object contact. Then the second of the two added fractals can be restored. And, finally, the second initial fractal can be carried out. This brings the organism back to the original point of origin, and creates a unique trajectory based upon the location of the two objects that are contacted during the course of one fractal. Since two points determine a line in a plane, the line representing the location and orientation of the axis of the trajectory of the animal is determined by the two contact points. So any rotation error in the axis of the trajectories would be corrected by the hefting process.

*3.10.1  Trajectory created by two contact points in one fractal*

Four fractals can create a trajectory that can heft when the animal encounters two objects during the time of one fractal, as shown in Figure 92.

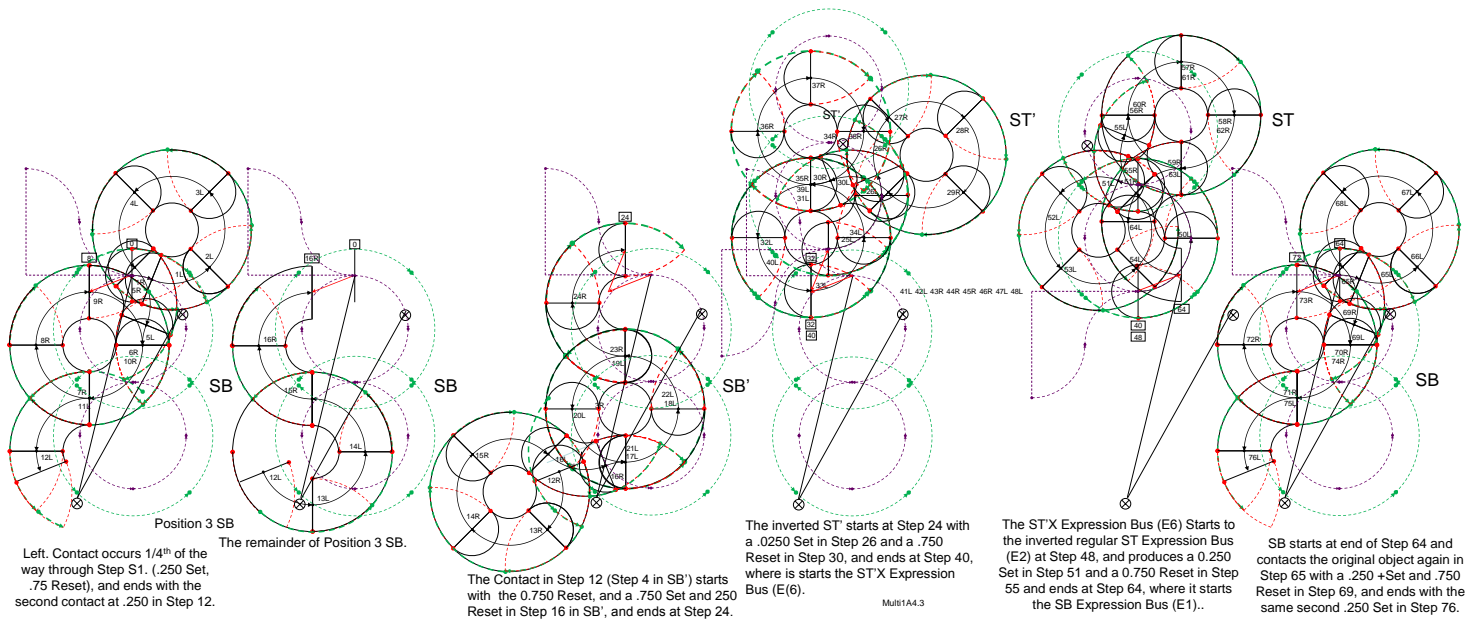

**Figure 92** *Four fractals work alternately* **to maintain the ability of the organism to heft in the presence of two objects in one fractal.**

This process is similar to the process used by a solitary 911 telephone operator, who answers and listens to an emergency call. If a second call comes in while the operator is dealing with the first call, the phone line is switched to the second caller, immediately. The operator deals with the second call, and then returns to the first call when through with the second call. This protocol is based upon the thought that the second call may be of greater urgency than the first. The principle here is that the organism must start to deal with the contact with any new object contact, immediately.

### 3.10.2 Combined trajectory

The trajectories shown above are combined and shown in Figure 93.

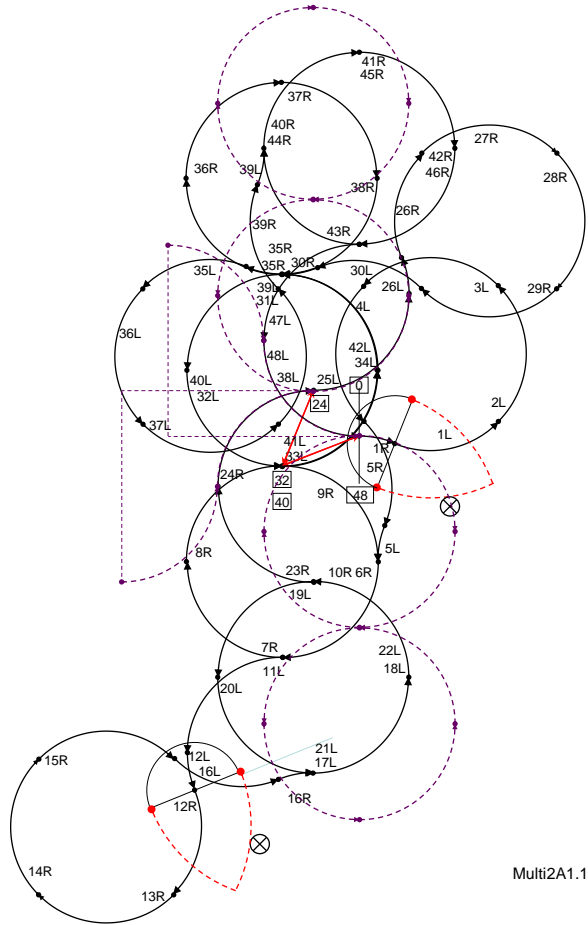

1. Contact occurs 1/4th of the way through Step SB1. (.250 Set, .750 Reset) In Position 3SB.
2. Contact occurs again in Step 12 (Step 4 in S') with a 0.250 set and a 0.750 reset at Step 16 in SB', and goes to Step 24.
3. The inverted ST' starts at Step 24 with a .0250 set in Step 26 and a reset in Step 30, and shifts to goes to ST in Step 35
4. The inverted ST starts at Step 35 using a 0.250 Set and a 0.750 Reset in Step 39, and goes to Step 48..

**Figure 93** *Four fractals are expressed sequentially* **to deal with the contact with two objects in the first fractal.**

The animal model returns to its starting point at the end of Step 48.

### 3.10.3 *Control system needed to deal with contact with two objects during the time of one fractal*

If contact is made with an object sometime in the first eight steps, a response is made by the SB matrix in the manner shown previously. If a second contact is made with an object before the completion of the sixteenth step of that fractal, a circuit is added that transfers expression to another fractal. This is done in a manner similar to that used in the poly-phase system, which can determine which fractal is expressed at a given moment in time. In operation, the Start Signal starts the SB fractal and energizes its E1 Expression Bus. This causes the actions of the SB actuators to be expressed, as shown in Figure 94.

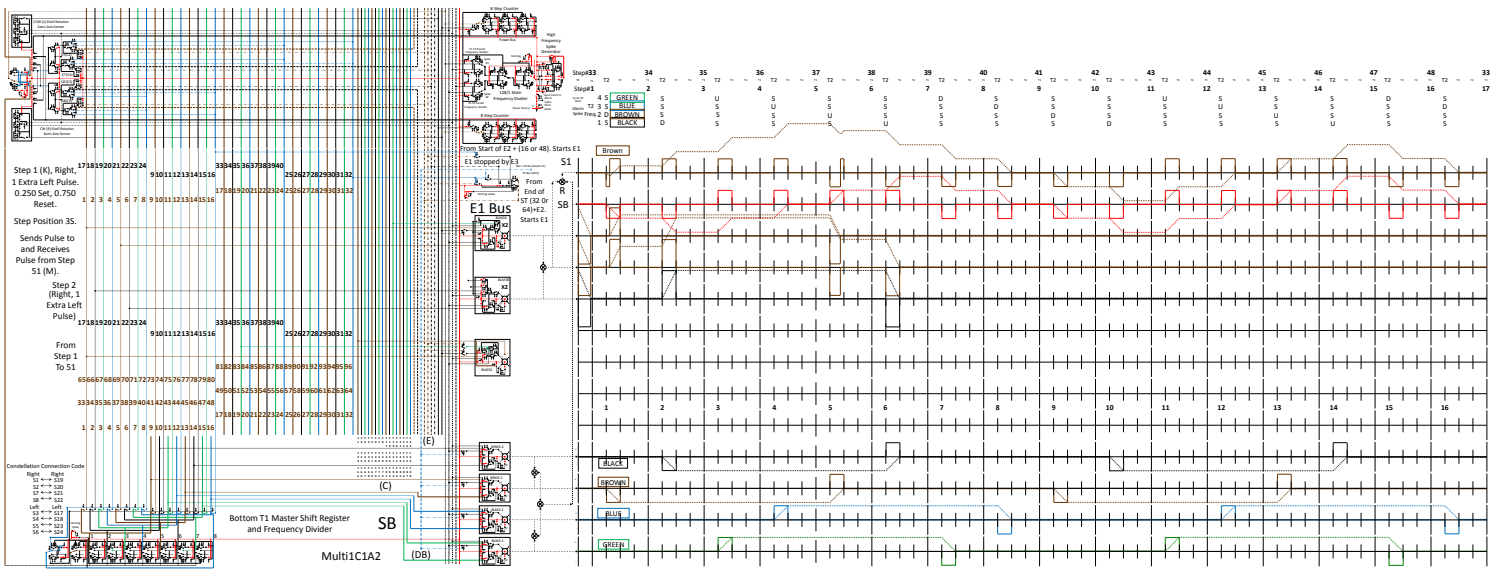

**Figure 94** *The first fractal (SB) starts to express itself* **after the Start Signal.**

Spike timing

In this example, contact is made in Step 1, and the reset pulse for the contact made in Step 1 is made four steps later in Step 5. If no other objects are encountered, the fractal will express itself until the end at Step 16. Then the inverted fractal (ST) will be started, as usual as shown in Figure 64, with the ST expression bus (E2) being energized by the SB expression bus E1 Bus plus Step 16. With no additional contact made, the ST fractal will produce a restoring pulse in Step 19, as shown in the trajectory in Figure 50, and the SB fractal will be restarted in Step 32.

If SB is interrupted by a second contact, its E1 Bus is deactivated. So it cannot activate E2 at the end of Step 16. This causes the sensor spike and E1 to activate the E3 Bus of SB' instead, as shown in Figure 95. The E1 Bus of SB can be re-activated by the E2 Bus at the end of ST, and E5 Bus and the end of ST at Step 32 or Step 64.

3.10.4 *Operation of the control system when a second object is encountered after the eighth step*

If a second object is to be encountered after the eighth step and before the sixteenth step, a second sixteen-step-fractal (SB') is needed. The SB' fractal is started at the eighth step of the original sixteen step fractal (SB), creating a fractal that is eight steps out of phase with the first fractal. But the E3 Buss of SB' is not express until contact is made with the second object (as in the example give in Step 12) between the eighth step and the sixteenth step, as shown in Figure 95.

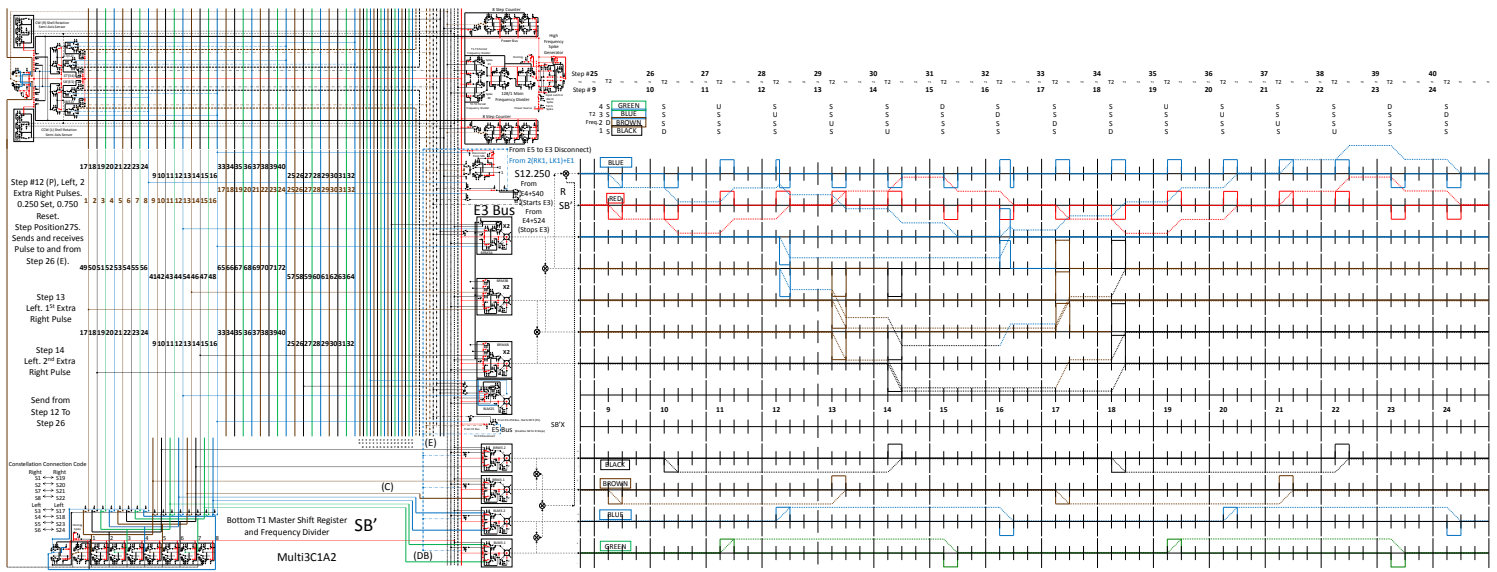

**Figure 95 The *SB' fractal expresses itself from the moment there is contact with a second object in Step 12*, and continues to express itself to the end of the fractal at Step 24.**

The SB' fractal does not start until Step 8 because the T1 pulse train is not sent to the Master Frequency Divider and Shift Register until the start of the 9$^{th}$ Step. It does not express itself until there have been two contacts made in the SB fractal as detected by the 2(RK1, LK1) + E1 Circuit. In the example shown, the second contact is made at 0.250 sec in Step 12. The restoring pulse for the contact made in Step 12 is made four steps later in Step 16. The SB' fractal fills in the remaining eight steps of the first fractal with actuation that can act during the time past the eighth step when the first fractal cannot act. The (SB') fractal completes itself at Step 24. When the E3 Bus of SB' is activated, it sends a pulse to the E1 disconnect of the SB fractal. This de-expresses the SB fractal, allowing it to finish and continue in its figure eight without influencing the output of the system.

### 3.10.5 Restoration of the SB' contact with an additional inverted ST' fractal

The multi-contact motion also requires the addition of an inverted fractal (ST') shown in Figure 96 to provide the restoring pulse needed for hefting after the pulse in SB'. So ST' is started by Step 24 and E3, and continues to express itself until Step 40 due to the activation of its E4 expression bus by the step spike in Step 40.

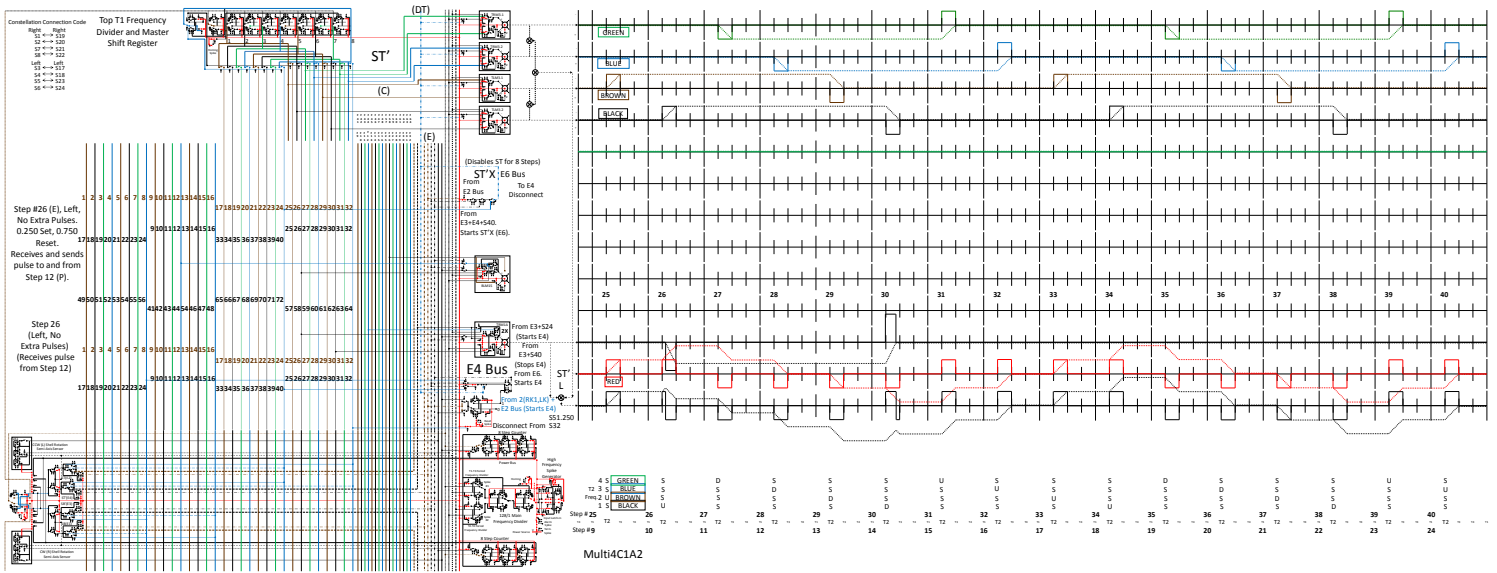

**Figure 96** The *inverted fractal (ST')* expresses the restoring motion starting in Step 26 due to the contact in SB' at Step 12. The reset pulse for the restoring pulse in Step 26 is made four steps later in Step 30. This restores the (') fractals. But the base fractal SB remains to be restored by ST.

### 3.10.6 A circuit that creates a pause

As stated, the first contact example time of 0.250 sec. made in Step 1 in SB still needs to be restored by ST. However, ST must start at the beginning of Step 16 or Step 48 to work with SB. So a pause of eight steps has to be made before ST starts at Step 48, as shown in Figure 100. In music, this would be a "rest" period at the end of a phrase in which nothing is expressed. Because SB is interrupted by a second contact, SB' is activated before ST'. So the E6 Bus of the top rest period ST'X is activated by the E4 Bus of ST' and Step 40, as shown in Figure 100. This activates the ST fractal in Step 48, which restores the trajectory back to its point of origin and causes the first example to be repeated. The rest periods SB'X and ST'X are created by the E5 or E6 Enable Bus at the end of SB' or ST'.

In a different example, if ST is interrupted by a second contact, as shown in Figure 104, ST' is activated before SB'. So the E5 Bus of the bottom rest period SB'X is activated by the E3 Bus and Step 56. This re-activates the SB fractal in Step 64, which restores the trajectory back to its point origin and causes the second example to be repeated. The logic circuits for these two operations are nearly symmetrical.

### 3.10.7  *Restoration of the original contact in SB*

After the top rest period ST'X, the inverted fractal ST and its Enable Bus E2 are started by Step 48 plus E6, and the restoring motion from the first contact in Step 1 from SB is expressed in Steps 51 thru 54 in the inverted fractal ST, as shown in Figure 97.

**Figure 97  The *top fractal (ST) restores the effects of the contact in (SB)* and ends at Step 64. Then expression is returned to the first bottom fractal (SB).**

If there is no slippage, and the object hasn't moved, contact will be made with the original object in the same time in Step 65. Then the whole process will be repeated.

### 3.10.8 Control system that includes all four control systems into one matrix.

The SB, SB', ST' and ST control systems can be combined, as shown in Figure 98.

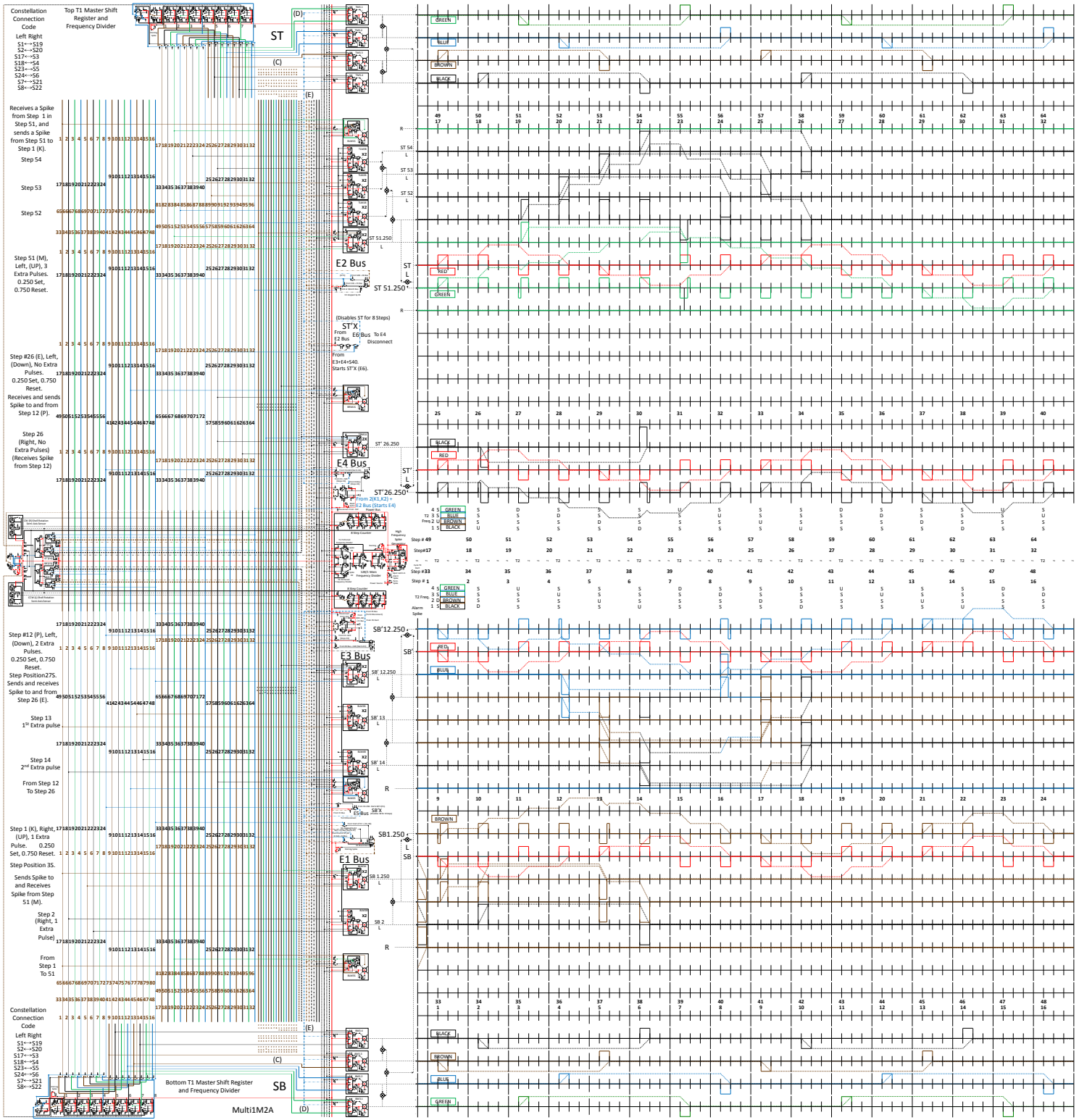

**Figure 98** The combined control system contains an *Enable Bus (E)* that allows each fractal to start and stop other fractals.

Because contact is made in Step 1 and in step 12 of the SB fractal, its Enable Bus E1 is followed by the E3 Enable Bus in SB', the E4 Enable Bus of ST', the E6 of the mute fractal ST'X, and the E2 Enable Bus of ST.

### 3.10.9 Overall Logic circuit that controls the sequence in which the fractals are expressed

The overall circuit that controls the sequence in which the fractals are expressed is shown in Figure 99.

Spike timing

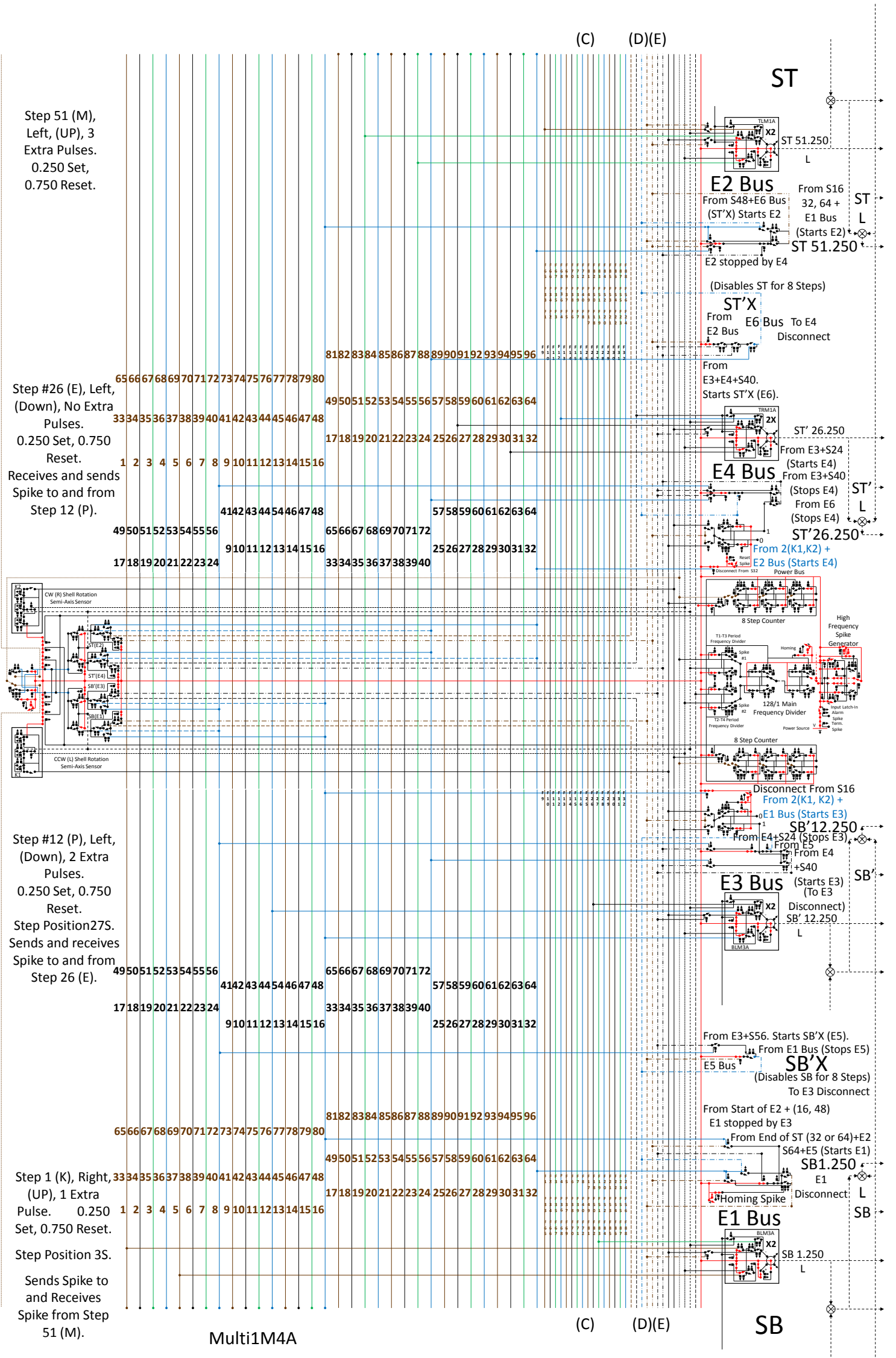

**Figure 99** The *sequence in which each fractal is expressed* is determined by the activation of the two sensors and the logic circuits driven by the six Enable Buses.

Spike timing

As stated previously, the Start (Alarm) signal starts SB and its Enable Bus E1. If contact is made in the second half of SB, SB' is activated. Then, ST' and ST are activated, as shown in Figure 100. ST is not activated after SB because its Enable Bus E1 is deactivated before Step 16.

3.10.10 *Layout of the SB' and ST' matrices*

The original fractals SB and ST have eight "free" steps after the first eight steps in which additional objects can be contacted. So expression can be transferred to SB' or ST'. Since there is an overlap between the expression of SB and SB', there has to be an "underlap" (gap) between Step 40 when ST' ends, and a gap in Step 48 when ST begins. So a rest period is created by the mute fractal ST'X (E6) in this example. The expression sequence is shown more clearly in Figure 100.

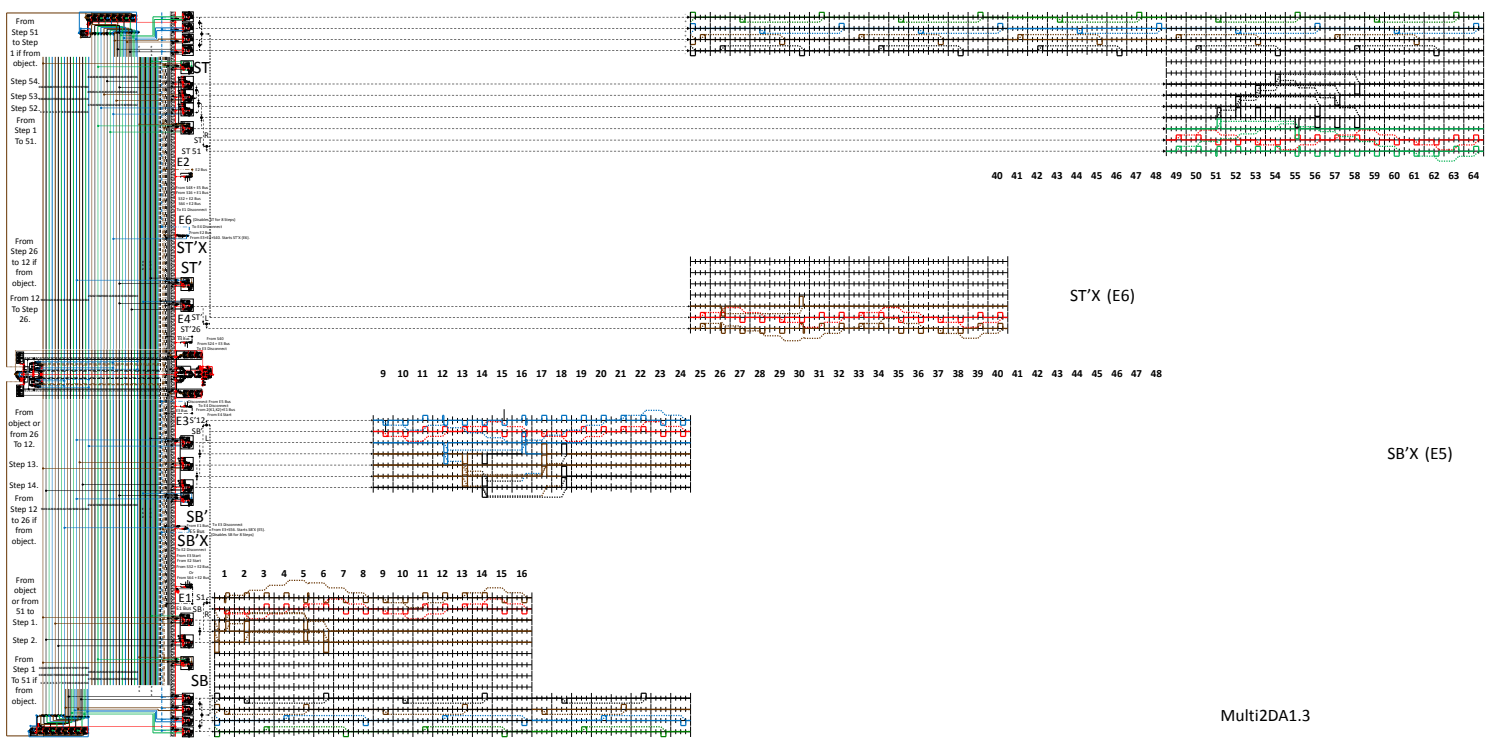

**Figure 100  The *path followed by the fractals* allows a response to be made for each contact with objects, and allows a restoring action to be made for each response. The response and restoration actions bring the path back to its starting point.**

The trajectory for this fractal path is shown in Figure 92 and Figure 93 with the first contact being made in Step 1 of the SB matrix. The first contact could be made in the second (ST) fractal as shown in the next sections.

### 3.10.11 *First contact made in the second fractal*

As stated previously, the first contact can be made in the second top (ST) fractal. This means that the second contact is made in the ST' fractal shown in Figure 101 instead of the SB' fractal shown in Figure 92.

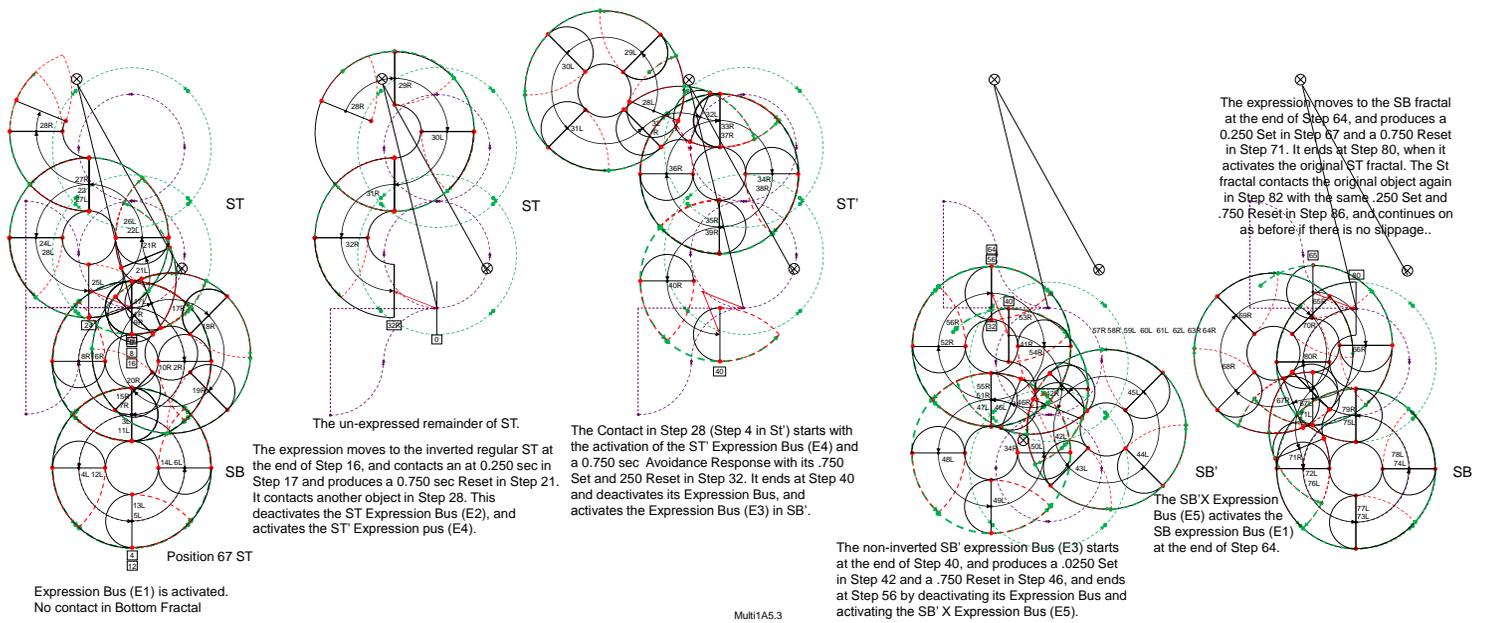

**Figure 101 *First Contact* is made in Step 17 in ST, and the second contact is made in Step 28 in ST'.**

The restoring pulses are made in Step 42 in the SB' fractal, and are made in Step 67 in the SB fractal.

### 3.10.12 *Combined trajectory*

These trajectories can be combined, as shown in Figure 102.

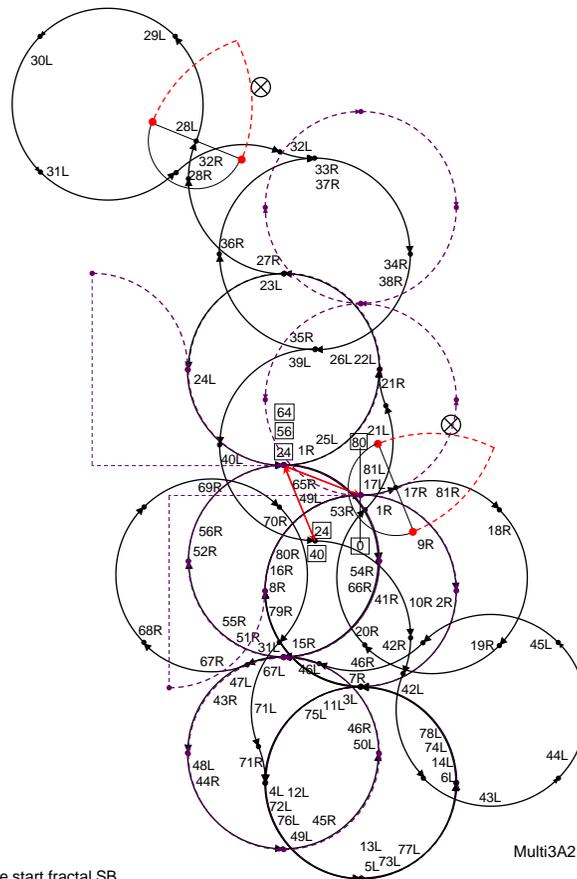

1. There is no contact in the start fractal SB.
2. Contact occurs 1/4th of the way through Step S17. (.250 Set, .750 Reset) In the inverted fractal ST (Position 67ST), and reset in Step 21.
3. Contact occurs again 0.250 into Step 28 in the inverted ST fractal that is completed in the inverted ST' fractal, with a 0.250 set and a 0.750 reset at Step 32 in the ST' fractal that goes to end of Step 40.
4. The non-inverted SB' starts at Step 40 and produces a .0250 set in Step 42 and a reset in Step 46, and goes to end of Step 56.
5. The non-inverted SB starts at the end of Step 64 and produces a 0.250 Set in Step 67 and a 0.750 Reset in Step 71, and goes to the end of Step 80.

**Figure 102 Pulses created in each of the four fractals cause the organism *to avoid the two objects and to return to the point of origin*.**

When the first two contacts can be made in the inverted fractal (ST) instead of (SB) in the example shown in Figure 92 and Figure 93, the (ST') would be expressed before (SB'), as shown in Figure 104.

3.10.13 *Control system that produces the trajectory when the first object is contacted in the top fractal (ST)*

A different arrangement of actuators is made active when the first object is contacted in the Top Fractal (ST), as shown in Figure 103.

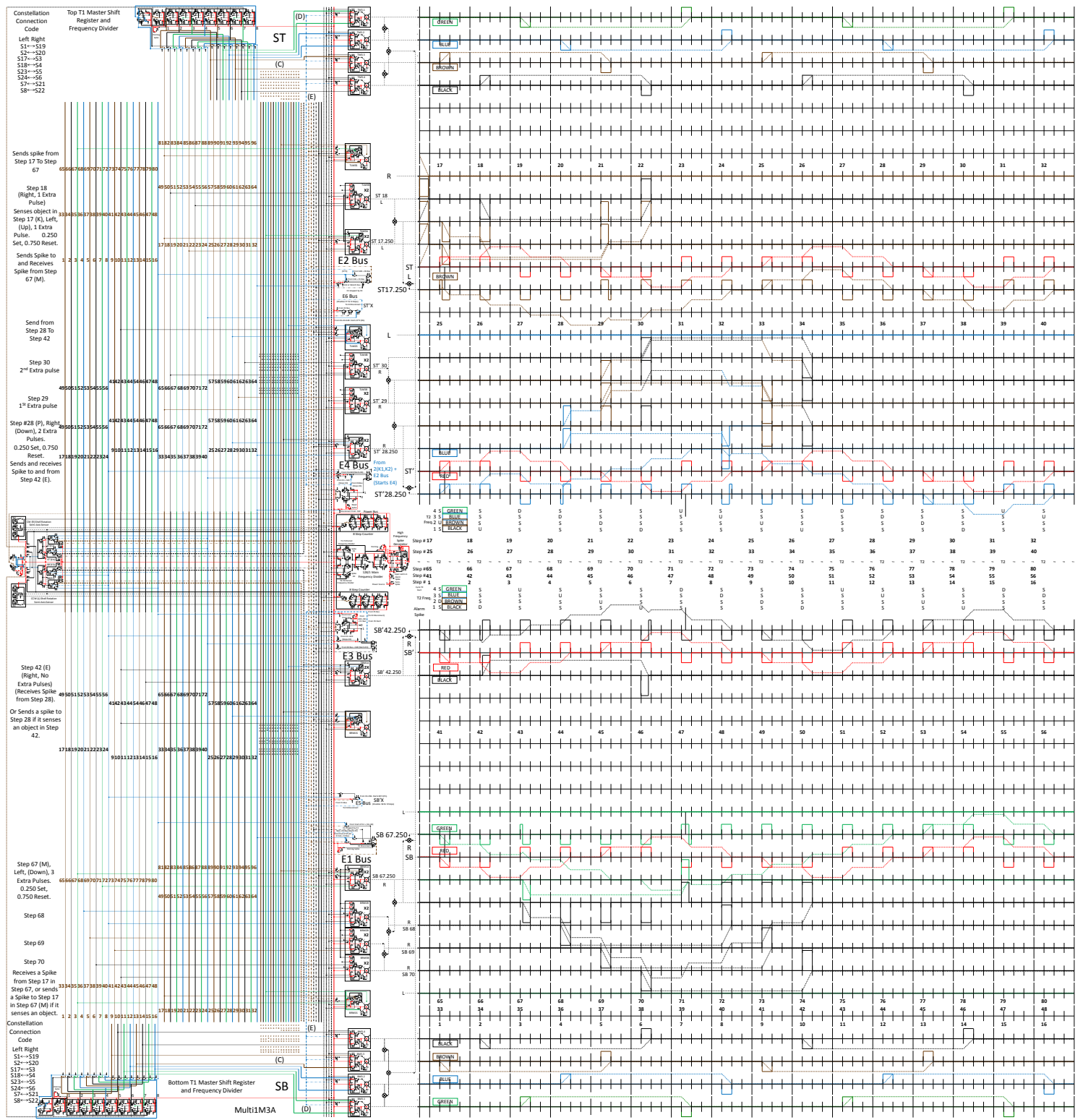

**Figure 103 The four fractals (SB, SB', ST', and ST) are generated in *one complete circuit*.**

When all four fractals are generated in one complete circuit, the Expression Bus (E) can be used to activate each fractal, as shown in Figure 103.

3.10.14 *Activation sequence*

The sequence in which each fractal is activated is shown below in Figure 104.

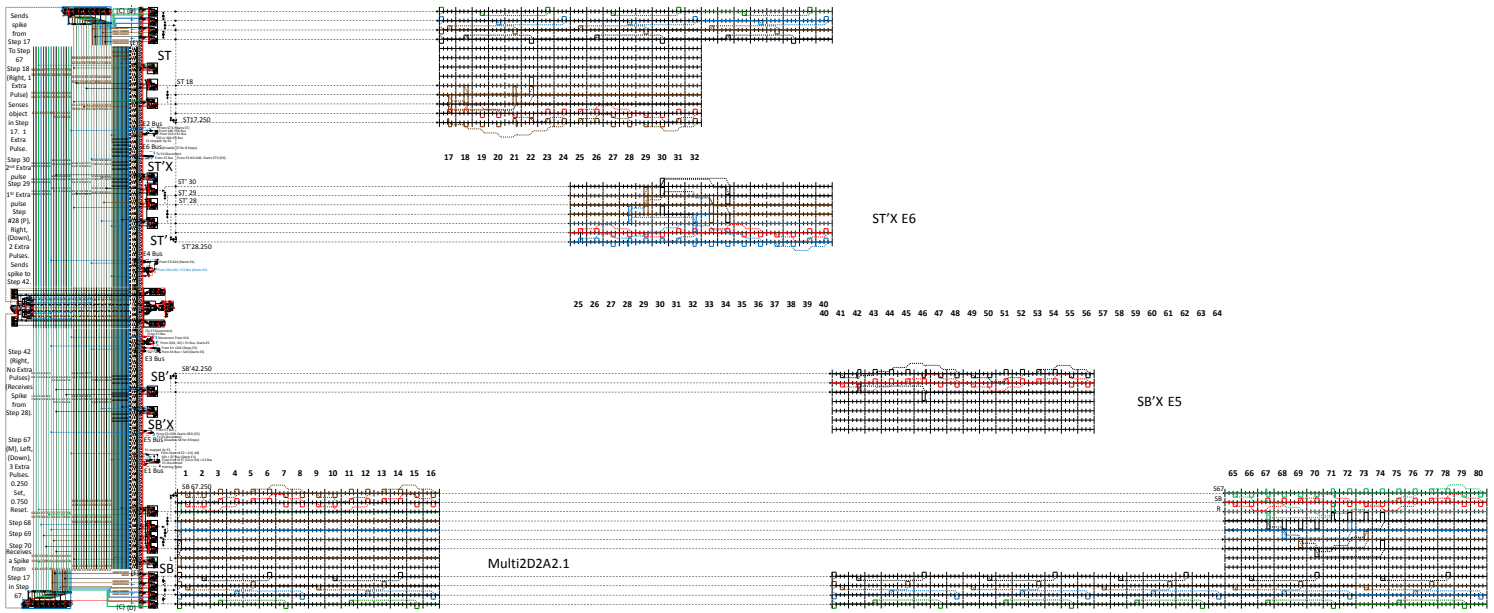

**Figure 104** The *path through the fractals* **is determined by when contact is made with two objects.**

When the two contacts are made in the ST fractal, the path of restoration goes through the SB'X rest period rather than the ST'X rest period.

The logic circuit in Figure 99 creates the different paths needed for any set of two objects contacted in either the SB or ST fractals. The process of activating a fractal and restoring the original location by activating a second fractal is similar to the process of activating a unique pulse upon contact with an object and restoring the orientation of the fractal by a restoring pulse in the opposite direction. What is turned on is turned off. What is done is undone. The system attempts to return to its original internal state. But in the process of doing this, it undergoes a unique activity in its environment.

3.10.15 *Contact made with more than two objects*

The (SB') and (ST') fractals also have eight "free" steps after their first eight steps. So, additional objects can be contacted if additional fractals are provided. These (SB") and (ST") fractals respond to and restore a third object while the first two processes are on hold. So up to sixteen objects can be contacted and dealt with successfully using the two, eight-step fractal shown in this study. The number of objects an animal can deal with is determined by the depth of the fractal nest. With a depth of eight complete systems, as shown in Figure 105, the animal model can make a unique hefting response to eight objects.

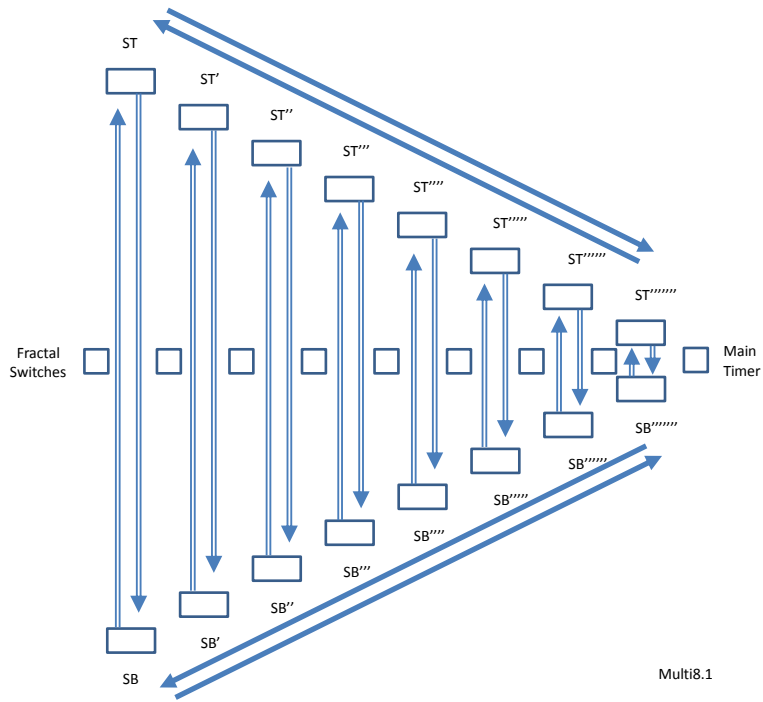

**Figure 105 Contact can be made with objects in *all eight steps* in a set of nested fractals.**

The sequence of activation goes deeper into the nested fractals as more objects are contacted and more fractals are put on-hold. The activation returns toward the beginning of the nest as more fractals are restored than new objects are encountered.

The fractals-on-hold can store information until they are restored. The queue can be used as a path for data retrieval. This process can be used by an animal to return its place of origin (reproductive migration), also.

The control systems shown in this section (dealing with multiple objects) are designed to produce hefting. So they are shown to use temporary working memories created by contacts with objects. Each fractal can consist of all of the cells and support cells needed for each of the eight non-inverted and eight inverted steps. These are shown in the control system in Figure 66. This means that objects can be dealt with in any of the sixteen steps according to the depth of the fractal nest. Even though only one step can be dealt with in a given fractal, other steps in that fractal could be used if needed later.

The control systems shown in the Section 3.7 that produce an attachment to a moving object (imprinting) require permanent memories. These permanent memories can be unmade (restored to zero) only by rebooting the system (applying the Termination and Alarm Spikes). Once a fractal controller has a value stored in it, another value cannot be stored in this fractal controller. So, the elements of the controller not being used by the stored value can be eliminated. These elements could be used to create a new fractal controller that can store any new value. So the process of creating a long-term memory may require the system to stimulate the growth of new controllers as each permanent memory is made.

3.10.16 *Section summary*

Making contact with a second object before the end of the sixteen-step fractal requires that there be a second fractal network that is 90 degrees out of phase with the first network. The second contact puts the first fractal on hold, and activates the second network, which completes its avoidance response and reactivates the first network so it can complete its avoidance response.

3.11  *Remembering the location of multiple objects*

The fixed objects in an environment have a specific relationship with the original position and orientation of the animal, and the basic fractal. If the systems can remember where and when these objects are encountered, it may be possible for the animal to move within that environment without contacting more than the first few objects belonging to the set of objects in a particular pattern. This frees up the sensors so they can deal with new or unexpected objects. This ability would be useful to a Limpet and other grazers. It would be particularly useful for farmers whose fields are littered with unexploded land mines. They could move freely through the fields if they know where the land mines are located. This section shows a process of learning the location of multiple objects in an environment using long-term, proximity memory co-cells, and then avoiding the objects without contacting them, sub sequentially.

3.11.1  *Proximity Sensor*

A proximity sensor is needed to detect an object before it makes contact with the shell. This can be done by adding another pair of contacts (RK1a and LK1a) shown in Figure 106 to the original contacts (RK1 and LK1) in the sensors shown in Figure 41.

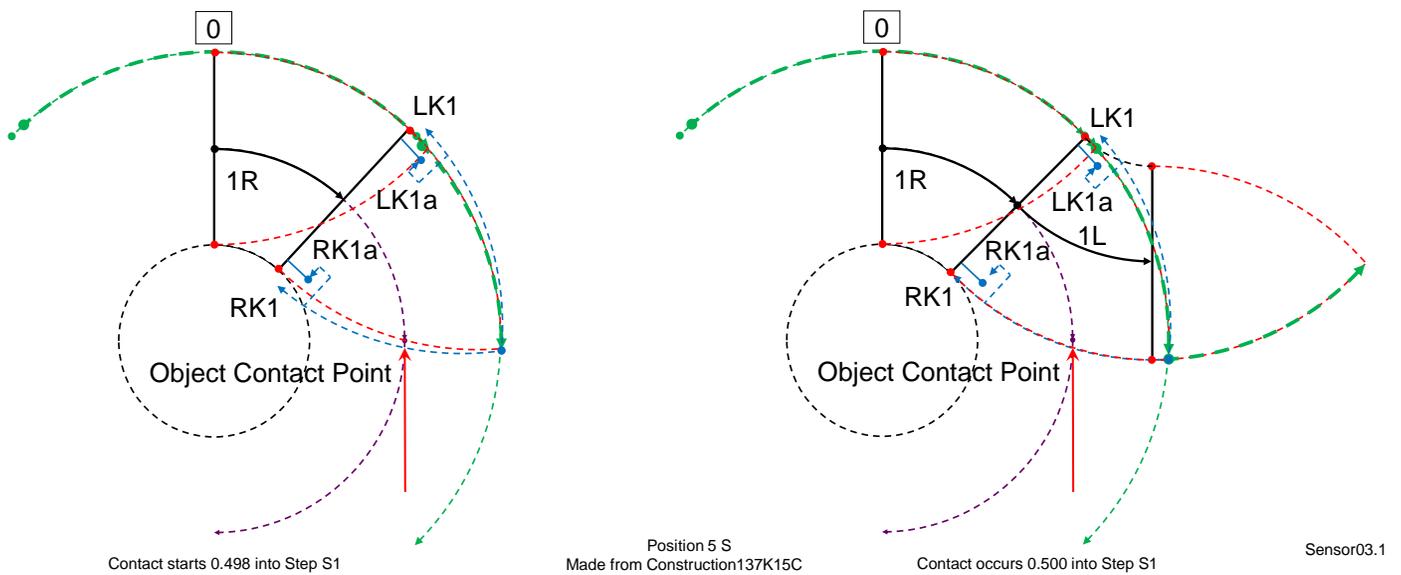

**Figure 106 A *proximity sensor* has a pair of normally closed contacts (RK1a and LK1a) that are broken (opened) by the presence of an object before the front shell makes contact with the object.**

When the front shell approaches an object for the first time, the contacts of (RKa) or (LKa) are opened. This signal is recorded by the Proximity Sensor Controlled Co-Cell (BLM3M) shown in Figure 108. The front shell continues to more toward the object until the contacts of RK1 or LK1 are closed. This causes the front shell to turn the animal away from the object in the usual manner. But the time measured by the break of the contacts at (RK1a) or (LK1a), which is before that actual contact, is used to turn the animal away from the object when it nears it a second time and subsequent times.

3.11.2  *Proximity sensor circuit*

The Proximity Sensor in Figure 107 uses a normally closed Output Contact (Key) [11] to hold open the normally closed Output Disconnect [10]. So there is no voltage on the Output Terminal [10], initially. When an object gets close enough to the Front Shell, the Output Contact Key [11] is opened. This releases the Output Disconnect allowing it to close. This starts the Output Spike [10], which closes the Output Disable Latch [8], shutting off (completing) the spike.

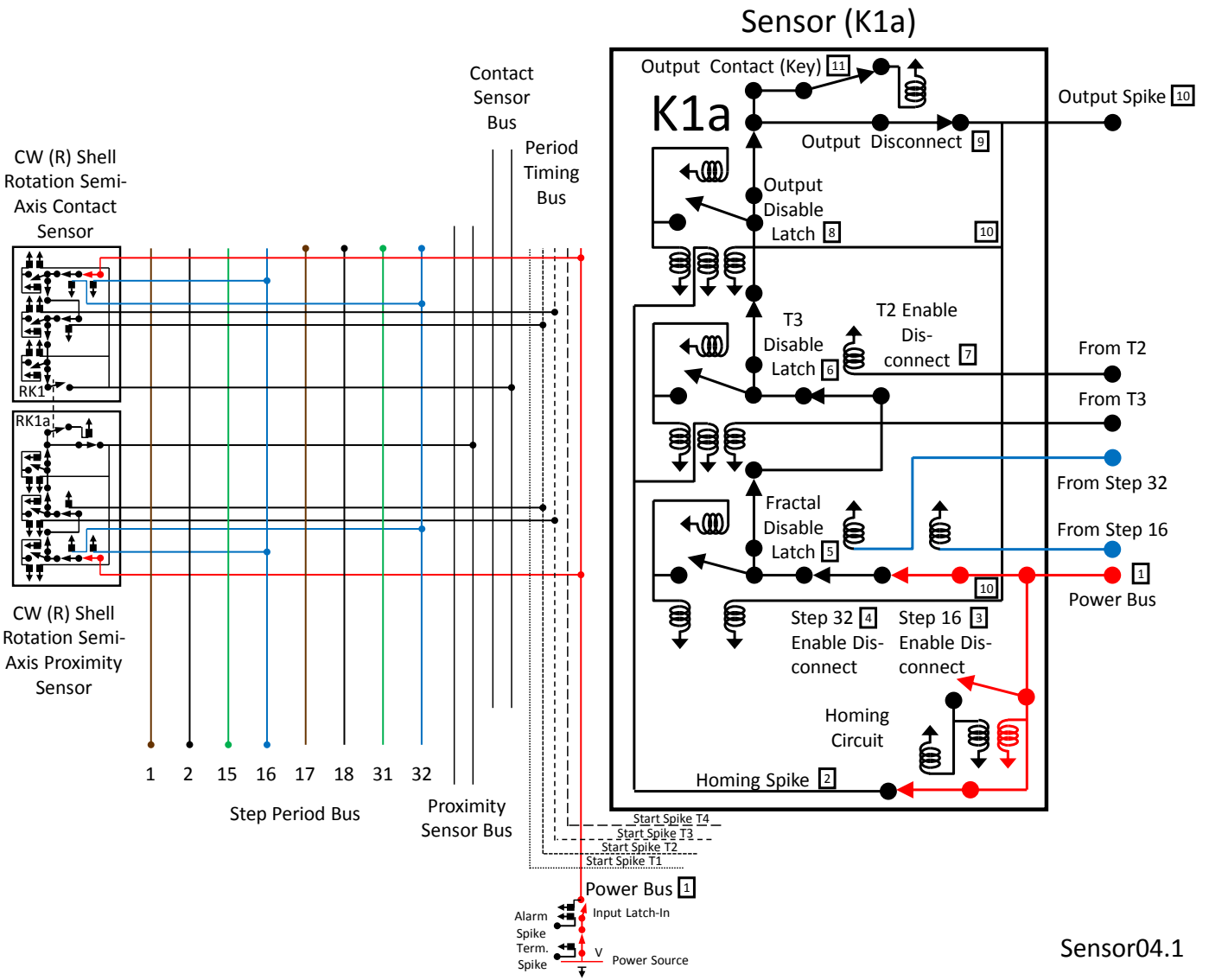

**Figure 107 The *proximity sensor circuit* produces a spike just before contact is made with an object.**

The Contact Output Key [11] of the Proximity Sensor is connected physically to the Contact Output Key of the Contact Sensor. So they both move together. The Proximity Sensor is homed when power is turned on to the Power Bus. This disconnects power to the Output Contact (Key) until the first T2 spike, which open the T3 Disable Latch [6]. The Output of the Proximity Sensor is connected to a separate Proximity Sensor Bus, as shown in Figure 107.

### 3.11.3 *A Proximity Sensor Controlled Memory Co-Cell*

The Proximity Sensor Controlled Memory Co-Cell in Figure 108 works like a Send Cell in Figure 61, but the contact spike time is not reset to zero. Instead, it remains in the position determined by the sensor spike time, and is advanced for one sec the next time that Step Period occurs. This causes the time value to pass through the Max Terminal, where it produces an output spike at the Max Output Terminal [21] at the same time in the period as the original contact time. Then is it reset for one sec during the next reset period, which brings it back to its original position. It does not produce an output spike on its way back past the Max Output Terminal. The Sensor Input Latch [12] in the Proximity Sensor Controlled Memory Co-Cell is connected to the Proximity Sensor (RK1a or LK1a) in Figure 107, as shown in Figure 108.

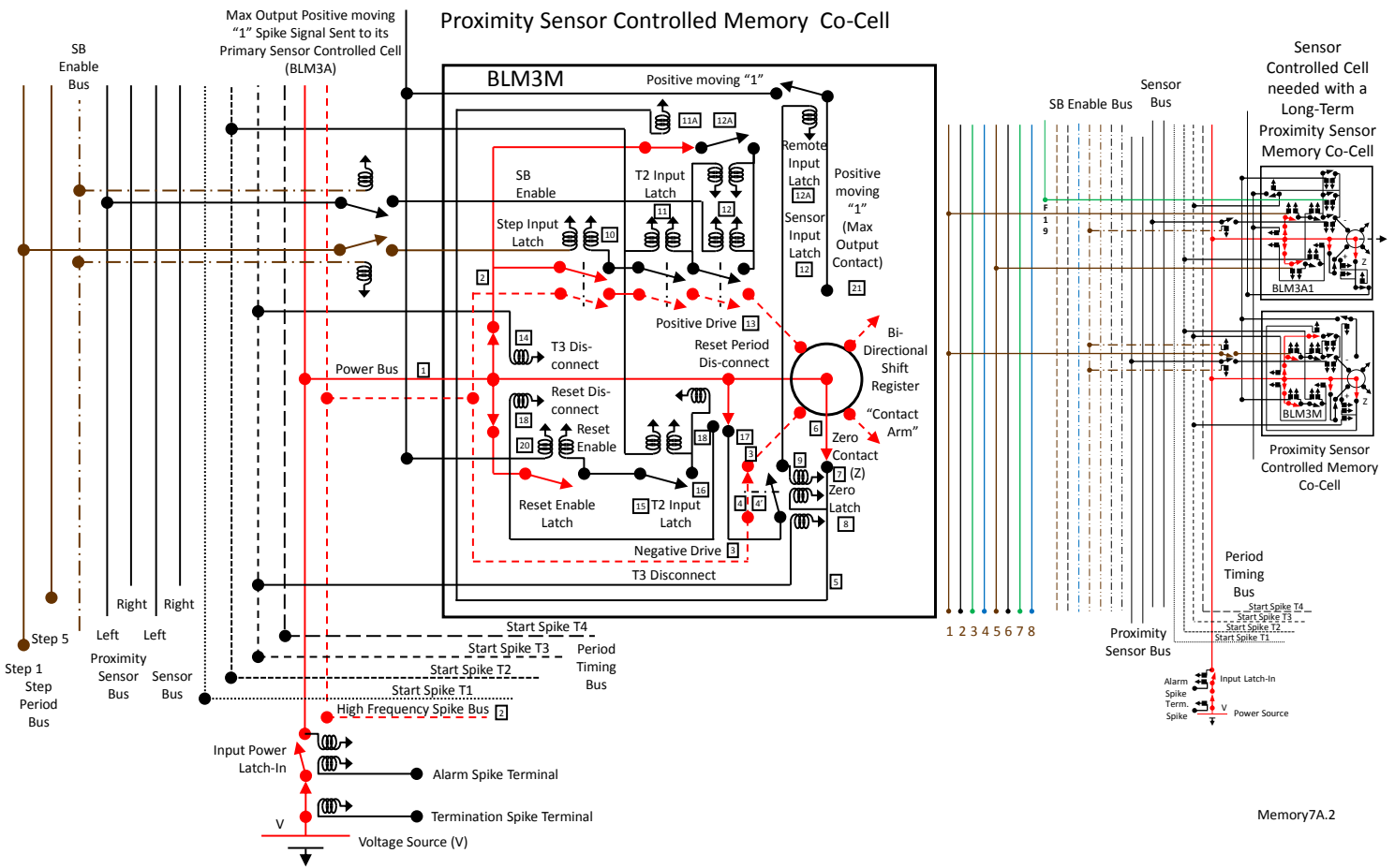

**Figure 108 The Proximity Sensor Controlled Memory Co-Cell records the time within the Step Period when the animal nears an object.**

The Proximity Sensor Controlled Memory Co-Cell uses the same kind of bidirectional shift register as the Send Cell (BLMA3S) shown in Figure 62, except that it needs twice the range of spike times, and it produces a Maximum Output Spike as the spike count passes through its midpoint, as shown Figure 109. A diagram of its action is shown in Figure 111. This Max Output Spike [21] is sent directly to the Sensor Controlled Cell (BLM3A). Since this spike takes place slightly before the object sensor spike, the animal turns away from the object before it contacts the object. So no signal is produced by the object sensor.

### 3.11.4 An extended range, bidirectional shift register

The time delays recorded in the Memory Co-Cell can be carried out by the Extended Range, Bi-Directional Shift Register shown in Figure 109.

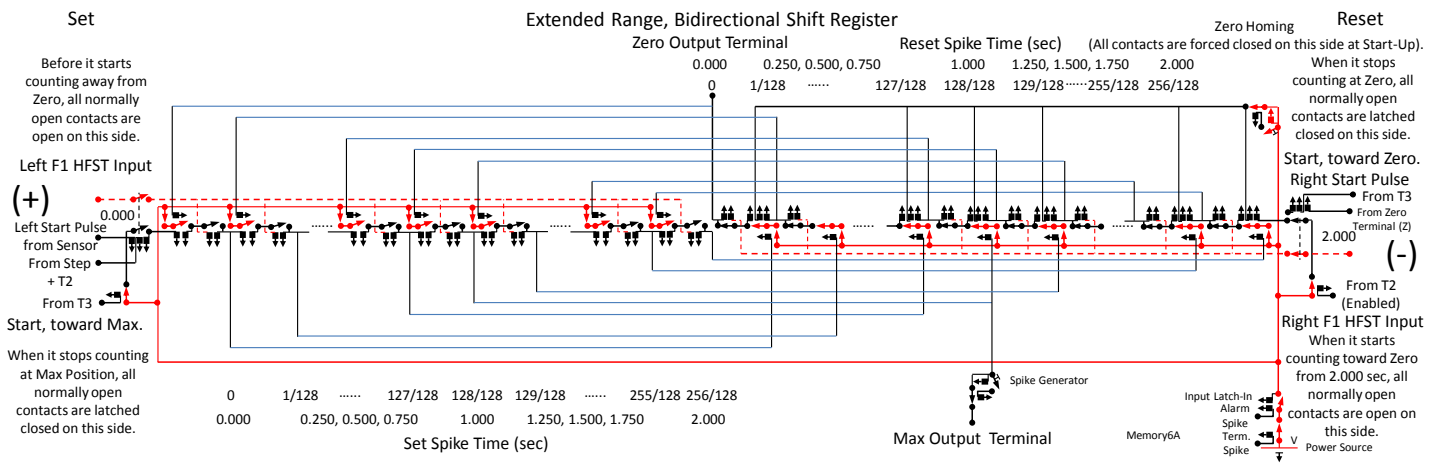

**Figure 109** The *extended range, bi-directional memory shift register* produces an output when it counts up through one second. The extended range, bi-directional shift register is similar to the bi-directional shift register shown in Figure 62. It also produces an output spike at the Max Output Terminal when the 128/128 (1.000sec) contact on the left side is closed, and produces an output spike at the Zero Output Terminal when the last contact on the right side is closed. However, the extended range, bi-directional memory shift register does not reset to zero after it produces an output. Instead, it is advanced to the end of the period, and reset one whole period. So it retains the original spike time information indefinitely. This information is deleted only if power to the Memory Co-Cell is interrupted.

The value of the memory pulse is not stored by just one logic unit, but is stored at the transition between open and closed contacts in the multiple logic units in the bi-directional shift register.

3.11.5 *Sensor Controlled Actuator needed with a Proximity Sensing Co-Cell*

Another Remote Input Latch is needed in the Sensor Controlled Actuator (BLM3A1) so it can respond to the Max Output Signal from the Proximity Sensor Controlled Co-Cell (BLM3M) in Figure 108, forming the Sensor Controlled Actuator (BLM3A2) shown in Figure 110.

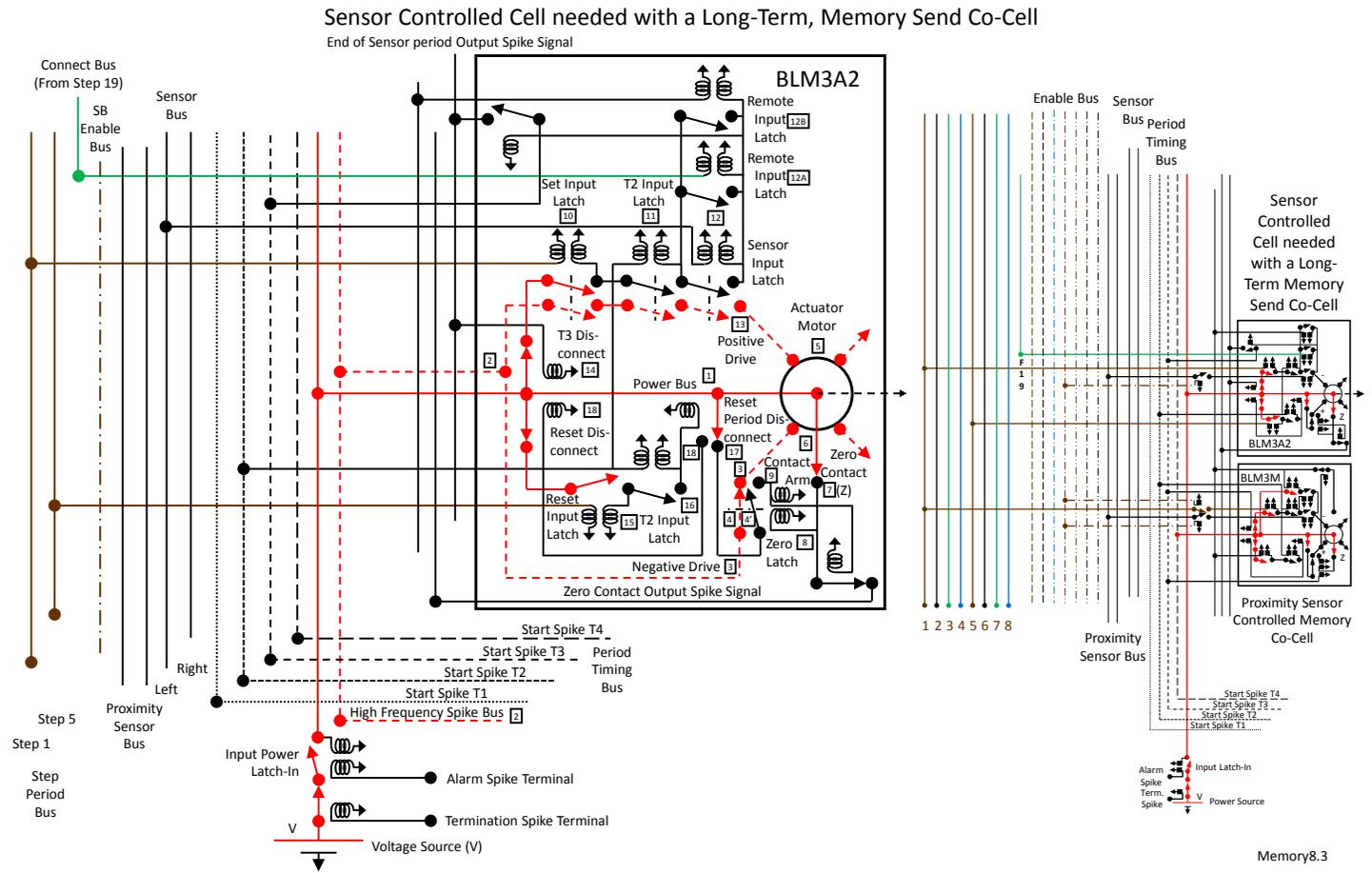

**Figure 110 Another Remote Input Latch [12B] is needed in the Sensor Controlled Actuator (BLM3A2) needed with the Proximity Sensor Controlled Co-Cell (BLM3M)**

The Sensor Controlled Actuator (BLM3A2) is started at 0.250 sec into Step 1 by the Contact Sensor (LKa) at the first contact with an object. In the second approach with the object it is started earlier by its Proximity Sensor Co-Cell (BLM3M), which is set at 0.247 sec into Step 1.

### 3.11.6 *The action of the long-term proximity sensor controlled memory co-cell*

The Long-Term Proximity Sensor Controlled Memory Co-Cell records and reproduces the time in which a spike occurs due to the contact with an object. The timer in the Long-Term Memory Cell is shown as the same symbol as the muscle motor/actuator and contact arm. But it uses the bi-directional memory shift register to read and write spike times instead of an actuator. The circuit of the cells in Figure 112 and Figure 113 is required to carry out this action shown in the diagram in Figure 111.

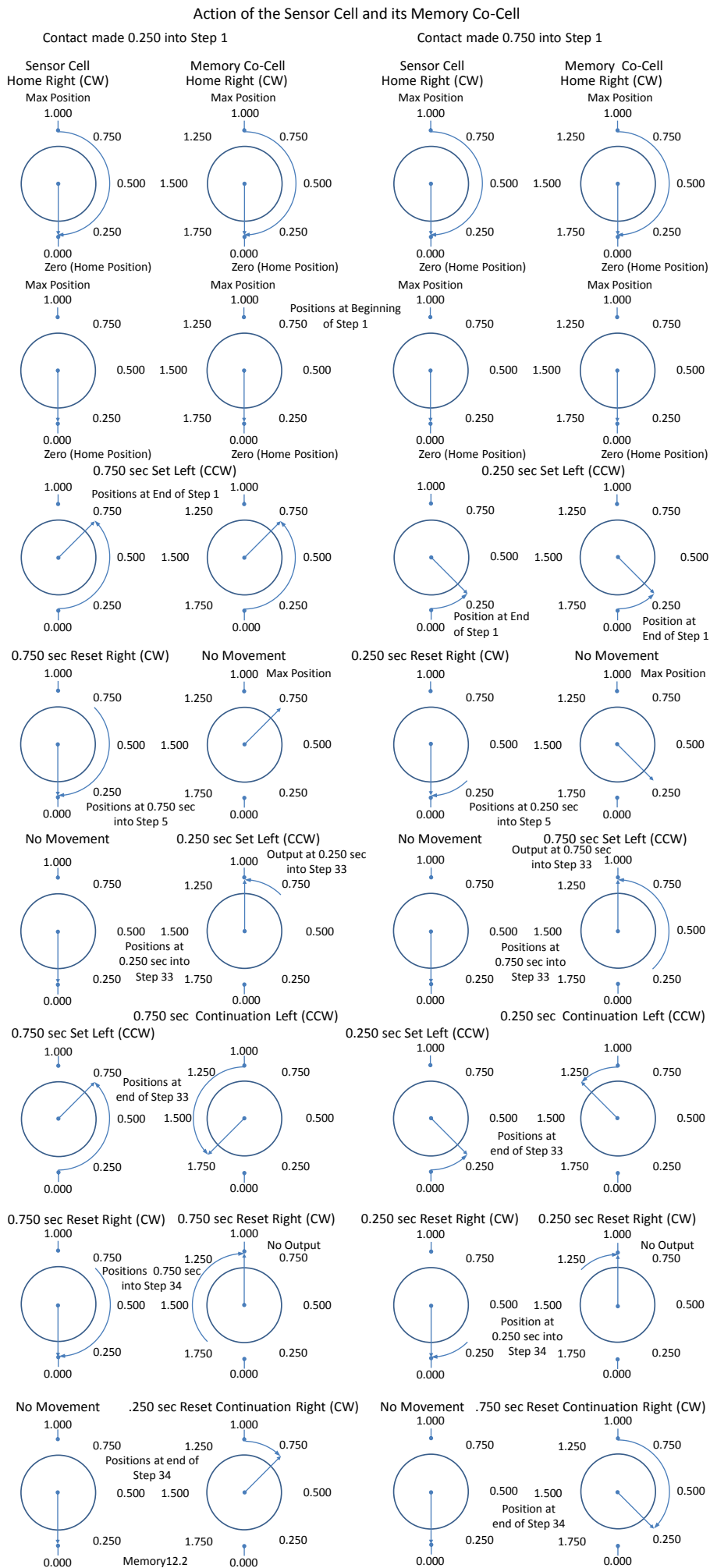

**Figure 111 The Memory Co-Cell** *reproduces the contact spike time* **previously encountered by its Action Cell.**

Spike timing

The memory Co-Cell is not reset to zero. Instead, it is advanced by one whole period (one sec). The time that it passes by the Max Terminal contains the information related to the original time of contact experienced by its Action Cell. And since it starts a second time at some point above Zero and advances for one sec, and resets for one sec, it never gets back to Zero. The information about its starting point can be changed by a different proximity time. But it is not erased except by a Termination Spike (power shutdown).

### 3.11.7 Close up view of the control system with the long-term memories

The Sensor Cells (BLM3A) and (TLM1A) receive spike time signals from a Sensor, a Send Cell (BLM3S) or (TLM1S) in the other fractal, and a Long-Term Memory Cell (BLM3M) or (TLM1M). The Sensor Cells also send a spike time signal to a Sensor Cell in the other fractal, and can send a start signal to Block Cells (BLM1B) or (TLM3B), as shown in Figure 112.

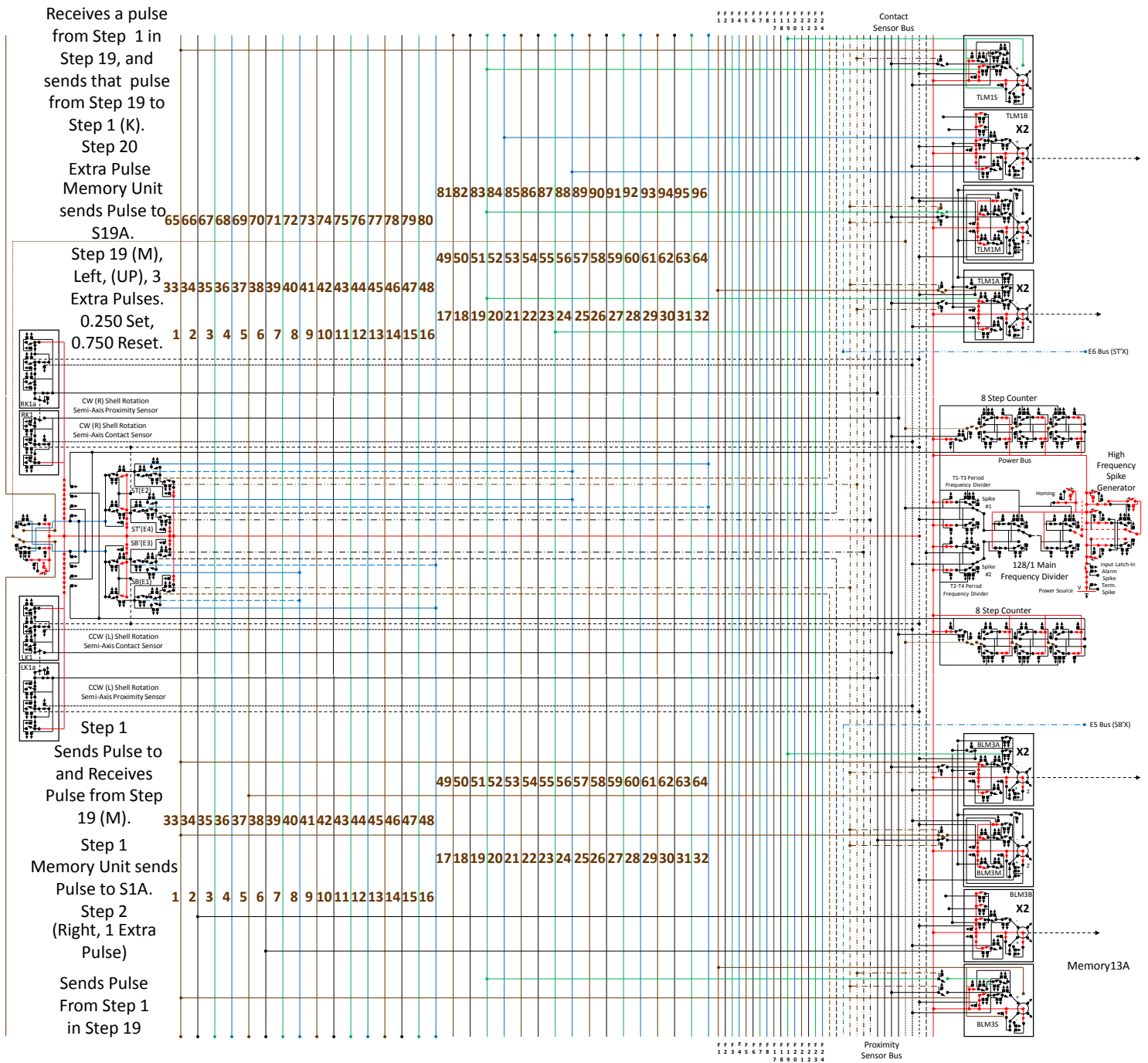

**Figure 112 The Long-Term Memory Cell (BLM3M) acts as if it is a Sensor (K).**

The Long-Term Memory Cell (BLM3M) sends a spike time signal to a Sensor Controlled Cell (BLM3A) slightly before contact is made with the object that caused the original contact spike time from the Sensor (K).

### 3.11.8 *The control system using a long-term memory co-cell*

A long-term memory co-cell can be connected to the sensor cell in each step, and connected into the control system that deals with multiple objects, as shown in Figure 113.

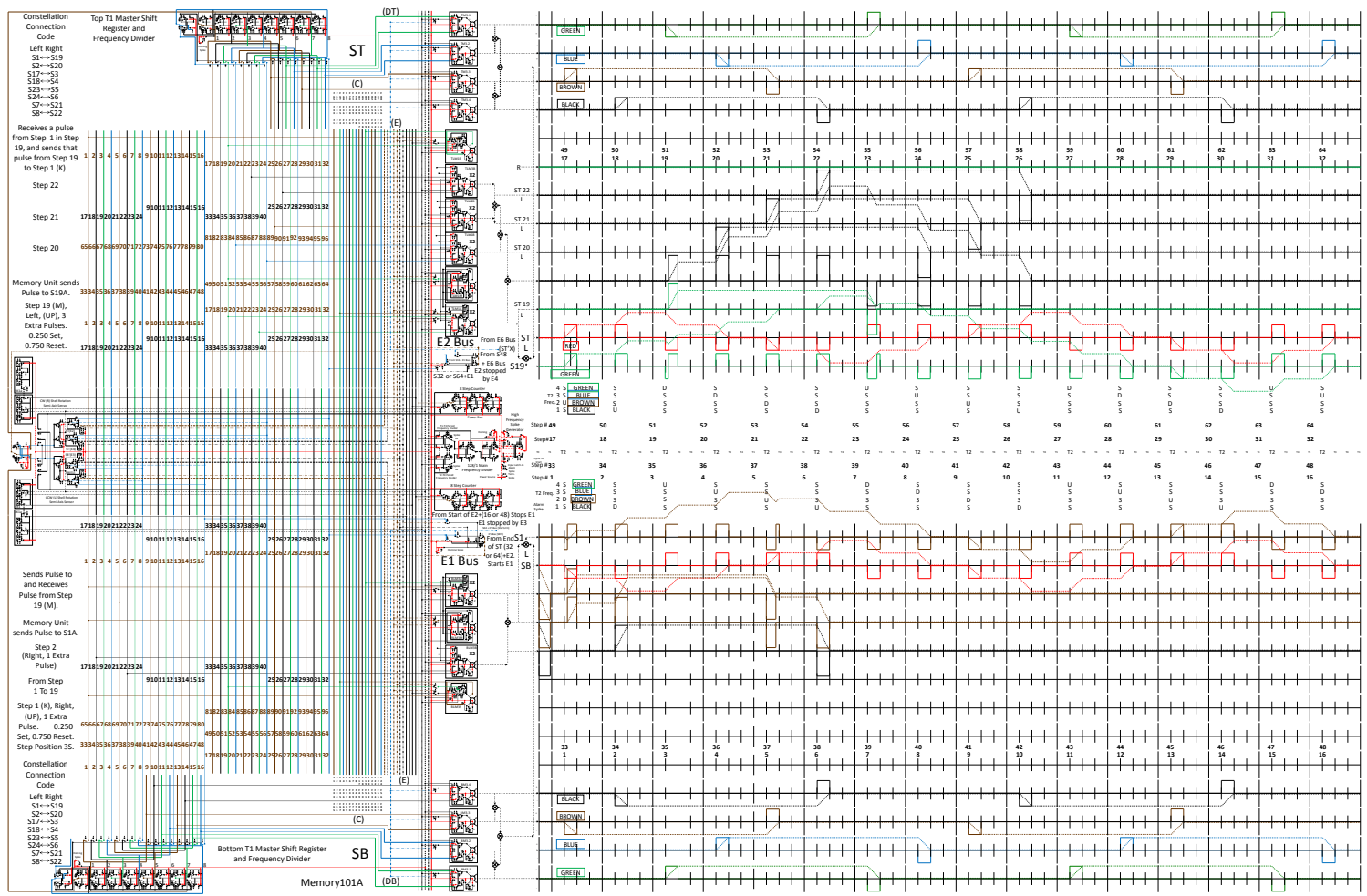

**Figure 113** A *long-term memory cell* can be used to record the time of contact with an object in a given step, and produce that contact time the next time the fractal arrives at that step.

The second co-cell (TLM1M) is connected to the same value of the Step Period Bus and the Sensor Bus (K) as the Sensor Cell (TLM1A). It records the time of contact in that step, but is not reset. The next time that step value is energized (after 32 steps) it is energized by the high frequency spike train for the time of one full step period (128 spikes). This causes it to reproduce an output spike when it passes the Max Terminal that is the same as the original time of contact in that period. The memory co-cell continues past the Max Terminal, and stops at a position past the Max Contact that is the complement of the original spike time. Then the co-cell is reset using the 128 spikes in a step period. This leaves the contact arm a distance away from the Max position that represents the original spike time.

### 3.11.9 *Choosing grazing paths using the Long-Term Proximity Memory Cell circuits*

Once multiple objects are contacted and their positions remembered, the organism follows a specific path based upon the location of these objects rather than just following the original figures-eight. It has acquired the ability of remembering a path through a Maze.

Since it does not contact these objects, its primary sensors are not employed. So it is free to contact and remember objects closer together. This increases the resolution of the sensor system. Since the new grazing paths created by memory create a more complex raster, the complexity of its behavior can be increased.

Now that the organism is able to produce new, specific grazing paths, the door is open for it to choose paths that have more food than other paths, and it can avoid competing with other organism for food. The chance of determining its own destiny emerges.

Spike timing

3.11.10 *Section summary*

An extended range, bi-directional shift register can store a permanent pulse memory by adding one unit pulse time to a sensor contact spike time so it cannot then be reset to zero in one time period. This allows the positions of multiple objects to be remembered indefinitely.

*3.12    High frequency, parallel, interlaced, spike trains*

The limiting factor in the production of a single high frequency spike train is the time required to produce and recover from a neuronal spike. The time of a neural spike is said to be around one m second, with the recovery time being around 4 m seconds. So a single neuron may not be able to repeat sooner than in 5 m seconds periods. This limits the frequency of a spike train created by a single, repeating neuron firing to 200 spikes per second. This limitation can be overcome by creating a spike train produced by multiple neurons connected in parallel. Using two parallel, interlaced spike trains can produce 400 spikes per second, and four parallel, interlaced spikes trains can produce 800 spikes per second. In theory, the frequency can be increase well beyond these values by using more parallel, interlaced spike trains.

3.12.1   *High frequency, multi-spike generator*

The High Frequency Spike Generator (F) shown in Figure 3 can be expanded to produce spikes that occur in sequence using different switch cells, as shown in Figure 114.

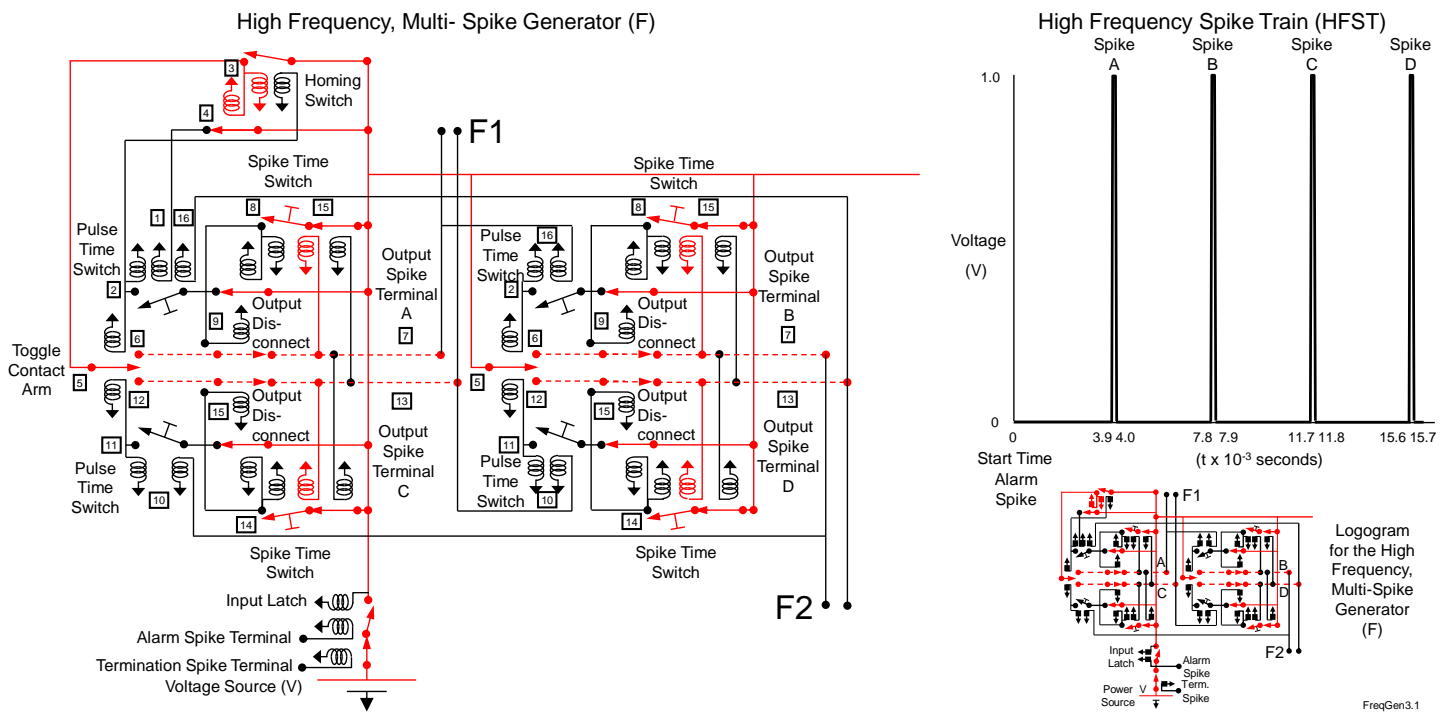

**Figure 114  Using a second circuit, the High Frequency, Multi-Spike Generator (F') produces the second spike on its F1' bus while the first circuit is recovering.**

Using two separate circuits to produce the spike train increases the firing rate of the spike train given the limitations of the repeat firing rate of the individual switches.

### 3.12.2 Multi-spike driven Muscle Actuator

The spike driven muscle actuator in Figure 6 can be operated by the two values of the multi-spike generator in Figure 114 using its F1' Spike Train, as shown in Figure 115.

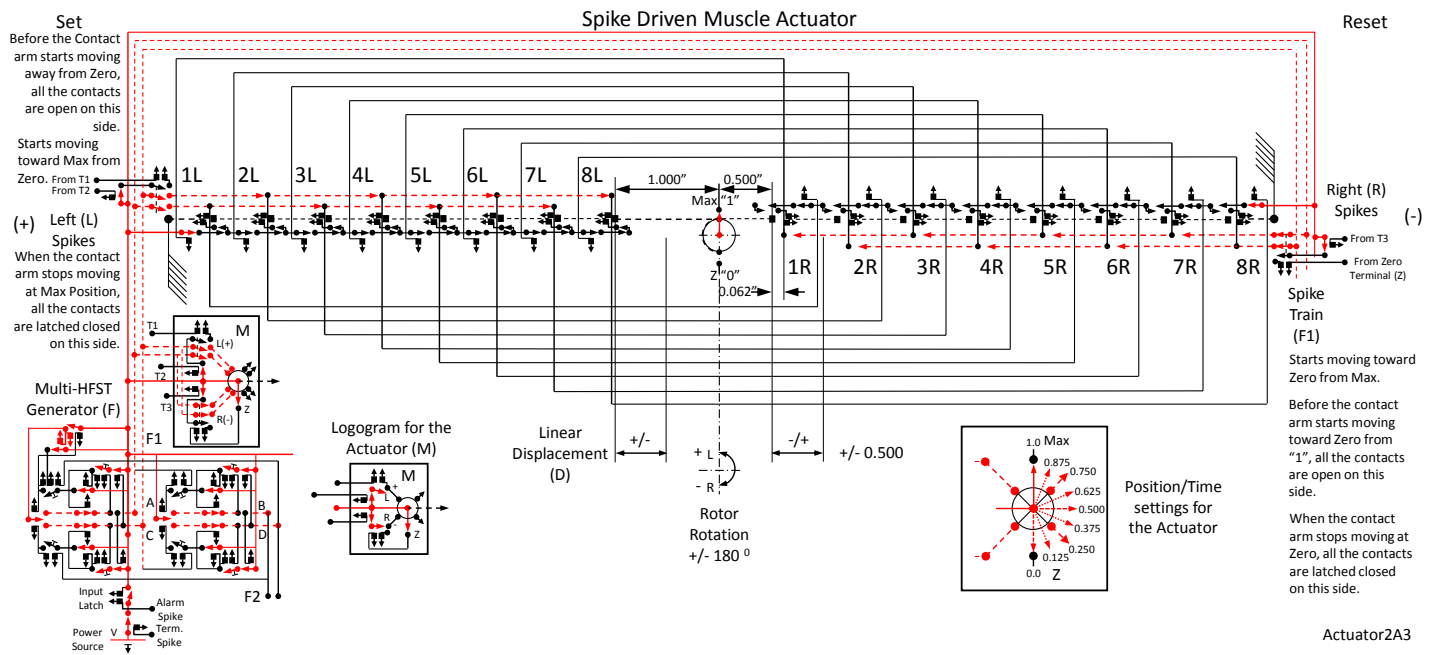

**Figure 115  Each cell (1-8) is connected to the F1 values (A or B) alternately. This doubles the spike frequency that can be applied to the muscle.**

The parallel multi-spike system allows the system to operate at a higher frequency (faster) than a single spike system.

### 3.12.3 A complete parallel, multi-spike system

The multi-spike frequency generator (F') shown in Figure 114 can replace the single spike generator (F) shown in the control system in Figure 7, as shown in Figure 116.

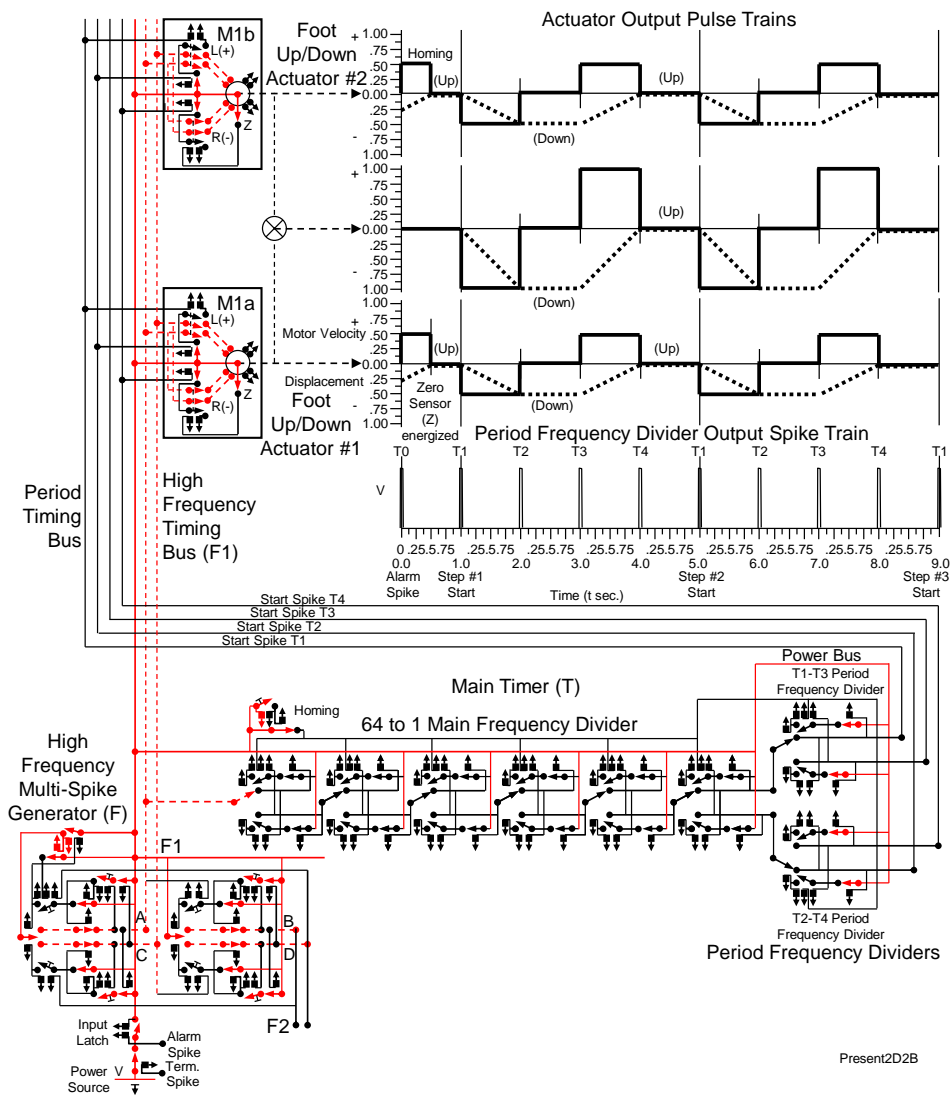

**Figure 116  The first stage of the Main Frequency divider operates at half of the frequency of the actuators.**

Spike timing

Using the multi-spike generator (F') allows the Main Frequency Divider to run at a lower frequency. So, fewer reductions are required in the Main Frequency Divider, as shown in Figure 116.

### 3.12.4  *Four stage frequency generator*

The action and reaction time of the first stage of the Frequency Generator and the first stage of the Main Frequency Divider limit the maximum frequency of the system. The frequency of the system can be increased by adding more timing circuits to the High Frequency, Multi-Spike Generator (F'), while reducing the frequency of the input to the frequency divider. Two more stages can be added to the multi-spike generator in Figure 116, as shown in Figure 117.

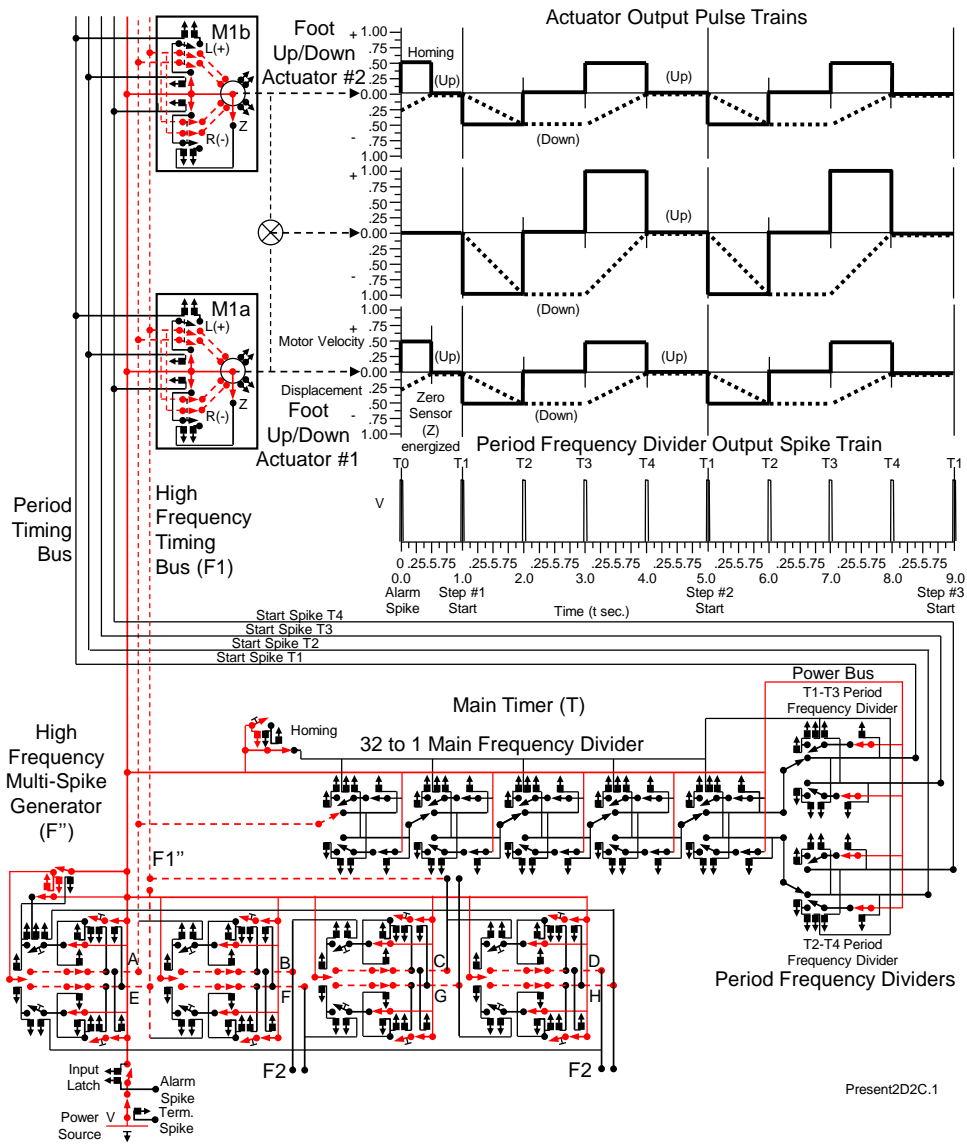

**Figure 117  Doubling the number of generator stages cuts the frequency to the frequency divider by one half.**

Cutting the frequency to the frequency divider by one-half reduces the number of stages in the frequency divider by one stage. So a 32-to-1 frequency divider is used instead of the 64-to-1 frequency divider used in Figure 116.

### 3.12.5 *An eight-stage frequency generator*

Doubling the number of spike generator stages cuts the frequency to the frequency divider in half again, as shown in Figure 118.

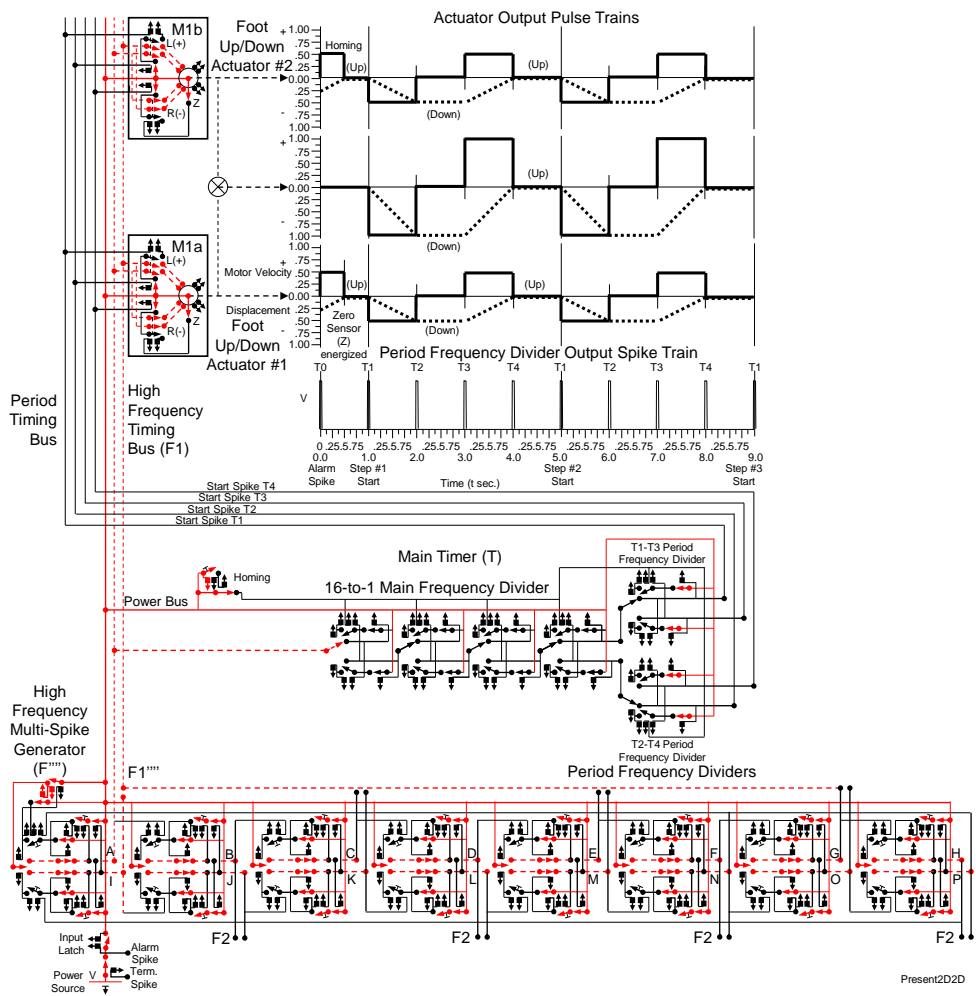

**Figure 118** An eight-stage frequency generator reduces the number of stages in the frequency divider by one stage.

So a 16-to-1 frequency divider is used instead of the 32-to-1 frequency divider used in Figure 117. Doubling the number of generator stages to 16 would reduce the number of stages in the frequency divider to three, providing an 8-to-1 frequency divider. Assuming that the maximum frequency of one neuron is 200 spikes per second, the sixteen stage generator increases the maximum spike frequency of the system by a factor of 16. This provides a maximum system frequency of 3,200 Hz in one parallel system. This provides a Step Period of about 1/3 of a second in which a full (90 degree) motion can be carried out with a resolution 1 part in 128, which is better than 1%. This is good performance, even for an industrial robot.

### 3.12.6 *Section summary*

The duty cycle of a single neuron may not be greater than 5 ms. This limits the frequency of a single spike generator to 200 Hz, which is too slow for advanced animal life forms. The High Frequency Spike Generator needed for the Limpet like proto-animal needs to operate at 256 spikes per second to provide the two phases, each running at 128 spikes per second. This provides 128 different possible pulse widths (movements) in each Pulse Time Period of one second. The Alpha waves of 8-12 Hz in our brain are fast enough to provide Pulse Periods of 1/10 of a second using PWM. Using 100 spikes per Pulse Period requires a High Frequency Spike Train of around 1000 Hz to generate PWM timing in the brain. This is close to the ultra-fast Gamma frequencies of 400-800 Hz. However, higher frequency provides higher information transfer rate, which leads to faster movements and greater intelligence. Information transfer rates increases exponentially with spike train frequency the PWM memory systems in this paper. Higher frequency would have a huge beneficial influence on survival. These higher frequencies can be achieved by using multiple, parallel frequency generators.

Spike timing

## 4. Methods and Materials

Relays, timing motors, voltage sources, and conductors in this paper are used to describe a temporal control process based upon pulse width modulation. Of course, other devices would be used in a biological system to carry out the same temporal process. But the purpose of this paper is to show how timers can be used to produce behavior useful in an animal model. So spike operated muscle actuators and bi-directional shift registers are used in this paper to create motion and store timing information using a control system based upon PWM.

Connections between neurons do not have to be made, broken, or changed in any way in the spike the timing system shown in this paper to change behavior. In fact, having a constant conductance at each connection is essential for the logic circuits in this paper to work correctly. Connections between logic elements are as important as the logic elements that make up the logic circuits needed to produce the spike processing system. One can assume that these connections are determined according to genetic information in the same way that the connection of the cells of bones, teeth, and other features of the body of an animal or plant are determined. But once the connections are made, and the groups of cells in the body are grown, there is no need to change these arrangements to produce changes in behavior. Changes in behavior in the spike timing model can be made by changes in spike timing, exclusively.

In this paper, the normally open contacts of the relays represent excitatory neurons, and the normally closed contacts represent inhibitory neurons. The actuators are timers that can express a physical movement. I show their movements being summed physically by the symbol for a differential in a manner similar to that used by Vannevar Bush's Differential Analyzer, since the linear movements of muscle cells are summed in the same manner as a differential when they are connected in series.

The circuits in this paper are drawn with the spike sources (encoders) such as sensors and the system timing bus on the left side of the page, the actuators (decoders) in the middle of the page, and the pulse trains (output of the decoders) on the right side of the page in the manner of the Bush Differential Analyzer. The spikes produced on the left hand side of the page must be equal to the onset and offset spikes in the pulses on the right hand side of the page, forming a balanced set of equations. These pulse trains define the fractal behavior expressed by the organism. Graphical analysis of the pulse trains using circular curve fitting produces the resultant trajectories (movements) of the organism based upon the kinematics of the animal model. All of the pulse trains and motion trajectories in the paper are based upon rigorous calculations based upon spike times, and their accuracy can be checked using the numerical date in the computer generated graphics.

## 5. Conclusion

The use of pulse width time as the source of the high resolution information in working, temporary, and long-term memories in an animal brain offers a fundamental change in way of thinking about how the brain works.

The paper shows how the high power, low energy, and great timing ability of spikes that can be used efficiently to start and stop timers in a rhythm based control process using pulse width modulation. It is hard to imagine how a rhythm based system like the animal brain could operate without timers. And timers are intrinsic memory devices. Many examples are shown of how the timers associated with pulse width modulation can be used as actuators and memory devices that produce useful behavior.

Another foundational element shown in this rhythm based system is the use of recurrent fractals to produce

Spike timing

cyclical motion and repeating internal states that act as scanners. The sensors in these scanners produce spike time information when they sense an object in their environment, and the PWM system uses these spike times to alter the fractals. These alterations in the fractals produce variations in the beat of the system that results in unique rhythms and behavior.

Fractals created using Pulse Width Modulation form dynamic waves that consist of an equal number of set pulses and reset pulses. Under some conditions, a pulse can be set, and not reset to zero. This leaves a permanent (static) pulse in the fractal that influences the properties of the fractal. These static pulses have the characteristics of a particle rather than a wave. A static pulse can act as a permanent (long-term) memory that influences behavior long after it is set.

Also, the paper shows a process that attempts to maintain the coherence seen in a stable recurrent fractal. The process achieves coherence by resetting itself in real time after responding to each sensed input, resulting in a zero sum relation of pulses. This restores its internal order by eliminating the buildup of higher than normal pulse values. This allows information to be stored and retrieved, and results in a changed relationship with its environment such as avoiding obstacles. These features underlay a greater understanding of navigation, echolocation, voice recognition using the principles of spectrometry, vision system using fractals as raster scanners, and the animal brain using spike timing.

The circuits shown in this paper can be used to create computers programs that produce PWM since computers have a built in high frequency generator (clock) that operates at a very high frequency. This clock frequency can be used to write and read very high resolution pulse time information into and from up/down counters in the computer.

Since the pulse value is not stored by just one logic unit, but is stored at the transition between open and closed contacts in the multiple logic units in these bi-directional shift registers, the system architecture shown in this paper expands the opportunities for explaining how the brain works.

Using spike operated muscle motors, shift register memories, and dual input shift registers to add (compute the sum of two spike pulses) provides the technology needed to deal with almost any kind of motion control (robotic) application. And the evolution and expansion of its use in biology may account for the success of intelligent animal life.

**Acknowledgements:**

I wish to thank Dapeng Li, Ph.D., Department of Bioengineering, University of Massachusetts, Dartmouth, for his interest in and encouragement of this effort.

And I wish to thank Konrad Kording, Northwestern University, for informing me of the Bahill paper on human eye movements, and his instructive suggestions and insights concerning the writing of this paper.